\documentclass[a5, 10pt]{book}  
\usepackage{graphicx}
\usepackage{epstopdf}
\usepackage{fancyhdr}  
\usepackage{amsmath, amsthm, amssymb}
\usepackage{booktabs}
\usepackage{subfigure}
\usepackage{color}
\usepackage{caption}
\usepackage{hyperref}
\usepackage{color}
\definecolor{DarkBlue}{rgb}{0.1,0,0.55}
\definecolor{olive}{rgb}{0.176,0.498,0.118}
\definecolor{royal}{rgb}{0.412, 0.114, 0.482}

\hypersetup{
   pdfauthor = {Matthias Ossadnik},
	colorlinks=true, citecolor=royal, linkcolor=DarkBlue, bookmarksnumbered = true,
}

\usepackage[numbers,sort&compress]{natbib} 
\usepackage{tocbibind}
\fancypagestyle{plain}{  
	\fancyhead{}  
	  
}

\def\openone{\leavevmode\hbox{\small1\kern-3.8pt\normalsize1}}%

\usepackage[printonlyused]{acronym}

\newcommand{\pdag}{{\phantom{\dagger}}}
\newcommand{\ham}[1]{\mathcal{H}_{\text{#1}}}

\newcommand{\braket}[2]{\left\langle #1\,\right|\left. #2\right\rangle}
\newcommand{\bra}[1]{\left\langle #1\,\right|}
\newcommand{\ket}[1]{\left|\, #1\right\rangle}

\newcommand{\al}{{\boldsymbol\alpha}}

\newcommand{\kettwo}[2]{\left| \begin{array}{c} #1 \\ #2\end{array}\right\rangle}

\begin{document}

\frontmatter
 \begin{titlepage}
  Diss.\ ETH No.\ 20039 \\
   \vspace{1cm}
  \begin{center}
     {\Large {\bf \sc A wave packet approach to interacting fermions}\\}
  \vspace{1.5cm}
    {\sc Abhandlung zur Erlangung des Doktor der Wissenschaften \\}
    {\sc der \\}
  \vspace{0.3cm}
    {\sc ETH Z\"urich\\}
    \vspace{0.3cm}
    
    \vspace{1.5cm}
  {vorgelegt von\\}
   \vspace{0.3cm}
    {\sc Matthias Ossadnik\\}
    \vspace{0.5cm}
    {Dipl. Phys., Julius-Maximilians-Universit\"at W\"urzburg (Germany) \\}
    {geboren am 20.03.1981\\}
    {Staatsangeh\"origkeit: Deutsch\\}
    \vspace{1cm}
    {Angenommen auf Antrag von\\}
    {Prof.\ Dr.\ M.\ Sigrist, examiner\\}
    {Prof.\ Dr.\ T.\ M.\ Rice, co-examiner\\}
    {Prof.\ Dr.\ C.\ Honerkamp, co-examiner\\}    
    \vspace{1.cm}
    {2012}
  \end{center}
\end{titlepage}

\thispagestyle{empty}
\cleardoublepage


\chapter{Abstract}

\noindent
The complexity of the cuprate superconductors continues to challenge physicists even 25 years after their discovery. At half-filling they are antiferromagnetic Mott insulators. Upon doping the antiferromagnetism vanishes, and at some finite doping superconductivity sets in. In between these two phases lies the so-called pseudogap phase, which features a gap for electronic excitations in the anti-nodal directions around the saddle points $(0,\pi)$ and $(\pi,0)$as well as a partial spin gap \cite{pseudogap_experiment}. At the same time, electronic excitations around the nodal points $(\pm \pi/2, \pm \pi/2)$ are gapless even for relatively small doping. For large doping, the superconductivity vanishes, and the cuprates behave like ordinary Fermi liquids.

\noindent
The Mott insulator and the Fermi liquid phase are understood very well, yet the intermediate pseudogap phase remains controversial. In order to tackle this problem theoretically, one may base the description either on the Mott insulator and consider the effect of doping, or on the Fermi liquid, and consider the partial truncation of the Fermi surface. 

\noindent
In this thesis, we use the latter approach, and study the breakdown of the Fermi liquid state using the renormalization group (RG) \cite{tnt}. The advantage of the method is that it is well suited for studying anisotropies in momentum space. Moreover, it treats $d$-wave pairing- and antiferromagnetic spin-fluctuations on an equal footing. 

\noindent
In the first part of this thesis, we use the RG approach in order to study anisotropic quasi-particle scattering rates. This undertaking is motivated by transport experiments on overdoped cuprates \cite{abdel}, which point towards a breakdown of the Fermi liquid phenomenology due to strong scattering at the saddle points, leading to a linear temperature dependence of the transport scattering rates. We show that a similar linear dependence arises from the renormalization group treatment down to low temperatures, thus providing an additional piece of evidence that the breakdown of the Fermi liquid phase is dominated by the saddle points.

\noindent
In the remainder of the thesis, we seek to extend the work on the crossover from the Fermi liquid state to the pseudogap phase \cite{rice_rvb}. In earlier works it has been argued that the RG flows in the so-called saddle point regime, where the Fermi surface lies close to the saddle points, are indicative of a transition to an insulating spin liquid state, which truncates the Fermi surface in the vicinity of the saddle points \cite{furukawarice, saddlepointregime, laeuchli}. Progress in the derivation of effective models for the conjectured spin liquid state has been hindered, however, by the difficulties involved in solving the strong coupling low energy Hamiltonian. We approach the problem by observing that the pseudogap phase is intermediate between a phase where electrons are localized in real space (Mott insulator) and a phase where they are localized in momentum space (Fermi liquid). We introduce an orthogonal wave packet basis, the so-called Wilson-Wannier (WW) basis \cite{wilsonbasis,wilsonbasis2}, that can be used to interpolate between the momentum space and the real space descriptions. Its main feature is that the basis functions are localized in phase space, which allows for a coarse grained description of the physics in both momentum space and real space at the same time. The price that is paid for this convenience is that the translational invariance of the lattice is explicitly broken from the onset.

\noindent
Nevertheless, the positive features of the WW basis appear to be very attractive for studying  the pseudogap regime because it is experimentally well established that nodal and anti-nodal states behave very differently in this regime, so that the anisotropy in momentum space is important. At the same time, scanning tunneling microscope measurements suggest that the physics of the anti-nodal states takes place in real space, rather than momentum space \cite{pseudogap_review}.

\noindent
In order to prepare the stage for the study of the saddle point regime, we develop the necessary ideas step by step, starting with the construction of the basis and the derivation simple approximate formulas for the transformation of the Hamiltonian. Since the approach is novel, these steps are performed in considerable detail.

\noindent
We then discuss the relation of the WW basis to the phenomenon of fermion pairing in both particle-particle and particle-hole channels in one and two dimensions. We use the example of simple mean-field Hamiltonians to show that when the size of the wave packets is chosen properly, the description of states with fermion pairing simplifies considerably. Moreover, we relate the geometry of the Brillouin zone in two dimensions to a separation of length scales between nodal and anti-nodal directions, which suggests that the anti-nodal states are locally decoupled from the nodal states. The phase space localization allows to include both of these aspects.

\noindent
In the remainder we show how to combine the WW basis with the RG, such that the RG is used to eliminate high-energy degrees of freedom, and the remaining strongly correlated system is solved approximately in the WW basis. 

\noindent
We exemplify the approach for different one-dimensional model systems, and find good qualitative agreement with exact solutions even for very simple approximations.

\noindent
Finally, we reinvestigate the saddle point regime of the two-dimensional Hubbard model. We show that the anti-nodal states are driven to an insulating spin-liquid state with strong singlet pairing correlations, thus corroborating earlier conjectures \cite{furukawarice, rice_rvb}. 

\noindent
Throughout, we limit ourselves to the simplest treatment of each model, so that all results are rather qualitative in nature. On the other hand, we hope that this allows to understand the workings of the method and the physical arguments that are derived from the phase space analysis of the interacting fermion system.

\tableofcontents
\mainmatter
\chapter{Introduction}

\begin{figure}
\centering
\includegraphics[width=8cm]{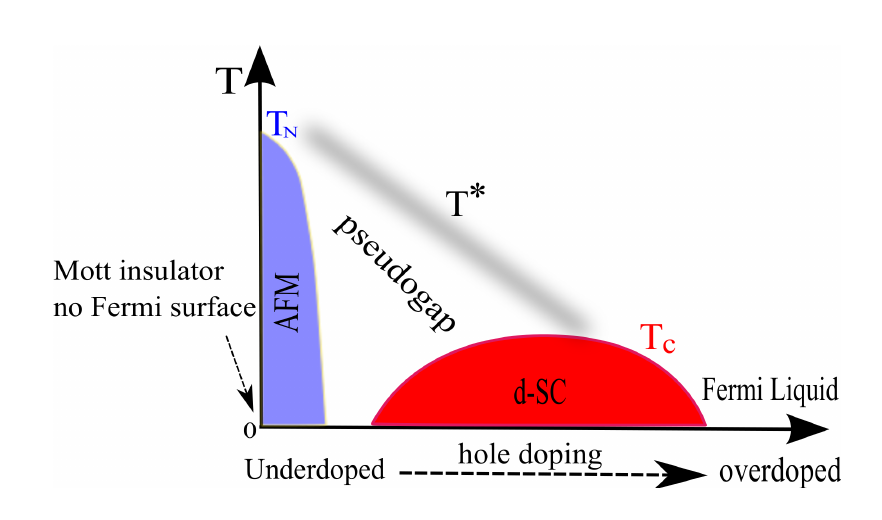}
\caption{Schematic phase diagram of the cuprates. (Figure reproduced from \cite{yrz})}
\label{fig:phase_diagram_cuprates}
\end{figure}

\noindent
Despite many efforts, the phenomenology of the cuprate superconductors still offers challenging problems \cite{pseudogap_review, pseudogap_experiment}. Their schematic phase diagram is shown in Fig.~\ref{fig:phase_diagram_cuprates}. At half-filling, they are antiferromagnetic Mott insulators. Upon doping, the antiferromagnetic order is destroyed rapidly, and the materials enter a phase known as the pseudogap phase that has a variety of exotic properties \cite{pseudogap_experiment}. Among them is a gap for electronic excitations around the so-called anti-nodal directions $(0,\pi)$ and $(\pi,0)$ that coexists with a truncated Fermi surface around the nodal directions. At even larger doping, they eventually become superconducting with a $d$-wave order parameter. The pseudogap gradually decreases with doping, until it merges with the superconducting gap around the optimal doping, where $T_c$ is maximal. As the doping is increased even more (overdoped region in Fig.~\ref{fig:phase_diagram_cuprates}), $T_c$ decreases, and the system behaves like a conventional Fermi liquid. 

\noindent
There are different routes that can be followed in order to increase our understanding of these complex materials theoretically. Either one starts from the Mott insulator and investigates the effect of doping, or one starts at the overdoped side and tries to understand the transition from a Fermi liquid to the unconventional pseudogap state around optimal doping. We follow the latter path and focus on the transition from a normal Fermi liquid phase to the pseudogap phase. Since the cuprates are Fermi liquids in the overdoped regime, we base our investigation on a weak to moderate coupling approach, the functional renormalization group \cite{tnt}. 

\noindent
 This method has been successfully used in the past in order to arrive at a phase diagram of the Hubbard model at moderate coupling \cite{saddlepointregime,zanchi,halboth,breakdown,ehdoping}. Similar to the phase diagram of the cuprates, antiferromagnetic and superconducting phases are obtained at half-filling and moderate doping, respectively. In between, one finds the so-called saddle point regime, which is characterized by a crossover between the two phases, with a strongly anisotropic scattering vertex and dominant correlations around the saddle points. 

It has been conjectured that the latter regime is the weak coupling analogue of the Fermi surface truncation that is observed in cuprates \cite{furukawarice,laeuchli,rice_rvb}. This conjecture is based on similarities of the flow to strong coupling in the saddle point regime and quasi-one dimensional ladder systems \cite{nlegladder, linbalents}. The latter systems can be solved exactly, and exhibit the so-called $d$-Mott phase at half-filling, with gaps for all excitations and strong singlet correlations, similar to the RVB states proposed by Anderson \cite{anderson, rice_rvb}. This analogy has been fruitfully used as a starting point for a phenomenological theory of the underdoped cuprates recently \cite{yrz}. 

\noindent 
In this thesis, we try to add some new aspects to these earlier works. It consists of two parts: The first part is very short, consisting only of Ch.~\ref{ch:rg}. In this part, we apply the renormalization group to study recent transport measurements on overdoped cuprates \cite{abdel}. In the experiment, superconductivity was suppressed using a magnetic field, and the transport scattering rate was determined from interlayer angle-dependent magnetoresistance (ADMR) measurements. Interestingly, it was found that the onset of superconductivity is accompanied by a strong anisotropic scattering rate with maxima in the anti-nodal directions. Moreover, the anisotropic part of the scattering rate shows a linear temperature dependence, whereas the isotropic part retained the usual quadratic temperature dependence. From the point of view of the saddle point regime found within the RG, the pronounced anisotropy is very natural since the scattering vertex itself is highly anisotropic, with the strongest scattering occurring at the saddle points. Hence we investigate the quasi-particle scattering rates using a band structure from \cite{abdel} in order to see whether the anisotropy and temperature dependence of the renormalized vertex can explain the observed phenomena. We find good qualitative agreement with the experimental results, including the approximately linear temperature dependence of the scattering rate.

\noindent 
The bulk of this work is contained in the second part, where we seek to establish a new approach for the approximate solution of the strong coupling fixed point found in the RG. Since the two parts are independent of each other, we present a separate introduction to this part in Ch.~\ref{ch:ww_intro}.

\chapter{Renormalization group calculation of angle-dependent scattering rates in the two-dimensional Hubbard model}
\label{ch:rg}

\section{Introduction}

\noindent
In this chapter we apply the functional renormalization group (RG) in order to compute quasi-particle life-times in the two-dimensional Hubbard model. The motivation for this study is provided by transport experiments by Abdel-Jawad et al. \cite{abdel}. In their experiment they found that the close to onset of superconductivity in overdoped cuprate superconductors, the quasi-particle scattering rate can be decomposed into two term: The first term is isotropic, does not depend strongly on doping and shows the usual $T^2$ dependence on temperature characteristic of Fermi liquids. The second term is anisotropic, with the maxima at the anti-nodal points $(\pi,0)$ and $(0,\pi)$. This term increases strongly at the onset of superconductivity as the superconducting dome is approached from the overdoped side. Strikingly, it shows linear temperature dependence. Since the cuprates appear to be ordinary Fermi liquids in the highly overdoped regime, the linear temperature dependence is a puzzling result. In the following, we use the functional renormalization group (FRG) investigate the problem. Since we are dealing with the overdoped regime, effects due to strong onsite interactions are not expected to be very important, so that it makes sense to employ the (weak coupling) renormalization group in the following. 

\noindent
We note that within the RG framework, it is natural to expect anisotropy along the Fermi surface due to an interplay of anisotropy of the Fermi velocity and the imperfect nesting of the Fermi surface in the regime under study. In fact, earlier investigations of the two-dimensional Hubbard model on a square lattice using the renormalization group found that $d$-wave pairing in the overdoped region of the phase diagram was driven by the appearance of a strongly anisotropic scattering vertex in the particle-particle and particle-hole channels  at low energies and temperatures \cite{zanchi,halboth,breakdown,katanin2loop}. Relatedly, it was shown that the self-energy is also anisotropic \cite{zanchi2,katanin,ehdoping,rohe}.

\noindent
In the following section, we describe the RG setup used, in particular the calculation of the self-energy, and the introduction of an (artificial) decoherence rate for the fermions, that we use to suppress the superconducting instability in our calculation.

\noindent
We then show how the anisotropic scattering vertex leads to the anisotropy of the quasi-particle scattering rate. Surprisingly, the renormalization of the vertex gives rise to a linear temperature dependence of the scattering rate due to the scale dependence of the vertex when the pairing divergence is suppressed, in good qualitative agreement with the experiments.

\section{Renormalization group setup}

\begin{figure}
 \centering
\includegraphics[width=7cm]{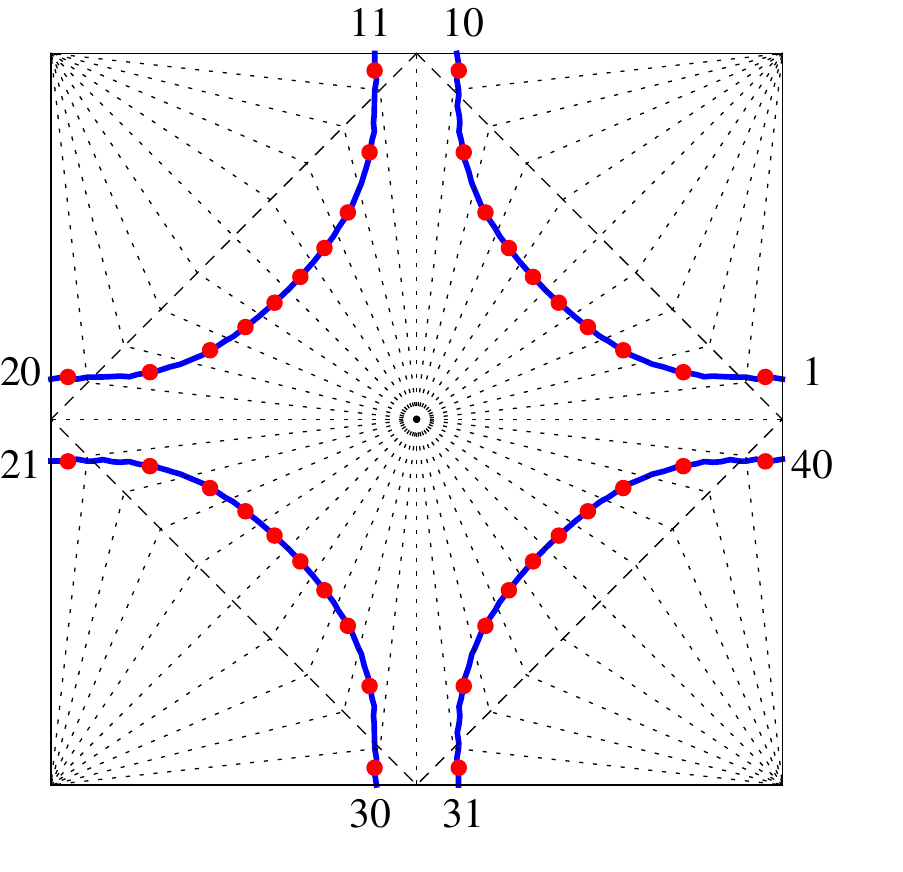}
\includegraphics[width=3cm]{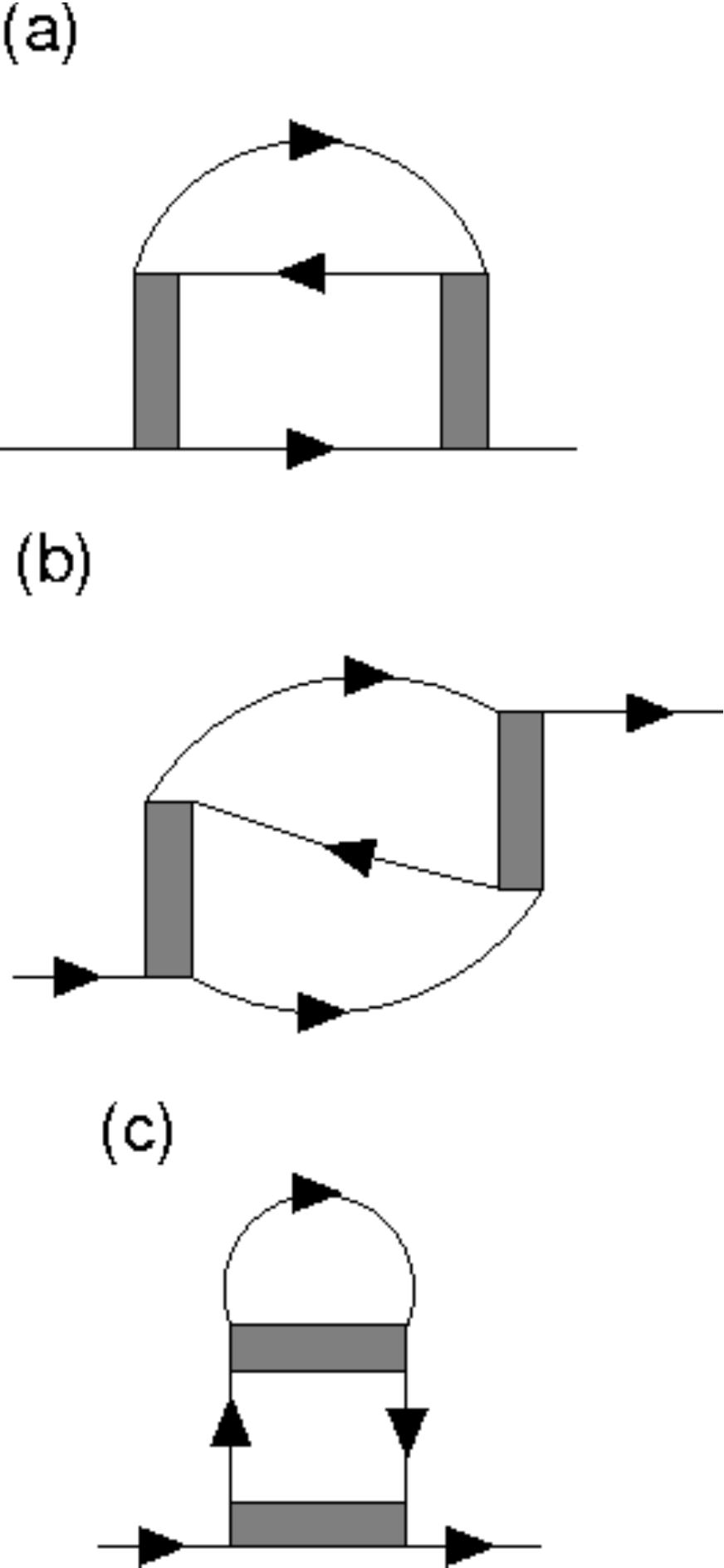}
\caption{Left: Fermi surface (solid line) and discretization of the BZ for $p=0.22$. The boundaries of the patches (labelled by $1,2,\ldots,40$) are indicated by the dashed lines. All vertices are evaluated at the points marked by the red dots, and are taken constant within each patch. Right: Two-loop diagrams contributing to the self-energy. Only diagrams a) and b) contribute to the scattering rate.}
\label{fig:patches}
\end{figure}

\noindent
Our approach relies on the functional RG equation for the one-particle irreducible (1PI) generating functional $\Gamma[\Phi]$, which is derived in \cite{tnt,breakdown}, and which leads to a hierarchy of coupled flow equations for the 1PI vertices after a suitable expansion of the functional. We use a Wilsonian flow scheme with a sharp momentum cutoff. This cutoff underestimates effects of small wavelength scattering \cite{tflow}, but these processes are important only if the FS is very close to a van Hove singularity. This is not the case in the doping regime studied here. To solve the flow equations, the hierarchy has to be truncated, and in the following we will use the standard truncation of neglecting all vertices with more than four legs. In this approximation, the only quantities appearing in the calculation are the self-energy $\Sigma_\Lambda(k)$ and the four-point vertices $V_\Lambda(k_1,k_2, k_3)$. The $k_i$ also contain the frequency, $k_i = \left(\omega_i,\mathbf{k}_i\right)$. All propagators contain a sharp infrared cutoff $\chi_\Lambda(\mathbf{k})=\theta(|\epsilon(\mathbf{k})|-\Lambda)$ in momentum space, where the flow parameter $\Lambda$ flows from $\Lambda=\infty$ to $\Lambda=0$ with the initial condition $V_\infty(k_1,k_2,k_3)=U$. We neglect the frequency dependence of all vertices and discretize their momentum dependence. The latter is done by dividing the Brillouin zone into elongated patches, cf. Fig. \ref{fig:patches}. The vertices are taken to be constant in each patch. Their values are calculated at a reference point in each patch, which we choose to lie where the FS crosses the center of the patch. 

\noindent
Typically, the truncated flows diverge at some energy scale $\Lambda>0$. The leading divergence can be interpreted as the dominant instability. The scale at which the divergence occurs gives an estimate of the corresponding $T_c$ \cite{breakdown}. In the regime of interest here, $d$-wave pairing is the leading instability, and in our approximation $T_c$ takes the values $T_c = 0.26 t_1$ for $p=0.15$, $T_c=0.22t_1$ for $p=0.22$, and $T_c=0.16t_1$ for $p=0.30$. These $T_c$s are way too high, mainly because we neglect self-energy corrections in the flow of the scattering vertex. Nevertheless, $T_c$ grows with decreasing hole doping reproducing qualitatively the experimental results \cite{abdel}. 


 \noindent
The experiments by Abdel-Jawad et al. \cite{abdel} were carried out on well characterized Tl$_2$Ba$_2$CuO$_{6+x}$ samples. The interlayer angle-dependent magnetoresistance (ADMR) provided detailed Fermi surface (FS) information which we use to fix the band parameters as follows: 
\begin{eqnarray}
\epsilon(k_x,k_y) &=&-2t_1\left(\cos k_x + \cos k_y\right) +4t_2\left(\cos k_x \cos k_y\right)\nonumber\\
 && + 2t_3\left(\cos 2k_x + \cos 2 k_y\right) +4t_4(\cos 2k_x\cos k_y 
\nonumber\\ &&+ \cos 2k_y\cos k_x) + 4t_5\left(\cos 2k_x\cos 2k_y\right),
\end{eqnarray}
with $t_1=0.181,\;t_2=0.075,\;t_3=0.004,\;t_4=-0.010,\text{ and } t_5=0.0013 (\text{eV})$. 

\begin{figure}
 \centering
\includegraphics[width=9cm]{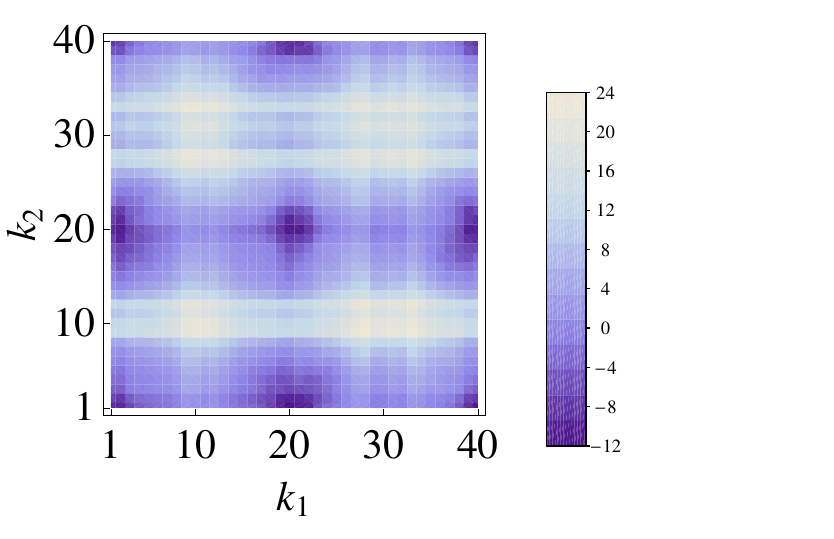} 
\caption{Characteristic momentum dependence of the renormalized vertex $V_\Lambda(\mathbf{k}_1,\mathbf{k}_2,\mathbf{k}_3)/t_1$ for $p=0.22$, $T=0.02 t_1$, $1/\tau_0=0.2t_1$ at $\Lambda=0$. In the figure, the dependence on the two ingoing wave vectors $(\mathbf{k}_1, \mathbf{k}_2)$ is shown, where the outgoing wave vector $\mathbf{k}_3$ is taken to lie in patch $1$ close to $(\pi,0)$ (cf. Fig. \ref{fig:patches}) and $\mathbf{k}_4$ is fixed by momentum conservation.}
\label{fig:vanis}
\end{figure}

\noindent
We use a moderate starting value of the onsite repulsion, $U$. Our goal is a qualitative rather than a quantitative description which would require a larger value of $U$ and multi-loop corrections to the RG flow equations. 
The experiments were carried out in a high magnetic field to suppress superconductivity and allow to access the normal state down to low $T$. However, including a magnetic field into our RG calculation is difficult, so that we choose to suppress superconductivity by introducing an isotropic scattering rate $1/\tau_0$ into the action. This smears out the Fermi distribution at the FS, which in turn regularizes the loop integrals and subsequently the flow of the four-point vertices. This scattering rate is included in the flow equation of the four-point vertex only, whereas the flow equation for the self-energy is left unaltered. We found that for our choice of $U=4t_1$ and the range of temperatures ($T\geq 0.004 t_1$) and dopings ($p \geq 0.15$), a scattering rate of $1/\tau_0 = 0.2 t_1$ is sufficient to suppress the divergences associated with superconductivity,
while leaving the one-loop integrals corresponding to other channels, e.g. the $\pi-\pi$-particle-hole diagram, basically unchanged. On average the vertices remain comparable to the bandwidth and the largest vertices do not grow larger than $\approx3\times$bandwidth (Fig. \ref{fig:vanis}). The renormalized vertex is used as an input into a standard lowest order calculation of the quasi-particle decay rate.

In Fig. \ref{fig:vanis} we display a typical result of our calculations for the four-point vertex $V_\Lambda(\mathbf{k}_1,\mathbf{k}_2,\mathbf{k}_3)$ at energy scale $\Lambda$ for fixed outgoing wavevector $\mathbf{k}_3$ close to $(\pi,0)$ as a function of the two incoming wavevectors $(\mathbf{k}_1,\mathbf{k}_2)$. The remaining outgoing wavevector is determined by momentum conservation allowing for umklapp processes. 
The strongest scattering processes occur for a momentum change of $(\pi,\pi)$ and the $\mathbf{k}_i$ near $(\pi,0)$ and $(0,\pi)$. 

\noindent
As we are only interested in the scattering rates at the FS, which are given by $\operatorname{Im}\Sigma(\mathbf{k}\in FS, \omega\rightarrow 0+i\delta)$, we will restrict the calculation of the self-energy to this quantity in the following. Obviously, the frequency-dependence of $\Sigma$ cannot be neglected in the calculation. On the other hand, if we neglect the frequency-dependence of the four-point vertices, it is clear from the structure of the flow equations that $\Sigma$ will also be frequency-independent, as only Hartree and Fock diagrams are included. This difficulty can be overcome by replacing the four-point vertex appearing in the self-energy flow equation by the integrated flow equation of the vertex \cite{ehdoping}, schematically,
\begin{equation}
 \Sigma_{\Lambda=0}=\int d\Lambda\left(\int d\bar{\Lambda} V_{\bar{\Lambda}} S_{\bar{\Lambda}} G_{\bar{\Lambda}} V_{\bar{\Lambda}}\right) S_\Lambda,
\label{eq:twoloop}
\end{equation}
where in our approximation the single-scale propagator $S_\Lambda$ \cite{breakdown,tnt} and the full propagator $G_\Lambda$ are related to the free propagator $G_0$ by $
 S_\Lambda = \dot{\chi}_\Lambda G_0 \text{ and } G_\Lambda = \chi_\Lambda G_0,
$
respectively. The RHS of eq. (\ref{eq:twoloop}) depends on $\Lambda$ only through the cutoff $\chi_\Lambda$. After a partial integration with respect to $\Lambda$ and after explicitly inserting a sharp cutoff $\chi_\Lambda(\mathbf{k}) = \Theta(|\epsilon(\mathbf{k})|-\Lambda)$ we have
\begin{eqnarray}
 \Sigma_{\Lambda=0}&=& \int d\Lambda \theta(|\epsilon(\mathbf{k}_1)|-\Lambda)\delta(|\epsilon(\mathbf{k}_2)|-\Lambda)\theta(\Lambda-|\epsilon(\mathbf{k}_3)|) \nonumber\\ & & \times V_\Lambda^2 G_0(k_1)G_0(k_2)G_0(k_3),
\label{eq:sigmafinal}
\end{eqnarray}
where summation and integration over internal momenta and Matsubara frequencies is implied. Inspection of (\ref{eq:sigmafinal}) shows that it amounts to evaluting the two-loop contribution to the self-energy, but with the flowing vertex $V_\Lambda$ instead of the initial interaction $U$.

\noindent
The diagrams corresponding to (\ref{eq:sigmafinal}) are shown in Fig. \ref{fig:patches}. As we are interested in the scattering rates at the FS, we need only consider diagrams a) and b), because the contribution of diagram c) is real for external frequencies $\omega+i\delta$. For a) and b), for external frequency $\lim_{\omega\rightarrow 0}\omega+i\delta$, we obtain an imaginary part $\propto \delta\left(\epsilon(\mathbf{k}_3)-\epsilon(\mathbf{k}_2)-\epsilon(\mathbf{k}_1)\right)$, reflecting energy conservation.

\noindent
Neglecting the flow of the four-point vertices, i.e. setting $V_\Lambda=U$ in eq. (\ref{eq:twoloop}), is equivalent to a second order perturbative calculation of the scattering rate, which gives a $T^2$ behavior away from van Hove singularities. All deviations from the Landau theory scaling form may be attributed to the renormalization of the four-point vertices.

\section{Results}

\begin{figure}
\centering
\includegraphics[width=8cm]{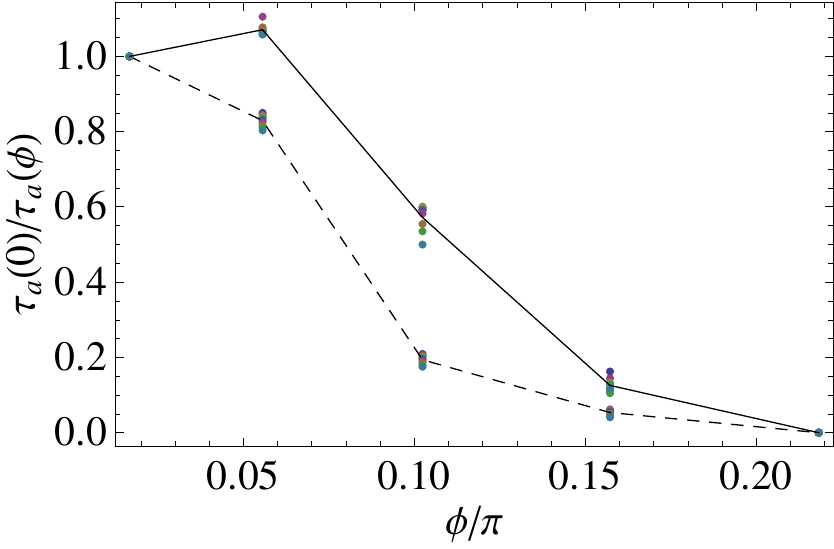}
\caption{Angle-dependence of the anisotropic component of the quasi-particle scattering rate on a Fermi surface segment for $p=0.15$ (dashed line) and $p=0.30$ (solid line). The scattering rates are normalized to unity in the antinodal direction. The dots are the values at different temperatures, the lines connect the temperature averages.}
\label{fig:angulardependence}
\end{figure}

\noindent
Based on eq.\,(\ref{eq:sigmafinal}) we calculate both the temperature and the doping dependence of the angle-resolved quasi-particle scattering rates at the Fermi surface. We find that the scattering rates are anisotropic for all choices of parameters. The precise shape of the angular dependence changes with doping, but does not change very much with temperature, as shown in Fig.~\ref{fig:angulardependence}. 
In general, we find that in the nodal direction ($\phi=\pi/4$) the scattering rates have a minimum, and increase towards the anti-nodal direction ($\phi=0$). The size of the anisotropy grows as doping is decreased. This parallels the increase of $T_c$ with lower doping in a calculation without the regularizing scattering rate. This is in accord with the results of Ref.~\cite{abdel}, where with decreasing doping both $T_c$ and the anisotropic part of the scattering rates increase, whereas the uniform component remains constant. 

\noindent
Separating the scattering rates into an isotropic and an anisotropic part, we write
\begin{equation}
 \frac{1}{\tau}(\phi,T)= \frac{1}{\tau_i}(T) + \frac{1}{\tau_a}(\phi,T),
\end{equation}
where $1/\tau_i \equiv \min_\phi 1/\tau(\Phi)$ so that $1/\tau_a(\phi) \geq 0$. We characterize the $T$-dependence of the anisotropic part by its average over the angle, $
\langle 1/\tau_a \rangle(T)$.
This makes sense as the angular dependence of $1/\tau_a$ is approximately $T$-independent (Fig.~\ref{fig:angulardependence}).
\begin{figure}
\centering
\includegraphics[width=7cm]{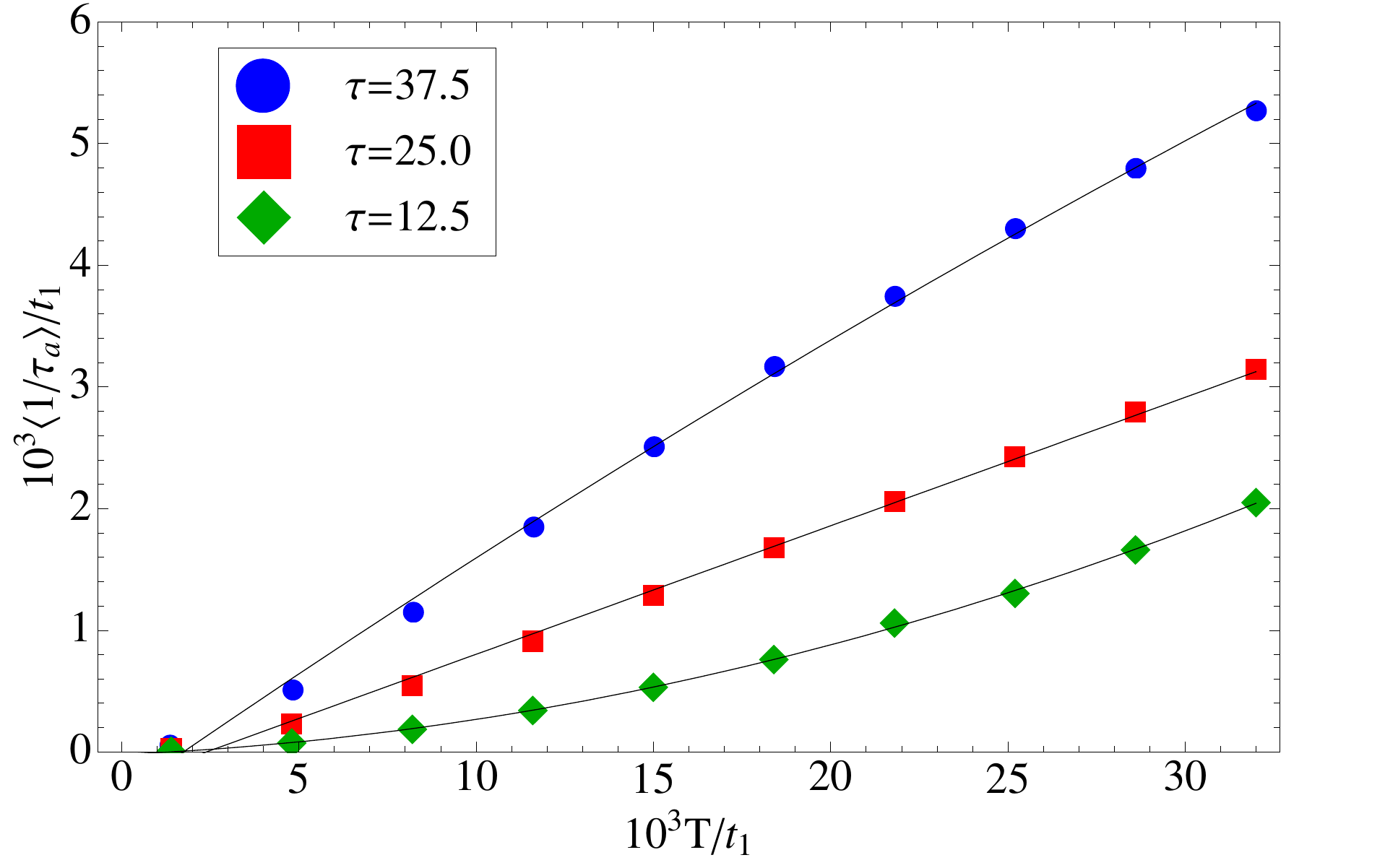}
\includegraphics[width=7cm]{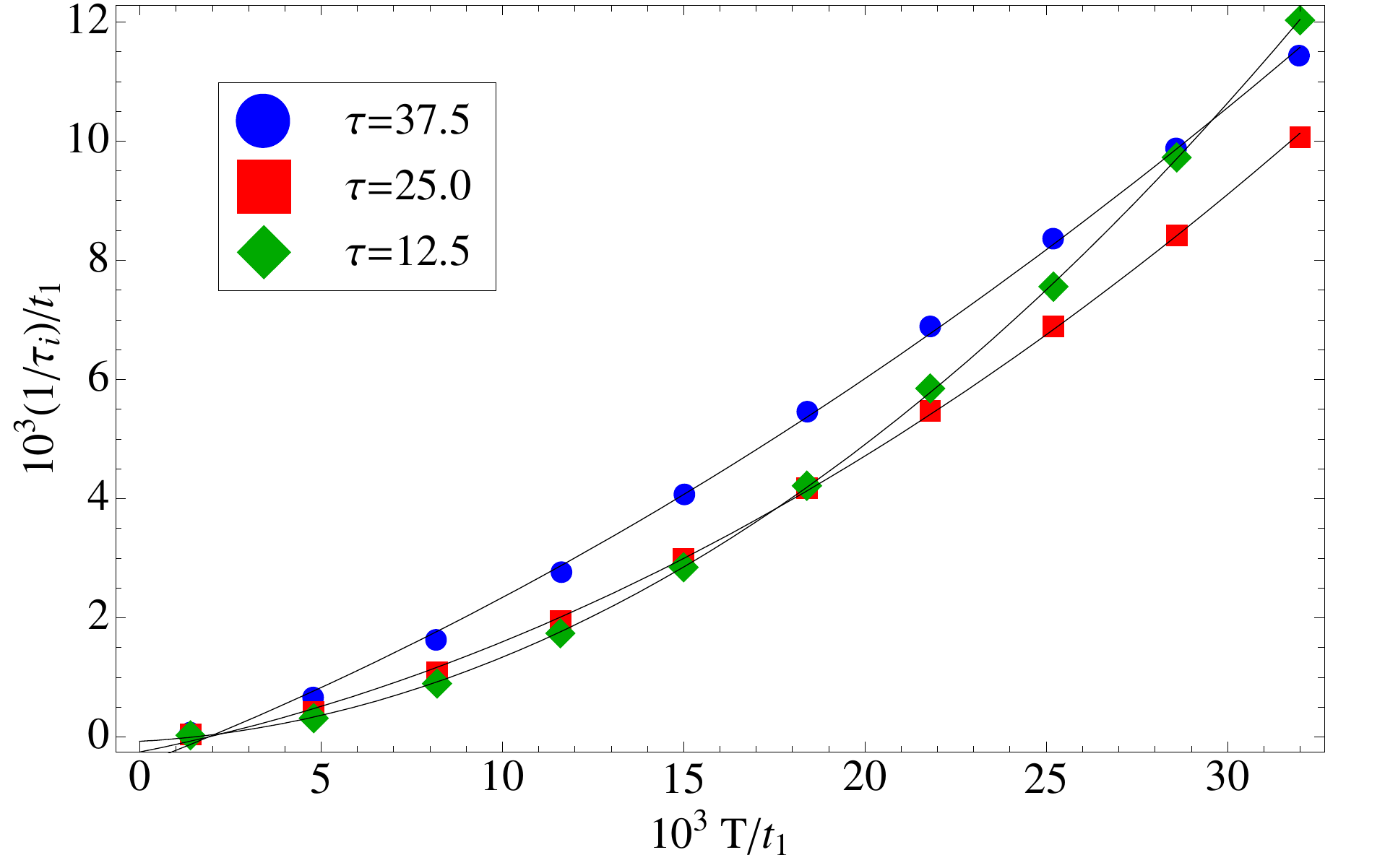}
\caption{Temperature dependence of (a) the anisotropic and (b) the isotropic component of the quasi particle scattering rate at the Fermi surface for different values of the hole doping. The solid lines are fits to quadratic polynomials in $T$. (Result of fits: $\frac{1}{\tau_a}=-0.35 + 0.20 T - 0.0007 T^2$ for $p=0.15$, $\frac{1}{\tau_a}= -0.17 + 0.09 T + 0.0004 T^2$ for $p=0.22$, and $\frac{1}{\tau_a}=-0.020 + 0.012 T + 0.0016 T^2$ for $p=0.30$}
\label{fig:Tdependence}
\end{figure}
Using these definitions, we find that the $T$-dependence of $1/\tau_i$ and $\langle 1/\tau_a\rangle$ can be fitted very well by a quadratic polynomial, as shown in Fig. \ref{fig:Tdependence}. In the same figure, one sees that for $p = 0.22$ and $p=0.15$, $\langle 1/\tau_a\rangle$ is \emph{linear} in $T$ with a coefficient which increases with decreasing hole doping. In contrast with this, the isotropic part $1/\tau_i$ has always a predominantly quadratic $T$-dependence and does not change much with doping. Thus our calculations reproduce the main features of the striking correlation between charge transport and superconductivity reported in Ref. \cite{abdel}.

\noindent
Note that in our theory, the {\em linear} relationship between $\langle 1/\tau_a\rangle$ at fixed $T$ and $T_c$ from Ref. \cite{abdel} is replaced by a slightly superlinear behavior that is consistent with $\langle 1/\tau_a\rangle=0$ for $T_c=0$. However, as mentioned earlier, our theoretical $T_c$'s are not reliable, and the experimental $T_c$s might be affected by additional effects like sample quality. In our view, the essential physical point of Ref. \cite{abdel} which is well reproduced by our theory is that $\langle 1/\tau_a\rangle$ and $T_c$ grow together, as they are both caused by the same interactions with wavevector transfer near $(\pi,\pi)$. 

\section{Conclusions}

\noindent
The results signal presented above point towards a clear breakdown of Landau-Fermi liquid behavior. The linear temperature dependence is not due to a proximity to the van Hove singularity at the saddle points of the band structure, as these points lie well below the Fermi surface at energies $\gg T$. Further, the increase in the linear term in $T$ with decreasing hole dopings occurs as the energy of the van Hove singularity moves further away from the Fermi energy. Recalling Eq. (\ref{eq:sigmafinal}), it is clear that deviations from the ordinary Fermi liquid behavior result from the scale-dependence of the scattering vertex, since this is the only quantity that differs from the perturbative calculation that leads to Fermi liquid behavior. Hence anomalous $T$-dependence of $\langle 1/\tau_a\rangle$ arises from the increase in the four-point vertex with decreasing temperature or energy scale. This increase is not restricted to the $d$-wave pairing Cooper channel since the divergence in this channel is suppressed in our calculations. Examination of the RG flows shows that several channels in the four-point vertex grow simultaneously, e.g. particle-hole and particle-particle umklapp processes, both with wavevector transfer near $(\pi,\pi)$ and with initial and final states in the anti-nodal regions. This phenomenon is not simply a precursor of $d$-wave superconductivity but rather signals that a crossover to strong coupling in several channels of the four-point vertex is responsible for the breakdown of the Landau-Fermi liquid. It will be challenging to find out more about the relation of this breakdown to the opening of the pseudogap at smaller doping levels. 
The simultaneous enhancement of several channels through mutual reinforcement was earlier identified as a key feature of the anomalous Fermi liquid in the cuprates and associated with the onset of resonant valence bond (RVB) behavior \cite{breakdown,laeuchli}. 

\noindent
In conclusion, our RG calculations suggest that the anomalous behavior of the inplane quasi-particle scattering rate revealed by the ADMR experiments \cite{abdel} on overdoped cuprates can be understood as an intrinsic feature of the doped Hubbard model that is already present at weaker interaction strengths. The positive correlation between $T_c$ and the anisotropic scattering rate shows up in the calculation as a general increase of correlations in the anti-nodal direction that is not restricted to the $d$-wave pairing channel. Our calculations are in agreement with earlier RG studies using different hopping parameters \cite{zanchi,halboth,breakdown,ehdoping} as far as the structure of the scattering vertex is concerned, so that we expect that our results hold in more general settings as well.

\chapter{Introduction to the wave packet approach to interacting fermions}
\label{ch:ww_intro}

\section{Introduction}

\noindent
In this chapter we give a general overview of the wave packet approach to interacting fermions that has been newly developed in this work. It is based on a description of electrons in terms of a complete orthogonal basis of phase space localized states - the wave packets. These states are intermediate between real and momentum space states in the sense that they have a finite extension in both spaces, similar to a Gaussian. As a consequence, a length scale $M$ is introduced into the problem from the beginning. Intuitively, this makes sense only when a physical length scale is present in the system under investigation. The typical example of the introduction of such a length scale is provided by systems with a gap for single particle excitations. Because of the gap, single particle correlations decay exponentially in space, $\left\langle c^\dagger(r)\,c^\pdag(0)\right\rangle \sim e^{-|r|/\xi}$, and in this case $\xi$ yields a natural length scale. 

\noindent
The two limiting case $\xi\rightarrow \infty$ and $\xi \rightarrow a$ (where $a$ is the lattice constant) are relatively well understood: The most celebrated example of the former is given by conventional, weakly coupled superconductors. These systems are known to be very well described by a mean-field approach, the BCS theory of superconductivity \cite{BCS}. In a superconductor, electrons are bound into pairs, and $\xi$ may be thought of as the pair size. The success of the BCS theory relies on the fact that the pairs are so large that many of them overlap, effectively eliminating quantum fluctuations \cite{superconductivity}. Corresponding to the large pair size in real space, the pairs are very localized in momentum space, and only a thin shell around the Fermi surface is correlated. The opposite limit of $\xi\rightarrow 0$ is exemplified by the strong coupling limit of a Mott insulator, where each lattice site is occupied by one electron and only local spin degrees of freedom remain, which are well separated from the charge sector. Alternatively, one may think of this state as a paired state as well, where each electron is bound to a hole. Since the pairs are localized, the pair size vanishes. Conversely, the pairs are very delocalized in momentum space and spread out over the whole Brillouin zone in this limit.

\noindent
Clearly, these two extreme cases are best described in the space where the fermion pairs are as local as possible, which allows to map the fermion problem to a tractable effective model. Our motivation is to obtain a similar description for the intermediate regime, where $\xi$ is neither small nor large, and hence momentum space concepts such as the Fermi sea and real space phenomena like the suppression of double occupancy both play a role. From this point of view it is quite natural to employ phase space localized basis functions: Due to their localization in real space some effects of local correlations can be taken into account, and their localization in momentum space allows to resolve certain features of the Brillouin zone, such as the approximate position of the Fermi surface. 

\noindent
The remainder of this chapter is organized as follows: First, we introduce the specific context of our study, the pseudogap phase of the cuprate superconductors. After a brief review of the part of the phenomenology that is relevant in the following, we discuss some theoretical studies that try to elucidate the opening of the pseudogap from a weak coupling point of view \cite{rice_rvb}, and the problems faced there related to the difficulties of treating renormalization flows that flow to strong coupling. Then we explain the wave packet approach and how it relates to the experimental and theoretical situation. Finally, we give an outline of the remaining chapters.

\section{The pseudogap phase of the cuprates and the saddle point regime of the Hubbard model}
\label{sec:intro_saddle}

\subsubsection{The pseudogap phase of the cuprates}

\begin{figure}
\centering
\includegraphics[width=10cm]{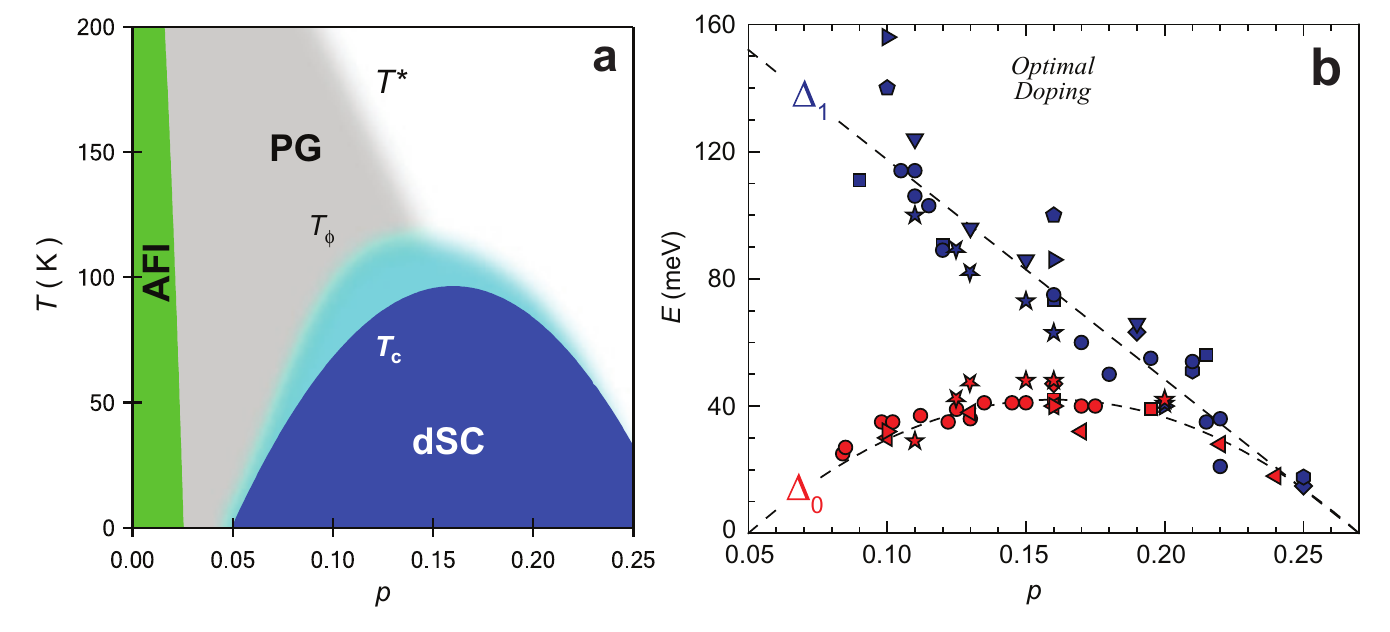}
\caption{Electronic excitations in the pseudogap phase. a) Schematic phase diagram of the cuprates. $T_c$ is the critical of $d$-wave superconductivity, $T_\phi$ is the maximal temperature at which superconducting phase fluctuations are detectable, and $T^\ast$ is the pseudogap temperature. b) Electronic excitations in the cuprates fall into two classes: The dome shaped curve ($\Delta_0$) corresponds to excitations above the superconducting state. The gap tracks $T_c$ and decreases for small doping. The curve labelled $\Delta_1$ separates from the superconducting gap in the underdoped regime, increasing towards half-filling. The different symbols correspond to different experimental techniques (Figure reproduced from \cite{pseudogapfig})}
\label{fig:pseudogapphase}
\end{figure}

\noindent
The pseudogap phase of cuprate superconductors is arguably one of the more puzzling aspects of their phenomenology. Here we highlight some aspects of this phase which are important for what follows, and refer to the review \cite{pseudogap_experiment} for a more detailed account. The phase lies between the Mott insulating state at zero doping, and the superconducting state at doping $p\approx 16\%$ as displayed schematically in Fig.~\ref{fig:pseudogapphase} a). Spectroscopic experiments, in particular ARPES measurements have shown that it is characterized by highly anisotropic electronic excitations, shown in Fig.~\ref{fig:pseudogapphase} b): In the nodal regions, close to $\left(\pi/2,\pi/2\right)$ a Fermi surface exists and electronic excitations are gapless. For large enough doping, a superconducting gap opens on these Fermi surface arcs. The corresponding gap for electronic excitations tracks the $T_c$ of the superconducting phase. At the same time, the gap for excitations at the saddle points stays large and increases as the doping is decreased \cite{pseudogap_review}. The gap for excitations at the saddle points persists up to the pseudogap temperature $T^\ast$, which is much larger than $T_c$ at low doping. NMR Knight shift measurements \cite{knightshift} indicate that a (partial) spin gap opens below $T^\ast$, which is generally taken as evidence for spin-singlet pairing. 

\subsubsection{The saddle point regime of the Hubbard model}

\begin{figure}
\centering
\includegraphics[width=7cm]{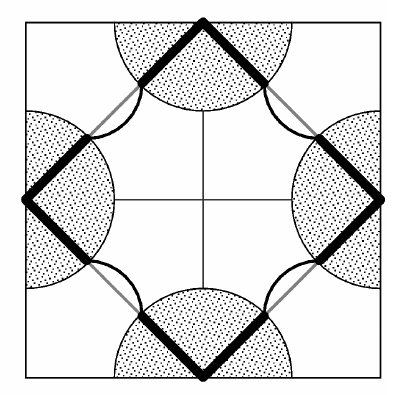}
\caption{Truncated Fermi surface due to strong RVB correlations. Figure taken from \cite{furukawarice}}
\label{fig:truncated_fermi_surface}
\end{figure}

\noindent
Despite the fact that the cuprates are often modeled as lightly doped Mott insulators, we have seen in Ch.~\ref{ch:rg} that the opposite approach using weak coupling renormalization group equations can yield valuable insights. The analysis of the RG equations for the full Hubbard model is still very complicated. Since the correlations are strongest in the vicinity of the saddle points, various researchers were led to study a reduced saddle point model instead of the full Hubbard model \cite{saddle1, saddle2, saddle3, saddle4,furukawarice}. Within the one-loop RG approach it has been established that the model has strong correlations at low energies, with the leading instablities occuring in the $d$-wave pairing and antiferromagnetic channels. However, it was found that at the same time, the uniform spin and charge susceptibilities are suppressed. Since the latter behavior is consistent with gaps for spin and charge excitations, Furukawa et al.~\cite{furukawarice} were lead to conjecture that the ground state for this model is an insulating spin liquid. The conjecture is based on an analogy to the physics of ladder systems \cite{nlegladder, linbalents}, where a similar RG flow leads to a spin liquid phase with gaps for all excitations. Later work using exact diagonalization of the low energy Hamiltonian on small clusters \cite{laeuchli} corroborated this view. However, it has proven difficult to derive an effective model for this problem, and to embed it into the full Hubbard model.

\section{Phase space localized basis functions}

\subsubsection{What is phase space localization?}

\begin{figure}
\centering
\includegraphics[width=10cm]{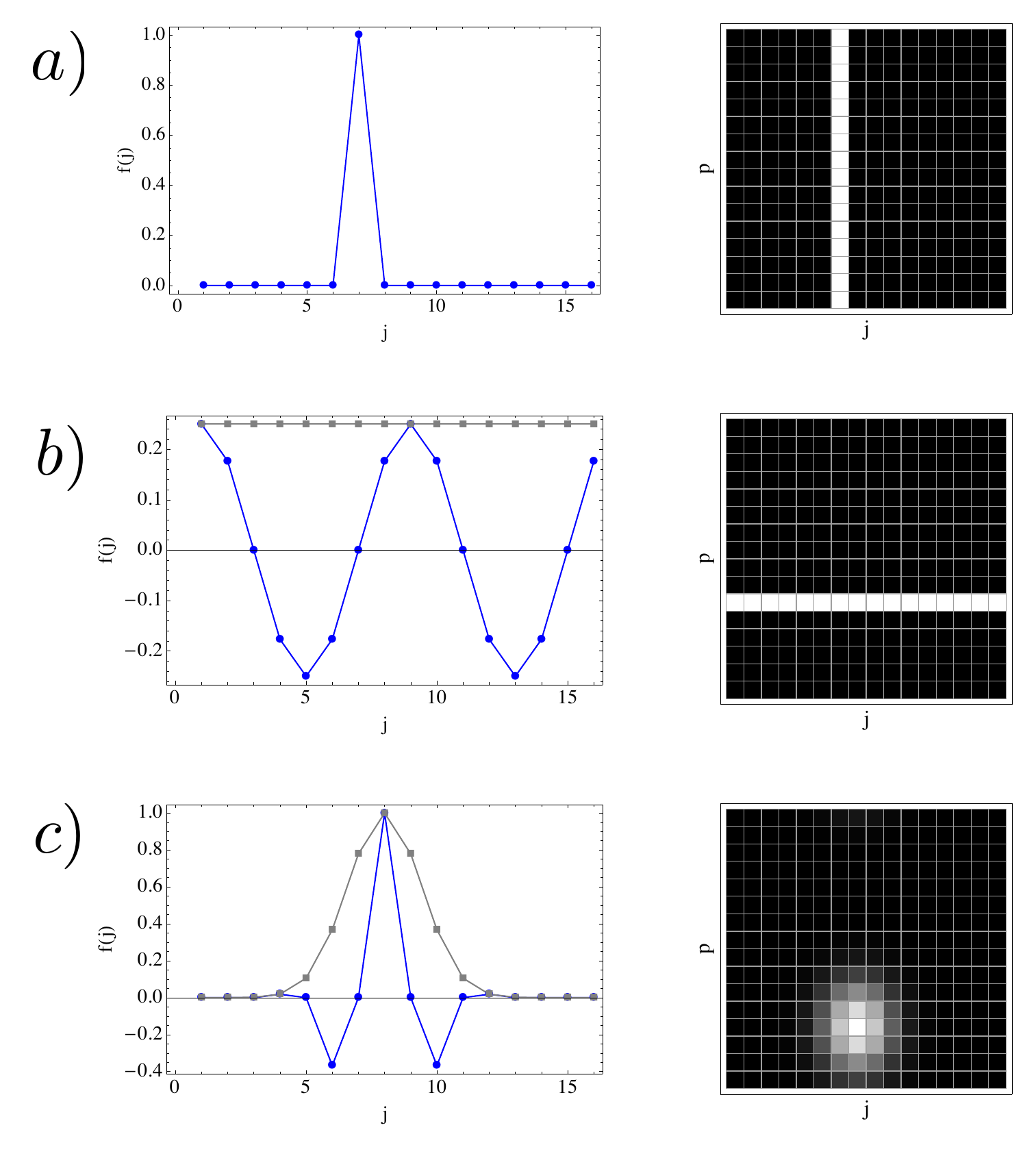}
\caption{Real space representation $f(j)$ (left) and phase space density $\left|f(j) \tilde{f}(p)\right|$ (right) for three different functions. a) Localized real space basis state, b) plane wave state, c) wave packet. The real space basis state and the momentum space basis state are fully localized in $j$ or $p$ directions, respectively, and fully delocalized in the other direction. The wave packet state on the other hand is localized in both real and momentum space.}
\label{fig:phasespacelocalization}
\end{figure}
\noindent
In this section we give a gentle introduction to phase space localized basis functions, the basic building block of out approach. As we have stated above, these functions are localized to some extent in both real space and momentum space at the same time. In order to contrast this with the usual real space and momentum space basis states, Fig.~\ref{fig:phasespacelocalization} compares the phase space density of three different functions. The phase space density can be defined in the following way: Take any function $f(j)$, that is defined on a one-dimensional lattice with $N$ sites, where the position is labelled by $j=0,\ldots,N-1$. Call its Fourier transform $\tilde{f}(p)$, where $p$ is the wave-vector, such that $f(j)=1/\sqrt{N} \sum_p e^{i p j} \tilde{f}(p)$. Define the phase space density to be $\rho(j, p) = \left| f(j) \tilde{f}(p)\right|$. The phase space density depends on both position and momentum variables, and is a neat way to visualize the localization of a function (or basis state) in real and momentum space simultaneously. Clearly, the real space basis state $f(j) = \delta_{ij}$ in Fig.~\ref{fig:phasespacelocalization} a) is localized in real space, but completely delocalized in momentum space, and thus represented by a vertical straight line in the phase space plot. The function can be used to generate a complete orthogonal basis by shifting it 'horizontally' in phase space, i.e. by shifting in real space, $f(j)\rightarrow f(j+1)$. Repeating this procedure $N$ times yields a complete orthogonal basis that is invariant under real space shifts. Similarly, the plane wave $f(j)=1/\sqrt{N} e^{ipj}$ in Fig.~\ref{fig:phasespacelocalization} is represented by a horizontal  line in phase space. A complete basis is generated by shifting it 'vertically', i.e. $f(j)\rightarrow e^{ij} f(j)$, and repeating the procedure $N$ times. Now consider the function in Fig.~\ref{fig:phasespacelattice}, which is a Gaussian. Due to its phase space localization, it looks more two-dimensional than the real space and momentum space basis states, in that it has both a definite mean position \emph{and} mean momentum, so that it appears like a smeared point in the phase space plot instead of an extended line. 

\subsubsection{How can one construct a nice basis from phase space localized functions?}

\begin{figure}
\centering
\includegraphics[width=6cm]{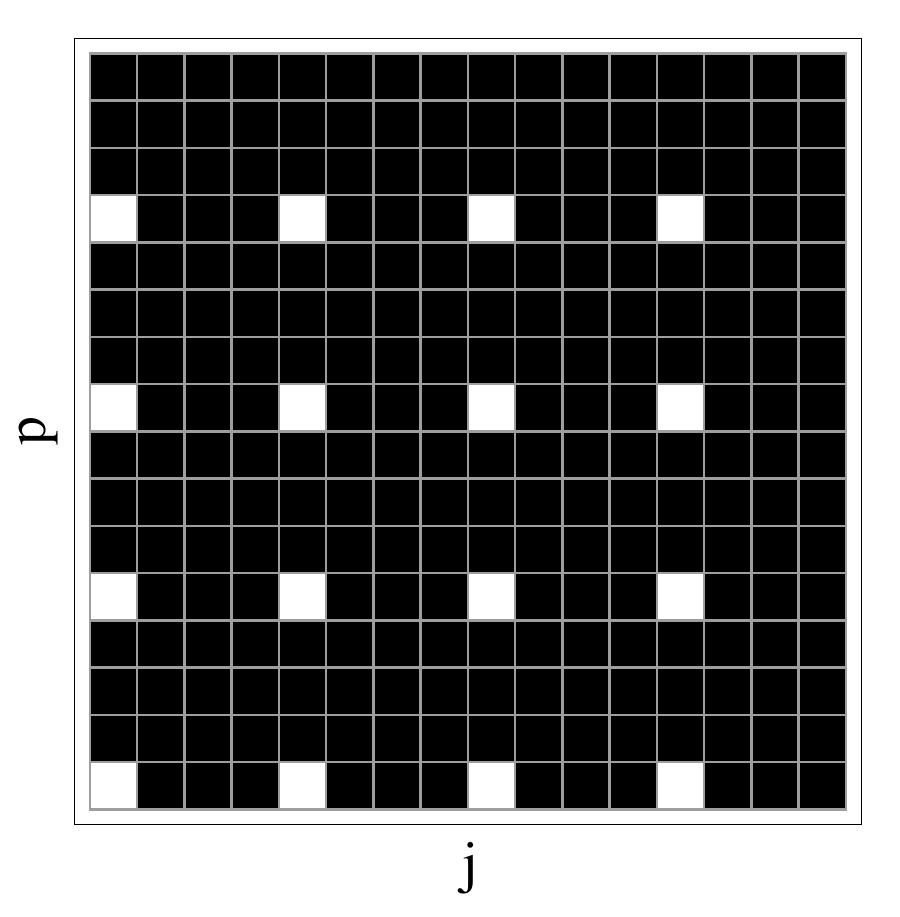}
\caption{Phase space positions of the basis states of a complete shift-invariant wave packet basis. The basis is created by applying real and momentum shifts to a single wave packet For a complete basis, the states generated lie on a lattice in phase-space. In order to have the correct number of states, the area of a unit cell must be $N$ for a lattice with $N$ sites.}
\label{fig:phasespacelattice}
\end{figure}

\noindent
The Gaussian is the type of state we intend to use as a basis function for the description of interacting electron systems. Thus the question arises how one can create a nice complete basis from this kind of wave function. By analogy with the examples above, the na\"ive approach is to use one such function, which we denote by $g(j)$. $g(j)$ is referred to as the \emph{window function} in the following. This function should be localized in phase space in the same way as the Gaussian, i.e.~it should have a maximum in real space, say at $j=0$, and a maximum in momentum space, at $p=0$. To be well localized, it should decay rapidly as one moves away from this maximum. To generate a basis that covers the full phase space, one shifts it around in both real space and momentum space, by defining
\begin{equation}
g_{mk}(j) = e^{i K k j} g(j-Mm),
\end{equation}
where $m$, $k$, and $M$ are integers. The mean position of the shifted window function $g_{mk}(j)$ is $Mm$, and its mean momentum is $Kk$. Counting the number of states that are obtained in this way, we see that $K$ must satisfy $K = 2\pi/M$. Note that the basis functions lie on a \emph{lattice} in phase space, as shown in Fig.~\ref{fig:phasespacelattice}. This basis has the following nice properties
\begin{enumerate}
\item Good phase space localization of the basis functions,
\item Shift invariance in real (with period $M$) and momentum space (with period $K$).
\end{enumerate}
Unfortunately, orthogonality is not among them. However, Wilson \cite{wilsonbasis} found a way to obtain a orthogonal basis that shares the good phase space localization with the na\"ive approach here, and (almost) retains its shift invariance properties. The construction principle is very similar, but involves a second step. Essentially one first generates a basis with $2N$ states by setting $K=\pi/M$ (instead of $K=2\pi/M$). One then forms linear combinations such as
\begin{equation}
\psi_{mk}(j) = \frac{1}{\sqrt{2}} \left[ g_{m,k}(j) \pm g_{m,-k}(j)\right],
\end{equation}
which are either even or odd under reflections (i.e. parity $\pm 1$). Finally, one discards half of the states, such that neighboring states in phase space have the opposite parity. In Ch.~\ref{ch:wwbasis} we present more details on the construction of this so-called \emph{Wilson-Wannier basis}. 

\subsubsection{How can phase space localization help to understand correlated electrons?}

\begin{figure}
\centering
\includegraphics[height=8cm]{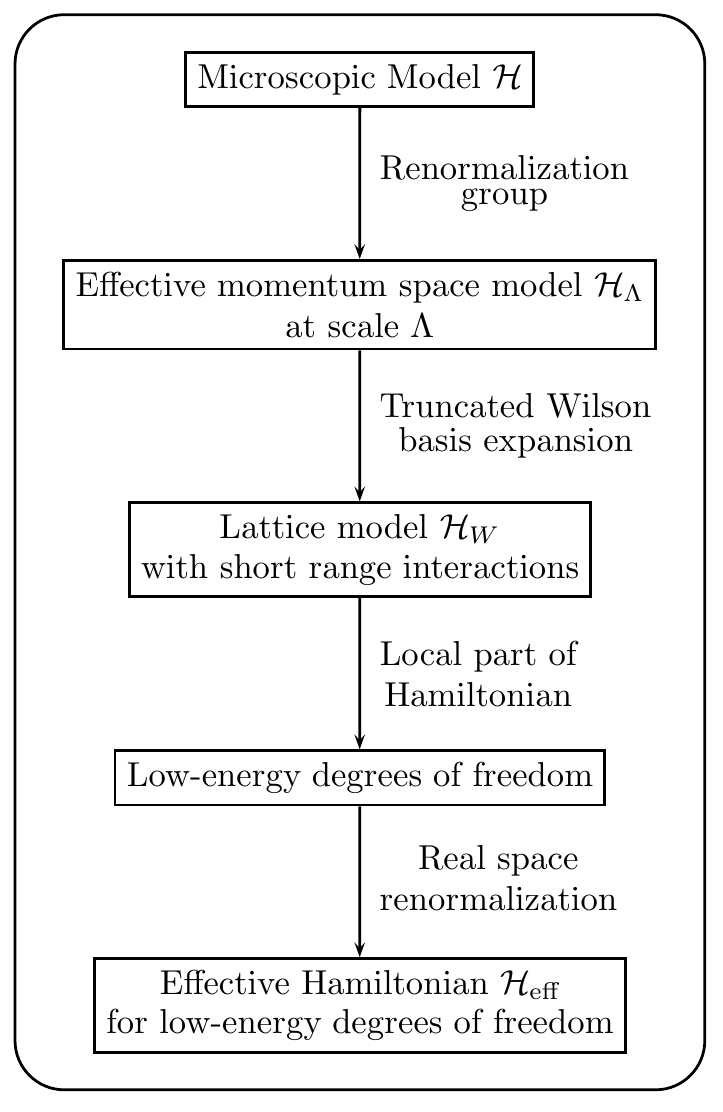}
\caption{The different steps in the wave packet approximation}
\label{fig:flowchart}
\end{figure}

\noindent
Most microscopic approaches to correlated electron systems can be loosely categorized into two classes: First, there are methods that are more naturally situated in momentum space such as the renormalization group \cite{tnt}. These methods are very effective in capturing longer range correlations, and relatedly are good at resolving features in the Brillouin zone, such as the anisotropy of quasi-particle life times on the Fermi surface in Ch.~\ref{ch:rg}. In addition, they often lead to simple physical pictures, that allow an intuitive understanding of complex physical problems, a feature that has merits on its own. On the other hand, it is rather difficult to treat strong short-range correlations, and more often than not uncontrolled approximations are necessary. The second class is usually defined in real space, and contains methods like exact diagonalization of small clusters \cite{exactdiagonalization}, strong coupling expansions \cite{strongcouplingexpansion}, or the dynamical mean-field theory \cite{dmft}. These methods are usually very effective when it comes to dealing with strong correlations on small length scales, but it is difficult to incorporate the build-up of correlations on longer length scales. Moreover, they often involve a high computational cost, and the sheer complexity of the calculations can make it hard to develop a physical understanding of the solution. 

\noindent
We try to find a middle ground between these two approaches, by combining features of both: We split the Brillouin zone into a low-energy part in the vicinity of the Fermi surface, and the remaining states which are at higher energies. It stands to reason that the states at the Fermi surface are more strongly correlated than states that are very far away from the Fermi surface. Hence we use the renormalization group to treat the high-energy problem perturbatively and obtain an effective Hamiltonian for the low energy degrees of freedom. Note that this only makes sense when the onsite interactions are not too strong. The remaining low-energy problem is then transformed to the Wilson-Wannier (WW) basis, and solved using real space methods, namely strong coupling approximations and real space RG \cite{morningstar}. The different steps are summarized in Fig.~\ref{fig:flowchart}.

\noindent
The usefulness of the momentum space localization lies in the fact that one can isolate the low-energy degrees of freedom simply by truncating the basis, retaining only those states whose mean moment lies close to the Fermi surface, or a part of the Fermi surface, such as the saddle points. At the same time, the real space localization is helpful because the effective Hamiltonian in the WW basis remains short-ranged, which makes it possible to analyze the strong coupling problem. 

\noindent
In closing, we would like to remark that the wave packet approach is not really a method to solve Hamiltonians, but rather a different way to view many-electron problems. Thus it is in principle compatible with many different methods that can be used to solve these problems.

\section{Outline of the remaining chapters}

\noindent
In the remainder of this thesis, we develop the above heuristic ideas in more detail, and discuss applications to one- and two-dimensional interacting fermion systems. Since our approach is novel, the exposition starts from scratch, gradually moving towards the saddle point regime that is the motivation for our work. We begin with two rather technical chapters where the basic formalism for dealing with the Wilson-Wannier basis is established.

\noindent
In Ch.~\ref{ch:wwbasis} we introduce the Wilson-Wannier (WW) basis states for one-dimensional lattices, following the exposition given in \cite{daubechies, discretewilson}. In particular, we show how the basis can be generated from a single window function (or wave packet) by applying shifts in real space and momentum space. We reformulate the construction by relating the construction principle to the point group of the lattice, which leads to a more compact and physically transparent form. Some elementary, yet lengthy mathematical derivations are involved in the construction. These are relegated App.~\ref{ch:windowfunction} to avoid interrupting the logical flow. In order to generate the basis, one must have a suitable window function first. We introduce a family of such window functions which allows to obtain many approximate analytical results. Finally, we extend the WW basis to the square lattice by taking the tensor product of one-dimensional basis functions.

\noindent
Ch.~\ref{ch:wwtrafo} deals with the transformation of operators from real or momentum space to the WW basis in one and two dimensions, respectively. For each case, we derive the general basis transformation formula first. Then we decompose it into two steps: First the operator is expanded into an overcomplete wave packet basis, which has the advantage that matrix elements are much simpler to understand than in the WW basis itself. This step naturally leads to a systematic and intuitively appealing approximation method for local operators (referred to as $1/M$-expansion, where $M$ is the size of a wave packet), similar to the gradient expansion in field theory. In the second step, the orthogonalization procedure for the WW basis is applied to the wave packet transform in order to arrive at the final form. We find our group theoretical formulation form Ch.~\ref{ch:wwbasis} very helpful in developing intuition and (relatively) simple formulas for this step. 

\noindent
In Chs.~\ref{ch:ww_pairing} and \ref{ch:wave_packets_rg} we use the results from the preceding sections to explore the physics of interacting fermions from the point of view of phase space localization, focussing on superconductivity and antiferromagnetism and the resulting Fermi surface instabilities. 

\noindent
In Ch.~\ref{ch:ww_pairing} we investigate correlations in the ground state of simple mean-field Hamiltonians in the WW basis in one and two dimensions. This exercise serves the purpose of relating the free parameter of the WW basis, namely the size $M$ of the generating wave packet, to physical length scales due to fermion correlations. In particular, the ground state of all Hamiltonians considered exhibits fermion pairing, where the pairs may consist of either two particles (superconductivity) or a particle and a hole (antiferromagnetism). In one dimension, the pair binding energy $\Delta$ defined by the symmetry-breaking mean-field corresponds to a length scale $\xi \sim 2\pi v_F/ \Delta$, which may be interpreted as the pair size. We discuss the appearance of local physics in the WW basis as the ratio $\xi/M$ is varied. We find a crossover between the two limits $\xi \ll M$, where all fermions are paired into bound states that are local in the WW basis, and $\xi \gg M$, where locally the system appears to be almost uncorrelated. This insight will be used in later chapters in order to map interacting fermion systems to bosonic systems with the paired fermions as new degrees of freedom. 

\noindent
Subsequently we investigate the changes that appear in two-dimensions, focussing on the saddle point regime of the two-dimensional Hubbard model. We show that no single length scale can be associated to the pair breaking energy because of the large anisotropy of the Fermi velocity. Instead, we observe a separation of length scales along the Fermi surface, similar to the crossover in one dimension as $\xi/M$ is varied. Due to their small Fermi velocity, states in the vicinity of the saddle points are effectively bound into pairs on very short length scales, whereas states in the nodal direction around $(\pi/2,\pi/2)$ are very weakly correlated at the same length scale. This leads us to conjecture that the states at the saddle points decouple from the nodal states at short length scales, corroborating the arguments made in earlier works that were discussed above in Sec.~\ref{sec:intro_saddle}. 

\noindent
From Ch.~\ref{ch:wave_packets_rg} on we leap from simple mean-field Hamiltonians to interacting models. As a preparation for the renormalization group based studies that follow, we clarify the relationship between wave packets and the RG. We start by relating the $1/M$-expansion from Ch.~\ref{ch:wwtrafo} to the scaling dimension of operators in the RG approach, which serves to understand the relative importance of different operators. In a short technical section, we discuss certain problems that occur due to the cutoff that is introduced by the RG, and point out a remedy for this issue. Finally, we pick up the discussion on the separation of length scales in the saddle point regime (Ch.~\ref{ch:ww_pairing}), and perform a similar analysis based on the geometry of the low energy phase space from the point of view of the renormalization group. We obtain similar results as before.

\noindent
After this long preparation, we finally study actual interacting fermion systems in Ch.~\ref{ch:ww1}, starting with one-dimensional systems at weak coupling, where exact solutions are available from bosonization \cite{giamarchi,gogolin}. The main goal is not to aim at numerical accuracy, but to see if and how the qualitative behavior at low energies is reproduced in the wave packet approach. Hence we analyze two kinds of systems that exhibit strong coupling fixed points with very different behavior: First we treat chains with repulsive and attractive interactions, where the model is known to show quasi long range order in the form of algebraic decays of various correlation functions for charge, spin, and singlet pair densities. The second kind of system is the two-leg ladder at half-filling. This model is known to become Mott insulating with gaps for all excitations at any coupling strength. Nevertheless, the ground state features pronounced $d$-wave pair and antiferromagnetic spin correlations, on short length scale, resembling the RVB states proposed in the context of cuprate superconductors \cite{anderson, rice_rvb}.

\noindent
We study these models following the approach outlined in Fig.~\ref{fig:flowchart} above. In each case we first introduce the relevant RG fixed point. Then we transform the low energy degrees of freedom to the WW basis. In this step, we limit ourselves to the simplest possible approximation, and keep only states at the Fermi points. Effectively, this maps the low energy sector of the chain (ladder) at weak coupling to another, strongly coupled chain (ladder) with a larger lattice constant. The resulting model is then analyzed step by step at strong coupling, using mappings to effective bosonic models. The major reason for this approach is that then the resulting Hamiltonians are relatively simple to analyze, thus allowing to understand why and how the differences in low energy physics come about. Despite the simplicity of the approximations, we find that the qualitative behavior of all systems is reproduced well. In particular, the difference between quasi long range order and the RVB-like short ranged correlations show up very clearly. Finally, we demonstrate that the different behavior of the two kinds of systems can be related to the structure of the respective local Hilbert spaces in the WW basis, a result that will be useful in two dimensions as well. More concretely, for the chain models we generically obtain locally degenerate ground states in the strong coupling limit, whereas for the ladder the ground state is unique, with large gaps for all excitations. We link this difference to the energy separation (or lack thereof) between single particle excitations and collective modes.

\noindent
In Ch.~\ref{ch:ww2} we return to the discussion of the saddle point regime of the Hubbard model, the main motivation for our study. In the same manner as in one dimension, we use the RG in order to obtain an effective Hamiltonian at low energies. Based on the arguments from Chs.~\ref{ch:ww_pairing} and \ref{ch:wave_packets_rg}, we use the separation of length scales inherent to the model in the saddle point regime to devise approximations. We only consider the simplest such approximation, and ignore all low-energy states except the ones in the vicinity of the saddle points. For these states, we show that in the WW basis the saddle point states are mapped to a bilayer, the two-dimensional analogue of the two-leg ladder system. Consequently, the local Hilbert space is the same as the one for the ladder systems. Moreover, from the RG flow for different model parameters we infer that the effective low-energy Hamiltonian is similar as well, in that its local part has a unique local ground state with large gaps for all excitations and strong $d$-wave pairing and antiferromagnetic correlations. The main difference to the ladder model is that there is no universal fixed-point of the RG, and that in particular we observe a crossover between AF and $d$SC dominated regimes as a function of doping. In order to assess the effect of the higher dimensionality on the stability of the local ground state, we diagonalize the effective model on small clusters, and map it to an effective bosonic model that is analyzed by means of variational coherent states. We find that the RVB-like ground state appears to be robust over a sizable parameter range, indicating spin-liquid behavior.

\noindent
Even though the results are robust within our approximations, we are reluctant to draw definite conclusions from the calculations at this point, since the approximations involved are quite drastic. However, since the wave packet approach is still in its early stages of development, there is a lot of space for improvements in different directions, some of which we point out in the conclusions. Moreover, the approximation is based on physical arguments that are fairly elementary, namely the separation of length scales due to the vicinity of the saddle points and the possibility to localize states due to umklapp scattering (manifested by the commensurability of the AF spin correlations), both of which are independent of the details of the calculation.

\noindent
Finally, we summarize our thesis in Ch.~\ref{ch:conclusions}, and give an outlook on possible future work.

\chapter{Wilson-Wannier basis for a finite lattice}
\label{ch:wwbasis}

\noindent
In this technical chapter we introduce the Wilson-Wannier (WW) basis \cite{wilsonbasis, daubechies, discretewilson}, an orthogonal basis whose basis states are wave packets that are localized in real space and bimodal in momentum space. The reason for using this type of basis is that even though there are no-go theorems on the localization in both momentum- and real space of orthonormal wave packet bases\cite{balian, low}, these can be circumvented if one allows wave packets to be localized around two points in momentum space. Moreover, the packets can be chosen such that in momentum space each wave packet is localized around the two momenta $\pm p$, so that for systems with inversion symmetry, they can still be used to resolve states that are close to the Fermi points.

\section{Wilson-Wannier basis in one dimension}
\subsection{Construction of the basis functions}

\subsubsection{Definition}

\noindent
In the following we consider a finite one-dimensional lattice of size $N$ with periodic boundary conditions. In order to introduce the Wilson basis, we assume that $N$ can be written as $N =  M L$, where both $M$ and $L$ are even. The basis is generated from a single window function (or wave packet) $g(j),\, 0\leq j < N$. We demand that it is exponentially localized in both real and momentum space with widths of order $M$ and $K\equiv\frac{\pi}{M}$, respectively. In order to generate the basis, we will need the shifted window function
\begin{equation}
g_{m k}(j) = \underbrace{e^{i K k j}}_{\text{momentum shift}} g\underbrace{\left(j - m M\right)}_{\text{position shift}} .
\label{eq:gshift}
\end{equation}
The shifted window functions are labelled by two coordinates, the position coordinate $m=0,1,\ldots,L-1$, and the momentum coordinate $-M<k\leq M$. $K$ is the step size of a momentum shift. These coordinates are connected to the mean position $\bar{j}$ and mean momentum $\bar{p}$ of the function by
\begin{equation}
 \bar{j} = M m, \quad \bar{p} =  K k,
 \end{equation} 
There are $2N$ shifted window functions $g_{mk}(j)$ in total for a lattice of size $N$. 

\noindent
We now use the $g_{mk}(j)$ to generate a complete and orthogonal basis for the lattice. The basis states are denoted by $\left|m,k\right\rangle$. Their relation to the real space states $\left| j\right\rangle$ is given by the wave function
\begin{equation}
\braket{j}{m,k} = \psi_{mk}(j).
\end{equation}

\noindent
Following \cite{daubechies, discretewilson}, the wave functions $\psi_{mk}(j)$ are given by
\begin{equation}
\psi_{m k}(j) = \left\{ \begin{array}{l l} 
g_{m,0}(j) & \qquad m \text{ even},\; k=0 \\
g_{m,M}(j) & \qquad m \text{ even}, \; k=M \\ 
  \frac{1}{\sqrt{2}}\left( g_{m,k}(j) + g_{m,-k}(j)\right)&\qquad 1 \leq k < M, \; m + k \text{ even} \\
  \frac{-i}{\sqrt{2}}\left( g_{m,k}(j) - g_{m,-k}(j)\right)& \qquad 1 \leq k < M,\; m + k \text{ odd}
\end{array}\right.
\label{eq:defpsi_g}
\end{equation} 
When the window function $g(j)$ satisfies certain orthogonality conditions to be stated below, the states $\ket{m,k}$ form a complete orthogonal basis. In addition, the functions $\psi_{m,k}(j)$ have the useful property of exponential localization in both real space and (around two points in) momentum space if $g(j)$ is chosen appropriately.

\noindent
The window function $g(j)$ has to satisfy certain conditions to make the $\psi_{mk}(j)$ orthogonal. The derivation of the conditions on $g(j)$ for the case of a finite lattice with an even number of sites is given in appendix \ref{sec:gconditions}. In real space, the conditions are 
\begin{equation}
\sum_{m=0}^{L-1} g\left(j - m M\right) g\left(j - M\left(m+ 2 l\right)\right) = \frac{1}{M}\delta_{l,0},
\label{eq:gconditions}
\end{equation}
which has to be satisfied for all $j$ with $0\leq j < N$ and $l$ with $0\leq l < L/2$. Conditions (\ref{eq:gconditions}) are in the form of a convolution. The convolution can be turned into a multiplication by means of the Zak transformation \cite{zaktransform}, so that suitable window functions can be readily constructed on a computer. The usage of the Zak transformation in this context is detailed in appendix \ref{sec:zaktransform}. 

\noindent
In Sec. \ref{sec:analytical_window_functions}, we show that restriction to a special class of window functions that are band limited in momentum space leads to considerable simplification \cite{daubechies}. In fact, analytical window functions can be easily constructed in this case.

\begin{figure}
\centering
\includegraphics[width=8cm]{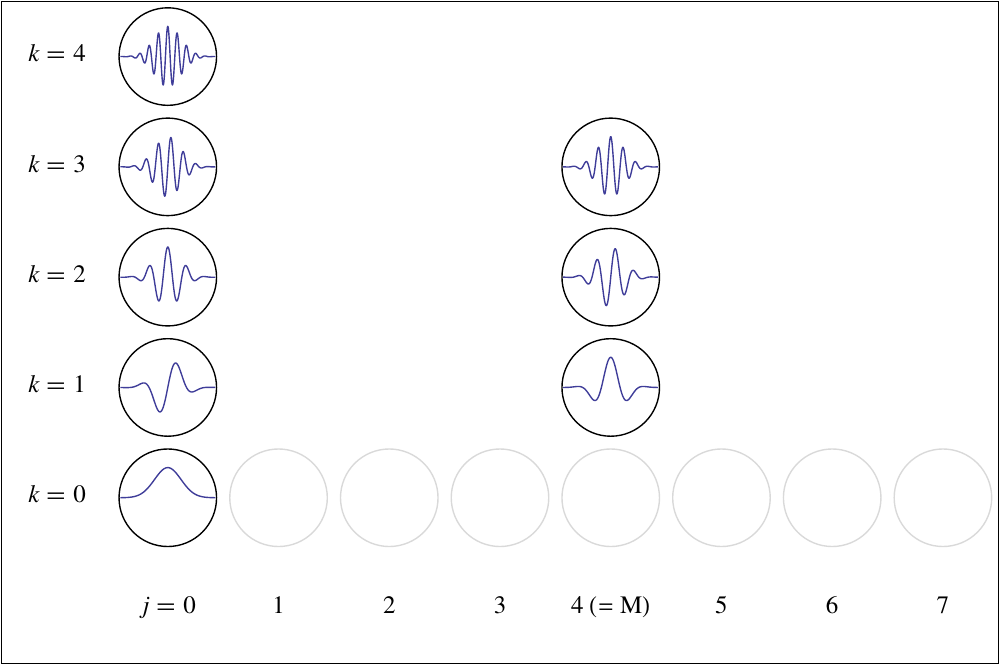}
\caption{Schematic representation of the relation of Wilson-Wannier functions to the real space lattice. The figure shows one unit cell of the Wilson basis for $M=4$. $j$ labels lattice sites in the real lattice, and gray circles represent these sites. The WW momentum is denoted by $k$ and runs in the vertical direction. The $2M = 8$ sites in the original lattice are replaced by two sets of states centered around $j=0$ and $j=M$ in the Wilson basis. Note that the two superlattice sites within one unit cell are inequivalent, which can be seen best from the fact that the states $k=0$ and $k=M$ exist only once per unit cell.}
\label{fig:wilson_basis}
\end{figure}

\begin{figure}
\centering
\includegraphics[width=8cm]{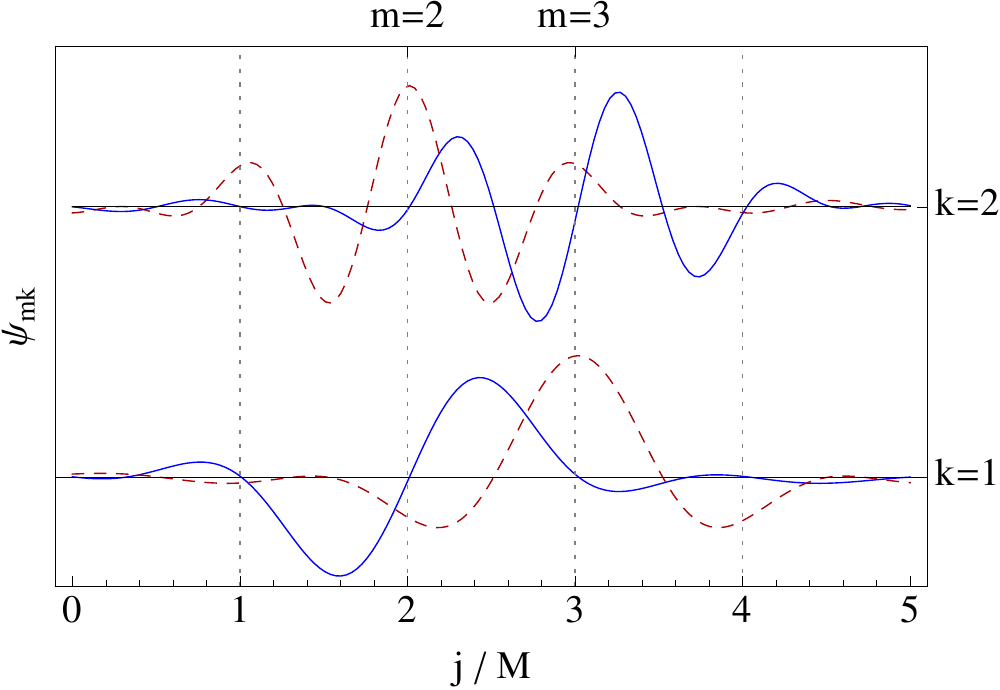}
\caption{A subset of the Wilson-Wannier basis functions within one unit cell of the basis for $N=2^{10}$ and $M=2^5$. The figure shows $\psi_{mk}(j)$ with $m=2,3$ and $k=1,2$. States with $k=2$ are offset vertically for sake of clarity. The centers of the wave packets in real space are marked by the dotted gray lines at positions $j = M m$. The parity of the states is given by $(-1)^{m+k}$ and hence follows a checkerboard pattern in the $m-k$-plane. States with even parity are shown in red (dashed line), states with odd parity in blue (solid line).}
\label{fig:wwbasisfunctions}
\end{figure}

\noindent
The choice of $M$ in the factorization $N=MK$ defines the length scale over which the basis functions are delocalized. The unit cell for the basis functions is $2M$ because of the phase factors $e^{\pm i \phi_{m+k}}$ in (\ref{eq:defpsi}) that are different on adjacent WW sites but identical on second nearest neighbor sites, corresponding to wave packets with even (odd) parity on even (odd) phase space lattice sites. The states with $k=0$ and $k=M$ are already parity eigenstates, so that they appear only once per unit cell, for all other $k$ there are two states per unit cell for even and odd parity (or $k<0$ and $k>0$). A schematic picture of the basis function in one unit cell is shown in Fig.~\ref{fig:wilson_basis}. From $N=ML$ one sees that there are $L/2$ unit cells in total. Note that this figure is intended to show how the states are rearranged in the new basis only, and that it does not reproduce the shape of the wave packets correctly. The real space form of the wave packets within one unit cell is shown in Fig.~\ref{fig:wwbasisfunctions}. The figure shows wave packets with $m=2,3$ and $k=1,2$. The parity of the states follows a checkerboard pattern in the $m-k$-plane, where nearest neighbors always have opposite parity. These findings suggest to use another definition of the WW basis function based on their symmetry properties.

\subsubsection{Improved definition based on group theory}

\noindent
The definition (\ref{eq:defpsi_g}) is awkward to work with. It can be simplified by the action of the point group $G$ of the lattice on a wave packet $g_{mk}(j)$. Since the lattice is one-dimensional, $G = \{\mathcal{E}, \mathcal{P}\}$, where $\mathcal{E}$ is the identity and $\mathcal{P}$ is the inversion. For practical purposes, we replace the elements of $G$ by their $1\times 1$ matrix representations when acting on momenta $p$. Note that the action on the mean momentum $k$ of a wave packet $g_{mk}(j)$ is the same as the one on the momentum $p = K k$. Hence we use the correspondence
\begin{eqnarray}
\mathcal{E}&\leftrightarrow& 1 \nonumber \\
\mathcal{P}&\leftrightarrow& -1 \nonumber \\
G &\leftrightarrow& \{1,-1\},
\end{eqnarray}
and define the point group action on a wave packet to be
\begin{equation}
A_\alpha \, g_{mk}(j) = g_{m,\alpha k},
\label{eq:groupaction_1d}
\end{equation}
where $\alpha = \pm 1$, and $A_\alpha \in G$ is its associated group element. Note that (\ref{eq:groupaction_1d}) implies that the center of mass of the wave packet is used as the origin of the lattice. For each $k$ we can find the stabilizer (or little group) $H_k \subseteq G$, which is the subgroup of $G$ that leaves $g_{mk}(j)$ invariant. We find
\begin{equation}
H_k = \left\{ \begin{array}{ccc} \{\mathcal{E}\} & \text{for} & 0<\left|k\right|<M \\ G & \text{for} & \left|k\right| = 0,M \end{array}\right.
\end{equation}
Finally, we denote the number of elements of a group $G$ by $\left| G\right|$. Then (\ref{eq:defpsi_g}) can be written as
\begin{equation}
\psi_{m, k}(j) =  \frac{1}{\sqrt{\left|G\right|  \left|H_k\right|}}\sum_{\alpha\in G} e^{-i\alpha\phi_{m+k}} g_{m,\alpha k}(j).
\label{eq:defpsi}
\end{equation}
The phase factor $\phi_{m+k}$ is given by
\begin{equation}
\phi_a = \left\{\begin{array}{l l} 0 & \qquad a \text{ even} \\ \frac{\pi}{2} & \qquad a \text{ odd}\end{array}\right.
\label{eq:def_phi}
\end{equation}
Hence
\begin{equation}
e^{-i\alpha \phi_{a}} = \left\{\begin{array}{ccc} 1 & \text{for} & a \text{ even} \\ \frac{\alpha}{i} & \text{for} & a \text{ odd}\end{array}\right.,
\end{equation}
and it is straightforward to show that the definitions (\ref{eq:defpsi_g}) and (\ref{eq:defpsi}) are equivalent.

\subsection{Relation to real space and momentum states}

\noindent
In the remainder of this work, we will frequently switch between momentum space, real space, and WW descriptions. This section summarizes the relation between states in all three bases.
We begin by summarizing the different variables and their meaning in Tab. \ref{tab:variables_1d}.
\begin{table}
\begin{center}
\begin{tabular}{| c | c | p{5cm} |}
\hline 
variable & range & meaning \\
\hline
$j$ & $0,\ldots, N$ & position in real space \\ 
\hline 
$p$ & $-\pi+\frac{2\pi}{N}, -\pi+2\frac{2\pi}{N},\ldots, \pi-\frac{2\pi}{N}, \pi$ & momentum \\
\hline
$m$ & $0,\ldots, L-1$ & WW position label for state with center position $\bar{j}=Mm$ \\
\hline
$k$ & $0,\ldots, M$ & WW momentum label for state with center momenta $\bar{p}=\pm K k$\\
\hline
$\alpha$ & $\pm 1$ & Representation of the point group $G$ on wave packets \\
\hline
$M$ & even & Real space shift length. Size of WW unit cell is $2M$ \\ 
\hline
$K$ & $\frac{\pi}{M}$ & Momentum shift length \\
\hline
$L$ & $\frac{N}{M}$ & Range of $m$ \\
\hline
\end{tabular}
\end{center}
\caption{Variables for the description of the spatial degrees of freedom in one dimension.}
\label{tab:variables_1d}
\end{table}

\noindent
We have already introduced the relation between WW basis states $\ket{m,k}$ and real space basis states $\ket{j}$, which defines the basis functions $\psi_{mk}(j)$,
\begin{equation}
\psi_{mk}(j) = \braket{j}{mk}.
\end{equation}
A brief glance at the definition (\ref{eq:defpsi_g}) of the basis functions reveals that they are real:
\begin{eqnarray}
\psi_{m,k}(j) &=& \frac{1}{\sqrt{2}} \left[e^{-i\phi_{m+k}}\,g_{m,k}(j) +e^{i\phi_{m+k}}\, g_{m,-k}(j)\right] \nonumber \\
&=& \frac{1}{\sqrt{2}} \left[ e^{i\left(K k  j-\phi_{m+k}\right)} \pm e^{-i\left(K k j-\phi_{m+k}\right)}\right] g\left(j - M m\right)\nonumber \\
&=& \sqrt{2} \cos\left[K k- \phi_{m+k}\right] \, g\left(j-M m\right).
\end{eqnarray}
As a consequence, we find that
\begin{equation}
\braket{m,k}{j} = \braket{j}{m,k}^\ast = \psi_{m,k}(j).
\end{equation}

\noindent
The relation to momentum states follows directly from
\begin{eqnarray}
\braket{p}{m,k} &=& \sum_j \braket{p}{j}\, \braket{j}{m,k} \nonumber \\
&=& \frac{1}{\sqrt{N}} \sum_j e^{-ipj} \psi_{m,k}(j) \nonumber \\
&=& \tilde{\psi}_{m,k}(p), 
\end{eqnarray}
where $\tilde{\psi}_{m,k}(p)$ denotes the Fourier transform of $\psi_{m,k}(j)$. Note that $\tilde{\psi}_{m,k}(p)$ is not real in general, so that $\braket{m,k}{p} = \tilde{\psi}_{m,k}(p)^\ast$. 

\subsection{Analytical window functions}
\label{sec:analytical_window_functions}

\noindent
In general, window functions that satisfy (\ref{eq:gconditions}) have to be constructed numerically. However, a special class of window functions can be readily constructed analytically. The key condition for the simplification is that the window function is band limited in momentum space. In order to make this notion more quantitative, we introduce the Fourier transform $\tilde{g}(p)$ of $g(j)$ via 
\begin{equation}
g(j) = \frac{1}{\sqrt{N}} \sum_p e^{i p j} \tilde{g}(p).
\end{equation}
Then we call the window function band limited when its Fourier transform satisfies
\begin{equation}
\tilde{g}(p) = 0 \text{ for } |p| \geq K.
\label{eq:condition_p_local}
\end{equation}
Condition (\ref{eq:condition_p_local}) states that only shifted window functions that are nearest neighbors in momentum space overlap, i.e.
\begin{equation}
\tilde{g}_{mk}(p)\,\tilde{g}_{m'k'}(p) = 0 \; \text{ for } |k-k'| > 1. 
\end{equation}
Moreover, from condition (\ref{eq:condition_p_local}) one sees it is more convenient to use the momentum space representation in order to specify $g(j)$, since the number of parameters needed to fix $\tilde{g}(p)$ is $N/2M=L/2$, which is independent of the wave packet size $M$. In appendix \ref{sec:gcondition_band_limited} we show that for a band limited window function the orthogonality conditions (\ref{eq:gconditions}) become
\begin{equation}
\left|\tilde{g}\left(p\right)\right|^2 + \left|\tilde{g}\left(K - p \right)\right|^2 = \frac{2M}{N} \;\text{ for } 0 \leq p \leq K.
\label{eq:orthogonality_condition_p_loc}
\end{equation}
This implies that the values 
\begin{eqnarray}
\tilde{g}(0) &=& \sqrt{\frac{N}{2M}}, \nonumber\\
\tilde{g}\left(K/2\right) &=& \frac{1}{2}\sqrt{\frac{N}{M}}
\end{eqnarray}
are fixed. For the remaining momenta, any value $\tilde{g}(p) \leq \sqrt{2M/N}$ can be chosen for $0<p<K/2$, the remaining values are fixed by (\ref{eq:orthogonality_condition_p_loc}) and (\ref{eq:condition_p_local}), and $\tilde{g}(p)=\tilde{g}(-p)$.

Window functions that satisfy (\ref{eq:condition_p_local}) are listed in Tab.~\ref{tab:windowfunction} for the cases $N/M=2,4$. Note that for $N/M=2,4$, the window function is unique, whereas for $N/M > 4$ it is not. In the following, we use the window function for $N=4M$ for most calculations.

\begin{table}
\centering
\begin{tabular}{| c | c | c | c | c |}
\hline 
$N/M$ & $\tilde{g}\left(0\right)$ & $\tilde{g}\left(\frac{2\pi}{N}\right)$    \\
\hline
2      &   1                                  &          0                                \\
\hline
4      &   $ \frac{1}{\sqrt{2}}$         &        $ \frac{1}{2}  $      \\ 
\hline
\end{tabular}
\label{tab:windowfunction}
\caption{Analytical window functions in momentum space for small lattices with $N/M=2,4$. The value of $\tilde{g}(p)$ for all other momenta is either zero or related to the ones given by symmetry.}
\end{table}

\noindent
It is noteworthy that for $N/M=2$ the WW basis states are simply standing waves with wave vector $p=K$, i.e.
\begin{equation}
\psi_{mk}(j) = \cos\left[K j - \phi_{m+k}\right].
\end{equation}
As a consequence, in order to resolve the full momentum dependence of matrix elements, the interactions in at least one WW unit cell have to be known, and the purely local matrix elements are insufficient to do so (except for $k=0,M$, where there is only one state per unit cell).

\section{Wilson-Wannier basis for the square lattice}

\noindent
The Wilson-Wannier (WW) basis for a square lattice is the tensor product of two one-dimensional bases. Each basis state $\left|\mathbf{m}, \mathbf{k}\right\rangle$ is labelled by two two-dimensional vectors: The mean position $\mathbf{m}$ and the WW momentum $\mathbf{k}$. The transformation from real- or momentum-space to the WW basis is hence no different from the transformation in one dimension. Hence we have for the state $\left| \mathbf{j}\right\rangle$ at lattice site $\mathbf{j} = \left(j_1, j_2\right)$, that
\begin{eqnarray}
\left\langle \mathbf{j} \right|\left. \mathbf{m},\mathbf{k}\right\rangle &=& \psi_{m_1,k_1}\left(j_1\right)\,\psi_{m_2, k_2}\left(j_2\right),\nonumber \\
&\equiv& \Psi_{\mathbf{m},\mathbf{k}}\left(\mathbf{j}\right),
\label{eq:ww_wavefunction_realspace}
\end{eqnarray}
with $\psi_{mk}(j)$ defined above in (\ref{eq:defpsi}). Thus the position labels $\mathbf{m}$ are integer vectors that define a square lattice with lattice constant $M$. The momentum label $\mathbf{k}=\left(k_1,k_2\right)$ lies in the first quadrant of the Brillouin zone since $0\leq k_i \leq M$. In a similar manner, we can define the two-dimensional window function as a tensor product of one-dimensional ones,
\begin{equation}
g_{\mathbf{m},\mathbf{k}}\left(\mathbf{j}\right) = g_{m_1,k_1}\left(j_1\right) \, g_{m_2, k_2}\left(j_2\right).
\end{equation}
In order to facilitate computations, it is again useful to consider point group actions on the wave packets $g_{\mathbf{m},\mathbf{k}}\left(\mathbf{j}\right)$ in order to simplify the definition of basis functions. Since we use a product of one-dimensional basis functions, we do not base the discussion on the point group $C_{2v}$ of the square lattice but on its subgroup
\begin{equation}
G = \{\mathcal{E}, \mathcal{P}_x, \mathcal{P}_y, \mathcal{P}_x \mathcal{P}_y\},
\end{equation}
consisting of reflections around the $x$ and $y$ axes. The action of a group element $A$ on a wave vector $\mathbf{k}$ is given by a $2\times2$ matrix, parametrized by two numbers $\al=(\alpha_1, \alpha_2)$:
\begin{eqnarray}
A_\al \, g_{\mathbf{m},\mathbf{k}}\left(\mathbf{j}\right) &=& g_{\mathbf{m}, A_\al \mathbf{k}}\left(\mathbf{j}\right) \nonumber \\
A_\al &=& \left(\begin{array}{cc} \alpha_1 & 0 \\ 0 & \alpha_2\end{array}\right).
\end{eqnarray}
In the following we use the vector $\al = \left(\alpha_1, \alpha_2\right)$ in order to label elements of $G$, similar to what we have done above in one dimension. The correspondence between group elements and vectors is summarized in Tab. \ref{tab:A_alpha_correspondence_2d}.
\begin{table}
\centering
\begin{tabular}{| r | c | c | c | c |}
\hline
group element & $\mathcal{E}$ & $\mathcal{P}_x$ & $\mathcal{P}_y $ & $\mathcal{P}_x \mathcal{P}_y$  \\
\hline
$\al$ & $\left(1,1\right)$ & $\left(-1,1\right)$  & $\left(1,-1\right)$ &$\left(-1,-1\right)$ \\
\hline
\end{tabular}
\caption{Correspondence between point group elements and the integer vectors $\al$.}
\label{tab:A_alpha_correspondence_2d}
\end{table}

\begin{figure}
\centering
\includegraphics[width=8cm]{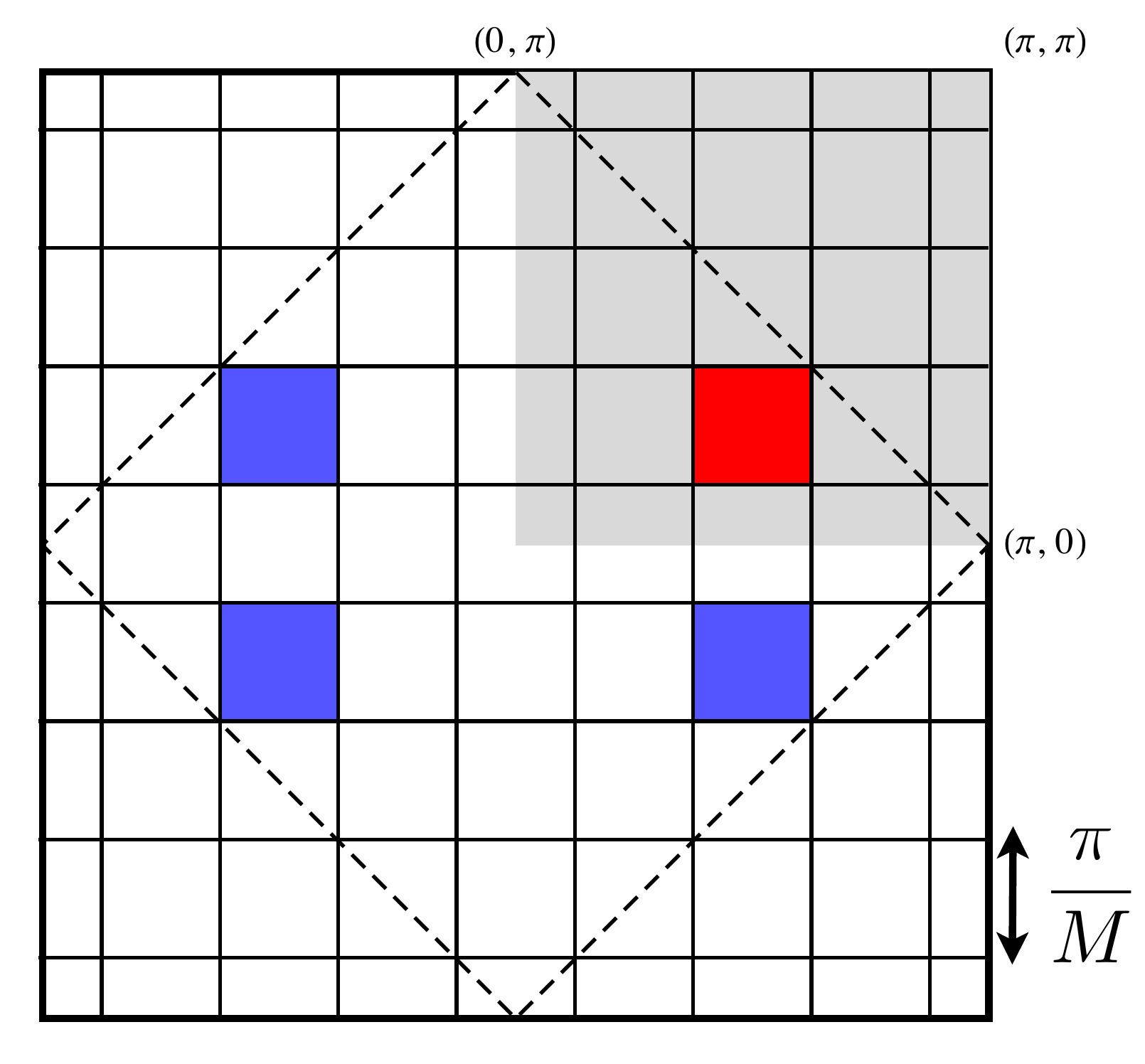}
\caption{Connection between wave packet momenta and WW momentum for the square lattice. Each WW basis state $\ket{\mathbf{m},\mathbf{k}}$ is a linear combination of wave packets with up to four different mean momenta $A_\al \mathbf{k}$, where $\al\in G$. All WW momenta lie in the first quadrant of the Brillouin zone (shaded area). The phase space cells for the are drawn as a grid. The three blue cells are obtained from the red cell by group transformations, and are thus represented by the same WW momentum $\mathbf{k}$.}
\label{fig:ww_basis_2d}
\end{figure}

\noindent
Now we can express the WW basis functions on the square lattice in a convenient way as
\begin{equation}
\Psi_{\mathbf{m}\mathbf{k}}\left(\mathbf{j}\right) = \frac{1}{\sqrt{\left| G\right| \,\left|H_{\mathbf{k}} \right|}}\sum_{\al \in G} e^{-i\al\cdot \mathbf{\Phi}_{\mathbf{m}+\mathbf{k}}} g_{\mathbf{m}, A_\al \mathbf{k}}(\mathbf{j}),
\label{eq:defpsi_2d}
\end{equation}
where $\left|G\right| = 4$ is the number of elements of $G$, and $\left| H_\mathbf{k}\right|$ is the number of elements of the stabilizer $H_{\mathbf{k}}$ of $\mathbf{k}$. $\left| H_{\mathbf{k}}\right|=1$ when $\pi/M \,\mathbf{k}$ lies in the interior of the first quadrant of the BZ, $\left| H_{\mathbf{k}}\right|=2$ when it lies on the boundary, and $\left| H_{\mathbf{k}}=4\right|$ when it lies on a corner. Thus in general each basis state is a linear combination of wave packet states with up to four different momenta that lie on the $G$-orbit of the WW momentum $\mathbf{k}$. The connection between WW momentum $\mathbf{k}$ and wave packet momenta is shown in Fig.~\ref{fig:ww_basis_2d}

\noindent
Since the WW basis functions for a square lattice derive directly from the one-dimensional variant, the analytical window functions from Sec.~\ref{sec:analytical_window_functions} can be directly used in two dimensions as well. For later convenience, the variables used in the different representations - real space, momentum space, and WW basis - are listed in Tab. \ref{tab:variables_2d}.
\begin{table}
\begin{center}
\begin{tabular}{| c | c | p{5cm} |}
\hline 
variable & range & meaning \\
\hline
$\mathbf{j} = \left(j_1,j_2\right)$ & $j_i = 0,\ldots, N$ & position in real space \\ 
\hline 
$\mathbf{p}=\left(p_1,p_2\right)$ & $p_i=-\pi+\frac{2\pi}{N}, -\pi+2\frac{2\pi}{N},\ldots, \pi-\frac{2\pi}{N}, \pi$ & momentum \\
\hline
$\mathbf{m} =\left(m_1,m_2\right)$ & $m_i=0,\ldots, L-1$ & WW position label for state with center position $\bar{\mathbf{j}}=M\,\mathbf{m}$ \\
\hline
$\mathbf{k}=\left(k_1,k_2\right)$ & $k_i = 0,\ldots, M$ & WW momentum label for state with center momenta $\bar{\mathbf{p}}=K A_\al\mathbf{k}$, where $\al \in G$\\
\hline
$\al=\left(\alpha_1,\alpha_2\right)$ & $\alpha_i =\pm 1$ & Representation of the point group $G$ \\
\hline
\end{tabular}
\end{center}
\caption{Variables for the description of the spatial degrees of freedom in two dimensions.}
\label{tab:variables_2d}
\end{table}

\chapter{Wilson-Wannier representation of operators}
\label{ch:wwtrafo}

\noindent
This chapter discusses the transformation of operators to the WW basis. We consider the one-dimensional case first, beginning with the general transformation formula for many-body operators in Sec.~\ref{sec:wwtrafo_1}. In Sec.~\ref{sec:wave_packet_transformation} we introduce a useful splitting of the general formula into two steps: A transformation into an overcomplete wave packet basis, and a second step to orthogonalize these basis states, similar to the construction in Ch.~\ref{ch:wwbasis}. We focus in particular on the transformation of local and almost local operators, leading to an expansion in $1/M$ for those operators that is analogous to gradient expansions in field theory. The subsequent sections apply these results to the most relevant cases, namely hopping and interaction operators. Finally we generalize all results to two dimensions in Sec.~\ref{sec:ww_trafo_2d}.

\section{General transformation formula in one dimension}
\label{sec:wwtrafo_1}

\subsubsection{Transformation of annihilation and creation operators}

\noindent
We consider transformation properties of the fermion creators and annihilators first. We denote the state with no fermions by $\ket{0}$, and states with one fermion in state $j$ by $\ket{j}$ etc. For sake of clarity, we omit spin indices in this section. The fermion annihilator (creator) for a fermion in state $\ket{m,k}$ is denoted by $\gamma_{m,k}$ ($\gamma^\dagger_{m,k}$). Using the resolution of the identity $\sum_j \ket{j}\bra{j}$, we the find
\begin{eqnarray}
\gamma^\dagger_{m,k}\ket{0} &=& \ket{m,k} \nonumber \\
&=& \sum_j \ket{j}\, \braket{j}{m,k} \nonumber \\
&=& \sum_j \psi_{m,k}(j) \, c^\dagger_j \ket{0}.
\label{eq:derivation_gamma_c_j}
\end{eqnarray}
Taking the Hermitian conjugate of (\ref{eq:derivation_gamma_c_j}) yields a similar relation for $\bra{0} \gamma_{m,k}$. Hence the transformation from real space to WW basis takes the form
\begin{eqnarray}
\gamma^\dagger_{m,k} &=& \sum_j \psi_{m,k}(j) c^\dagger_j,\nonumber \\
\gamma_{m,k} &=& \sum_j \psi_{m,k}(j) \, c_j.
\label{eq:c_j_to_gamma}
\end{eqnarray}

\noindent
Using the resolution of the identity $\sum_p \,\ket{p}\bra{p}$ instead of $\sum_j \ket{j}\bra{j}$, the analogous transformation from momentum space to WW basis is obtained:
\begin{eqnarray}
\gamma^\dagger_{m,k} &=& \sum_p \tilde{\psi}_{m,k}(p) \, c^\dagger_p \nonumber \\
\gamma^\pdag_{m,k} &=& \sum_p \tilde{\psi}_{m,k}(p)^\ast \, c^\pdag_p.
\label{eq:c_p_to_gamma}
\end{eqnarray}

\subsubsection{Transformation of arbitrary operators}

\noindent
Eqns.~(\ref{eq:c_j_to_gamma}, \ref{eq:c_p_to_gamma}) can be used to transform all many-body operators from real (momentum) space to the WW basis. This is done by applying the transformation rule for the fermion operators to each operator separately. Consider a general operator $\hat{O}$. It can be expanded in any of the three bases. The transformation rule for the expansion coefficients follows from the fact that the operator is independent of the particular representation chosen. We assume that the real space expansion is given by
\begin{equation}
\hat{O} = \sum_{j_1\cdots j_{2n}} O\left(j_1,\ldots, j_{2n}\right) c^\dagger_{j_1}\cdots c^\dagger_{j_n} \, c^\pdag_{j_{n+1}}\cdots c^\pdag_{j_{2n}}.
\label{eq:O_real_representation}
\end{equation}
The same operator in the WW representation can be written as
\begin{equation}
\hat{O} = \sum_{m_1k_1\cdots m_{2n}k_{2n}} O\left(m_1k_1,\ldots,m_{2n}k_{2n}\right)\;\gamma^\dagger_{m_1k_1}\cdots \gamma^\dagger_{m_nk_n}\,\gamma^\pdag_{m_{n+1}k_{n+1}}\cdots \gamma^\pdag_{m_{2n}k_{2n}}.
\label{eq:O_WW_representation}
\end{equation}
Using (\ref{eq:c_j_to_gamma}), the relation between the two expansions is given by
\begin{eqnarray}
O\left(m_1k_1,\ldots,m_{2n}k_{2n}\right) &=& \sum_{j_1\cdots j_{2n}} O\left(j_1,\ldots,j_{2n}\right)\,\left[\prod_{l=1}^{2n}\, \psi_{m_l,k_l}\left(j_l\right)\right].
\label{eq:O_j_to_WW}
\end{eqnarray}

\noindent
In a similar manner, the momentum space representation can be written as
\begin{equation}
\hat{O} = \frac{1}{N^{n-1}}\sum_{p_1\cdots p_{2n}} O\left(p_1,\ldots,p_{2n}\right) c^\dagger_{p_1}\cdots c^\dagger_{p_n} \, c^\pdag_{p_{n+1}}\cdots c^\pdag_{p_{2n}}.
\end{equation}
It is related to the real space representation (\ref{eq:O_real_representation}) by Fourier transformation in each index.
From the WW representation (\ref{eq:O_WW_representation}) and the transformation rule (\ref{eq:c_p_to_gamma}) it follows that 
\begin{equation}
\begin{split}
O\left(m_1k_1,\ldots,m_{2n}k_{2n}\right) =&\frac{1}{N^{n-1}} \sum_{p_1\cdots p_{2n}} O\left(p_1,\ldots,p_{2n}\right)\\ &\;\;\;\;\times \;\;\left[\,\prod_{l=1}^{n}\, \tilde{\psi}_{m_l,k_l}\left(p_l\right)\right]\left[\,\prod_{l=n+1}^{2n}\, \tilde{\psi}_{m_l,k_l}\left(p_l\right)^\ast\right].
\end{split}
\label{eq:O_p_to_WW}
\end{equation}

\noindent
Eqns. (\ref{eq:O_j_to_WW}, \ref{eq:O_p_to_WW}) can be used directly in order to obtain the WW representation of any operator. On the other hand, it is not very intuitive due to the  rather complicated definition of the $\tilde{\psi}_{mk}(p)$. In the next section, we bring (\ref{eq:O_p_to_WW}) into a more easily understandable (if less compact) form. 

\section{Wilson-Wannier basis and wave packet transformation}
\label{sec:wave_packet_transformation}

\noindent
The goal of this section is to split the transformation (\ref{eq:O_p_to_WW}) into two steps, where the first step involves sums over products of window functions. This step yields matrix elements between different wave packet states with wave function $g_{mk}(j)$, and hence we refer to it as \emph{wave packet transformation}. The second step involves the sum over the point group $G$ that takes the overcomplete wave packet states to the WW basis states. Simplifications occur because the wave packet states are shift invariant under general phase space shifts, so that only a few matrix elements have to be evaluated. Moreover, the wave packet matrix elements are easier to understand, since each wave packet is localized around one point in phase space (instead of two points for the WW basis states). In the following we focus on the wave packet part of the WW transformation, since this part contains most of the physical information. The effect of the symmetrization is discussed in Secs. \ref{sec:ww_trafo_hopping_1}-\ref{sec:ww_trafo_interaction_1} below.

\noindent
We restrict ourselves to the transformation from momentum space to the WW basis, since this transformation will be used most of the time in the remainder of this work. The corresponding formulas in real space are completely analogous.

\noindent
First, we use the definition (\ref{eq:defpsi}) of $\psi_{mk}(j)$ (and hence $\tilde{\psi}_{mk}(p)$) in terms of the wave packets, and split each sum over $p$ in two parts as follows:
\begin{eqnarray}
\sum_p \Big( \cdots \Big) \, \tilde{\psi}_{mk}(p) &=& \underbrace{\frac{1}{\sqrt{\left|G\right| \left| H_k\right|}} \sum_{\alpha\in G} e^{-i\alpha \phi_{m+k}}}_{\text{orthogonalization}} \,\times\,\underbrace{ \left[\sum_p \Big(\cdots\Big) \tilde{g}_{m,\alpha k}(p)\right] }_{\text{wave packet transformation}}
\label{eq:ww_trafo_split_p}
\end{eqnarray}
Now we exploit the shift invariance of the window function in order to replace all $\tilde{g}_{mk}(p)$ by $\tilde{g}(p)$, using Eq.~(\ref{eq:g_sum_simplification}) that is derived in App.~\ref{sec:window_function_gymnastics}. It states that for an arbitrary function $f(p)$ the identity
\begin{eqnarray}
\sum_p f(p)\tilde{g}_{mk}(p) = \sum_p f\left(p+Kk\right) e^{-iMmp} \tilde{g}\left(p\right)
\label{eq:g_sum_simplifcations_chap}
\end{eqnarray}
holds.
\noindent
 Applying (\ref{eq:ww_trafo_split_p}) and (\ref{eq:g_sum_simplifcations_chap}) to (\ref{eq:O_p_to_WW}), we find
\begin{eqnarray}
O\left(m_1k_1,\ldots,m_{2n}k_{2n}\right) &=& \frac{1}{\left|G\right|^n}\frac{1}{\sqrt{\prod_{l=1}^{2n} \left|H_{k_l}\right|}}\\ &&\times\sum_{\alpha_1\cdots \alpha_{2n}} \exp\left\{-i\sum_{l=1}^{n} \left(\alpha_l \,\phi_{m_l+k_l}-\alpha_{n+l} \phi_{m_{n+l}+k_{n+l}}\right) \right\} \nonumber\\
&& \times \bar{O}\left(m_1\alpha_1k_1,\ldots,m_{2n}\alpha_{2n}k_{2n}\right),
\label{eq:O_p_to_ww_final}
\end{eqnarray}
where the wave packet transform $\bar{O}\left(m_1\alpha_1k_1,\ldots,m_{2n}\alpha_{2n}k_{2n}\right)$ is given by
\begin{equation}
\begin{split}
\bar{O}\left(m_1\alpha_1k_1,\ldots,m_{2n}\alpha_{2n}k_{2n}\right)=&\frac{1}{N^{n-1}} \sum_{p_1\ldots p_{2n}} O\left(p_1+\alpha_1Kk_1,\ldots,p_{2n}+\alpha_{2n}Kk_{2n}\right) \\
& \times \exp\left\{-i M \sum_{l=1}^n\left( m_l p_l-m_{n+l}p_{n+l}\right)\right\} \left[\prod_{l=1}^{2n} \tilde{g}\left(p_l\right)\right].
\end{split}
\label{eq:O_wave_packet_transform}
\end{equation}

\subsubsection{Wave packet transformation for local operators}

\noindent
Eqns.~(\ref{eq:O_p_to_ww_final}-\ref{eq:O_wave_packet_transform}) appear to be rather formidable, and it is indeed tedious to discuss their properties in full generality. Hence we will consider only a special case here, which is nevertheless instructive. To be specific, we consider the wave packet transform of a local operator,
\begin{equation}
O\left(p_1,\ldots,p_{2n}\right) = O_{\text{loc}} \, \delta\left(\sum_{i=1}^n \left[p_i - p_{i+n}\right]\right),
\label{eq:O_local_p}
\end{equation}
where $\delta\left(\cdots\right)$ is the delta function modulo $2\pi$ that enforces momentum conservation. We also define the wave packet momentum mismatch
\begin{equation}
Q = K\sum_{i=1}^{2n} \left[ \alpha_i k_{i} - \alpha_{i+n} k_{i+n}\right],
\end{equation}
which specifies the amount of violation of conservation of $k$, where $Q=0$ when $k$ is conserved modulo $2M$. Inspection of (\ref{eq:O_wave_packet_transform}) shows that the delta function in (\ref{eq:O_local_p}) now enforces the condition
\begin{equation}
\sum_{i=1}^n \left[p_i - p_{i+n}\right] = -Q \mod 2\pi.
\end{equation}
Since this is the only dependence on the variables $\alpha_i$ and $k_i$, these variables influence the matrix element only through $Q$. Thus we may set $k_1 = Q/K$, $k_i = 0$ for $i>1$, and $\alpha_i=1$ without loss of generality. In order to keep the notation readable, we use $p_1 \equiv -Q - p_{n+1} + \sum_{i=2}^{n} \left[p_{i} - p_{n+i}\right]$ as a short hand. Then (\ref{eq:O_wave_packet_transform}) becomes
\begin{eqnarray}
\bar{O}\left(m_1,1,Q; \ldots; m_{2n},1,0\right) &=& O_{\text{loc}} \frac{1}{N^{n-1}}\sum_{p_2\cdots p_{2n}} \prod_{i=1}^{2n} \tilde{g}\left(p_i\right)\\
&& \times\, \exp\{-iM\sum_{i=1}^{n} \left[ m_i p_i - m_{n+i}p_{n+i}\right]\}\nonumber,
\label{eq:O_local_wave_packet}
\end{eqnarray}
which depends only on $\tilde{g}$, on $Q$, and on the $m_i$. By virtue of translational invariance of the wave packet states, the expression is shift invariant in real space with period $2M$ when $Q\neq 0$, and with period $M$ when $Q=0$.

\noindent
It is important to realize that the assumption that $\hat{O}$ is local is not as restrictive as it might seem. Since the presence of the window functions restrict all sums over $p$ to $-K<p<K$, in fact the local case is a valid approximation whenever
\begin{equation}
O\left(p_1 + K \alpha_1k_1,\ldots,p_{2m}+ K\alpha_{2m}k_{2m}\right) \approx C \, \delta\left(\sum_{i=1}^{2n} \left[ p_i + K\alpha_i k_i - p_{n+i} - K \alpha_{n+i} k_{n+i}\right]\right)  
\end{equation}
for some constant $C$ and $-K<p_i<K$. This translates into the statement that the spatial range of $\hat{O}$ should be small compared to $M$. This short range spatial dependence is then contained in the dependence on $\alpha_i k_i$ in the wave packet transform, and $k_i$ in the WW basis. As a consequence, only few matrix elements need to be evaluated, and the evaluation is very fast because (\ref{eq:O_local_wave_packet}) can be computed once and tabulated for later use. 

\noindent
When greater precision is necessary (for instance for longer ranged interactions), or when matrix elements vanish in the local approximation (which we will see to be the case for the hopping operator), it is straightforward to improve the approximation by Taylor expanding $O\left(p_1,\ldots,p_{2m}\right)$ around the maximum of the product of wave packets. Since the window functions restrict the summation over the momenta to $|p| \lesssim \pi/M$, the Taylor expansion leads effectively to an expansion in $1/M$. This $1/M$-expansion leads to expansion coefficients that are universal in the sense that they depend on the window function and a few parameters only. We will pursue this procedure for the hopping matrix elements below in Sec.~\ref{sec:ww_trafo_hopping_1}, but stick to the local approximation for interaction terms. We discuss the relation of the wave packet transform of an operator to the notion of scaling dimension of operators that is used in the renormalization group below in Ch.~\ref{ch:wave_packets_rg}.

\section{Transformation of hopping and interaction operators}

\subsection{Hopping}
\label{sec:ww_trafo_hopping_1}

\noindent
In this section we discuss the WW representation of the kinetic energy operator
\begin{equation}
\ham{kin} = \sum_p \epsilon(p) c^\dagger_p\, c^\pdag_p,
\end{equation}
where we assume $\epsilon_p = -2t\cos p$ in the following, but the treatment below applies to any dispersion. $t$ is the hopping rate for nearest neighbor hopping, the bandwidth is $W=4t$. We denote the WW transform of $\epsilon(p)$ by $T(mk,m'k')$. From the transformation formula (\ref{eq:O_p_to_WW}) we obtain
\begin{equation}
T\left(mk,m'k'\right) = \sum_p \epsilon(p) \tilde{\psi}_{mk}(p) \, \tilde{\psi}_{m'k'}\left(p'\right)^\ast.
\end{equation}
Since $\tilde{g}(p)$ is band limited (cf. Sec. \ref{sec:analytical_window_functions}), $T\left(mk,m'k'\right) = 0$ if $\left|k-k'\right| > 1$ because of vanishing overlap of the WW basis functions in momentum space. In the following we focus on the case $k'=k$, which, again by virtue of momentum space localization, is the dominant term in the expansion. We express $\tilde{\psi}_{mk}(p)$ in terms of $\tilde{g}(p)$ using (\ref{eq:O_p_to_ww_final}) as discussed above:
\begin{eqnarray}
T\left(mk,m'k\right) &=& \frac{1}{\left|G\right| \left| H_k\right|} \sum_{\alpha,\alpha'} e^{-i\left(\alpha \phi_{m+k} -\alpha'\phi_{m'+k}\right)} \; \delta_{\alpha k,\alpha' k}\\
&&\;\times\; \sum_p \epsilon\left(p+\alpha K\right) e^{-iMp(m-m')} \, \left|\tilde{g}\left(p\right)\right|^2 \nonumber \\
&=& \frac{1}{2} \sum_{\alpha} e^{-i\alpha \left(\phi_{m+k}-\phi_{m'+k}\right)} \;\sum_p \epsilon\left(p+\alpha K\right) e^{-iMp(m-m')} \, \left|\tilde{g}\left(p\right)\right|^2,\nonumber
\end{eqnarray}
where we have used $\left|G\right|=2$. Note that the factor $1/H_k$ cancels against one of the sums over $\alpha$ for the case $k=0$, where the Kronecker delta does not restrict the summation over $\alpha$.

\subsubsection{Wave packet transformation and $1/M$-expansion around the local limit}

\noindent
We note that the factor $\left|\tilde{g}\left(p\right)\right|^2$ restricts the sum over $p$ to $-\pi/M < p < \pi/M$, so that it makes sense to expand $\epsilon\left(p + \alpha q\right)$ around $p=0$:
\begin{equation}
\epsilon\left(p + \alpha K\right) \approx \underbrace{\epsilon\left(Kk\right)}_{\text{mean kinetic energy}} + \underbrace{\alpha \epsilon'\left(Kk\right)}_{\text{group velocity}} p+  \underbrace{\epsilon''\left(Kk\right)}_{\text{dispersion}} \frac{1}{2}p^2 + \ldots,
\end{equation}
where $\epsilon'\left(Kk\right)=\partial_p \epsilon\left(Kk + p\right)\big|_{p=0}$ etc. 

\noindent
We treat the kinetic energy contribution term by term, starting with the first term. Since this term is a constant, the corresponding contribution in the WW basis is diagonal by virtue of the orthogonality of the basis states. Hence the contribution is $\sum_{mk} \epsilon\left(Kk\right) \gamma^\dagger_{mk}\,\gamma^\pdag_{mk}$ 

\noindent
For the remaining contributions, we focus on the leading terms only. These are the terms that are diagonal in $k$ and act over the shortest distance. We use the wave packet transformation first in order to estimate the magnitude of the different terms, and apply the orthogonalization afterwards. Moreover, we use the analytical window function for $N/M=4$ from Tab. \ref{tab:windowfunction} in order to obtain analytical estimates. For the group velocity term the leading contribution arises from nearest neighbor hopping with $m'-m = \pm 1$, with magnitude 
\begin{equation}
\epsilon'\left(K k\right) \sum_p p e^{\pm iMp} \,\tilde{g}^2(p) = \pm i \frac{\pi}{4M} \epsilon'\left(Kk\right),
\end{equation}
which is $O(1/M)$ as announced above. The term proportional to $p^2$ is dominated by hopping to second nearest neighbors, $m'-m=\pm2$, and its contribution is
\begin{equation}
\frac{1}{2}\epsilon''\left(K k\right) \sum_p p^2 e^{\pm i2Mp} \,\tilde{g}^2(p) = -\frac{\pi^2}{16M^2} \epsilon'' \, \left(Kk\right).
\end{equation}

\subsubsection{Orthogonalization}

\noindent 
In order to arrive at the WW representation of the hopping operator, its wave packet transform has to be orthogonalized. This eliminates some term that are finite in the wave packet transform. For example, consider the first term of the expansion of $\epsilon\left(Kk + p\right) \approx \epsilon\left(Kk\right)$. In the wave packet transform, matrix elements between wave packets that are adjacent in phase space are finite, since they involve the scalar product of two wave packets,
\begin{equation}
\epsilon\left(Kk\right) \sum_p \tilde{g}_{mk}\left(p\right) \, \tilde{g}_{m'k'}\left(p\right)^\ast \neq 0,
\end{equation}  
and the wave packet states are not orthogonal. The WW basis states, on the other hand, are orthogonal, so that their scalar product is $\braket{mk}{m'k'} = \delta_{mm'}\delta_{kk'}$. As a consequence, the constant term in the expansion of $\epsilon(p)$ leads to a diagonal contribution to the WW transform. 

\noindent
In a similar manner, other matrix elements that are finite in the wave packet transform vanish in the WW basis due to the symmetry of the basis states (recall that the orthogonalization amounts to linear combining wave packet states such that they fall into irreducible representations of the point group). 

\subsubsection{Hopping Hamiltonian in WW-representation}

\noindent
In summary, the hopping Hamiltonian in the WW basis is dominated by the $k$-diagonal part, however, in principle the other hopping matrix elements are not negligible. Contributions can be evaluated using an expansion around the local approximation, which yields an expansion of $\ham{kin}$ in powers of $1/M$. For the analytical window function for $N/M=4$, the analytical estimate of the kinetic energy Hamiltonian is
\begin{eqnarray}
\ham{kin} &\approx &\sum_k \sum_{m,m'} \gamma^\dagger_{mk}\,\gamma^\pdag_{m'k} \Big[ \epsilon\left(K k\right) \delta_{m,m'} + \left(-1\right)^{m} \frac{\pi}{4M} \epsilon'\left(Kk\right) \delta_{m+1,m'}\nonumber \\&&  \qquad\qquad+ \left(\frac{\pi}{4M}\right)^2 \epsilon''\left(Kk\right) \left(\delta_{m,m'}-\delta_{m+2,m'}\right)  \Big] \; + \; \text{ h.c.}
\label{eq:ham_kin_ww_1}
\end{eqnarray}
Since the goal of this work is to use the WW basis in order to obtain new approximation schemes to interacting fermion systems rather than to aim at the highest precision, we will confine ourselves to the approximate kinetic energy (\ref{eq:ham_kin_ww_1}) in the remainder of this work.

\subsection{Interaction}
\label{sec:ww_trafo_interaction_1}

\noindent
In this section we consider the WW transform of a short ranged (compared to $M$) interaction Hamiltonian 
\begin{equation}
\ham{int} = \frac{1}{2N} \sum_{p_1\cdots p_4} \delta\left(p_1+p_2-p_3-p_4\right) \, \tilde{U}\left(p_1,p_2;p_3,p_4\right) \, \hat{J}\left(p_1,p_3\right) \, \hat{J}\left(p_2,p_4\right),
\end{equation}
where $\hat{J}\left(p_1,p_2\right) = \sum_s c^\dagger_{p_1,s}\,c^\pdag_{p_2,s}$. $\tilde{U}\left(p_1,p_2;p_3,p_4\right)$ contains the spatial dependence of $\ham{int}$, for a purely local interaction with strength $U$ we have $\tilde{U}\left(p_1,p_2;p_3,p_4\right)= U$.

\subsubsection{Wave packet transformation}
\noindent
Since we assume that $\ham{int}$ is short ranged, we apply the local approximation (\ref{eq:O_local_wave_packet}) to the wave packet transformation. Moreover, we consider $k$-conserving matrix elements only, i.e. matrix elements for which $\alpha_1 k_1 + \alpha_2 k_2 - \alpha_3 k_3 - \alpha_4 k_4 = 0\mod 2M$. In order to obtain analytical estimates, we employ the analytical window function from \ref{tab:windowfunction} with $N/M=4$. 

\noindent
The first step is to make use of the local approximation by transforming (\ref{eq:O_local_p}) to real space. This leads to the wave packet transform
\begin{eqnarray}
\bar{U}\left(m_1\alpha_1k_1,\ldots,m_4\alpha_4k_4\right) &=& \frac{\tilde{U}\left(\alpha_1k_1,\ldots,\alpha_4k_4\right)}{2N} \sum_{p_1\cdots p_4} \delta\left(p_1+p_2-p_3-p_4\right)\nonumber \\&& \times \exp \Big\{ -iM\left(m_1p_1 + m_2p_2-m_3p_3-m_4p_4\right)\Big\} \prod_{i=1}^4\tilde{g}\left(p_i\right)\nonumber \\
&=& \frac{\tilde{U}\left(\alpha_1Kk_1,\ldots,\alpha_4Kk_4\right)}{2} \sum_j\left[ \prod_{i=1}^4 g\left(j-Mm_i\right)\right]
\label{eq:U_wavepacket_trafo_1}
\end{eqnarray}
It follows that the spatial dependence of the wave packet matrix elements is determined by
\begin{equation}
V\left(m_1,\cdots,m_4\right) \equiv \sum_j\left[ \prod_{i=1}^4 g\left(j-Mm_i\right)\right].
\label{eq:def_V_1}
\end{equation}

\noindent
The dominant matrix elements arise for $m_1=m_2=m_3\neq m_4$, where three operators reside on one site, and $m_1=m_2 \neq m_3=m_4$, where two operators are located at the same site. Note that only the relative positions matter due to the residual translational invariance of the wave packet states. Approximate analytical values of $V\left(m_1,\ldots,m_4\right)$ for these cases are tabulated in Tab. \ref{tab:V_values}. Interactions decay rapidly, with spatial separation, so that $V(0,0,0,2)/V(0,0,0,0) \approx 1/32$. Consequently, we will take only nearest neighbor interactions into account. The table also shows the corresponding value for the case that the window function for $N/M=2$ is used. Since these are very similar, we will use the latter in the following for analytical calculations, and the former for numerical ones.

\begin{table}
\centering
\begin{tabular}{| c |  c | c | c | c |}
\hline
$N/M$ & $m$ & $0$ & $1$ & $2$ \\
\hline
4 & $V\left(0,0,0,m\right)$ & $\frac{17}{32M}$ & $\frac{8}{32M}$ & $-\frac{1}{64M}$ \\
\hline 
4 & $V\left(0,0,m,m\right)$ & $\frac{17}{32M}$ & $\frac{7}{32M}$ & $\frac{1}{64M}$ \\
\hline
2 & $V\left(0,0,0,m\right)$ & $\frac{1}{2M}$ & $\frac{1}{4M}$ & $0$ \\
\hline 
2 & $V\left(0,0,m,m\right)$ & $\frac{1}{2M}$ & $\frac{1}{4M}$ & $0$ \\
\hline
\end{tabular}
\caption{Approximate analytical values of $V\left(m_1,\ldots,m_4\right)$ (see Eq. (\ref{eq:def_V_1})) for the dominant matrix elements in the wave packet transform of a local interaction. In the following, we us the matrix elements for $N/M=2$.}
\label{tab:V_values}
\end{table}

\subsubsection{Orthogonalization}

\noindent
In order to obtain the WW representation of the interaction, the wave packet states have to be orthogonalized (see Sec.~\ref{sec:wave_packet_transformation}). The orthogonalization leads to cancellations between some non-local terms. We focus on weakly coupled systems, so that the states at $k=0,M$ can be ignored, which implies $\left|H_k\right| = 1$ in the following. Plugging the wave packet transform (\ref{eq:U_wavepacket_trafo_1}) into the orthogonalization formula (\ref{eq:O_p_to_ww_final}), we obtain
\begin{eqnarray}
U\left(m_1k_1,\ldots,m_4k_4\right) &=& \frac{V\left(m_1,\ldots,m_4\right)}{2}\,\frac{1}{2} \sum_{\alpha_1\cdots\alpha_4} \, \tilde{U}\left(\alpha_1Kk_1,\ldots,\alpha_4Kk_4\right)\nonumber\\
&& \;\; \times \;e^{-i\left(\alpha_1\phi_{1} +\alpha_2\phi_{2} - \alpha_3\phi_{3} - \alpha_4 \phi_{4}\right)},
\label{eq:U_wp_to_ww_1}
\end{eqnarray}
where we have introduced the short hand $\phi_i = \phi_{m_i+k_i}$. 

\noindent
At low energies, the interaction Hamiltonian that couples states at the Fermi points can be reduced to a small set of coupling constants. This will be considered in Ch.~\ref{ch:ww1}.

\section{Two-dimensional square lattice}
\label{sec:ww_trafo_2d}

\noindent
Having discussed the transformation of operators to the WW basis for the one-dimensional case, we now turn to the two-dimensional square lattice. All the steps from the one-dimensional calculation can be essentially repeated in the same manner. Therefore the discussion in this section is somewhat shorter, highlighting the differences to the one-dimensional case. Only the transformation from momentum space to WW basis will be treated. All the WW basis related quantities can be found in Tab. \ref{tab:variables_2d}.

\subsection{General transformation formula, wave packet transform, and local approximation}
\subsubsection{Transformation of annihilation and creation operators}

\noindent
The transformation formula follows directly from the definition (\ref{eq:defpsi_2d}) of the basis functions $\Psi_{\mathbf{m}\mathbf{k}}(\mathbf{j}$ and their Fourier transform $\tilde{\Psi}_{\mathbf{m}\mathbf{k}}\left(\mathbf{p}\right)$, where now all position related coordinates are vectors, e.g. $\mathbf{j} =\left(j_1,j_2\right)$. Thus we have
\begin{eqnarray}
\gamma^\dagger_{\mathbf{m}\mathbf{k}} &=& \sum_{\mathbf{p}} \tilde{\Psi}_{\mathbf{m}\mathbf{k}}\left(\mathbf{p}\right) \, c^\dagger_\mathbf{p}\nonumber,\\
\gamma^\pdag_{\mathbf{m}\mathbf{k}} &=& \sum_{\mathbf{p}} \tilde{\Psi}_{\mathbf{m}\mathbf{k}}\left(\mathbf{p}\right)^\ast \, c^\pdag_\mathbf{p}.
\end{eqnarray}

\subsubsection{Transformation of arbitrary operators}

\noindent
In the same way as in one dimension, an arbitrary many-body operator $\hat{O}$ can be expanded in any single-particle basis. The transformation rules for the coefficients in the expansion follow from the fact that the operator is independent of the basis. For sake of readability, we continue to suppress spin indices in this section. We assume that $\hat{O}$ contains $n$ creators and $n$ annihilators. Hence its momentum representation can be written as
\begin{equation}
\hat{O} = \frac{1}{N^{2n-2}}\sum_{\mathbf{p}_1\cdots \mathbf{p}_{2n}} O\left(\mathbf{p}_1,\ldots,\mathbf{p}_{2n}\right)\,c^\dagger_{\mathbf{p}_1}\cdots \,c^\pdag_{\mathbf{p}_{2n}}.
\label{eq:O_p_representation_2d}
\end{equation}
The WW representation can be similarly written as
\begin{equation}
\hat{O} = \sum_{\mathbf{m}_1\mathbf{k}_1\cdots \mathbf{m}_{2n}\mathbf{k}_{2n}} O\left(\mathbf{m}_1\mathbf{k}_1,\cdots,\mathbf{m}_{2n}\mathbf{k}_{2n}\right) \, \gamma^\dagger_{\mathbf{m}_1\mathbf{k}_1}\cdots\,\gamma^\pdag_{\mathbf{m}_{2n}\mathbf{k}_{2n}}.
\label{eq:O_WW_representation_2d}
\end{equation}
Equating the right hand sides of (\ref{eq:O_p_representation_2d}) and (\ref{eq:O_WW_representation_2d}), we obtain the general transformation rule for the square lattice:
\begin{equation}
\begin{split}
O\left(\mathbf{m}_1\mathbf{k}_1,\ldots,\mathbf{m}_{2n}\mathbf{k}_{2n}\right) =& \frac{1}{N^{2n-2}}\sum_{\mathbf{p}_1\cdots\mathbf{p}_{2n}} \,O\left(\mathbf{p}_1,\ldots,\mathbf{p}_{2n}\right) \\& \;\;\;\times\;\; \left[\prod_{l=1}^n \tilde{\Psi}_{\mathbf{m}_l \mathbf{k}_l}\left(\mathbf{p}_l\right) \right]\left[\prod_{l=n+1}^{2n} \tilde{\Psi}_{\mathbf{m}_l \mathbf{k}_l}\left(\mathbf{p}_l\right)^\ast \right]
\end{split}
\label{eq:O_trafo_p_ww_2d}
\end{equation}

\subsubsection{WW basis and wave packet transformation}

\noindent
In order to arrive at a more manageable form of (\ref{eq:O_trafo_p_ww_2d}), we use the two-dimensional version of the wave packet transformation (cf. Sec. \ref{sec:wave_packet_transformation}) and write
\begin{eqnarray}
\sum_\mathbf{p} \Big( \cdots \Big) \, \tilde{\Psi}_{\mathbf{m}\mathbf{k}}(p) &=& \underbrace{\frac{1}{\sqrt{\left|G\right| \left| H_\mathbf{k}\right|}} \sum_{\al\in G} e^{-i\al\cdot\mathbf{\Phi}_{m+k}}}_{\text{orthogonalization}} \,\times\,\underbrace{ \left[\sum_\mathbf{p} \Big(\cdots\Big) \tilde{g}_{\mathbf{m},A_\al \mathbf{k}}(\mathbf{p})\right] }_{\text{wave packet transformation}}\nonumber\\
\label{eq:ww_trafo_split_p_2d}
\end{eqnarray}

\noindent
We only treat $\mathbf{k}$-conserving matrix elements in the following. In the local approximation, the wave packet transform of $O\left(\mathbf{p}_1,\ldots,\mathbf{p}_{2n}\right)$ is given by
\begin{equation}
\begin{split}
\bar{O}\left(\mathbf{m}_1\al_1\mathbf{k}_1,\ldots,\mathbf{m}_{2n}\al_{2n}\mathbf{k}_{2n}\right) \approx&\;\; O\left(K A_{\al_1}\mathbf{k}_1,\ldots,K A_{\al_{2n}} \mathbf{k}_{2n}\right) \\ &\times\;\;\sum_{\mathbf{j}} \left[\prod_{i=1}^{2n} g\left(\mathbf{j}-M\mathbf{m}_i\right) \right]
\end{split}
\end{equation}

\noindent
The WW representation of $\hat{O}$ is then obtained from the wave packet transform using the orthogonalization part of the WW transformation. This yields
\begin{equation}
\begin{split}
O\left(\mathbf{m}_1\mathbf{k}_1,\ldots,\mathbf{m}_{2n}\mathbf{k}_{2n}\right) =&\; \frac{1}{\left|G\right|^n \sqrt{\prod_{i=1}^{2n} \left|H_{\mathbf{k}_i}\right|}}\sum_{\al_1\cdots \al_{2n}} e^{-i \sum_{l=1}^n \left[ \al_l\cdot\mathbf{\Phi}_l - \al_{n+l}\cdot\mathbf{\Phi}_{n+l}\right] } \\
& \quad\times\quad \bar{O}\left(\mathbf{m}_1\al_1\mathbf{k}_1,\ldots,\mathbf{m}_{2n}\al_{2n}\mathbf{k}_{2n}\right)
\end{split}
\label{eq:O_p_to_ww_2d}
\end{equation}
where we have used the short hand $\mathbf{\Phi}_{i} \equiv \mathbf{\Phi}_{\mathbf{m}+\mathbf{k}}$. In analogy to the one-dimensional case, the local approximation can be improved by expanding$O\left(\mathbf{p}_1,\ldots,\mathbf{p}_{2n}\right)$ around $\mathbf{p}_i=KA_{\al_i}\mathbf{k}_i$, leading to an expansion in $1/M$.

\subsection{Hopping}

\noindent
Wo consider the hopping operator 
\begin{equation}
\ham{kin} = \sum_\mathbf{p} \,c^\dagger_{\mathbf{p}}\,c^\pdag_{\mathbf{p}}\, \underbrace{\left[-2t\left( \cos p_x + \cos p_y \right) + 4t' \cos p_x \cos p_y -\mu \right]}_{=:\,\epsilon\left(\mathbf{p}\right)}
\end{equation}
which includes nearest and next-to-nearest neighbor hopping. First note that when $t'=0$, the hopping Hamiltonian consists of two one-dimensional hopping terms, so that in this case the results from one dimension can be reused without any changes for each direction. In order to take the $t'$ term into account, we abbreviate $\mathbf{Q} = K A_\al \mathbf{k}$ and expand $\epsilon\left(\mathbf{Q}+\mathbf{p}\right)$ to leading order around $\mathbf{p} = 0$. To the first order we obtain
\begin{equation}
\epsilon\left(\mathbf{Q} + \mathbf{p}\right) \approx \epsilon\left(\mathbf{Q}\right) + \mathbf{v}_g\left(\mathbf{Q}\right)\cdot \mathbf{p} + \ldots,
\end{equation}
where the group velocity $\mathbf{v}_g\left(\mathbf{Q}\right)$ is given by
\begin{eqnarray}
\mathbf{v}_g\left(\mathbf{Q}\right) &=& \left(\begin{array}{c} 2 t \sin Q_{x} \left[1-\frac{2t'}{t} \cos Q_y \right]  \\ 2 t \sin Q_{y} \left[1-\frac{2t'}{t} \cos Q_{x}\right]\end{array}\right).
\end{eqnarray}
Thus we can still use the one-dimensional result without changes up to $O(1/M)$, and the $t'$ term merely leads to a $\mathbf{Q}$-dependent multiplicative correction to the nearest neighbor hopping term. At order $O(1/M^2)$, a mixed term $\propto p_x p_y$ appears, which leads to diagonal hopping in the WW basis. However, taking into account the leading order only, this term can be neglected unless the first order term vanishes. This is the case whenever $\mathbf{v}_g\left(\mathbf{Q}\right)=0$, i.e. at the band edges $\mathbf{Q}=\left(0,0\right)$ and $\mathbf{Q}=\left(\pi,\pi\right)$, and at the saddle points $\mathbf{Q}=\left(\pi,0\right)$ and $\mathbf{Q}=\left(0,\pi\right)$. In both cases, the group velocity vanishes because of the higher symmetry around these points which demands $\epsilon\left(\mathbf{Q}+\mathbf{p}\right) = \epsilon\left(\mathbf{Q}-\mathbf{p}\right)$. However, the same symmetry forbids the mixed term $p_xp_y$ in the expansion of the kinetic energy, so that the second order contribution is a sum of two one-dimensional terms as well in this case. 

\noindent
Summarizing the above, to the leading order in the $1/M$-expansion the hopping operator for the square lattice is the sum of two one-dimensional hopping operators with a $\mathbf{k}$ and direction dependent hopping rate
\begin{eqnarray}
t _x &\rightarrow& t\left(1 - \frac{2t'}{t} \cos K k_y\right)\nonumber\\
t_y &\rightarrow&t\left(1 - \frac{2t'}{t} \cos K k_x\right)
\end{eqnarray}

\subsection{Interaction}

\noindent
We consider the WW representation of the interaction operator
\begin{equation}
\ham{int} = \frac{1}{2N^2}\sum_{\mathbf{p}_1\cdots\mathbf{p}_4} \delta\left(\mathbf{p}_1+\mathbf{p}_2-\mathbf{p}_3-\mathbf{p}_4\right)\, \tilde{U}\left(\mathbf{p}_1,\ldots,\mathbf{p}_4\right) J\left(\mathbf{p}_1,\mathbf{p}_3\right)\, J\left(\mathbf{p}_2,\mathbf{p}_4\right),
\end{equation}
where $J\left(\mathbf{p},\mathbf{p}'\right) = \sum_s c^\dagger_{\mathbf{p},s}\,c^\pdag_{\mathbf{p}',s}$. Similarly, in the WW basis we write
\begin{equation}
\ham{int} = \frac{1}{2}\sum_{\mathbf{m}_1\mathbf{k}_1\cdots\mathbf{m}_4\mathbf{k}_4} U\left(\mathbf{m}_1\mathbf{k}_1,\ldots,\mathbf{m}_4\mathbf{k}_4\right) J\left(\mathbf{m}_1\mathbf{k}_1,\mathbf{m}_3\mathbf{k}_3\right)\, J\left(\mathbf{m}_2\mathbf{k}_2,\mathbf{m}_4\mathbf{k}_4\right),
\end{equation}
where $J\left(\mathbf{m}\mathbf{k},\mathbf{m}'\mathbf{k}'\right) = \sum_s \gamma^\dagger_{\mathbf{m}\mathbf{k},s}\,\gamma^\pdag_{\mathbf{m}'\mathbf{k}',s}$.

\noindent
We follow the argumentation of Sec. \ref{sec:ww_trafo_interaction_1} and focus on matrix elements that are $\mathbf{k}$-conserving, i.e. $A_{\al_1}\mathbf{k}_1+A_{\al_2}\mathbf{k}_2 = A_{\al_3}\mathbf{k}_3 + A_{\al_4}\mathbf{k}_4$ in the wave packet transform $\bar{U}\left(\mathbf{m}_1\al_1\mathbf{k}_1,\ldots,\mathbf{m}_4\al_4\mathbf{k}_4\right)$ of $\ham{int}$. Moreover, we use the local approximation for the interaction matrix elements in momentum space. The general transformation formula (\ref{eq:O_p_to_ww_2d}) then reduces to
\begin{equation}
\begin{split}
U\left(\mathbf{m}_1\mathbf{k}_1,\ldots,\mathbf{m}_{4}\mathbf{k}_{4}\right) =& \frac{1}{\left|G\right|^n \sqrt{\prod_{i=1}^{4} \left|H_{\mathbf{k}_i}\right|}}\sum_{\al_1\cdots \al_{4}}\bar{U}\left(\mathbf{m}_1\al_1\mathbf{k}_1,\ldots,\mathbf{m}_{4}\al_{4}\mathbf{k}_{4}\right) \\
& \qquad\qquad\qquad\;\;\;\;\;\;\times\;\; e^{-i \left(\al_l\cdot\mathbf{\Phi}_l+\al_2\cdot\mathbf{\Phi}_2 - \al_{3}\cdot\mathbf{\Phi}_{3} - \al_4\cdot\mathbf{\Phi}_4\right)}.
\end{split}
\end{equation}
Similarly to Sec. \ref{sec:ww_trafo_interaction_1}, the wave packet transform $\bar{U}\left(\mathbf{m}_1\al_1\mathbf{k}_1,\ldots,\mathbf{m}_{4}\al_{4}\mathbf{k}_{4}\right)$ is given by
\begin{equation}
\begin{split}
\bar{U}\left(\mathbf{m}_1\al_1\mathbf{k}_1,\ldots,\mathbf{m}_{4}\al_{4}\mathbf{k}_{4}\right) =&\; \frac{1}{2} \tilde{U}\left(KA_{\al_1}\mathbf{k}_1,\ldots,KA_{\al_4}\mathbf{k}_4\right)\\ & \;\times\;\delta_{A_{\al_1}\mathbf{k}_1+A_{\al_2}\mathbf{k}_2,A_{\al_3}\mathbf{k}_3+A_{\al_4}\mathbf{k}_4} \\&\;\times\;V\left(m_{1,x},\ldots,m_{4,x}\right) V\left(m_{2,y},\ldots,m_{4,y}\right) 
\end{split}
\end{equation}
with the same function $V\left(m_1,\cdots,m_4\right) = \sum_j \prod_{i=1}^4 g\left(j-Mm_i\right)$ as in the one-dimensional case. Thus the decay properties are the same as in one dimension, and we restrict ourselves to nearest neighbor interactions only, neglecting all but the leading matrix elements. It is noteworthy that individual interaction matrix elements $U\left(\mathbf{m}_1\mathbf{k}_1,\ldots,\mathbf{m}_{4}\mathbf{k}_{4}\right)$ are proportional to $1/M^2$ because $V(m_1,\ldots,m_4) \propto 1/M$ (cf. Tab. \ref{tab:V_values}), whereas they are $\propto 1/M$ in one dimension. The scaling of operators and its connection to their relevance in the RG sence is considered in Ch.~\ref{ch:wave_packets_rg}. We conclude this chapter with the remark that since the low-energy region of the Brillouin zone in two dimensions extends over the whole Fermi surface, there is no universal low-energy parametrization of interactions as it was the case in one dimension. Instead, the full momentum dependence of the interaction along the Fermi surface has to be taken into account in general. Whether approximations are admissible has to be decided based on the shape of the Fermi surface and the length scale $M$ under consideration. This is discussed in Ch.~\ref{ch:ww2}.

\chapter{Wave packets and fermion pairing}
\label{ch:ww_pairing}

\noindent
Symmetry breaking in fermion systems can often be understood as a transition from free to paired fermions. The best known example is superconductivity \cite{superconductivity}, where electrons bind into pairs which form the condensate that characterizes the superconducting state. However, spin and charge density waves may also be viewed as pairing of electrons and holes, so that a wide variety of states falls into the class of paired fermion states. Such a paired state introduces an energy scale $\Delta$, given by the fermion gap, and a length scale, the pair size $\xi$. In the weak coupling limit, we can estimate
\begin{equation}
\xi \approx \frac{2\pi v_F }{\Delta}
\end{equation}
on dimensional grounds. 

\noindent
In this section we discuss fermion pairing in the context of the WW basis. Since the WW basis states are localized on the length scale $M$, one expects that for $M >\xi$ pairs are (predominantly) local in the WW basis, whereas for $M <\xi$ they are non-local. On the other hand, the pair correlations in the BCS state decay as one moves away from the Fermi surface, and the corresponding width in momentum space is $2\pi/\xi \sim \Delta/ v_F$. Hence, we expect states within a distance less than $\Delta / v_F$ to the Fermi surface to be strongly correlated, whereas they should be weakly correlated when they are far away from the Fermi surface. 

\noindent
These estimates suggest that it is possible to replace fermionic degrees of freedom by pairs that are local (in real space) in the WW basis when $M/\xi$ is chosen large enough. In this way the low energy problem may be bosonized. They also suggest that only about $\xi/M$ states in the direction perpendicular to the Fermi surface are strongly correlated, whereas the remainder may be treated perturbatively. Thus pairs may be considered to be local in momentum space (in the direction perpendicular to the Fermi surface) when $M/\xi$ is made small enough. It is natural to expect that for $M\sim \xi$, pairs are reasonably localized in both momentum and real space, hence allowing for a simplified description of the low energy physics in terms of relatively few (by momentum space localization) bosonic degrees of freedom that are local in real space (by virtue of real space localization) in the WW basis.

\noindent
The remainder of this chapter elaborates on these heuristic considerations. To this end, we compute the WW transform of several mean-field Hamiltonians, discussing the relevance of different terms and their dependence on $\xi/M$. Moreover, we compute single-particle correlations of the respective ground states in the WW basis. Motivated by applications in later chapters, we focus on superconductivity and antiferromagnetism. In Sec.~\ref{sec:ww_mf_one_dimension} we treat one-dimensional systems, the extension to the square lattice follows in Sec. \ref{sec:ww_mf_two_dimensions}.

\section{One dimension}
\label{sec:ww_mf_one_dimension}

\subsection{Superconductivity}
\noindent
We consider the BCS mean-field Hamiltonian in one dimension, which is given by
\begin{equation}
\ham{BCS} = -t\sum_{j}\sum_s \left[ c^\dagger_{j,s} \, c^\pdag_{j+1,s} + c^\dagger_{j+1,s}\,c^\pdag_{j,s} - \mu c^\dagger_{j,s}\,c^\pdag_{j,s}\right] + \Delta \sum_j \left[c^\dagger_{j,\uparrow}\,c^\dagger_{j,\downarrow} + c^\pdag_{j,\downarrow} \, c^\pdag_{j,\uparrow} \right].
\label{eq:BCS_Hamiltonian}
\end{equation}
$\ham{BCS}$ can be easily diagonalized by applying a Bogoliubov transformation to the fermion operators \cite{superconductivity}, so that in particular its ground state and correlation functions can be computed exactly. We first transform $\ham{BCS}$ to the WW basis and investigate the relative importance of the hopping and the pairing term as a function of $\xi/M$. Afterwards we compute single-particle correlations of the exact ground state in the WW basis. Both results illustrate that unpaired fermions become less and less important as $\xi/M$ goes to zero.

\subsubsection{WW representation of the BCS mean-field Hamiltonian}

 We consider the WW representation of $\ham{BCS}$. The hopping term has already been discussed above in Sec. \ref{sec:ww_trafo_hopping_1}, so we will turn directly to the anomalous term $\Delta \sum_j \left(c^\dagger_{j,\uparrow}\,c^\dagger_{j,\downarrow} + \text{h.c.}\right)$. Its WW representation can be obtained using Eq. (\ref{eq:c_j_to_gamma}):
\begin{eqnarray}
\Delta \sum_j \left(c^\dagger_{j,\uparrow}\,c^\dagger_{j,\downarrow} + \text{h.c.}\right)&=& \Delta \sum_{mk,m'k'} \left(\gamma^\dagger_{mk,\uparrow}\,\gamma^\dagger_{m'k',\downarrow} + \text{h.c.}\right) \, \sum_j \psi_{mk}(j) \psi_{m'k'}(j) \nonumber \\
&=& \Delta\sum_{mk} \left(\gamma^\dagger_{mk,\uparrow}\,\gamma^\dagger_{mk,\downarrow} + \text{h.c.}\right).
\end{eqnarray}
The pairing mean-field term is local in the WW basis, hence it acts independently on each WW orbital. All non-vanishing matrix elements are $O(1)$. In conjunction with the WW transform of the hopping operator, Eq.~(\ref{eq:ham_kin_ww_1}), we obtain in the limit of large $M$ that
\begin{equation}
\ham{BCS} = \sum_{mk}\sum_{s} \epsilon\left(K k\right) \gamma^\dagger_{mk,s}\,\gamma^\pdag_{mk,s} + \Delta \sum_{mk} \left(\gamma^\dagger_{mk,\uparrow}\,\gamma^\dagger_{mk,\downarrow} + \text{h.c.}\right) + O\left(\frac{1}{M}\right)
\label{eq:ham_BCS_ww_loc}
\end{equation}
In order to estimate the effect of the neglected $O(1/M)$ hopping terms, we first obtain the single particle gap $E_k$ for each WW orbital from the local Hamiltonian. It is given by
\begin{equation}
E_k = \sqrt{\epsilon\left(K k\right)^2 + \Delta^2}.
\label{eq:E_k_BCS}
\end{equation}
Now we compare the single particle energy with the band width $4t_k$, where the hopping rate $t_k \sim \frac{\pi}{4} v_F/M $ is given by the $k$-diagonal nearest-neighbor hopping matrix element $T(m,k;\, m+1,k)$ (cf. Sec. \ref{sec:ww_trafo_hopping_1}). This yields the dimensionless ratio
\begin{eqnarray}
\frac{4t_k}{E_k} &\sim& \frac{\pi}{M} \frac{v_F}{\sqrt{\epsilon\left(Kk\right)^2+\Delta^2}} \nonumber \\
&\sim& \frac{\xi}{2M}\frac{1}{\sqrt{\frac{\epsilon\left(Kk\right)^2}{\Delta^2}+1}},
\end{eqnarray} 
where we have used $\xi \sim 2\pi v_F /\Delta$. It is clear that the importance of the hopping term decreases as one moves away from the Fermi points since $\epsilon\left(K k\right) \sim K v_F \left(k - p_F/K\right)$. Thus we consider the states at the Fermi points, $k \approx p_F/K$ to estimate the importance of the hopping term. When the gap $E_k$ exceeds the band width $4 t_k$, the system can be considered to be strongly coupled in the sense that the hopping term leads to corrections that can be treated perturbatively and decay over distances of about $M$. On the other hand, when $E_k < 2t_k$, the energy gain from delocalizing an electron is large enough to overcome the single particle gap locally. In this case perturbation theory around the local Hamiltonian is not expected to converge rapidly (if at all).

\noindent
In summary, the hopping term can be treated perturbatively in the mean-field Hamiltonian $\ham{BCS}$ when we choose $M$ such that
\begin{equation}
\frac{\xi}{M} \lesssim \frac{1}{2}.
\label{eq:strong_coupling_condition}
\end{equation}

\subsubsection{One particle correlations in the BCS mean-field state}

\noindent
In the following, we will be interested in one-body equal time correlation functions only. Since the calculation is elementary (see e.g. \cite{superconductivity}), we state the results directly. There are two different correlation functions, the normal part $\tilde{G}_p := \left\langle c^\dagger_{p,s}\,c_{p,s}\right\rangle$, which yields the fermion distribution function, and the anomalous part $\tilde{F}_p := \left\langle c_{p,s}\,c_{-p,-s}\right\rangle$, that contains pair correlations. They are given by
\begin{eqnarray}
\tilde{G}_p &=& \frac{E_p - \epsilon_p}{2 E_p} \\
\tilde{F}_p &=& \frac{\Delta}{E_p},
\end{eqnarray}
where $\epsilon_p = -2t\cos p-\mu$ is the kinetic energy, and $E_p = \sqrt{\epsilon_p^2+\Delta^2}$ is the energy of a Bogoliubov quasi-particle. The corresponding correlation function in the WW representation is obtained using Eq. (\ref{eq:c_p_to_gamma}). For the normal part $G_{mk,m'k'}$ this yields
\begin{eqnarray}
G_{mk,m'k'} &=& \left\langle \gamma^\dagger_{mks}\,\gamma^\pdag_{m'k's}\right\rangle \nonumber \\
&=& \sum_p \tilde{\psi}_{mk}(p)\,\tilde{\psi}_{m'k'}(p)^\ast \, \left \langle c^\dagger_{p,s} \,c^\pdag_{p,s}\right\rangle\nonumber \\
&=& \sum_p \tilde{\psi}_{mk}(p)\,\tilde{\psi}_{m'k'}(p)^\ast \, \tilde{G}_p.
\end{eqnarray}
Similarly, for the anomalous part we obtain
\begin{eqnarray}
F_{mk,m'k'} &=& \left\langle \gamma^\pdag_{mk,s}\,\gamma^\pdag_{m'k',-s}\right\rangle\nonumber \\
&=& \sum_p \tilde{\psi}_{mk}(p)^\ast \, \tilde{\psi}_{m'k'}(-p)^\ast \, F_p.
\end{eqnarray}
\begin{figure}
\centering
\includegraphics[width=7cm]{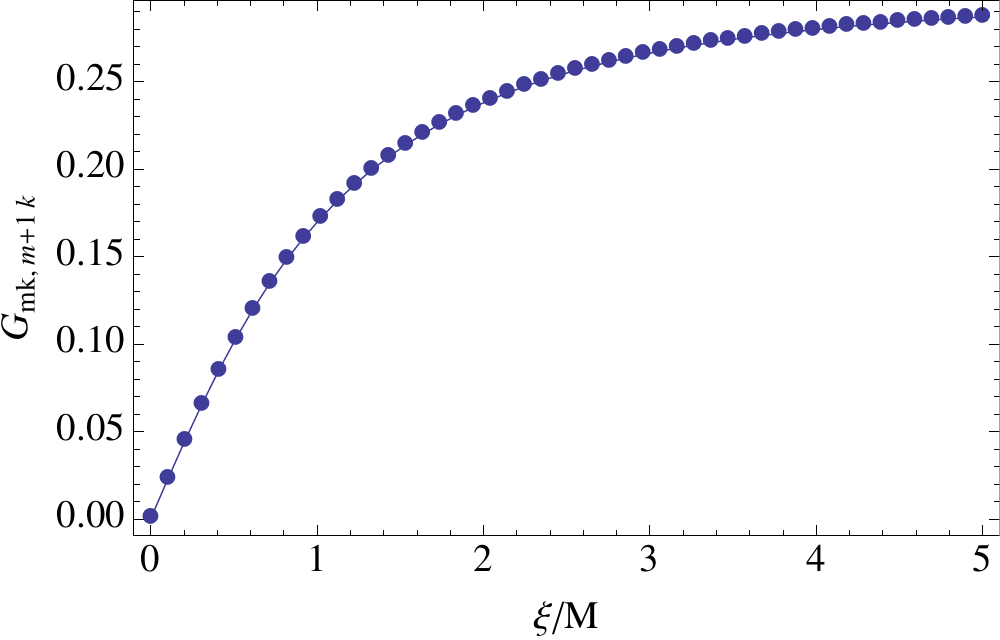}
\includegraphics[width=7cm]{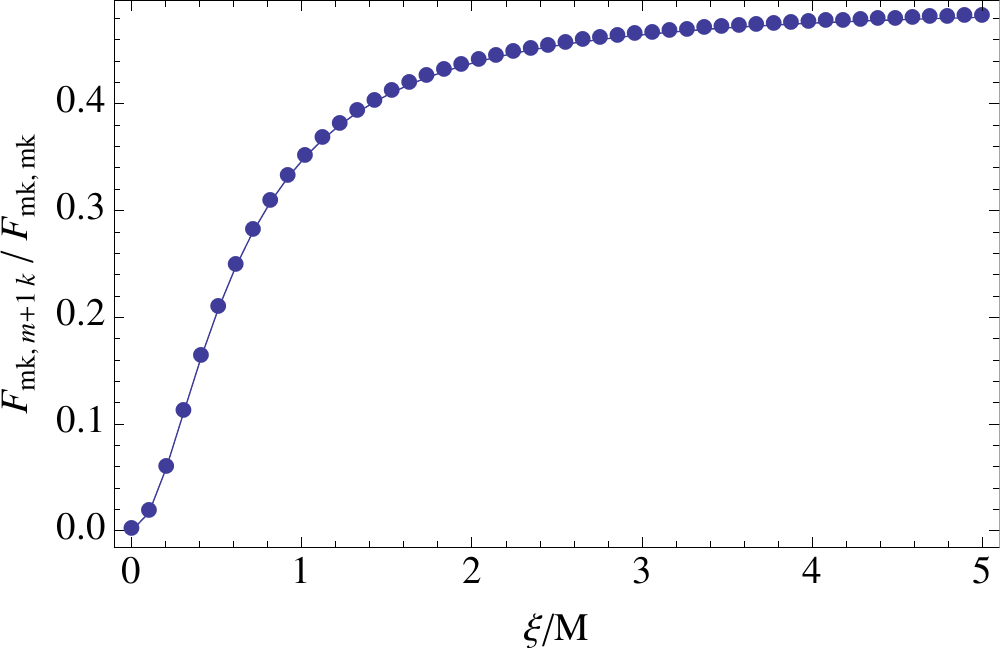}
\caption{Dependence of nearest neighbor correlations on $\xi/M$. The left panel shows the normal correlator, which measures the strength of hopping of unpaired fermions. The right panel displays the ratio of nearest-neighbor pair correlations to onsite pair correlation. Both non-local correlations are seen to decrease with decreasing $\xi/M$.}
\label{fig:sc1_nearest_neighbor}
\end{figure}

\noindent
In the following, we focus on two key quantities. First, we evaluate $G_{mk,m+1,k}$, which non-local fermion correlations, and $F_{mk,m+1,k}/F_{mk,mk}$, which measures the decay of pair correlations with distance from the Fermi surface. Since correlations are strongest at the Fermi surface, we set $k=k_F=p_F/K$. A plot of these ratios as a function of $\xi/M$ is shown in Fig.~\ref{fig:sc1_nearest_neighbor}. It can be seen that the non-local correlators vanish when the pair size $\xi$ is much smaller than $M$. 

\begin{figure}
\centering
\includegraphics[width=7cm]{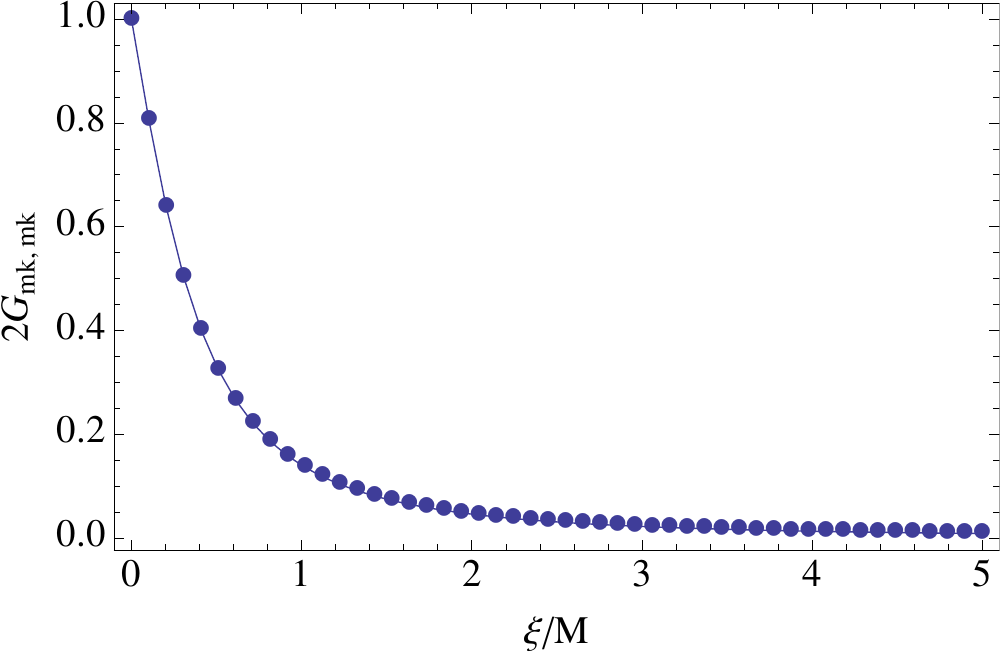}
\includegraphics[width=7cm]{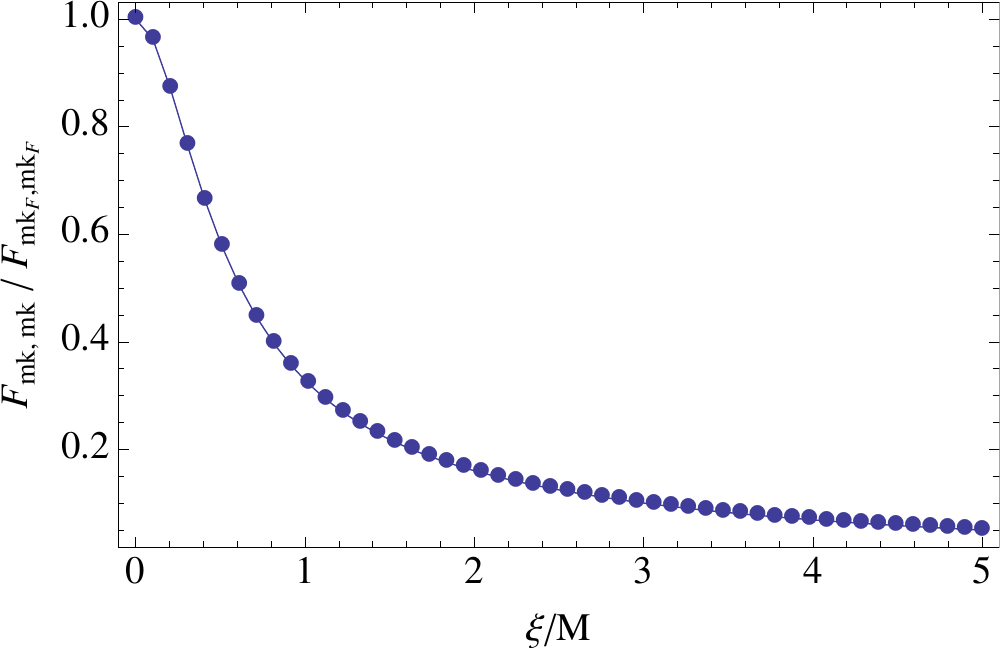}
\caption{Decay of correlation with separation from the Fermi surface as a function of $\xi/M$. Left panel: Decay of the occupation number at $k=k_F+1$. Note that the plot shows two times the occupation number. Right panel: Ratio of local pair correlations at $k=k_F+1$ and $k=k_F$. Both quantities decrease with increasing $\xi/M$.}
\label{fig:sc1_k_decay}
\end{figure}

\noindent
Next we consider the decay of correlations with distance from the Fermi surface. To this end, we evaluate the ratios $2G_{mk,mk}$ for $k = 1+k_F$. This is a measure of the occupation of WW orbitals next to the Fermi surface. Analogously, we evaluate the anomalous correlator $F_{mk,mk}/F_{m\,k_F,m\,k_F}$ at $k=1+k_F$, which measures the decay of pair correlations as one moves away from the Fermi surface. This is shown in Fig.~\ref{fig:sc1_k_decay}.

\begin{figure}
\centering
\includegraphics[width=8cm]{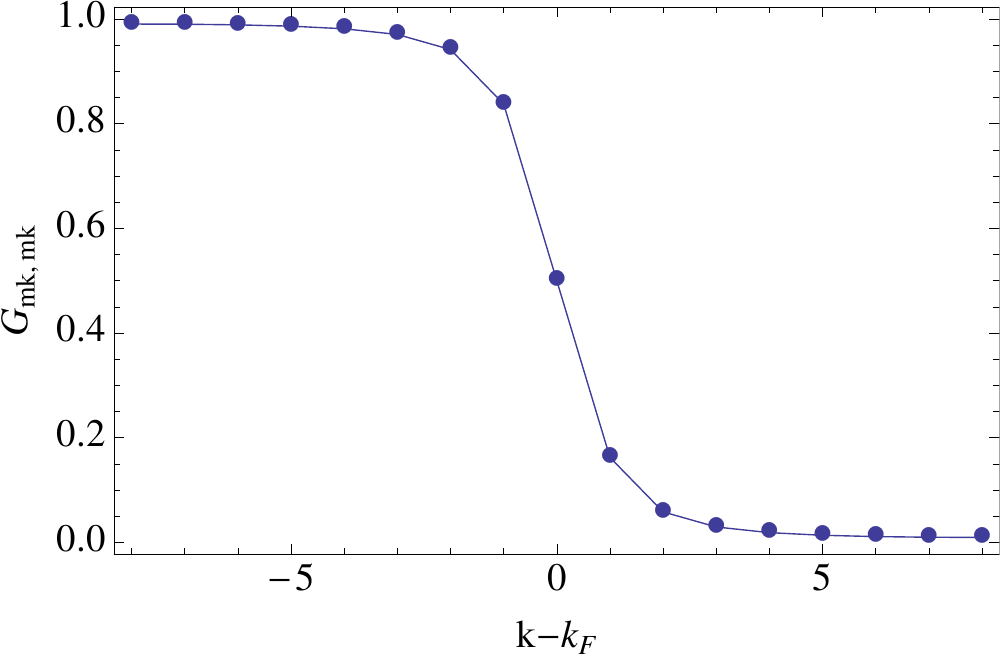}
\caption{WW orbital occupation number as a function of $k-k_F$ for $\xi/M=1/2$. All orbitals except the ones at the Fermi surface and their nearest neighbors deviate only weakly from their non-interacting occupation number.}
\label{fig:sc1_G_loc}
\end{figure}

\noindent
We are interested in the case that the states at the Fermi points are strongly coupled, i.e. $\xi/M \approx 1/2$ (cf. Eq.~\ref{eq:strong_coupling_condition}), the non-local correlations are strongly suppressed. At the same time, from Fig.~\ref{fig:sc1_k_decay} we see that the WW orbitals next to the Fermi surface have a mean occupancy of about $1/6$ above the Fermi surface ($5/6$ below the Fermi surface). For larger $\left|k-k_F\right|$, the occupation number approaches its non-interacting value apart from small deviations, as shown in Fig.~\ref{fig:sc1_G_loc}. This suggests that most of the WW orbitals can be treated perturbatively, since they deviate only slightly from their non-interacting ground state, where all orbitals below (above) the Fermi surface are filled (empty).

\subsection{Antiferromagnetism}	

\noindent
The antiferromagnetic case is relevant when the system half-filled, which implies that $p_F = \pi/2 = K M/2$ and $\mu=0$. The mean-field Hamiltonian is given by
\begin{equation}
\ham{AF} = -t\sum_{j}\sum_s \left[ c^\dagger_{j,s} \, c^\pdag_{j+1,s} + c^\dagger_{j+1,s}\,c^\pdag_{j,s} \right] + \Delta \sum_j \left(-1\right)^j\,  c^\dagger_{j,s}\, \sigma^z_{ss'}\,c^\pdag_{j,s'},
\label{eq:AF_Hamiltonian}
\end{equation}
where $\Delta$ is the mean-field for the staggered magnetization, which we take to point into the $z$-direction. In many respects, $\ham{AF}$ leads to results similar to the BCS case above, so that the discussion will be focussed on obtaining the WW transform of the AF mean-field operator.

\subsubsection{WW representation of the AF mean-field Hamiltonian}

\noindent
The WW transform  of the mean-field term is given by
\begin{eqnarray}
\Delta \sum_j \left(-1\right)^j\,  c^\dagger_{j,s}\,\sigma^z_{ss'}\, c^\pdag_{j,s'} &=&\Delta \sum_{mk,m'k'} \gamma^\dagger_{mk,s}\,\sigma^z_{ss'} \,\gamma^\pdag_{m'k',s'} \sum_j \left(-1\right)^j \psi_{mk}(j) \psi_{m'k'}(j)\nonumber \\ \label{eq:AF_ww_step_1}
\end{eqnarray}
Now we use that $\left(-1\right)^j = e^{i\pi j}$ to obtain
\begin{eqnarray}
\left(-1\right)^j \psi_{mk}(j) &=& \frac{1}{\sqrt{|G|\left|H_k\right|}}\sum_{\alpha} e^{-i\alpha\phi_{m+k}} \, e^{i\pi j} g_{m,\alpha k}\left(j\right) \nonumber \\
&=& \frac{1}{\sqrt{|G|\left|H_k\right|}}\sum_{\alpha} e^{-i\alpha\phi_{m+k}} \, g_{m,-M+\alpha k}(j)\nonumber \\
&=& \frac{1}{\sqrt{|G|\left|H_k\right|}}\sum_{\alpha} e^{-i\alpha\phi_{m+k}} \, \left(-1\right)^{m+k}\, g_{m,M-\alpha k}(j)\nonumber \\
&=& \left(-1\right)^{m+k} \, \psi_{m,M-k}(j),
\end{eqnarray}
where we have used that $e^{-i\phi_{m+k}} = \left(-1\right)^{m+k} e^{+i \phi_{m+k}}$ and $0\leq M-k = -M+k \mod 2M$ in the third line. Plugging the result into (\ref{eq:AF_ww_step_1}) we obtain
\begin{equation}
\Delta \sum_j \left(-1\right)^j\, \sigma^z_{ss'}\, c^\dagger_{j,s}\, c^\pdag_{j,s'} = \Delta \sum_{mk} \left(-1\right)^{m+k}\, \gamma^\dagger_{mk,s}\,\sigma^z_{ss'}\, \gamma^\pdag_{m,M-k,s'}.
\end{equation}
In general, the staggered magnetization couples the two WW orbitals $\ket{m,k}$ and $\ket{m,M-k}$. However, at the Fermi points we have $k=M/2=M-k$, so that only one orbital is involved. 

\noindent
The AF mean-field Hamiltonian in the WW representation is thus given by
\begin{equation}
\ham{AF} = \sum_{mk}\left[ \delta_{ss'} \epsilon\left(Kk\right) \gamma_{mk,s}^\dagger\, \gamma^\pdag_{mk,s'} +\Delta\,\sigma^z_{ss'} \, \gamma^\dagger_{mk,s} \,\gamma^\pdag_{m,M-k,s'}\right] + O\left(\frac{1}{M}\right)
\label{eq:ham_af_1d}
\end{equation}

\noindent
Note that the single particle gap is exactly the same as for the BCS case above, Eq. (\ref{eq:E_k_BCS}). Thus the ratio $t_k/E_k$  and the single particle correlations show exactly the same behavior as a function of $\xi/M$ and $k$. 


\section{Two dimensions: Effect of anisotropy}
\label{sec:ww_mf_two_dimensions}

\noindent
In this section we apply a similar analysis to the case of the two-dimensional square lattice. Since we will be interested mainly in the case that the Fermi surface lies close to the saddle points $\left(0,\pi\right)$ and $\left(\pi,0\right)$, we use the kinetic energy
\begin{equation}
\ham{kin} = -2t \sum_\mathbf{p} \sum_s\left(\cos p_x + \cos p_y\right) \, c^\dagger_{\mathbf{p},s}\,c^\pdag_{\mathbf{p},s}
\label{eq:ekin_2_wwrg}
\end{equation}
and set the chemical potential $\mu=0$. We consider two types of mean-fields, $d$-wave superconductivity ($d$SC) and antiferromagnetism (AF). The major novelty compared to the one-dimensional case is that now several WW orbitals lie close to the Fermi surface. Moreover, the number of orbitals at the Fermi surface depends on the wave packet scale $M$, with $M$ orbitals at the Fermi surface for the model considered here. In addition, the Fermi velocity along the Fermi surface is strongly anisotropic. As one moves from $\left(\pi,0\right)$ to $\left(0,\pi\right)$ along the Fermi surface, i.e. $p_y = \pi-p_x$, its $x$-component (the $y$-component is the same) varies as
\begin{equation}
v_{F,x}\left(\mathbf{p}\right) = 2t \sin p_x.
\end{equation}
As a consequence, no simple relation between the gap magnitude $\Delta$ and the pair size $\xi$ exists that holds for the whole Fermi surface even in the case of an angle-independent mean-field. 

\noindent
Since correlations are expected to be strongest where $v_F$ is smallest, we use estimates of the pair size at the saddle points in the following, and adjust $M$ accordingly. In the vicinity of the saddle point $\left(\pi,0\right)$, the kinetic energy (\ref{eq:ekin_2_wwrg}) can be approximated as
\begin{equation}
\epsilon\left(\mathbf{p} - \left(\pi,0\right)\right) \approx t \left(p_y^2 - p_x^2\right),
\end{equation}
and similarly for the other saddle point $\left(0,\pi\right)$. The natural length scale $\xi$ for a pair gap of size $\Delta$ is then given by
\begin{equation}
\xi = \pi \sqrt{\frac{t}{\Delta}}.
\label{eq:def_xi_2d}
\end{equation}

\begin{figure}
\centering
\includegraphics[width=8cm]{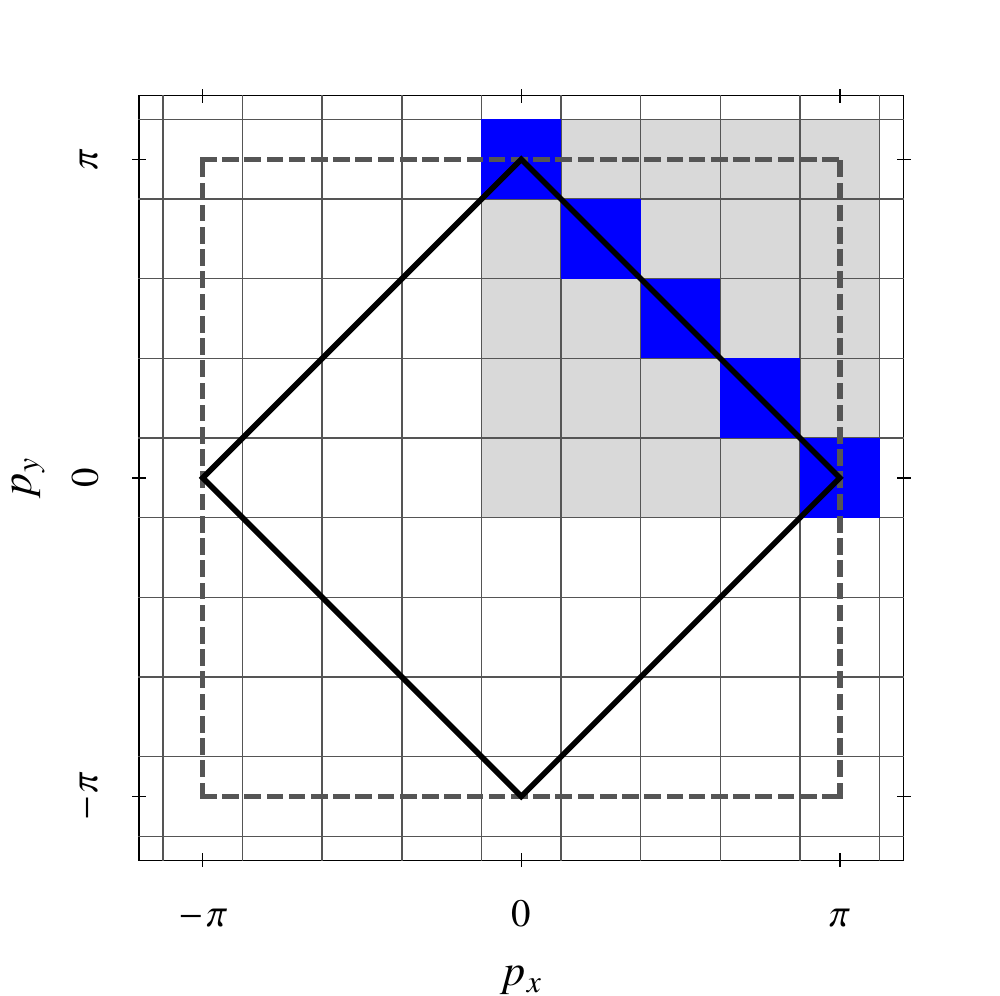}
\caption{Low energy WW orbitals (blue cells) for the two-dimensional square lattice with nearest neighbor hopping at half-filling for $M=4$. The Fermi surface (solid black line), and the phase space cells (gray grid) are also shown.}
\label{fig:cell_mf2} 
\end{figure}
\noindent
We are mainly interested in cases where pairing occurs at relatively small scales. We hence fix $M=4$ for the remainder of this chapter. The phase space cells and the states at the Fermi surface for this case are shown in Fig.~\ref{fig:cell_mf2}. 

\noindent
Since at the saddle points the distance between two WW states in real space is $2M$ (cf. Ch.~\ref{ch:wwbasis}), $\xi$ should be compared to $2M$. For $\xi = 2M$ we obtain
\begin{equation}
\Delta \sim  \frac{\pi^2}{4M^2}\,t = \frac{\pi^2}{64}\,t,
\end{equation}
which evaluates to $\Delta \approx 0.15 \,t$. 

\noindent
In the following two sections we discuss pairing in the AF and $d$SC channels. We will show that even when the order parameter is constant along the Fermi surface, the strength of correlation is much larger at the saddle points than for the other states at the Fermi surface. This is due to the large anisotropy of the Fermi velocity. In order to measure the strength of correlation for different angles, we compute the ratio
\begin{equation}
\frac{\left\langle E_{\text{kin}} \right\rangle_{\mathbf{k}}}{\left\langle E_{\text{pair}}\right\rangle_{\mathbf{k}}} = \frac{\sum_{\mathbf{m'}}T\left(\mathbf{m}\mathbf{k},\mathbf{m'}\mathbf{k}\right)\; G\left(\mathbf{m}\mathbf{k},\mathbf{m}'\mathbf{k} \right)}{\sum_{\mathbf{m'} }\Delta\left(\mathbf{m}\mathbf{k}, \mathbf{m}' \mathbf{k}'\right) \, F\left(\mathbf{m}\mathbf{k}, \mathbf{m}' \mathbf{k}'\right)}
\label{eq:ekin_epair_ratio}
\end{equation}
where $T$ and $\Delta$ are the WW transforms of the hopping Hamiltonian and the order parameter, respectively. $\mathbf{k}'$ in the denominator of the right hand side is either $\mathbf{k}'=\mathbf{k}$ for $d$SC, or $\mathbf{k}'=\left(M,M\right)-\mathbf{k}$, as discussed below. $F$ and $G$ are the WW transforms of the normal and anomalous correlators, cf. Sec.~\ref{sec:ww_mf_one_dimension} above. Hence (\ref{eq:ekin_epair_ratio}) yields the strength of correlation of the WW orbitals $\ket{\mathbf{m}\mathbf{k}}$ in the mean-field ground state in the sense that when the ratio is large, the energy is dominated by the kinetic energy, so that electrons in the orbital can be considered as almost uncorrelated at scale $M$. On the other hand, when it is small, the pairing energy (AF or $d$SC) dominates over kinetic energy, so that the orbital can be approximated by almost local fermion pairs.

\subsection{Antiferromagnetism}

\noindent
We start with the AF case, since this is more similar to the one-dimensional examples in that the order parameter is constant along the Fermi surface. The WW transform of the AF mean-field Hamiltonian can be obtained in exactly the same way, as in one dimension, Eq.~(\ref{eq:ham_af_1d}). Since in two dimension the wave vector of the antiferromagnetic ordering is $\mathbf{Q}=\left(\pi,\pi\right)$, the WW orbitals $\ket{\mathbf{m},\mathbf{k}}$ and $\ket{\mathbf{m}, \mathbf{Q}/K - \mathbf{k}}$ are coupled. The WW transform of the staggered AF magnetization is
\begin{eqnarray}
\ham{AF} &=& \Delta \sum_\mathbf{j} \left(-1\right)^{j_x + j_y}\sigma^z_{ss'} c^\dagger_{\mathbf{j},s}\,c^\pdag_{\mathbf{j},s'} \nonumber \\&=& \Delta \sum_{\mathbf{m}\mathbf{k}} \left(-1\right)^{m_x+m_y+k_x+k_y}\sigma^z_{ss'} \, \gamma^\dagger_{\mathbf{m}\mathbf{k},s}\,\gamma^\pdag_{\mathbf{m},\mathbf{Q}/K-\mathbf{k},s'}.
\end{eqnarray}

\begin{figure}
\centering
\includegraphics[width=8cm]{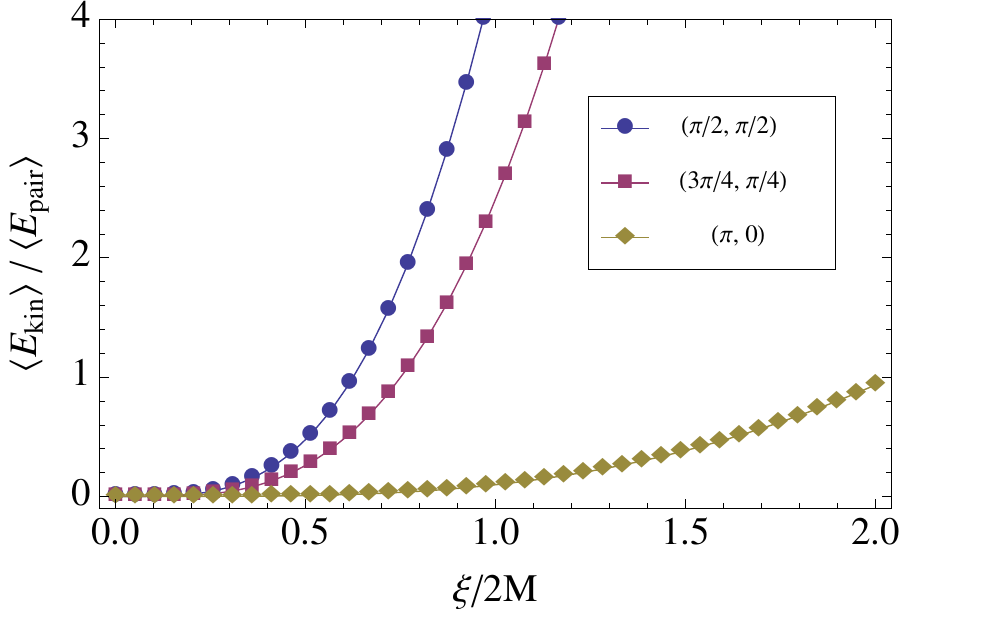}
\caption{Strength of correlations for the WW orbitals $\ket{\mathbf{m}\mathbf{k}}$ at the Fermi surface. There are three inequivalent classes of states, with mean momenta $K \mathbf{k} =\left(\pi/2,\pi/2\right)$, $\left(3\pi/4,\pi/4\right)$, and $\left(\pi,0\right)$, respectively. The strength of correlation is determined by applying Eq.~\ref{eq:ekin_epair_ratio} to the ground state of the mean-field Hamiltonian $\ham{kin}+\ham{AF}$. The pair size at the saddle points $\xi$ is connected to the size of the order parameter via Eq.~(\ref{eq:def_xi_2d}). Local correlations are strong when $\left\langle E_\text{kin}\right\rangle/\left\langle E_{\text{pair}}\right\rangle < 1$. }
\label{fig:af_mf2_ekin_epair}
\end{figure}
\noindent
We use Eq. (\ref{eq:ekin_epair_ratio}) in order to estimate the strength of correlation in different orbitals, dependent on the ratio $\xi/M$. This ratio is shown in Fig.~\ref{fig:af_mf2_ekin_epair}. It is evident that for $\xi \sim 2M$, the states at the saddle points are very strongly correlated, with $\left\langle E_\text{kin}\right\rangle/\left\langle E_{\text{pair}}\right\rangle \approx 0.1$. The other states are much less correlated, the corresponding ratios are given by $\left\langle E_\text{kin}\right\rangle/\left\langle E_{\text{pair}}\right\rangle \approx 2.7$ for the states with $K\mathbf{k} = \left(3\pi/4, \pi/4\right)$, and $\left\langle E_\text{kin}\right\rangle/\left\langle E_{\text{pair}}\right\rangle \approx 4.8$ for the nodal states with $K\mathbf{k} = \left(\pi/2,\pi/2\right)$. 
\begin{figure}
\centering
\includegraphics[width=8cm]{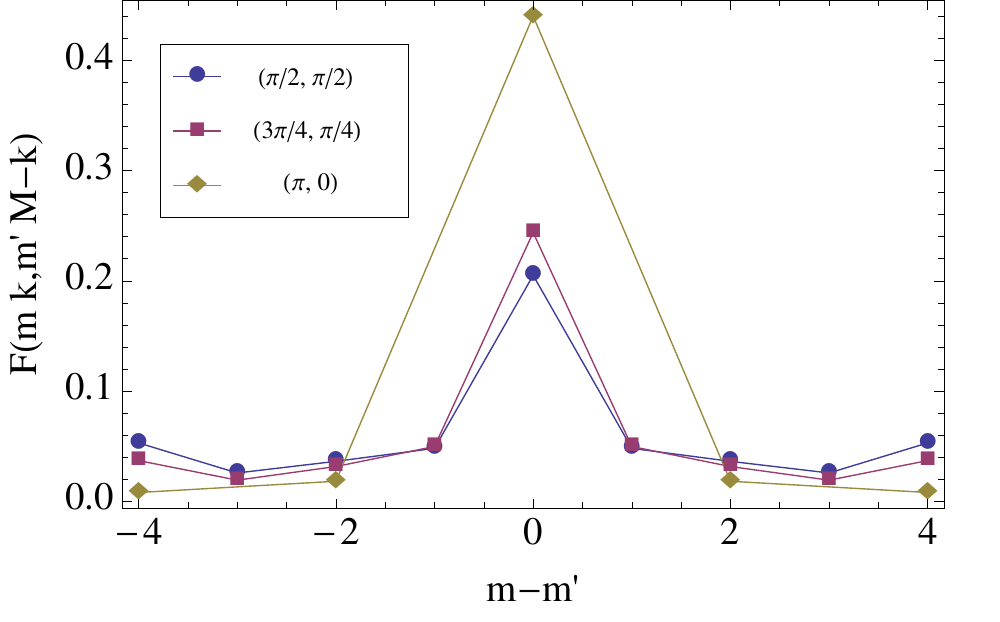}
\caption{Spatial decay of the AF pair correlations for WW states at the Fermi surface for $M=4$, $\xi = 2M$. The figure shows the anomalous correlator $F\left(\mathbf{m},\mathbf{k},\mathbf{m}', \mathbf{M}-\mathbf{k}\right)$ (where $\mathbf{M} = \left(M,M\right)$) for $\mathbf{k}$ on the Fermi surface, and $\mathbf{m}' = \mathbf{m} = m' \left(1,1\right)$, the slowest decay direction.}
\label{fig:af_mf2_pairsize}
\end{figure}

\noindent
The spatial decay of the pair correlations for the states at the Fermi surface is shown in Fig.~\ref{fig:af_mf2_pairsize}. The pair correlations are strongest at the saddle points. Surprisingly, the pair correlations decay very rapidly even for the nodal states, which are weakly correlated. This is probably connected to the fact that the system is half-filled and thus perfectly nested, so that even the uncorrelated Fermi sea contains relatively strong density-density correlations at the wave vector $\mathbf{Q}$.

\subsection{$d$-wave superconductivity}

\noindent
In this section, we consider $d$-wave superconductivity. The corresponding mean-field Hamiltonian is given by
\begin{equation}
\ham{\emph{d}SC} = \ham{kin} + \frac{\Delta}{2} \sum_{\mathbf{j}} \epsilon_{ss'} \left[ c_{\mathbf{j},s}\, c_{\mathbf{j}+\mathbf{x},s'} - c_{\mathbf{j},s}\,c_{\mathbf{j}+\mathbf{y},s'}\right] \;+\; \text{h.c.},
\end{equation}
where $\mathbf{x}$ and $\mathbf{y}$ are unit vectors in the $x$- and $y$-directions, respectively. 

\subsubsection{WW representation of the Hamiltonian}

\noindent
The transformation of this Hamiltonian to the WW basis can be obtained from the transformation of the hopping operator (\ref{eq:ham_kin_ww_1}), noting that in real space $c_{\mathbf{j},s}$ and $c^\dagger_{\mathbf{j},s}$ transform in the same way because the basis functions are real. Thus we obtain to leading order in $1/M$
\begin{equation}
\ham{\emph{d}SC} = \sum_{\mathbf{m}\mathbf{k}}\sum_s \epsilon\left(K\mathbf{k}\right) \gamma^\dagger_{\mathbf{m}\mathbf{k},s}\,\gamma^\pdag_{\mathbf{m}\mathbf{k},s} + \Delta \sum_{\mathbf{m}\mathbf{k}} \left(\sin K k_x - \sin K k_y\right) \gamma^\dagger_{\mathbf{m}\mathbf{k},\downarrow}\,\gamma_{\mathbf{m}\mathbf{k},\uparrow} \;+\;\text{h.c.}.
\label{eq:ham_sc_mf2}
\end{equation}
As a consequence, the $d$-wave pairing term for the nodal states vanishes to leading order in $1/M$, so that the pairs are always non-local there. By symmetry, the corrections at the saddle point are $O(1/M^2)$ and hence relatively small. Naturally, the correlations are strongest at the saddle points because of the $d$-wave symmetry of the order parameter.

\subsubsection{Correlations in the mean-field ground state}

\begin{figure}
\centering
\includegraphics[width=8cm]{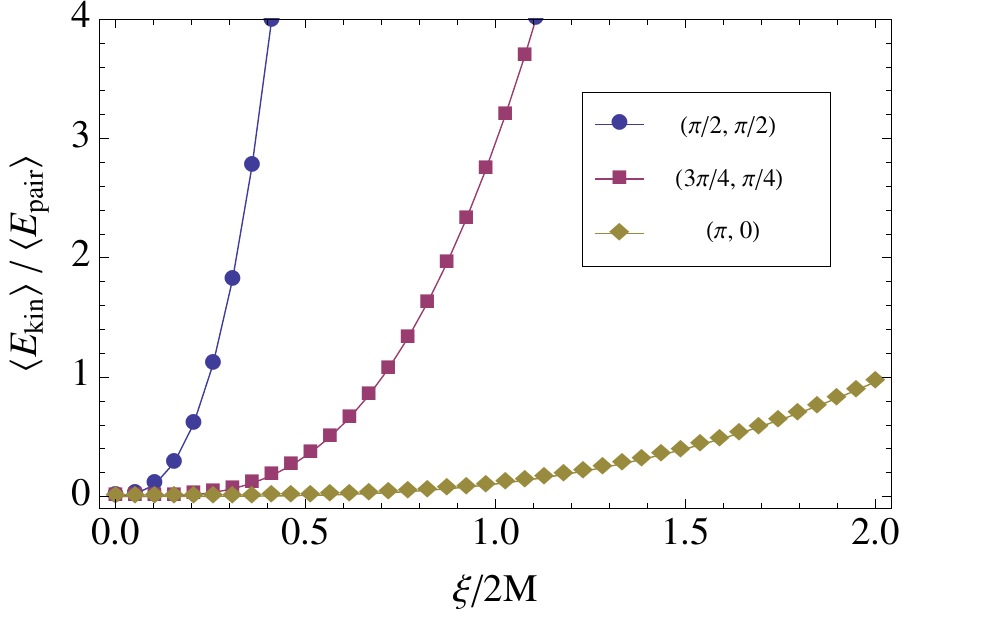}
\caption{Amount of correlation for the WW orbitals $\ket{\mathbf{m}\mathbf{k}}$ at the Fermi surface for the $d$SC mean-field state. There are three inequivalent classes of states, with mean momenta $K \mathbf{k} =\left(\pi/2,\pi/2\right)$, $\left(3\pi/4,\pi/4\right)$, and $\left(\pi,0\right)$, respectively. The strength of correlation is determined by applying Eq.~\ref{eq:ekin_epair_ratio} to the ground state of the mean-field Hamiltonian (\ref{eq:ham_sc_mf2}). The pair size $\xi$ at the saddle points is connected to the size of the order parameter via Eq.~(\ref{eq:def_xi_2d}). Local correlations are strong when $\left\langle E_\text{kin}\right\rangle/\left\langle E_{\text{pair}}\right\rangle < 1$. }
\label{fig:sc_mf2_ekin_epair}
\end{figure}
\noindent
Similar to the AF case above, we calculate the ratio $\left\langle E_\text{kin}\right\rangle/\left\langle E_{\text{pair}}\right\rangle$ for all states at the Fermi surface. In the $d$SC case, the correlations decay faster towards the nodal direction because of the angle dependence of the order parameter. We compute the pair correlations to order $1/M$, so that the pair amplitude is finite for the nodal states. The results are shown in Fig.~\ref{fig:sc_mf2_ekin_epair}. The behavior at the saddle points is almost the same as in the AF case, namely, the saddle points are strongly correlated for $\xi \lesssim 4M$, with $\left\langle E_\text{kin}\right\rangle/\left\langle E_{\text{pair}}\right\rangle \approx 0.11$ at $\xi/2M=1$. The states at $\left(3\pi/4, \pi/4\right)$ are less correlated, with $\left\langle E_\text{kin}\right\rangle/\left\langle E_{\text{pair}}\right\rangle \approx 3.2$ for $\xi/2M = 1$. This is a slightly larger ratio than for the AF case, but still similar. The nodal states, on the other hand have $\left\langle E_\text{kin}\right\rangle/\left\langle E_{\text{pair}}\right\rangle \approx 40$, so that correlations can be neglected at scale $M$.

\begin{figure}
\centering
\includegraphics[width=8cm]{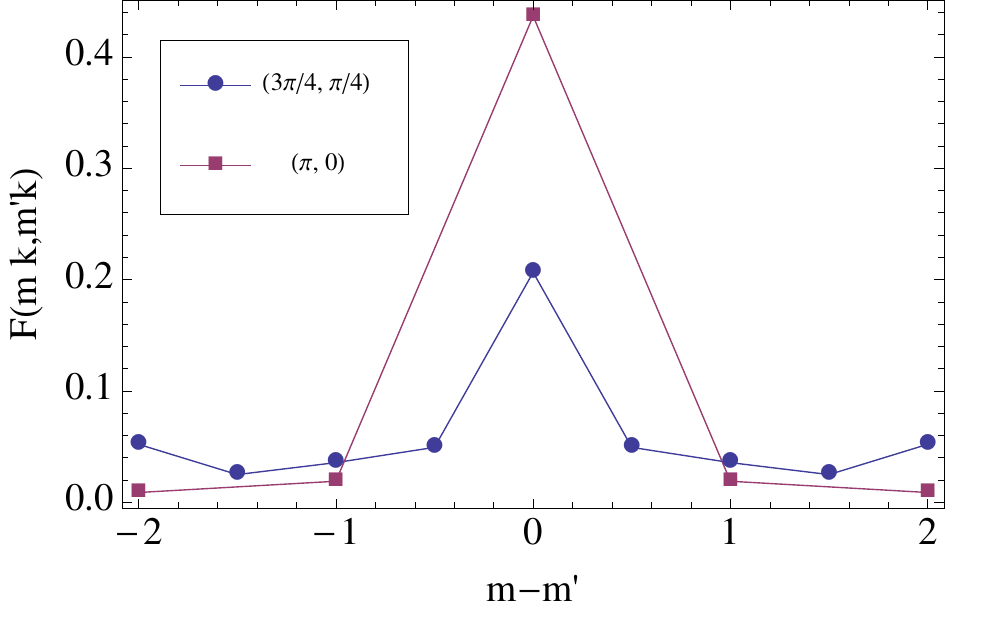}
\caption{Spatial decay of the $d$SC pair correlations for WW states at the Fermi surface for $M=4$, $\xi = 2M$. The figure shows the anomalous correlator $F\left(\mathbf{m},\mathbf{k},\mathbf{m}' \mathbf{Q}/K-\mathbf{k}\right)$ for $\mathbf{k}$ on the Fermi surface, and $\mathbf{m}' = \mathbf{m} = m' \left(1,1\right)$, the slowest decay direction.}
\label{fig:sc_mf2_xi}
\end{figure}
\noindent
The spatial decay of the pair correlations at $\xi/2M=1$ is different from the AF case because of the angular dependence of the order parameter. Since pair correlations for the nodal states are small, we only show the decay of correlations for the other states. This is displayed in Fig.~\ref{fig:sc_mf2_xi}. The increase of the correlation function for the states at $K\mathbf{k} =\left(3\pi/4,\pi/4\right)$ is again an artifact of the window function that is used, which is obtained for $N/M=8$ (cf. Ch.~\ref{ch:wwbasis}). The main result is that correlations at the saddle points decay very rapidly, similar to the AF case.

\section{Conclusions}

\noindent
In this chapter we have used the WW basis states in order to analyze the correlations of AF and $d$SC mean-field states in one and two dimensions. We found that the ratio $\xi/M$ of the (particle-hole or particle-particle) pair size to the size of the wave packets that define the WW basis controls a crossover between locally almost uncorrelated fermions for $\xi \gg M$ and tightly bound fermions for $\xi \ll M$. Even though this result is based on simple mean-field Hamiltonians, it is reasonable to expect that locally the physics is similar at scale $M$ independently of the behavior at larger scales. This insight will be used in the following sections to analyze strong coupling fixed points of the renormalization group in one and two dimensions. The key idea from this section is that if the wave packet size $M$ is chosen appropriately, the fermions can be replaced by effective bosonic degrees of freedom corresponding to the paired fermions. The interactions between the new degrees of freedom determine the physical behavior at larger distances and low energies. 

\noindent
In two dimensions, we have seen that even in simple mean-field states the physics involves multiple length scales when the Fermi velocity is strongly anisotropic. In particular, when the Fermi surface lies in the vicinity of the saddle points, the wave packets there are easily localized and bound into pairs. On the other hand, states in the nodal direction remain essentially uncorrelated at the same length scale. This separation of length scales forms the basis of the treatment of the saddle point regime of the two-dimensional Hubbard model in Ch.~\ref{ch:ww2}.

\chapter{Wave packets and the renormalization group}
\label{ch:wave_packets_rg}

\noindent
In the last chapter we have discussed the manifestation of fermion pairing in the WW basis for simple mean-field Hamiltonians. The type of pairing and its associated length scale $\xi$ have been put in by hand. It is clear, however, that microscopic Hamiltonians rarely are of the mean-field variety, instead it is a complex task by itself to obtain an effective Hamiltonian that eventually leads to pair formation (or more complicated orderings) from the microscopic interactions. For weak to moderate initial interactions, the renormalization group is one of the standard methods for obtaining effective Hamiltonians for the low-energy degrees of freedom of a many-fermion system from the microscopic interactions \cite{shankar}. 

\noindent
In this chapter we seek to establish the connection between the renormalization group for interacting fermions and the WW basis. We begin in Sec.~\ref{sec:scaling_dimensions} by relating the $M$-dependence of the wave packet transform of an operator to its na\"ive scaling dimension. This serves to discuss the relevance of different operators for large $M$. In Sec.~\ref{sec:cut} we introduce very briefly the method of continuous unitary transformations \cite{wegner,cut1,cut2}, a Hamiltonian formulation of the renormalization group, which avoids certain problems connected with the cutoff in the renormalization group. Since this topic is rather technical, we relegate the bulk of the material to App.~\ref{ch:wegner}. Finally, in Sec.~\ref{sec:bz_geometry} we discuss the implications of the geometry of the Brillouin zone and the Fermi surface for the treatment of low-energy problems. In particular, we show that in the proximity of the van Hove singularity, the problem can be simplified due to a separation of scales. This is an ingredient to the treatment of the saddle point regime of the Hubbard model in Ch.~\ref{ch:ww2}.

\section{Scaling dimensions}
\label{sec:scaling_dimensions}

\noindent
In the following we elaborate on the relation between the so-called na\"ive scaling dimension of operators in the RG approach \cite{shankar} and the dependence of matrix elements on the wave packet scale $M$ in the wave packet transform of operators. In particular, we will see that the latter is given by the scaling dimension of the operator under consideration.

\subsubsection{Scaling dimensions in the RG approach}

\noindent
The scaling dimension within the renormalization group arises as follows: In its original formulation \cite{shankar}, each renormalization step consists in integrating out degrees of freedom with energy $ \Lambda/s < E < \Lambda$, where $s$ is close to one. Afterwards, the cutoff $\Lambda$ is lowered to $\Lambda/s$, and all length scales are rescaled, such that in the new units the cutoff is again $\Lambda$. All field operators are rescaled such that the kinetic energy part remains unchanged, which allows to compare the relative growth of the interaction part compared to the kinetic energy. In addition to the rescaling, perturbative corrections arise from integrating out degrees of freedom, which completes the RG step. The na\"ive scaling dimension of an operator is related to the rescaling of lengths, and is thus obtained by omitting the perturbative renormalization. It measures the importance of operators at low energies. An operator that becomes asymptotically more important than the kinetic energy is called relevant, it is called marginal if it scales in the same way as the kinetic energy, and irrelevant if it decreases at low energies. We follow the discussion in \cite{shankar} for the RG part.

\noindent
For interacting fermion systems, the programme is implemented by imposing an upper cutoff on all momentum space integrals as follows
\begin{equation}
\int d^d p \rightarrow \int_{-\Lambda/v_F}^{\Lambda/v_F} dp_\perp \int d^{d-1}p_\parallel,
\end{equation}
where $p_\perp$ measures the distance to the Fermi surface, $p_\perp = \left|\mathbf{p}\right| - p_F$ with the Fermi momentum $p_F$. $v_F$ is the Fermi velocity which is assumed to be constant. The remaining integral $\int d^{d-1}p_\parallel$ runs over fixed energy shells, with energy $v_F p_\perp$. The action of spinless free fermions in the vicinity of the Fermi surface (i.e. with cutoff $\Lambda \ll v_F p_F$) is then given by
\begin{equation}
S= \int d\omega \int_{-\Lambda/v_F}^{\Lambda/v_F} dp_\perp \int d^{d-1}p_\parallel \bar{\psi}\left(\omega,\mathbf{p}\right) \Big[ -i\omega + v_F p_\perp\Big] \psi\left(\omega,\mathbf{p}\right).
\end{equation}
This action is a fixed point of the RG transformation $\Lambda \rightarrow \Lambda/s$ when $\omega, p_\perp$, and $\psi\left(\omega,\mathbf{p}\right)$ are rescaled as
\begin{eqnarray}
\omega &\rightarrow& \omega/s \nonumber \\
p_\perp &\rightarrow&  p_\perp/s \nonumber \\
\psi\left(\omega, \mathbf{p}\right) &\rightarrow& s^{3/2} \psi\left(\omega,\mathbf{p}\right).
\end{eqnarray} 
$\bar{\psi}\left(\omega,\mathbf{p}\right)$ is rescaled in the same way as $\psi\left(\omega,\mathbf{p}\right)$. This definition of the rescaling ensures that the kinetic energy part of the action remains invariant, so that the effect of rescaling on other operators measures the change in their importance when the cutoff is lowered. Relevant (irrelevant) operators are proportional to some positive (negative) power of $s$ after rescaling, whereas marginal ones are unchanged. On the other hand, other types of operators are affected by the rescaling, for example one may add a quadratic term $\propto p_\perp^2$ to the kinetic energy. Under rescaling (omitting the angular integrals over $p_\parallel$ which do not play a role), this term changes as
\begin{eqnarray}
\int d\omega \int p_\perp \; \bar{\psi}\left(\omega,\mathbf{p}\right) p_\perp^2 \psi\left(\omega,\mathbf{p}\right) &\rightarrow& \int \frac{d\omega}{s} \int \frac{dp_\perp}{s}  s^{3/2}\bar{\psi}\left(\omega,\mathbf{p}\right) \left(\frac{p_\perp}{s}\right)^2 s^{3/2}\psi\left(\omega,\mathbf{p}\right)\nonumber \\
&=&\frac{1}{s} \int d\omega \int dp_\perp \; \bar{\psi}\left(\omega,\mathbf{p}\right) p_\perp^2 \psi\left(\omega,\mathbf{p}\right),
\end{eqnarray}
so that it decreases as the cutoff is lowered. It is clear that additional powers of $p_\perp$ make the decrease even faster. In a similar manner, one finds that interactions that involve the full Fermi surface when evaluated at $p_\perp=0$ are marginal, whereas all others are irrelevant. Since it is quite lengthy to show this, we refer the reader to \cite{shankar}. Instead, we show how the scaling dimensions of operators follow directly from their wave packet transform. 

\subsubsection{Scaling dimensions from wave packets}

\noindent
The dependence of the wave packet transform of an operator on $M$ can be found using the transformation rules from Ch.~\ref{ch:wwtrafo}. Instead of reiterating them, we give intuitive arguments why they are true in this section. We first consider the kinetic energy, and consider its wave packet transform. We focus on $k$-conserving matrix elements that involve wave packets with $m' = m+1$. Since the states are wave packets with a mean momentum $Kk$, they move at the group velocity $v_g = \epsilon'\left(K k\right)$. The distance between two adjacent sites is $M$, hence we find for the wave packet hopping rate
\begin{eqnarray}
\text{hopping rate} &\sim& \frac{\text{group velocity}}{\text{distance}}\nonumber \\
&\sim& M^{-1}.
\end{eqnarray}
Rescaling the Hamiltonian by $M$, the leading part of the kinetic energy is hence independent of $M$. Consequently, operators that decrease faster (slower) than $1/M$ as $M\rightarrow \infty$ are irrelevant (relevant), and operators that scale like $1/M$ are marginal. Expanding the kinetic energy to higher order around $Kk$ leads to higher powers of $p$ that are integrated against the wave packets. Since the wave packets are narrow in $p$-space, with a width $\sim 1/M$, each additional power of $p$ contributes an additional power of $1/M$, so that corrections to the leading term are less and less important as $M\rightarrow \infty$. 

\noindent
For states at points of high symmetry in the Brillouin zone, the behavior is different. Because of the symmetry, $\epsilon'\left(Kk\right)$ vanishes, and the kinetic energy is $O(1/M^2)$. Consequently, the kinetic energy at these points can be neglected compared to generic points on the Fermi surface for large $M$.

\noindent
The scaling of local (at scale $M$) interactions can be understood as follows: The matrix element of a local interactions between two pairs of wave packet states is of the form
\begin{eqnarray}
\text{interaction strength} &\sim& \underbrace{\left(\text{density of wave packet}\right)^2}_{\sim M^{-2d}} \, \times \, \underbrace{\text{volume of wave packet}}_{\sim M^d} \nonumber \\ &\sim& M^{-d}.
\end{eqnarray}
Note that the interaction need only be local compared to the wave packet scale $M$ for the argument to hold, since all $k$-conserving matrix elements transform in the same way, regardless of the momenta involved. 

\begin{figure}
\centering
\includegraphics[width=8cm]{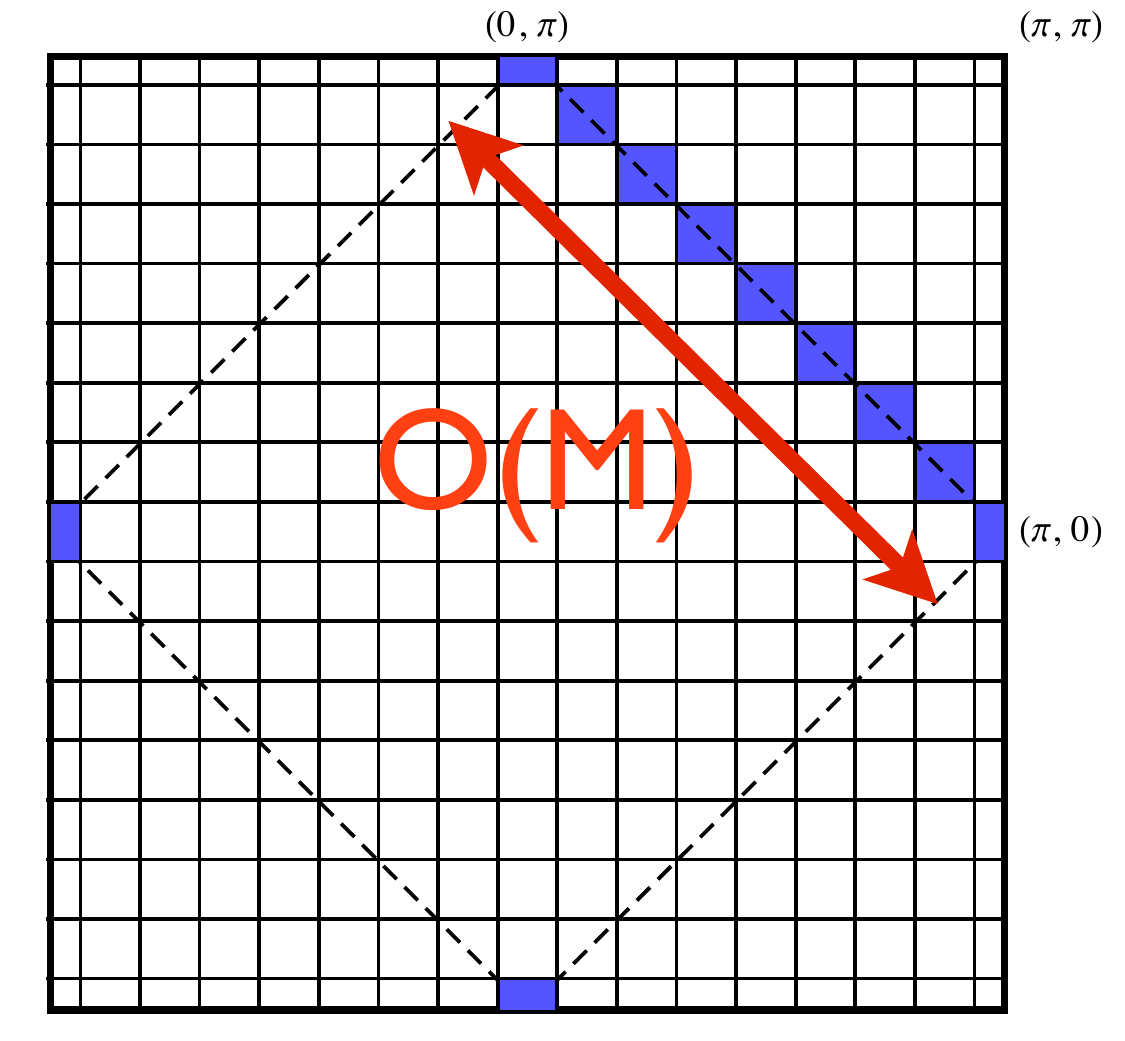}
\caption{WW basis states at the Fermi surface in two dimensions. There are $O(M)$ states at the Fermi surface in total.}
\label{fig:om_states}
\end{figure}
\noindent
In order to link this result to the scaling dimensions from the RG approach, we implement the cutoff by restricting the WW basis states to states that lie at the Fermi surface. There is one such state per WW lattice site in one dimension, regardless of $M$. Since the interaction matrix elements for this state scale like $1/M$, the interaction is marginal in one dimension. The matrix elements in two dimensions scale like $M^{-2}$ and may thus appear to be irrelevant. However, the number of states at the Fermi surface scales linearly with $M$, since the Fermi surface has a fixed length in momentum space, cf. Fig.~\ref{fig:om_states}. Thus each state can couple to $M$ other states, and interaction terms that involve a finite fraction of the states at the Fermi surface (i.e. with all four momenta at the Fermi surface) scale like $1/M$ instead. In the same manner as for the kinetic energy, when the interaction is not local but contains additional powers of $p$, each power of $p$ contributes an additional power of $1/M$ in the wave packet transform, so that for large enough $M$ interactions can be assumed to be local to high accuracy. 

\noindent
The argument that interactions have to couple many states at the Fermi surface is not new, and has been used by various investigators to facilitate the analysis of weakly coupled fermion systems by means of a $1/N$-expansion, where $N$ is the number of states at the Fermi surface with discretized angles \cite{shankar,fradkinboso,oneovern2,oneovern3,oneovern4}.

\noindent
It should be noted, that for the states at the saddle points (or other points of high symmetry) the interaction in two dimensions always scales in the same way as the hopping terms. This suggests that when the Fermi surface touches the saddle points, the states there may become strongly coupled even when they are not coupled to the rest of the Fermi surface. 

\section{One-loop RG via continuous unitary transformations}
\label{sec:cut}

\noindent
In this section we introduce the continuous unitary transformations (CUT) \cite{wegner,cut1,cut2}, which are the Hamiltonian equivalent of the action-based RG flow equations \cite{shankar, tnt}. Whereas the RG is based on integrating out degrees of freedom, and thus decreases the number of the degrees of freedom in the system, the CUT method merely decouples states with different kinetic energies, so that the number of degrees of freedom stays the same. The decoupling is achieved by a sequence of infinitesimal canonical transformations that is applied to the Hamiltonian of the system. We denote the generator of the infinitesimal transformation by $\eta\left(B\right)$, where $B$ is the flow parameter related to the renormalization scale. The flowing Hamiltonian $\mathcal{H}\left(B\right)$ obeys the equation
\begin{equation}
\frac{d}{dB}\mathcal{H}\left(B\right) = \left[ \eta\left(B\right), \mathcal{H}\left(B\right)\right],
\label{eq:flow_equation_ch}
\end{equation}
which is just the first order expansion of the unitary transformation $e^{-\eta(B)}\mathcal{H} e^{\eta(B)}$. In principle, the flow equation is exact, but in practice one has to resort to approximations. In App.~\ref{ch:wegner} we show that for a suitable generator $\eta(B)$, the flow equation (\ref{eq:flow_equation_ch}) leads to a set of coupled flow equations for a set of auxiliary coupling constants $F\left(\mathbf{p}_1,\cdots,\mathbf{p}_4\right)$. These flow equations are structurally very similar to the usual RG equations when the frequency dependence is neglected. The major difference to the RG equations is that there is no cutoff. Instead, the physical coupling constants $U\left(\mathbf{p}_1,\cdots,\mathbf{p}_4\right)$ are obtained from the auxiliary ones via 
\begin{equation}
U\left(\mathbf{p}_1,\cdots,\mathbf{p}_4\right) = e^{-\left[\epsilon\left(\mathbf{p}_1\right) + \epsilon\left(\mathbf{p}_2\right) - \epsilon\left(\mathbf{p}_3\right) -\epsilon\left(\mathbf{p}_4\right) \right]^2/16\Lambda^2} F\left(\mathbf{p}_1,\cdots,\mathbf{p}_4\right),
\label{eq:F_to_U}
\end{equation}
where $\Lambda$ is the RG energy scale. The exponential on the right hand side contains the kinetic energy differences of the initial and final states of an interaction matrix elements. As a consequence, states with kinetic energy larger than $\Lambda$ decouple from the low energy states and can be neglected when $|\epsilon| \gg \Lambda$. The advantage of this scheme within the wave packet approach is that the wave packet states may lie partially below and partially above the cutoff. In the RG scheme it is difficult to obtain an effective Hamiltonian in this situation, whereas in the CUT method, Eq.~\ref{eq:F_to_U} gives a simple rule for all cases. Since the CUT flow equations in the one-loop approximations are essentially equivalent to the RG one-loop equations, we do not discuss them any further. The only CUT related equation that is needed in the remainder of this work is Eq.~\ref{eq:F_to_U}.

\noindent
More details can be found in App.~\ref{ch:wegner} and many more in the book by Kehrein \cite{kehrein}.

\section{The geometry of the low-energy states in the Brillouin zone}
\label{sec:bz_geometry}

\noindent
In this section we discuss the influence of the shape of the low-energy phase space on the low-energy physics of the Hubbard model in two dimensions. In particular, we look at the influence of strong anisotropy in the Fermi velocities along the Fermi surface. The most extreme case of anisotropy is realized when the Fermi surface touches the saddle points, where the van Hove singularity in the density of states manifests itself through a vanishing Fermi velocity at the saddle points. We will show that in this case, slow and fast parts of the Fermi system at a fixed energy scale $\Lambda$ live on different length scales, which leads to an approximate decoupling of these two kinds of states.

\subsubsection{van Hove singularities}

\begin{figure}
\centering
\includegraphics[width=6cm]{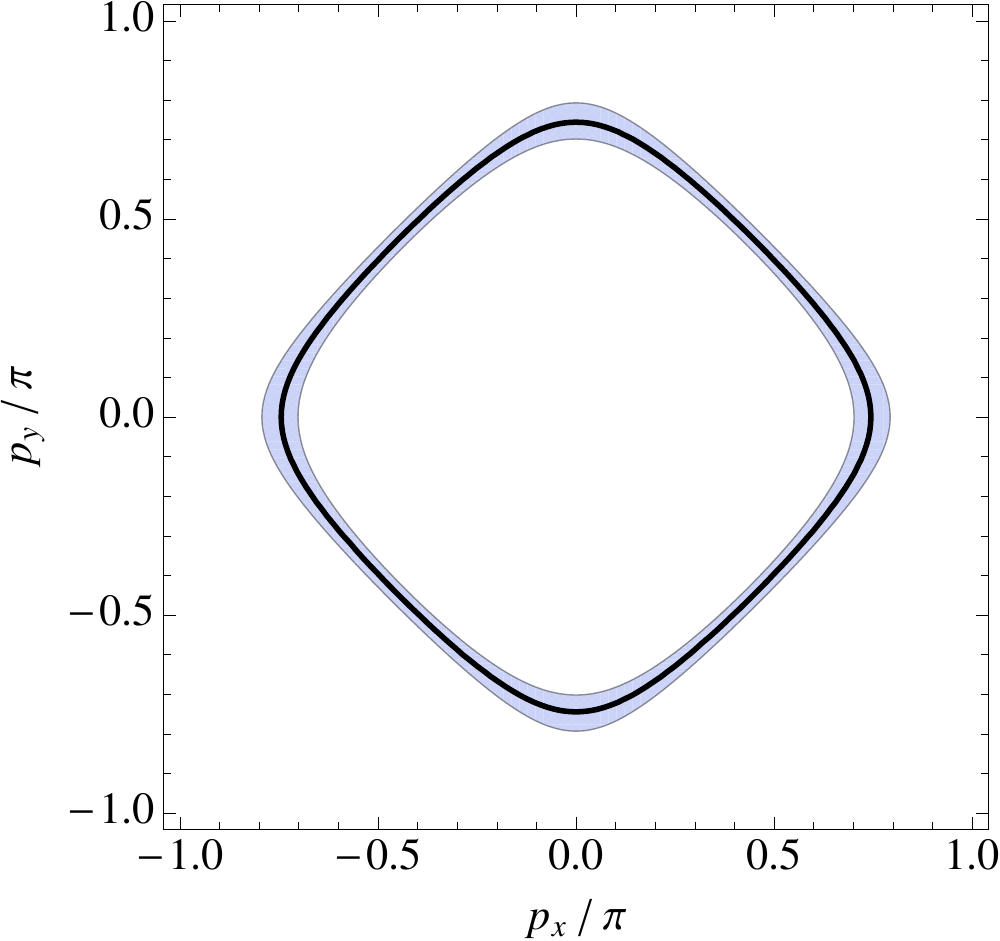}
\includegraphics[width=6cm]{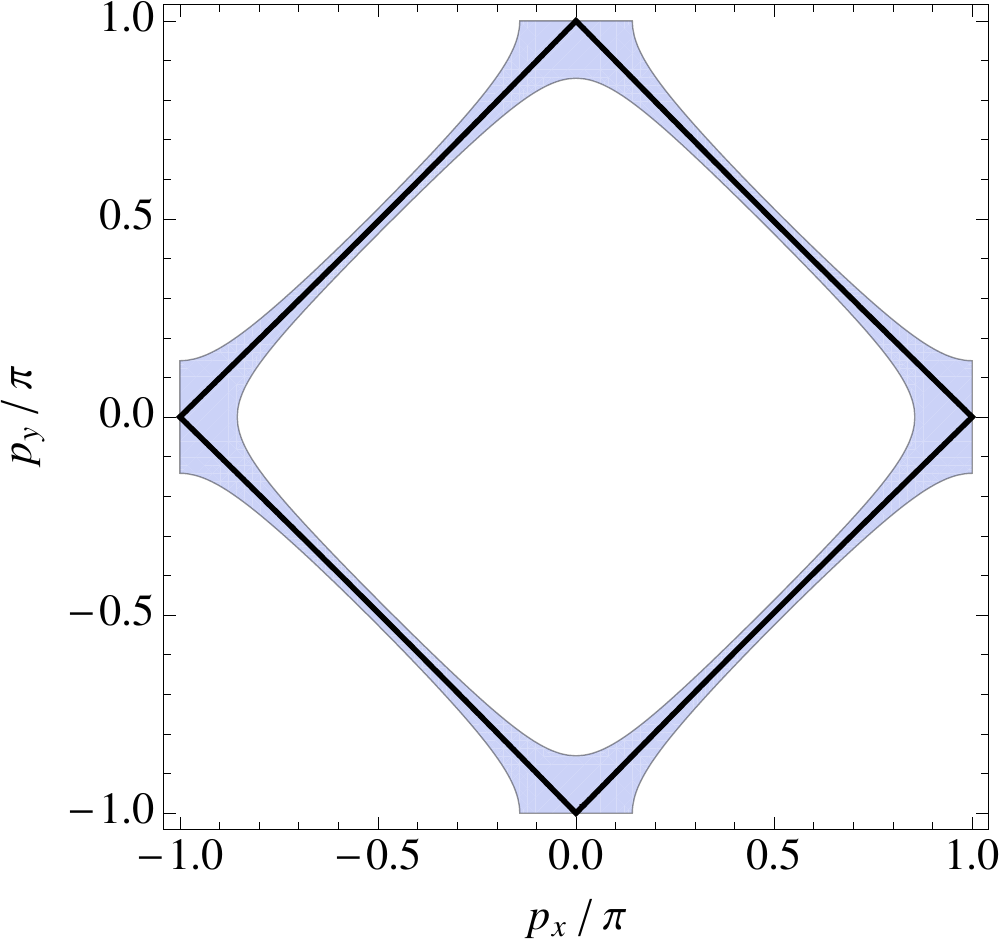}
\caption{Low energy phase-space for the two-dimensional square lattice with nearest neighbor hopping. The shaded area is the region where $\left|\epsilon\left(\mathbf{p}\right)\right|<0.1t$. The left panel shows the generic situation, where the Fermi surface is far away from the saddle points. The width of the tube around the Fermi surface is approximately independent of the angle. In the right panel, on the other hand, the Fermi surface touches the saddle points. Due to the van Hove singularity, the tube is much broader in the vicinity of the saddle points than around the rest of the Fermi surface.}
\label{fig:tubes}
\end{figure}

\noindent
The influence of the latter can be seen in Fig.~\ref{fig:tubes}, which shows the tube of states with energy $|\epsilon| < \Lambda$ for $\Lambda=0.1t$. In the case of a generic Fermi surface, the Fermi velocity is approximately constant along the Fermi surface. Since the energy close to the Fermi surface is $\epsilon \sim v_F p_\perp$, the tube has width $\Delta p \sim 2\epsilon/v_F$ everywhere. On the other hand, when the Fermi surface is in the vicinity of the saddle points, the Fermi velocity becomes strongly anisotropic (and vanishes at the saddle points). In the vicinity of the saddle points the width of the tube is asymptotically $\Delta p\sim 2 \sqrt{\Lambda/t}$. 

\begin{figure}
\centering
\includegraphics[width=8cm]{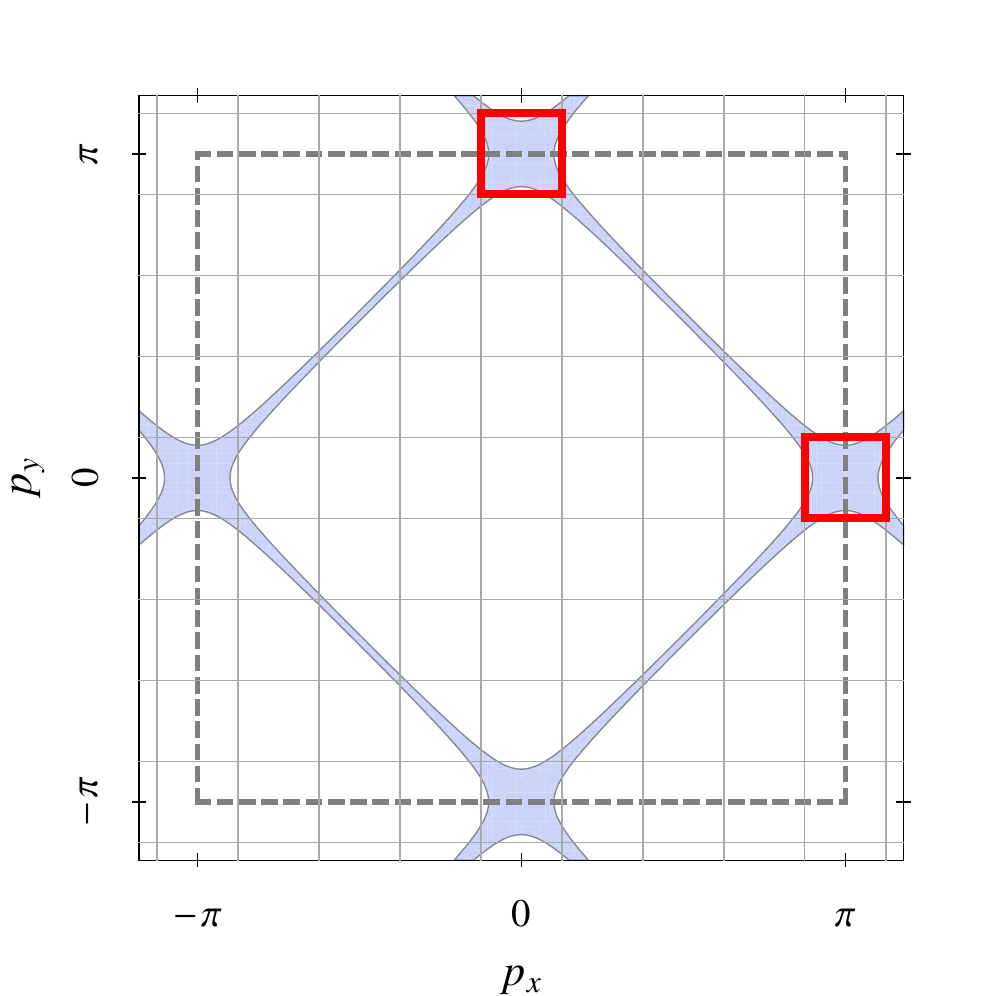}
\caption{Low energy phase space for the two-dimensional square lattice with nearest neighbor hopping at half-filling. $\Lambda=0.1t$, the region with $\left|\epsilon\right| < \Lambda$ is shaded. The phase space cells of the WW basis with $M=4$ are also shown. At the saddle points, the size $\pi/4$ of the cells matches the width of the low energy tube, but the cells are too large for the remaining part of the Fermi surface.}
\label{fig:tube_and_patches_wwrg}
\end{figure}

\noindent
We will be mainly interested in the latter case in the following. Due to the large anisotropy of the Fermi velocity, there is no unique length scale associated to the problem, so that there exists no single wave packet size $M$ that fits for all angles. However, a second glance at Fig.~\ref{fig:tubes} reveals that most of the low energy phase space is concentrated around the saddle points, so that it appears reasonable to adjust the size of the wave packets to this part of the Brillouin zone. Accordingly, the wave packets are too small (in real space) for the remaining part of the Fermi surface. This is shown in Fig.~\ref{fig:tube_and_patches_wwrg} for $\Lambda=0.1t$ and $M=4$. At the saddle points the phase space cell covers the low energy region neatly, but the tube is much narrower than a phase space cell as one moves into the nodal direction. The fact that the cells are too large in momentum space is also reflected in the WW transform of the interactions.

\subsubsection{Approximate decoupling of fast and slow states}
\begin{figure}
\centering
\includegraphics[width=8cm]{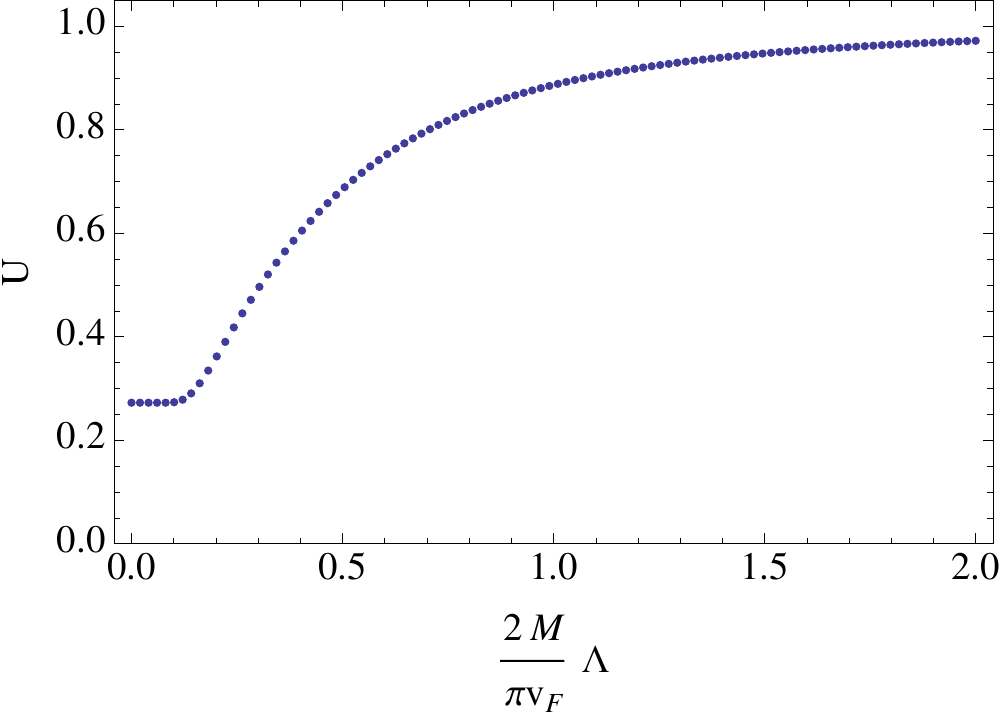}
\caption{Reduction of the local interaction due to size mismatch between low energy tube and phase space cells. The local part of the wave packet transform is shown as a function of $\frac{2M}{\pi v_F}\Lambda$. For $\frac{2M}{\pi v_F}\Lambda<1$, the tube is narrower than the phase space cell, and the exponential cutoff (\ref{eq:F_to_U}) reduces the interaction in the WW basis. A window function with $N/M=8$ is used. For very narrow tubes ($\Lambda\rightarrow 0$), the momentum resolution of this function is not sufficient to resolve the decreasing width of the tube.}
\label{fig:tube_mismatch}
\end{figure}

\noindent
At this point we can appreciate the CUT approach from Sec.~\ref{sec:cut} above, as it treats states below and above the cutoff on the same footing. From Eq.~\ref{eq:F_to_U} we see that states above the cutoff decouple exponentially, whereas states below the cutoff are relatively unaffected by the exponential suppression of the energy transfer. Therefore the WW transform for interactions at the saddle points is essentially the same whether the flowing auxiliary coupling $F\left(\mathbf{p}_1,\ldots,\mathbf{p}_4\right)$ (which is the pendant of the usual RG couplings) or the physical coupling $U\left(\mathbf{p}_1,\ldots,\mathbf{p}_4\right)$ is used. In particular, the coupling can be transformed using the local approximation from Sec.~\ref{sec:ww_trafo_2d}. When the tube is much narrower than a cell, however, the exponential suppression of interactions effectively restricts all the momentum space summations in the wave packet transform to the area of the tube. As a rule of thumb, we have
\begin{equation}
\text{interaction strength} \propto \frac{\text{area in momentum space}}{\text{area of BZ}}.
\end{equation}
Thus we expect the interactions in the nodal directions to decrease in the WW basis representation when the tube is chosen too small. This expectation is born out, and the decrease of the local interaction as a function of $\frac{2M}{\pi v_F}\Lambda$, which measures the size mismatch between the tube and the cells, is shown in Fig.~\ref{fig:tube_mismatch}. The decrease of the local effective interactions leads to a partial decoupling of the slow parts of the Fermi surface from the fast parts, where slow and fast refer to the Fermi velocity compared to $\Lambda$, i.e. a region is fast when $v_F > 2M/(\pi\Lambda)$ and slow otherwise. This effect is increased because of the kinetic energy in the WW basis. The key observation is that the RG flow does not enter the strong coupling regime for all parts of the Fermi surface simultaneously. In particular, for the states at the saddle points the band width in the WW basis is $W_{\text{sp}}\approx 8t/M^2 \sim \Lambda$, where $8t$ is the full band width of the model. The flow goes to strong coupling at the saddle points when the interaction strength $U \approx W_{\text{sp}} \sim \Lambda$. In the nodal region, on the other hand, the band width is $W_{\text{nod}}\sim 16t/M = 2 M W_{\text{sp}}$. Thus, even for the relatively small value $M=4$ we have $W_{\text{nod}} \approx 8 W_{\text{sp}}$. As a consequence, the fast regions behave almost like non-interacting fermions at scale $M$, and correlations involve very delocalized states only.

\noindent
This separation of scales justifies to treat the region around the saddle point in isolation when it becomes strongly coupled. The strong correlations imply that at larger length scales this region should be modeled in terms of the low energy degrees of freedom that emerge from the strongly correlated problem at scale $M$. The coupling to the remaining states at larger length scales involves these new degrees of freedom only. This route is pursued in Ch.~\ref{ch:ww2} for the saddle point regime of the two-dimensional Hubbard model. It should be emphasized that the decoupling of scales does not rely on the fact that the Fermi surface touches the saddle points exactly. Indeed, it is clear that for strong coupling problems at finite $\Lambda$, the width of the tube with $\left|\epsilon\right|<\Lambda$ limits the sensitivity to the precise position of the Fermi surface. For $\Lambda=0.1t$, the width at the saddle points is $\pi/4$, so that the same reasoning should hold when the Fermi surface is dislocated from the saddle points by less than about half this distance.

\noindent
Finally, it is noteworthy that the idea of a decoupling of fast and slow fermions is not new. In fact, it is a well established effect for multi-band systems, where the most celebrated example is probably the Kondo-lattice model of heavy fermion systems \cite{kondolattice1,kondolattice2}, where slow $f$-electrons are modeled as localized spins (i.e. effective low-energy degrees of freedom), whereas the fast $s$-electrons are treated as non-interacting. Another example that is more closely related to the problem here is given by $N$-leg ladders at weak coupling, where a similar decoupling of slow and fast bands has been observed within the renormalization group \cite{nlegladder}.

\chapter{Wave packets and effective Hamiltonians in one dimension}
\label{ch:ww1}

\section{Introduction}

\noindent
In this chapter we apply the WW basis states to three strongly coupled fixed points of RG flows for one dimensional systems: Chains with attractive interactions away from half-filling (Sec.~\ref{sec:attractive_chain}), chains with repulsive interactions at half-filling (Sec.~\ref{sec:repulsive_chain}), and the two-leg ladder with repulsive interactions at half-filling (Sec.~\ref{sec:ladder}). The low-energy phenomenology of these systems is very well understood (see e.g. \cite{giamarchi, gogolin}), so that we can compare the results obtained from the wave packet approach with exact solutions that are obtained from bosonization and Bethe ansatz \cite{giamarchi, gogolin}. We do not aim at quantitative results, and merely seek to obtain qualitative features of the low-energy physics. The main concern in this respect is the distinction between algebraic decays and exponential decays of correlation functions. Since the WW basis breaks the translational invariance of the system, it is not obvious that power-law correlations can be obtained at all.

\noindent
The qualitative nature of the study is reflected in the approximations used: Throughout, we discard all basis states except the ones at the Fermi points, with $k = p_F/K$, where $p_F$ is the Fermi momentum. We use the fixed point Hamiltonians obtained from one-loop RG for the interaction, and expand around the strong coupling limit. Somewhat surprisingly, we recover the nature of the dominant correlations in the ground state in all three cases, despite the simplicity of the approximations. 

\noindent
More explicitly, we will see that within the WW approach, the long range physics at the strong coupling fixed points is to a large extent determined by the structure of the local (at scale $M$) Hilbert space and its ground state degeneracy. This motive is picked up in the study of the two-dimensional Hubbard model in the next chapter.

\section{Renormalization group and wave packets for chains}
\label{sec:rg}

\noindent
In this section we briefly review the one-loop renormalization group equations for the one-dimensional Hubbard model at weak coupling (see e.g.~\cite{giamarchi}). Based on the weak coupling assumption we take into account only interaction terms that are allowed in the vicinity of the Fermi points. Momentum conservation can then be used to  parametrize the interaction in terms of four coupling constants $u_1,\ldots, u_4$ following the so-called g-ology scheme \cite{giamarchi}. 

\subsubsection{Low energy Hamiltonian in momentum space}
\noindent
We begin with the interaction part of the Hamiltonian, and introduce the current operators
\begin{equation}
J_{\alpha,\alpha'}\left(p',p'\right) = \sum_s c^\dagger_{\alpha p_F + p,s}\,c^\pdag_{\alpha' p_F + p'}.
\end{equation}
The indices $\alpha,\alpha'$ label right- and left-movers and take the values $\alpha = \pm1$. Note that this use of $\alpha$ coincides with the one in the definition of the WW basis functions, Ch.~\ref{ch:wwbasis}. 

\begin{figure}
\centering

\caption{Scattering processes corresponding to the four coupling constants $u_1,\ldots,u_4$.}
\label{fig:g-ology_1}
\end{figure}

\noindent
In the spirit of the renormalization group we assume that the interaction does not depend on the momenta relative to the Fermi points, i.e. we set
\begin{eqnarray}
\ham{int} &=& \sum_{\alpha_1\cdots \alpha_4} \tilde{U}\left(\alpha_1 p_F,\ldots, \alpha_4 p_F\right) \delta_{\alpha_1 p_F+\alpha_2 p_F,\alpha_3 p_F+\alpha_4 p_F}, \nonumber \\
&& \; \times \;\frac{1}{N}\sum_{p_1\cdots p_4} \delta_{p_1+p_2,p_3+p_4}\; J_{\alpha_1\alpha_2}\left(p_1,p_2\right) \, J_{\alpha_3\alpha_4}\left(p_3,p_4\right),
\end{eqnarray}
where $\tilde{U}\left(p_1,\ldots,p_4\right)$ is the interaction in momentum representation. Note that this approximation is analogous to the local approximation in the wave packet transformation introduced in Ch.~\ref{ch:wwtrafo}. Momentum conservation restricts the values of the $\alpha_i$, so that one can parametrize
\begin{eqnarray}
 \tilde{U}\left(\alpha_1 p_F,\ldots,\alpha_4 p_F\right) &=& \phantom{+\;}u_1\;\delta_{\alpha_1,-\alpha_3} \delta_{\alpha_2,-\alpha_4}\delta_{\alpha_1,-\alpha_2} \,  \nonumber \\
&& +\; u_2\;\delta_{\alpha_1,\alpha_3} \delta_{\alpha_2,\alpha_4}\delta_{\alpha_1,-\alpha_2} \nonumber \\
&& +\; \frac{u_3}{2}\;\delta_{\alpha_1,-\alpha_3} \delta_{\alpha_2,-\alpha_4}\delta_{\alpha_1,\alpha_2}  \nonumber \\
&& +\; \frac{u_4}{2}\;\delta_{\alpha_1,-\alpha_3} \delta_{\alpha_2,-\alpha_4}\delta_{\alpha_1,\alpha_2}.
\label{eq:g_ology_couplings}
 \end{eqnarray}
Note that $u_3$ is present at half-filling only, when umklapp scattering is allowed at low energies because of $p_F = \pi/2$. The coupling $u_4$ will be neglected in the following, since it does not influence the flow to strong coupling.

\noindent 
The prefactors in (\ref{eq:g_ology_couplings}) are chosen such that for the case of an onsite interaction $U$ the coupling constants have the value
\begin{equation}
u_i = U,
\end{equation}
which is used as initial condition for the renormalization group.

\noindent
The kinetic energy can be linearized around the Fermi points at weak coupling and is given by
\begin{equation}
\ham{kin} = \frac{2\pi v_F}{N}\sum_p \sum_\alpha \,\alpha\,p \; J_{\alpha\alpha}\left(p,p\right).
\end{equation}

\subsubsection{Renormalization group equations and their fixed points}

\noindent
The one-loop RG equations for the coupling constants $u_i$ are given by \cite{solyom, kopietz}:
\begin{eqnarray}
\dot{u}_1 &=& -\frac{1}{\pi v_F} u_1^2 \nonumber \\
\dot{u}_2 &=& -\frac{1}{2\pi v_F} \left( u_1^2 - u_3^2\right)\nonumber \\
\dot{u}_3 &=& -\frac{1}{2\pi v_F} \left(u_1 - 2 u_2\right) u_3, 
\label{eq:rgflow}
\end{eqnarray}
where the dot is shorthand for the logarithmic scale derivative $\frac{d}{ds} = -\frac{1}{\Lambda}\frac{d}{d\Lambda}$, so that $s= e^{-\Lambda/W}$. $\Lambda$ is the renormalization scale, and $W$ is the initial bandwidth. We will be interested in two cases, both involving finite scale singularities. The first one is the half-filled repulsive Hubbard model, for which the strong coupling fixed-point of (\ref{eq:rgflow}) is given by
\begin{eqnarray}
2 u_2 &=&  u_3 = u_{\text{AF}} > 0\nonumber \\
u_1 &=& 0.
\label{eq:AF_fixedpoint}
\end{eqnarray}
This fixed point is characterized by a diverging staggered spin-susceptibility.

\noindent
The second one is the attractive Hubbard model away from half-filling (i.e. $g_3=0$), for which the fixed point couplings are
\begin{equation}
u_1 = 2u_2 = u_{\text{SC}} < 0.
\label{eq:SC_fixedpoint}
\end{equation}
This fixed point is characterized by diverging singlet pairing correlations. 
 
 \noindent
The couplings $u_{\text{AF}}$ and $u_{\text{SC}}$ diverge at a critical scale $\Lambda_c$, that depends on the initial interaction strength $U$ and the full band width $W$. Here we are not interested in the magnitude of this scale. It is sufficient to know that at some scale the couplings exceed the band width below the cutoff, at which point the perturbative approach breaks down.

\noindent
In order to obtain a qualitative picture of the fixed point behavior, we assume in the following that the couplings define the largest energy scale in the problem, and investigate the strong coupling limit. This approach is similar to the semi-classical approximation to the sine-Gordon problem that describes the low-energy physics of one-dimensional chains in the bosonization method \cite{giamarchi,gogolin}. In order to study the fixed points, we need the WW representation of the fixed point Hamiltonian first.

\noindent
The kinetic energy part has been discussed in Sec.~\ref{sec:wwtrafo_1}, and will not be important in what follows.

\subsubsection{WW representation of $\ham{int}$}

\begin{figure}
\centering
\includegraphics[width=8cm]{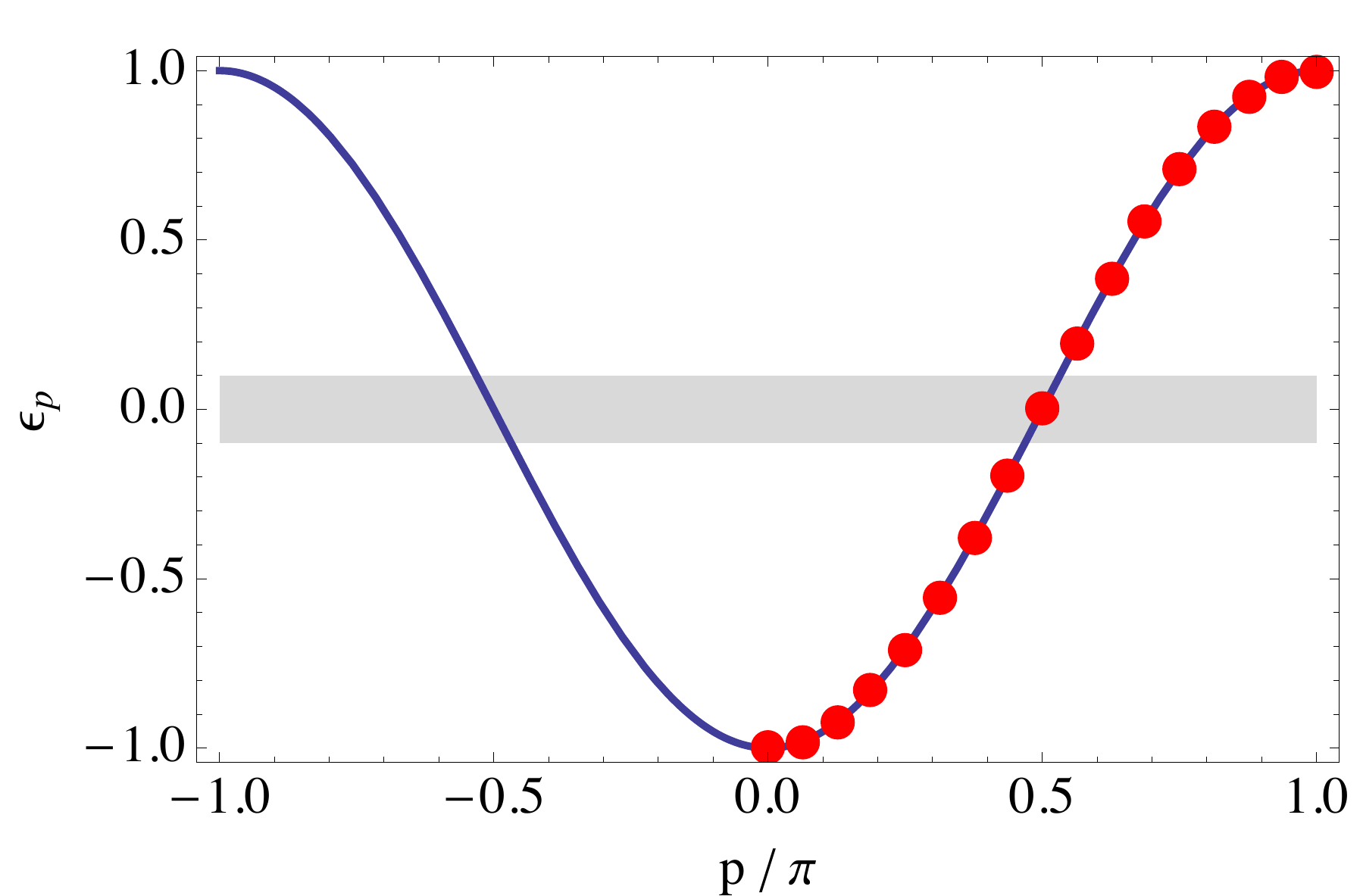}
\caption{The simplest wave packet approximation for chains: Only one WW momentum $k = p_F/K$ is kept, the remainder is discarded. It is assumed that $M$ is chosen such that only this state lies below the cutoff (shaded region). WW basis states are marked by red dots at momentum $Kk$, and drawn on top of the dispersion of the chain.}
\label{fig:ww_chain_cutoff}
\end{figure}

\noindent
Moreover, we will restrict the WW basis to states $\ket{mk}$ with $k=p_F/K \equiv k_F$, and assume that $M$ is chosen such that only this state lies below the cutoff, as indicated in Fig.~\ref{fig:ww_chain_cutoff}. In order to investigate the properties of the fixed points, we first transform the fixed point interactions to the WW basis using the methods from Sec.~\ref{sec:ww_trafo_interaction_1}, in particular Eq.~(\ref{eq:U_wp_to_ww_1}), which we restate here for convenience. Since we set all the $k_i$ to $p_F/K$, we suppress the WW momentum index $k$ in the following. Then Eq.~(\ref{eq:U_wp_to_ww_1}) becomes
\begin{eqnarray}
U\left(m_1,\ldots,m_4\right) &=& \frac{V\left(m_1,\ldots,m_4\right)}{2}\,\frac{1}{2} \sum_{\alpha_1\cdots\alpha_4} \, \tilde{U}\left(\alpha_1p_F,\ldots,\alpha_4p_F\right)\nonumber\\
&& \;\; \times \;e^{-i\left(\alpha_1\phi_{1} +\alpha_2\phi_{2} - \alpha_3\phi_{3} - \alpha_4 \phi_{4}\right)},
\label{eq:U_trafo_chain}
\end{eqnarray}
where $\phi_i = \phi_{m_i}$, and $\phi_a = \pi/2$ ($\phi_a=0$) for $a$ even (odd). The values of $V\left(m_1,\ldots,m_4\right)$ depend on the window function, the values we use are tabulated in Tab.~\ref{tab:V_values}.

 \noindent
Plugging the g-ology couplings (\ref{eq:g_ology_couplings}) into the right hand side of (\ref{eq:U_trafo_chain}), we observe that the Kronecker deltas can be used to perform three of the four sums over the $\alpha_i$. We evaluate the remaining sum for the $u_1$ term (the other terms being similar) only, and state the results for the other terms. We note that $\alpha_1=-\alpha_2=-\alpha_3=\alpha_4=\alpha$, and find
 \begin{equation}
\frac{1}{2} \sum_{\alpha} e^{-i\alpha\left(\phi_{1} - \phi_{2} + \phi_{3}-\phi_{4}\right)} = \cos \left[\phi_{1}+\phi_{3}-\phi_{2} - \phi_4 \right]
\end{equation}
Now recall from Eq. (\ref{eq:def_phi}) that $\phi_i$ can take on the values $0$ (for $m+k$ even) and $\pi/2$ (for $m+k$ odd) only. Then the $\cos$ vanishes when an odd number of operators acts on an odd (i.e. $m+k$ odd) WW orbital, since its argument is either $\pi/2$ or $3\pi/2$. In particular, terms of the form $m_1=m_2=m_3= m_4\pm 1$ vanish, since there is always an even number of odd $k_i$ (for $k$-conserving matrix elements). Since the interactions decay rapidly with distance, the contribution from this type of interaction is thus strongly suppressed in one dimension, and we neglect it in the following. Now we turn to the type $m_1=m_2\neq m_3=m_4$. Focussing on nearest neighbor interactions and setting all $k = p_F/K$, we find that the cosine contributes $\pm 1$, depending on which of the $m_i$ are even (odd). Thus there are no terms that vanish exactly in this case. Finally, when all sites involved are even (odd), the cosine always evaluates to $1$.

\noindent
The general form of the Hamiltonian for the states at the Fermi points, $k_i = p_F/K$ can now be written down using similar considerations for the other parts of the interaction. In order to simplify the notation, we define
\begin{equation}
\hat{J}\left(m_1,m_2\right) \equiv \sum_s \, \gamma^\dagger_{m_1,p_F/K,s} \, \gamma^\pdag_{m_2,p_F/K,s}.
\end{equation}
Then the WW Hamiltonian for the states at the Fermi points is given by
\begin{eqnarray}
\ham{int}\Big|_{k_i=p_F/K} &\approx& \frac{1}{4M} \sum_{m} \left(u_1+u_2+u_3/2+u_4/2\right) \, \hat{J}\left(m,m\right) \, \hat{J}\left(m,m\right) \\
&&\; + \; \frac{1}{8M}  \sum_{\langle m,m'\rangle} \Big[\left(-u_1+u_2-u_3/2+u_4/2\right) \hat{J}\left(m,m\right)\,\hat{J}\left(m',m'\right)\nonumber \\
&&\; + \; \left(u_1-u_2-u_3/2+u_4/2\right) \hat{J}\left(m,m'\right)\,\hat{J}\left(m',m\right)\nonumber \\
&& \; + \; \left(u_1+u_2-u_3/2-u_4/2\right) \hat{J}\left(m,m'\right)\,\hat{J}\left(m,m'\right) \,\Big]
\label{eq:u_i_trafo_chain}
\end{eqnarray}
where we have used the values of $V(0,0,m,m)$ from Tab.~\ref{tab:V_values}. The WW transforms of the fixed point couplings can be obtained in a straightforward manner from (\ref{eq:u_i_trafo_chain}). This is used in the following two sections to analyze the two fixed points above.

\section{Chain with repulsive interactions at half-filling}
\label{sec:repulsive_chain}

\noindent 
In this section we discuss the fixed point $2 u_2 =  u_3 = u_{\text{AF}} > 0$ for the half-filled chain with repulsive interactions. We use Eq.~(\ref{eq:u_i_trafo_chain}) in order to obtain the WW representation of the interaction. We proceed by diagonalizing the local part of the Hamiltonian, and find that charge degrees of freedom are gapped, so that locally only the spin degree of freedom survives. As a consequence, the ground state of the local interaction Hamiltonian has two-fold degeneracy per site, corresponding to the two possible spin states. In the next step, we show that the non-local part of the interaction is of the Heisenberg form, so that it leaves the local low-energy subspace invariant. The Heisenberg model can be solved using the Bethe ansatz \cite{giamarchi}, so that the asymptotic form of the spin-spin correlation functions can be obtained. The correlation functions of the effective model in the WW basis are then transformed back to momentum space, revealing a power-law form of the spin-spin correlation function for momenta close to $p = 0$ and $p=\pi$. This result is in qualitative agreement with the bosonization solution of the chain problem \cite{giamarchi,gogolin}.

\subsubsection{Fixed point interaction in WW representation}

\noindent
The fixed point interaction can be transformed to the WW basis by plugging $2u_2 = u_3 = u_{\text{AF}}$ into (\ref{eq:U_trafo_chain}). The result is 
\begin{equation}
\ham{AF} = \underbrace{\frac{1}{4M} u_{\text{AF}} \sum_{m} \hat{J}\left(m,m\right)\,\hat{J}\left(m,m\right)}_{\text{local interaction}} -\underbrace{\frac{1}{4M} u_\text{AF} \sum_m \hat{J}\left(m,m+1\right)\,\hat{J}\left(m+1,m\right)}_{\text{nearest neighbor interaction}}.
\end{equation}
Using the identity 
\begin{equation}
\sum_{i=1}^3 \sigma^i_{ab}\,\sigma^i_{cd} = \frac{1}{2}\delta_{ad}\,\delta_{bc} - \frac{1}{4} \delta_{ab}\,\delta_{cd}
\end{equation}
the non-local part of the interaction can be rewritten as
\begin{equation}
-\sum_m \hat{J}\left(m,m+1\right)\,\hat{J}\left(m+1,m\right) = 2\sum_m \hat{\mathbf{S}}\left(m\right) \cdot \hat{\mathbf{S}}\left(m+1\right) + \frac{1}{2} \hat{n}\left(m\right)\,\hat{n}\left(m+1\right),
\label{eq:af_chain_hnn}
\end{equation}
where the $\hat{S}^i\left(m\right)$ and $\hat{n}\left(m\right)$ are spin and charge operators at site $m$, respectively. Similarly, the local part can be rewritten as
\begin{equation}
\ham{loc} = \frac{1}{4M} u_{\text{AF}}\sum_{m} \hat{J}\left(m,m\right)\,\hat{J}\left(m,m\right) =\frac{1}{4M} u_{\text{AF}}\sum_m \hat{n}\left(m\right) \,\hat{n}\left(m\right).
\label{eq:af_chain_hloc}
\end{equation}

\subsubsection{Local Hilbert space and effective spin model}

\noindent
In order to arrive at suitable low-energy degrees of freedom, we investigate the local part (\ref{eq:af_chain_hloc}) of the interaction first. The local Hilbert space consists of the four states $\ket{0}$, $\ket{\uparrow}$, $\ket{\downarrow}$, and $\ket{\uparrow\downarrow}$. Since the wave packets lie at the Fermi surface, there is one fermion per site, and after adjusting the chemical potential accordingly, the energies of the four states are
\begin{eqnarray}
\ham{loc}\ket{\uparrow} &=& 0\nonumber \\ \ham{loc}\ket{\downarrow} &=& 0 \nonumber \\ \ham{loc} \ket{0} &=& \frac{1}{4M} u_{\text{AF}}\ket{0} \nonumber \\ \ham{loc}\ket{\uparrow\downarrow} &=& \frac{1}{4M} u_{\text{AF}}\ket{\uparrow\downarrow}.
\end{eqnarray}
Hence the low energy sector of the local Hamiltonian consists of the two states $\ket{\uparrow}, \ket{\downarrow}$. Since these states are degenerate, the effective Hamiltonian to leading order is found by simply projecting each WW site down to the spin sector. 

\noindent
The action of the non-local interactions in the spin sector can be inferred from (\ref{eq:af_chain_hnn}). The charge operators $\hat{n}\left(m\right)$ have no effect on the spin sector. Therefore we are left with the nearest-neighbor spin-spin interactions. Hence we arrive at the effective Hamiltonian
\begin{equation}
\ham{eff} = J \sum_m \hat{\mathbf{S}}\left(m\right)\cdot \hat{\mathbf{S}}\left(m+1\right),
\end{equation}
where $J = \frac{1}{2M}u_{\text{AF}} > 0$. The effective model is simply an antiferromagnetic $S-1/2$ Heisenberg model. 

\subsubsection{Asymptotic spin-spin correlation function}

The properties of the Heisenberg model in one dimension are well known, and the spin-spin correlation function has been derived using the Bethe ansatz \cite{luther,giamarchi}:
\begin{equation}
\left\langle \hat{\mathbf{S}}\left(m\right)\cdot \hat{\mathbf{S}}\left(m'\right)\right\rangle = C_1 \frac{1}{\left(m-m'\right)^2} + C_2 \left(-1\right)^{m-m'} \frac{1}{|m-m'|},
\label{eq:spincorrelation}
\end{equation}
where the $C_i$ are constants. It follows that there are soft excitations at the points $0$ and $\pi$ of the Brillouin zone of the superlattice defined by the WW basis states. 
 \begin{figure}
\centering
\includegraphics[width=6cm]{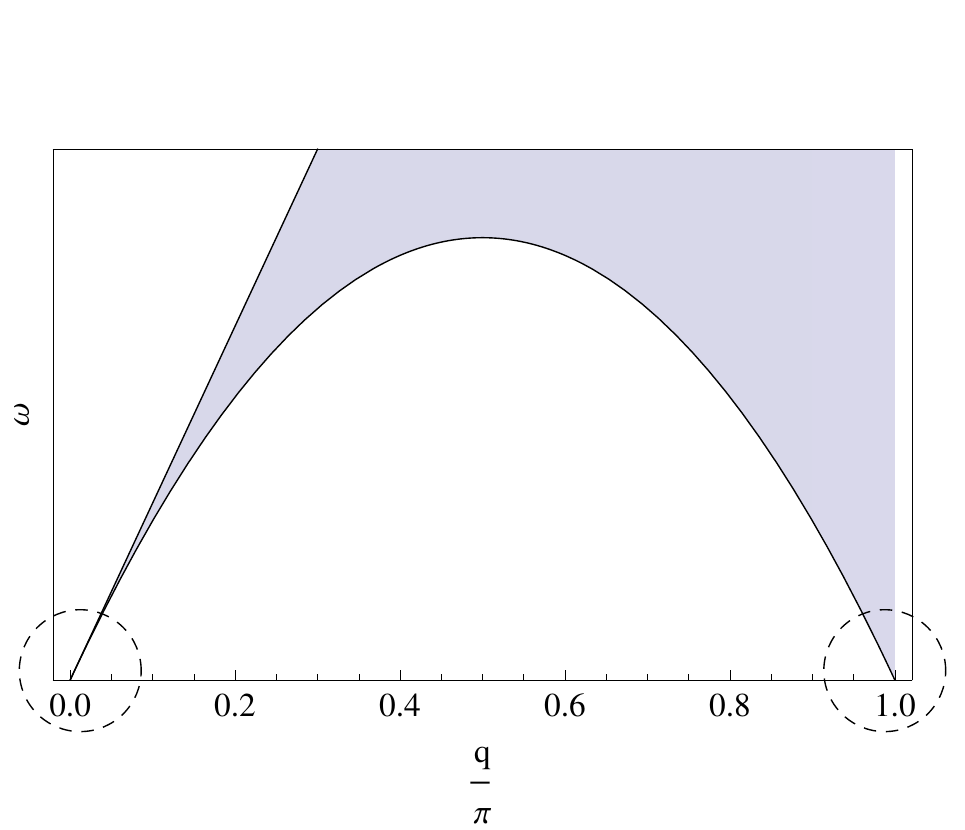}
\caption{Schematic spectral weight of spin-density excitations of the repulsive Hubbard model at half-filling. Low energy excitations exist at $q\approx0$ and $q\approx\pi$. The spin-density excitations in the truncated WW basis are confined to the encircled regions when transformed back into momentum space. At the points $q=0, \pi$ the correspondence is exact (cf. (\ref{eq:spintrafo_af_chain})), but for general momenta excitations around $q=0$ and $q=\pi$ are mixed because translational invariance is broken.}
\label{fig:spin_dispersion}
\end{figure}

\noindent
Now we transform this correlator back to momentum space in order to see what these results mean. We denote the momentum representation of the spin density by $\tilde{s}\left(q\right)$, where $q$ is the momentum transfer. Since the WW basis states are localized in momentum space around $\pm \pi/2$, the momentum transfer $q$ of two-fermion operators such as the $\hat{S}^i\left(m\right)$ is restricted to the vicinity of the two points $q=0$ and $q=\pi$, as indicated in Fig.~\ref{fig:spin_dispersion}. In Ch.~\ref{ch:ww_pairing} we have found that the staggered magnetization $\tilde{s}^i\left(q=\pi\right)$ is given by
\begin{eqnarray}
\tilde{s}^i\left(q=\pi \right) &=& \sum_j \left(-1\right)^j \sigma^i_{ss'} c^\dagger_{j,s}\,c^\pdag_{j,s'} \nonumber \\
&=& \sum_{mk} \left(-1\right)^{m+k} \sigma^i_{ss'} \gamma^\dagger_{mk,s} \, \gamma^\pdag_{m,M-k,s'} \nonumber \\
&\approx& \sum_{m} \left(-1\right)^m \sigma^i_{ss'} \gamma^\dagger_{m,M/2,s},\gamma^\pdag_{m, M/2,s'}\nonumber \\
&=& \sum_{m}\left(-1\right)^m \, \hat{S}^i\left(m\right),
\label{eq:spintrafo_af_chain}
\end{eqnarray}
in the WW basis. Therefore the power law in the staggered magnetization in the WW basis corresponds to the same power law in the staggered magnetization in real space. A similar line of reasoning for the uniform spin-density, $\tilde{s}^i\left(q=0\right)$ shows that in the same manner the uniform spin density in the WW basis corresponds to the uniform spin density in the original lattice. The momenta that lie far away from the center and boundary of the Brillouin zone of the WW basis are in general superpositions of momenta close to $0$ and $\pi$, reflecting the broken translational invariance.

\noindent
In summary, the asymptotic behavior of the spin-spin correlation function in the real space lattice inferred from the effective model is the same as in eq. (\ref{eq:spincorrelation}), with $m-m'$ replaced by $j-j'$. The wave velocity of excitations has to be rescaled because the unit cell is larger in the WW basis by a factor of $M$. We conclude that our approximation produces algebraically spin-density correlations round $q=0$ and $q=\pi$, and gaps for all charged excitations in agreement with bosonization treatments. However, the exponents of the power law decays, which are influenced by the Luttinger liquid physics are not recovered. This is not surprising given that the Luttinger liquid physics has its origins in the asymptotically linear fermion dispersion \cite{luttinger}, a feature that is hard to conserve in a cluster approximation (not to mention that the Fermi velocity appears nowhere in the present treatment). Nevertheless, we emphasize that the qualitative features of the model including algebraic decays are recovered, which is in our view a non-trivial result, in particular when taking into account the simplicity of the approximation.

\section{Chain with attractive interactions}
\label{sec:attractive_chain}

\noindent
Now we consider the fixed point for attractive interactions at arbitrary filling, characterized by $u_1 = 2u_2 = -u_{\text{SC}} < 0$. We follow the same step as in Sec.~\ref{sec:repulsive_chain} above. First we obtain the WW representation of the interaction Hamiltonian. We analyze the local part of the interaction and find that the low energy sector contains only singlet pairs of fermions. Then we map the projected Hamiltonian to a spin problem and deduce the asymptotic correlation function from the Bethe ansatz solution.

\subsubsection{Fixed point interaction in WW representation}

\noindent
The WW transform of the interaction Hamiltonian is again obtained using (\ref{eq:u_i_trafo_chain}). The local part is given by
\begin{eqnarray}
\ham{loc} &=& -\frac{3}{4M} u_{\text{SC}} \sum_m \hat{J}\left(m,m\right)\,\hat{J}\left(m,m\right) \nonumber \\
&=& -\frac{3}{4M} u_{\text{SC}} \sum_m \hat{n}\left(m\right)\,\hat{n}\left(m\right).
\end{eqnarray}

\noindent
For the non-local part, we introduce the additional pair annihilation operators
\begin{equation}
\hat{\Delta}\left(m\right) = \gamma_{m,p_F/K,\downarrow}\, \gamma_{m,p_F/K,\uparrow}
\end{equation}
and their hermitian conjugates $\hat{\Delta}^\dagger\left(m\right)$. In terms of the spin-, charge- and pair-operators the non-local interactions are given by
\begin{eqnarray}
\ham{n.n.} &=& \frac{3}{4M} u_{\text{SC}} \sum_m \Big[-\hat{\Delta}^\dagger\left(m\right) \, \hat{\Delta}\left(m+1\right) - \hat{\Delta}\left(m\right) \, \hat{\Delta}^\dagger\left(m+1\right)  \nonumber \\ &&\qquad\qquad\qquad\;\;+ \frac{1}{2} \hat{n}\left(m\right)\hat{n}\left(m+1\right) + \hat{\mathbf{S}}\left(m\right)\cdot \hat{\mathbf{S}}\left(m+1\right) \Big]
\label{eq:hnn_sc_chain}
\end{eqnarray}

\subsubsection{Local Hilbert space and effective spin model}

\noindent 
The analysis of the local Hilbert space and Hamiltonian is the same as in Sec.~\ref{sec:repulsive_chain}, except that now the interactions are attractive, so that the low-energy states are $\ket{0}$ and $\ket{\uparrow\downarrow}$. Again, the ground state is two-fold degenerate per site so that we obtain the effective Hamiltonian by projection onto the degenerate subspace. The spin operators in the second line of (\ref{eq:hnn_sc_chain}) do not contribute since in the low energy sector all spins are bound into singlet pairs. In order to treat the remaining interactions we note that the local low energy subspace can be mapped onto a spin system with $S=1/2$, similar to the repulsive chain above. 

\noindent
The mapping is accomplished by identifying
\begin{eqnarray}
\hat{S}^z\left(m\right) &\equiv& \frac{1}{2}\left(\hat{n}\left(m\right) - 1\right) \\
\hat{S}^-\left(m\right) &\equiv& \hat{\Delta}\left(m\right). 
\label{eq:spin_mapping_sc_chain}
\end{eqnarray}
In terms of the spin operators, the projection of the interaction Hamiltonian to the low energy subspace becomes
\begin{equation}
\begin{split}
\ham{eff} =&\;   \sum_m \Big[-\frac{J}{2}\left(\hat{S}^+\left(m\right) \hat{S}^-\left(m+1\right) + \hat{S}^-\left(m\right) \hat{S}^+\left(m+1\right)\right) \\&\;\;\;\;\;\;\;\;\;+ J\hat{S}^z(m)\,\hat{S}^z\left(m+1\right)\Big],
\end{split}
\label{eq:hameff_sc_chain}
\end{equation}
where $J=\frac{3}{8M}u_{\text{SC}}$. The effective Hamiltonian is hence an XXZ spin chain model. Note that the sign of the first term can be reversed by a gauge transformation on the spin states, e.g. $\ket{\uparrow}\rightarrow -\ket{\uparrow}$ on every second site without changing the model. Consequently, we find that the effective model is again an antiferromagnetic $S-1/2$ Heisenberg model. Note however, that the emergent $SU(2)$-symmetry is due to the approximation, and that in general the symmetry is $U(1)$.

\subsubsection{Asymptotic correlation functions}

\noindent
Since the effective Hamiltonian (\ref{eq:hameff_sc_chain}) is again of the Heisenberg type, we do not need to discuss the asymptotic correlations in the WW basis, for the results we refer to Eq.~(\ref{eq:spincorrelation}). Instead we discuss the meaning of the different correlators in real space. In the same way as before, the uniform spin-densities in the WW basis correspond to uniform spin-densities in real space. The staggered spin-density in the WW basis corresponds to spin-density in real space that oscillates with wave vector $p_F$ (instead of $\pi/2$). Translating the spin operators back to the fermion operators (using Eq.~\ref{eq:spin_mapping_sc_chain}), we obtain that the charge density (corresponding to $\hat{S}^z$) becomes soft at the momenta $q=0$ and $q=2p_F$. The pair correlations (corresponding to $\hat{S}^\pm$) show algebraic decay for the same momenta. 

\noindent
Comparison with the solution from bosonization \cite{giamarchi} shows that the slow decay of pair correlations with momentum $2p_F$ is an artifact of our approximation. The other power laws are in qualitative agreement, however. For the same reasons as for the repulsive chain, it is clear that the precise power laws connected to Luttinger liquid physics can not be recovered.

\section{Two-leg ladder at half-filling}
\label{sec:ladder}

\noindent
In this section we discuss the low energy behavior of the SO(5) symmetric two-leg ladder at half-filling. Since the Hubbard ladder has two legs, the diagonalization of the hopping term leads to two bands. The bands will be labelled by indices like $a,b,c,\ldots = 1,2$ in the following. There are four Fermi points in total which are all equivalent in the sense that the Fermi velocity is the same (because of particle-hole symmetry). Compared to the single chains considered so far, this opens up the possibility of competition between different order parameters. Indeed, it is well established that the Hubbard ladder has a Mott insulating ground state at half-filling, regardless of the interaction strength. In this so-called $d$-Mott phase \cite{linbalents}, $d$SC and AF correlations are strong, but decay exponentially, and all excitations are gapped. Remarkably, the single particle excitations at weak coupling are of the same order of magnitude as the bosonic spin- and pair-excitations.

\noindent
The renormalization group flow for the Hubbard ladder at half-filling has been derived by Lin and Balents \cite{linbalents}. There it was also shown that the system generically flows to strong coupling, and that the fixed point displays an enhanced SO(8) symmetry. The resulting low-energy Hamiltonian turns out to be exactly solvable, with a so-called $d$-Mott ground state, which has gaps for all excitations. When the high-energy Hamiltonian is SO(5) symmetric, the system retains this symmetry, which shows up as a degeneracy of $d$-wave pair excitations and AF spin-excitations. Since the interplay between superconductivity and antiferromagnetism is our main interest, we focus on the SO(5) case. 

\begin{figure}
\centering
\includegraphics[width=8cm]{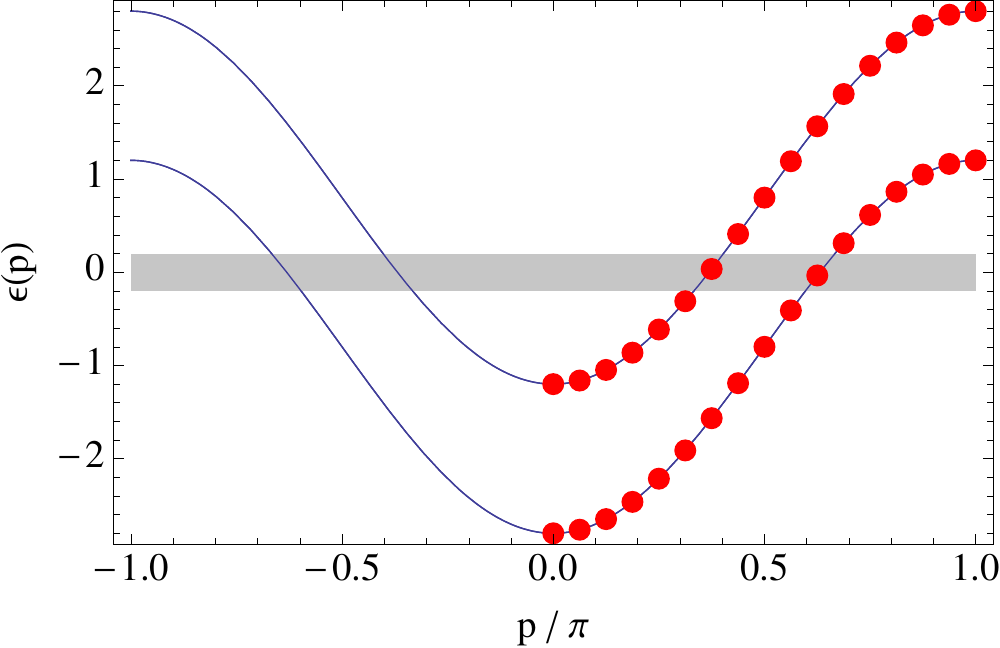}
\caption{The simplest wave packet approximation for the two-leg ladder. The blue lines show the kinetic energy of the two bands as a function of $p$. The red dots indicate the mean momenta and mean kinetic energy of the WW basis states. Since there are two bands, there are two families of WW basis states as well. In the simplest approximation to the low energy problem, we truncate the basis and keep only states at the Fermi points.}
\label{fig:band_ladder}
\end{figure}

\noindent
The existence of more than one band poses no problems for the transformation to the WW basis. Since the problem is weakly coupled, we transform each band to the WW basis separately, and label all WW quantities by an additional band index. The position of the WW states in the Brillouin zone is illustrated in Fig.~\ref{fig:band_ladder}. In the following, we use the same approximations as before, namely, we truncate the WW basis and keep only the states at the Fermi points. We use the fixed point couplings of the RG to define the interaction Hamiltonian, and solve it in the strong coupling limit.

\noindent
We will see that within the WW basis, the presence of two bands leads to a local problem with a non-degenerate ground state. This is to be contrasted with the generic case of degenerate ground states for chains above. The ground state features pronounced $d$-wave SC and AF correlations. The lowest excited states fall into two classes: There are eight degenerate fermionic excitations, and five degenerate bosonic excitations that correspond to the vector bosons above. Both types of excitations have comparable gaps. We find that this situation is robust against the perturbation by non-local interactions and fermion hopping, hence reproducing the exact solution of this problem qualitatively.

\subsubsection{Effective Hamiltonian}

\noindent
The SO(5) symmetry at low energies is realized in terms of local fermion bilinears in the continuum limit. The Lie algebra is generated by 10 operators: The particle number, the three spin operators and six so-called $\pi$-generators which create and annihilate triplet pairs with total momentum $\left(\pi,\pi\right)$. Because of the high symmetry of the kinetic energy part of the Hamiltonian, all these operators commute with the kinetic energy. The Lie algebra generators will not be important in the following, so that we refer the interested reader to the review by Zhang \cite{zhang98}. In addition to the generators, there are five bilinears that transform in the vector representation of SO(5). In order to avoid amassing more indices than necessary, we omit position labels, and write $R_{as}$ and $L_{as}$ instead of $c_{p\pm\alpha p_F, a,s}$. With this notation, the vector operators are given by
\begin{eqnarray}
\tilde{\Delta}^x &=& \frac{1}{2}\sigma^z_{ab}\,\epsilon_{ss'}  \left( R^\pdag_{a s} \, L^\pdag_{b s'} + R^\dagger_{a s'} \,L^\dagger_{b s}\right) \nonumber \\
\tilde{\Delta}^y &=& \frac{i}{2}  \sigma^z_{ab} \,\epsilon_{ss'}\left( R^\pdag_{a s} \, L^\pdag_{b s'} - R^\dagger_{a s'} \, L^\dagger_{b s}\right) \nonumber \\
\tilde{A}^i &=& \sigma^x_{ab}\,\sigma^i_{ss'} \left( R^\dagger_{a s} \, L^\pdag_{b s'} + L^\dagger_{a s}\, R^\pdag_{b s'}\right).
\end{eqnarray}
The effective interaction can be written in terms of these operators. Due to the symmetry, it is convenient to introduce the vector $\tilde{\mathbf{B}} = \left(\tilde{\Delta}_x, \tilde{\Delta}_y, \tilde{A}^x, \tilde{A}^y, \tilde{A}^z\right)$.The fixed-point interaction is given by \cite{linbalents}
\begin{equation}
\ham{int} = -u \int dr \, \tilde{\mathbf{B}}(r)\cdot\tilde{\mathbf{B}}(r)
\label{eq:hint_ladder}
\end{equation}
in the continuum limit, where $u>0$. The operators $\tilde{A}^i$ create AF spin-fluctuations with momentum $\left(\pi,\pi\right)$. Note that this involves a transition from one band to the other. The operators $\tilde{\Delta}^i$ create $d$-wave pair excitations with both fermions of a pair in the same band, and a relative minus sign of the phase between the bands.

\subsubsection{WW representation of the effective Hamiltonian}

\noindent
We now turn to the transformation of the interaction (\ref{eq:hint_ladder}) to the WW basis. Surprisingly, the transformation is simpler for the ladder than for the chain systems. The reason is that the interaction is written as a scalar product of operators with definite parity. In fact, all components of the vector $\mathbf{B}(r)$ are even under parity. Since the parity of a bilinear is the product of the parity of its constituting fermions, the fact that the parity of the WW basis states is opposite on neighboring sites can be expected to lead to cancellations of terms in the WW transformation.

\noindent
We show the highlights of the transformation for the $d$SC operators only, since the AF operators behave essentially in the same way. We use the local approximation as before, so that we only need the matrix elements $\bar{U}_\Delta\left(a_1,\alpha_1,s_1;\ldots;a_4,\alpha_4\,s_4 \right)$ at the Fermi points:
\begin{equation}
\bar{U}_\Delta\left(a_1,\alpha_1,s_1;\ldots;a_4,\alpha_4,s_4\right) = -u \; \underbrace{\frac{1}{4}\epsilon^T_{s_1,s_2} \epsilon_{s_3,s_4}}_{\text{spin singlet}} \;\underbrace{\delta_{a_1a_2}\,\delta_{a_3a_4}}_{\text{pair within one band}} \; \underbrace{\delta_{\alpha_1, - \alpha_2}\; \delta_{\alpha_3,-\alpha_4}}_{\text{pair momentum }=0},
\end{equation}
where we have omitted the $p_F$'s on the left hand side. The important point is that there are only two Kronecker deltas involving the $\alpha_i$, and that the first pair of coordinates is indpendent of the second pair of coordinates, because the of the factorization into bilinears. In the orthogonalization formula (\ref{eq:U_trafo_chain}) the summation over $\alpha_i$ then leads to
\begin{eqnarray}
U\left(m_1,\ldots,m_4\right) &\propto& \frac{1}{4}\sum_{\alpha_1\cdots \alpha_4} e^{-i \left(\alpha_1\phi_1 + \alpha_2 \phi_2 - \alpha_3 \phi_3 - \alpha_4 \phi_4\right)} \delta_{\alpha_1,-\alpha_2}\,\delta_{\alpha_3,-\alpha_4} \nonumber \\
&=& \left[\frac{1}{2}\sum_{\alpha_1} e^{-i \alpha_1\left(\phi_1 - \phi_2\right)} \right]\,\times\, \left[\frac{1}{2}\sum_{\alpha_2} e^{-i\alpha_2\left(\phi_3-\phi_4\right)}\right]  \nonumber \\
&=& \cos\left[\phi_1 - \phi_2\right]  \cos\left[\phi_3-\phi_4\right]\phantom{\Big|}.
\end{eqnarray}
Recalling that $\phi_i = \phi_{m_i}$ at the Fermi points, we see that the contribution is finite for $m_1 = m_2 \mod 2$ and $m_3 = m_4\mod 2$. Since we keep nearest neighbor interactions only, this leads to $m_1=m_2$, $m_3=m_4$. In the SC case this implies that interactions involve local pairs only, which may interact locally or hop to the neighboring site. The same conclusion holds for the operators $\tilde{A}^i$, which involve local particle-hole pairs only.

\noindent
In summary, the local Hamiltonian can be expressed conveniently in terms of the operators
\begin{eqnarray}
\hat{\Delta}^-\left(m\right) &=& \frac{1}{\sqrt{2}}\sigma^z_{ab}\,   \gamma^\pdag_{a, \downarrow} \, \gamma^\pdag_{b \uparrow}  \nonumber \\
\hat{\Delta}^+\left(m\right) &=& \frac{1}{\sqrt{2}}\sigma^z_{ab}\,   \gamma^\dagger_{a, \uparrow} \, \gamma^\dagger_{b \downarrow}  \nonumber \\
\hat{A}^i\left(m\right) &=& \sigma^x_{ab}\,\sigma^i_{ss'} \, \gamma^\dagger_{a s} \, \gamma^\pdag_{b s'}
\end{eqnarray}
and the corresponding local SO(5) vector operator
\begin{equation}
 \mathbf{B}\left(m\right) = \left((\hat{\Delta}^-(m)+\hat{\Delta}^+(m))/2, i(\hat{\Delta}^-(m)-\hat{\Delta}^+(m))/2, \hat{A}^x(m), \hat{A}^y(m), \hat{A}^z(m)\right). 
\end{equation}

\noindent
The interaction is given by
\begin{equation}
\ham{int} = -\frac{1}{4M}u \sum_m \left(\mathbf{B}(m)\cdot\mathbf{B}(m) +\frac{1}{2} \mathbf{B}(m)\cdot\mathbf{B}(m+1)\right).
\end{equation}
As usually, the numerical prefactors depend on the choice of the window function.

\subsubsection{Local Hilbert space and effective quantum rotor model}

\noindent
Following the same routine as before, we start by investigating the local Hilbert space of the problem, focussing on the bosonic states. The ground state $\ket{0}$ is non-degenerate, and consists of one $d$-wave pair per site, 
\begin{equation}
\ket{0} \equiv \kettwo{\uparrow\downarrow}{0} - \kettwo{0}{\uparrow\downarrow}.
\end{equation}
In addition, there are five degenerate bosonic excited states that are obtained by applying the components of the vector operator $\mathbf{B}$ to the ground state, 
\begin{eqnarray}
\ket{A^+} &=& \kettwo{\uparrow}{\uparrow} \;\; \times\; 3, \nonumber \\
\ket{\Delta^+} &=& \kettwo{\uparrow\downarrow}{\uparrow\downarrow},\; \text{and}\\
\ket{\Delta^-} &=& \kettwo{0}{0},
\end{eqnarray}
where $\ket{A^i} = \hat{A}^i\ket{0}$ etc. Finally, there are eight equivalent single fermion excitations. The energies of these two types of excitations are are very close to each other. In units of the WW basis coupling, they are given by 
\begin{eqnarray}
E_{\text{fermion}} &=& \frac{15}{4} \nonumber \\
E_{\text{boson}} &=& 4.
\end{eqnarray}

\noindent
This situation is quite different from the behavior of the single chains, where the bosonic (spin or pair) excitations are in the degenerate ground state manifold of the local Hamiltonian, and thus far below the single particle spectrum. In the ladder case, the bosonic and fermionic excitations are almost degenerate locally, and in fact the bosonic excitations lie slightly \emph{above} the fermionic ones. 

\subsubsection{Results}

\begin{figure}
\centering
\includegraphics[width=7cm]{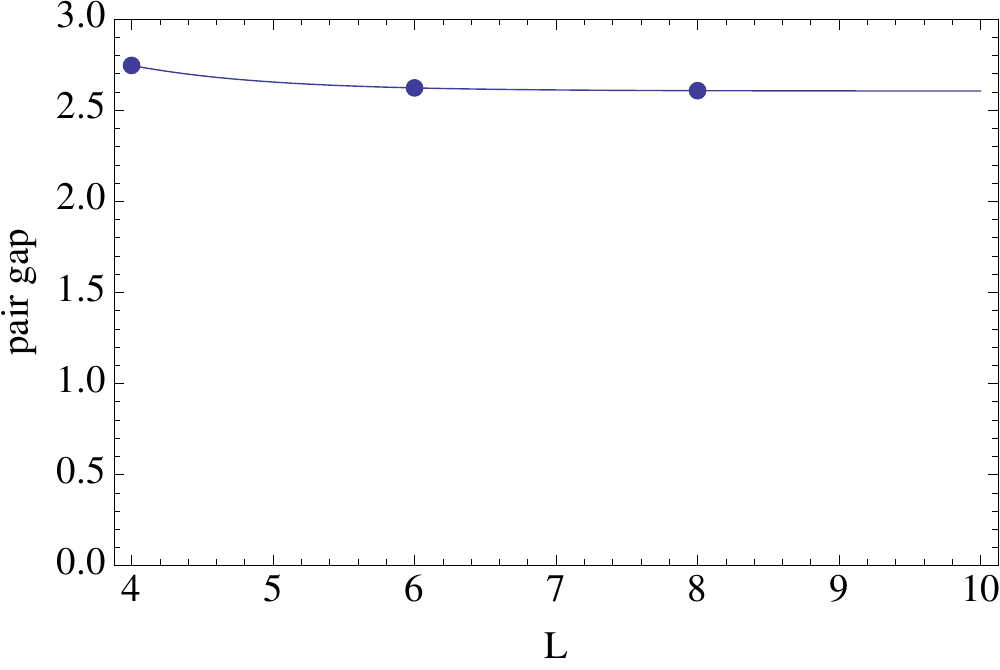}
\caption{Finite size scaling of the gap for SO$(5)$ vector excitations in the two-leg ladder. The gap quickly converges due to its infinite system value.}
\label{fig:ladder_gap_scaling}
\end{figure}

\noindent
Since the ground state is non-degenerate, we may expect it to be of the RVB form \cite{anderson, rice_rvb}, with strong but short ranged $d$-wave pairing and AF correlations. In order to check the robustness of these findings with respect to non-local perturbations, we use the CORE algorithm \cite{morningstar} to integrate out the fermionic degrees of freedom. A short description of this algorithm can be found in App.~\ref{sec:core}. The resulting bosonic model is then diagonalized on a cluster. Results of finite size scaling of the bosonic gap and the inferred value at infinite size are shown in Fig.~\ref{fig:ladder_gap_scaling}. The gap remains practically unchanged, and we conclude that the system is in a RVB state, in agreement with the exact solution \cite{linbalents}. 

\noindent
In addition to the results at half-filling, we can infer the behavior when the system is doped away from half-filling. From the spectrum of the local Hamiltonian it follows directly that doped holes (or electrons) are bound into pairs, since the pair binding energy is of the same order as the single particle gap. As soon as hole pairs are introduced, the ground state of the system is degenerate, and a low-energy model is obtained by projecting the Hamiltonian onto the degenerate ground state. This leads to a hardcore boson problem at small filling, with the number of bosons (i.e. hole pairs) equal to $\delta/2$, where $\delta$ denotes the doping level. We conclude that the system turns into a superfluid with superfluid density $\propto \delta$.

\section{Conclusions}

\noindent
In this section we have applied the simplest possible wave packet approximation to a variety of one-dimensional systems. The approach consists of first applying the renormalization group to the microscopic model to find possible fixed points of the flow. The fixed point Hamiltonians are then solved in the strong coupling limit for the set of WW basis states that lie at the Fermi surface. Due to the fact that the WW basis states always involve at least nearest neighbor interactions, the resulting effective Hamiltonians are non-trivial even when the kinetic energy is neglected. 

\noindent
A surprising amount of information about the nature of the ground state has been shown to be present in the local interaction and the excitation spectrum of the local problem in the WW basis. We have seen two radically different cases: For the chain problems, the 'order parameter' decouples from the fermionic degrees of freedom locally, in the sense that the corresponding bosonic excitations are ungapped, whereas fermions have a gap. This kind of behavior is very much in line with the assumptions made in deriving effective field theories for slow order parameter fields that separate from the gapped fermions. For the ladder case, on the other hand, we have seen that fermions and 'order parameter' excitations exist on the same energy scale, due to strong short ranged singlet correlations, leading to an RVB state that has very little in common with the paradigm of effective theories for slow variables.

\noindent
In the next chapter, we will apply a similar methodology to the two-dimensional Hubbard model, where we will find a similar behavior at the saddle points. 

\noindent
In closing this chapter, we would like to explain briefly how the admittedly oversimplified approach presented here can be extended. Based on our findings of correlations in mean-field states in Ch.~\ref{ch:ww_pairing}, it is clear that it is inconsistent to use the strong coupling limit for states at the Fermi surface while neglecting states away from the Fermi surface. The reason is that correlations of the neglected states are significant when the coupling at the Fermi points is strong. A straightforward extension of the approach considered here is two include more WW momenta, so that the transition from weak-coupling RG to strong-coupling wave packet approach becomes smoother. However, in the presentation here we have chosen simplicity over numerical accuracy, and leave this exploration for future work.

\chapter{Saddle point regime of the two-dimensional Hubbard model}
\label{ch:ww2}

\section{Introduction}
\label{sec:saddle_point_intro}

\noindent
In this chapter we apply the methodology developed in the previous chapter to the two-dimensional Hubbard model at moderate coupling. We focus on the so-called saddle point regime, where the Fermi surface lies in the vicinity of the saddle points $\left(\pi,0\right)$ and $\left(0,\pi\right)$. As discussed in Ch.~\ref{ch:ww_intro}, earlier RG studies \cite{saddlepointregime, furukawarice} in this regime have shown that both $d$-wave pairing ($d$SC) and antiferromagnetic (AF) correlations are strong in this region of the phase diagram, and especially the exact diagonalization study by L\"auchli et al.~\cite{laeuchli} points to the possibility that an insulating RVB is realized in the vicinity of the saddle points, which would lead to a natural explanation for the truncated Fermi surface of the cuprate superconductors \cite{pseudogap_experiment, rice_rvb} that does not rely on symmetry breaking. This type of scenario requires a 'phase separation in the Brillouin zone', in that electrons at the saddle points are localized on relatively short length scales, whereas the electrons in the nodal directions are delocalized over large distances. 

\noindent
In Ch.~\ref{ch:ww_pairing} we have seen that the corresponding separation of length scales naturally appears even at the mean-field level in the saddle point regime due to the vastly different Fermi velocities in the nodal and anti-nodal directions. Relatedly, we have shown in Ch.~\ref{ch:wave_packets_rg} that in this situation RG flows to strong coupling lead to a strongly correlated problem at the saddle points, whereas the faster states in the nodal directions behave almost like free particles at the same length scale (at moderate coupling). As a consequence, the slow states decouple from the fast states, and may be bound into pairs at relatively small length scales more or less independently of the what the fast states in the nodal direction do. 

\noindent
These findings in conjunction with the experimental phenomenology discussed in Ch.~\ref{ch:ww_intro} motivate us to apply the wave packet method to states in the vicinity of the saddle points. Following the procedure we have already used for one-dimensional systems, we first compute an effective Hamiltonian using the one-loop renormalization group. 
Details of the implementation can be found in Sec.~\ref{sec:rg_2d}. In order to monitor the flow of correlations at the saddle points, we compute the WW transform of the flowing interaction vertex at each step, as discussed in Sec.~\ref{sec:effective_hamiltonian_saddle_points}. In particular, we compute the single particle gap from the local interactions, and stop the flow when the gap exceeds the bandwidth of the wave packet states at the saddle points. The analysis of this local problem is dealt with in Sec.~\ref{sec:saddle_point_local}. We find a striking similarity between the local (in the WW basis) behavior of the saddle point states and the low-energy problem for the two-leg ladder from Sec.~\ref{sec:ladder}. In particular, the system has the same local ground state, and large gaps for all excitations when the flow approaches the strong coupling region. In order to assess the stability of the local ground state to the non-local couplings, we diagonalize the effective saddle point Hamiltonian on a plaquette in Sec.~\ref{sec:sp_flow_cluster}. Depending on the parameters, we find cooperon- and spin triplet-modes emerging below the single particle continuum. In order to overcome the limitations of small clusters, we compute an effect model for the $d$-wave pairs at the saddle points using the CORE algorithm \cite{morningstar} (see App.~\ref{sec:core}), and estimate the stability of the insulating RVB state against pair fluctuations with a variational ansatz in Sec.~\ref{sec:sp_quantum_rotor}. In this section we also discuss the general structure of the low energy states and relate it to the phenomenological SO(5) theory proposed by Zhang \cite{zhang98}.

\section{The microscopic model and its renormalization group treatment}
\label{sec:rg_2d}

\subsubsection{Model}

\noindent
We investigate the two-dimensional Hubbard model on a square lattice with nearest and next-to-nearest neighbor hopping. The Hamiltonian is given by
\begin{eqnarray}
\ham{} &=& \sum_{\mathbf{p}}\sum_s \epsilon\left(\mathbf{p}\right)\, c^\dagger_{\mathbf{p},s}\,c^\pdag_{\mathbf{p}} + \frac{U}{2N} \sum_{\mathbf{p}_1\cdots\mathbf{p}_4} \delta\left(\mathbf{p}_1+\mathbf{p}_2-\mathbf{p}_3-\mathbf{p}_4\right)\, \hat{J}\left(\mathbf{p}_1,\mathbf{p}_3\right)\,\hat{J}\left(\mathbf{p}_2,\mathbf{p}_4\right),\nonumber \\
\end{eqnarray}
where $\hat{J}\left(\mathbf{p},\mathbf{p}'\right) = \sum_s c^\dagger_{\mathbf{p},s}\,c^\pdag_{\mathbf{p}',s}$. The kinetic energy part is given by
\begin{equation}
\epsilon\left(\mathbf{p}\right) = -2t \left(\cos p_x + \cos p_y\right) + 4t' \cos p_x \cos p_y - \mu.
\end{equation}
We focus on the case that the Fermi surface touches the saddle points, which fixes $\mu = -4t'$. The interaction part describes onsite repulsion of strength $U$. 

\subsubsection{Renormalization group}
\noindent
The setup of the RG in this chapter is different from the one in Ch.~\ref{ch:rg}. Based on the findings of Ch.~\ref{ch:wave_packets_rg}, we prefer to use the Hamiltonian flow equations (or continuous unitary transformations \cite{wegner, cut1, cut2}) that are derived in App.~\ref{ch:wegner}. The main reason is that it is unclear in the usual RG approach \cite{tnt} how the transformation to the WW basis is to be performed when the cutoff cuts through a phase space cell. We recall, however, that the Hamiltonian flow equations are essentially the same as the RG equations when both are performed in the leading one-loop approximation as shown in App.~\ref{ch:wegner}.

\begin{figure}
\centering
\includegraphics[width=8cm]{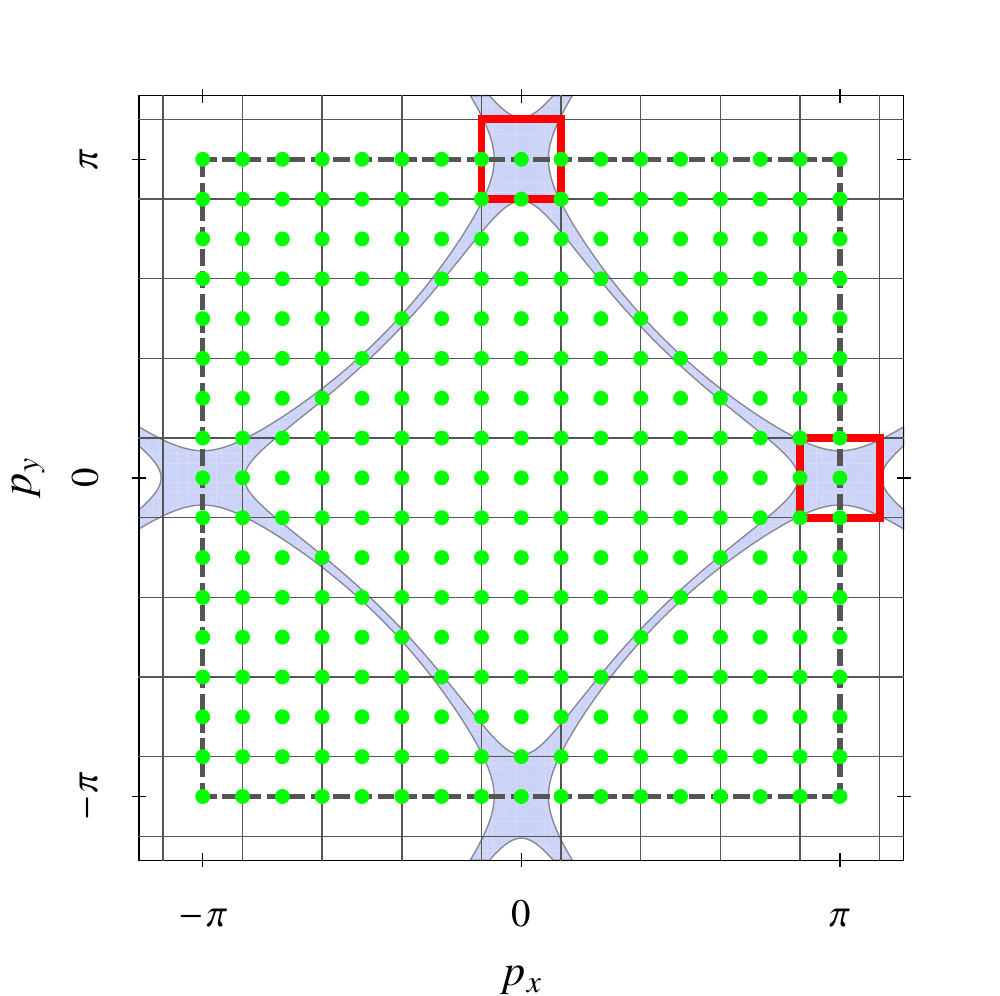}
\caption{Discretization of the flowing vertex. The vertex is computed on a discrete set of points that lie on a lattice with lattice constant $2\pi/16$ in the Brillouin zone (green dots). The shaded area contains all states with energy $|\epsilon\left(\mathbf{p}\right)| < 0.1 t$ for $t'=0.2t$. WW phase space cells for $M=4$ are also shown. The size of the $\pi/M$ of the cells is adjusted to the saddle point regions such that the cells around each saddle point (marked in red) contain approximately all low energy states there. The states that are kept in the saddle point truncation are marked red.}
\label{fig:tube_and_patches}
\end{figure}

\noindent
Moreover, we use a different discretization of the flowing vertex. This is connected to our aim to use the WW basis states in order to treat the low-energy problem. As discussed in Sec.~\ref{sec:ww_trafo_2d}, this is facilitated when the interaction vertex is known on a discrete set of $p$-points that lie on a square lattice. The lattice spacing (in the Brillouin zone) should be $\Delta p = \pi/(n M) = K/n$ for some integer $n$ in order to be able to transform the interaction to the WW basis using the formulas from Sec.~\ref{sec:ww_trafo_2d}. Since such a discretized Brillouin zone corresponds to a lattice of linear extension $N = 2nM$, the analytical window function for $N/M = 2n$ (cf. Sec.~\ref{sec:analytical_window_functions}) can be used in order to transform the discretized interaction numerically. In practice, we use $M=4$ and $N=4M=16$ throughout, so that the vertex is computed for 256 different momenta in total. The discretization is shown in Fig.~\ref{fig:tube_and_patches}.

\noindent
The restriction to $M=4$ has different reasons. First, this value corresponds to a strong coupling scale $\Lambda\approx 0.1t$, which is in the range where the unconventional RG flows have been observed in earlier studies \cite{saddlepointregime}. In addition, we think it is advantageous to use a relatively large value for the minimum energy scale $\Lambda$ at which the RG flow is stopped, since for small values the results depend very strongly on the precise shape of the Fermi surface, whereas we are more interested in the effect of short-ranged correlations which build up at higher energies and are therefore relatively insensitive to the detailed geometry of the Fermi surface.

\noindent
Fixing $M$ implies that we do not use a fixed value for $U$. Instead, we adjust the value of $U$ such that the flow goes to strong coupling at $\Lambda \approx 0.1t$. We estimate the onset of the strong coupling regime by computing the local fermion gap at the saddle points in the WW basis (see Sec.~\ref{sec:saddle_point_local}), and stop the flow when the gap exceeds the band width of the states there. All calculations are done at zero temperature. Since $U$ is fixed, the only parameter that is varied is $t'$. We consider the cases $t'/t = 0.1, 0.2$ and $0.3$ in the following. The corresponding values of $U$ are $U=2.5, 3,$ and $3.7$ respectively.

\noindent
The flows are very similar to the ones reported in earlier works \cite{saddlepointregime}, with the dominant correlations in the $d$SC and AF channels. The AF channel is dominant for small $t'$, but the $d$SC channel becomes increasingly important for larger $t'$. Since the general behavior of the RG flows in this regime has been dealt with extensively in the literature \cite{zanchi,halboth,breakdown,ehdoping}, we focus on the analysis of the saddle point states in the following.

\section{Effective Hamiltonian for the saddle points}
\label{sec:effective_hamiltonian_saddle_points}

\noindent
In the following we restrict the WW basis to the states right at the saddle points $(0,\pi)$ and $(\pi,0)$. Accordingly, we restrict the WW momentum coordinates $\mathbf{k}$ to the two vectors $\mathbf{k}^{(1)}=\left(M, 0\right)$ and $\mathbf{k}^{(2)} = \left(0,M\right)$.

\subsubsection{Interaction Hamiltonian in momentum space}

\noindent
Since we focus on the region around the saddle points in the following, only a fraction of the renormalized couplings is needed. It is useful to parametrize the couplings that act on states in the vicinity of the saddle point such that short- and long-range behavior are separated. This is achieved by shifting $\mathbf{p}\rightarrow K \mathbf{k}^{(a)}$, where $a=1,2$ labels the saddle points. Since all momentum space sums are effectively restricted to a square with side length $2\pi/M$ around the saddle points, there is no risk of double counting.

\noindent
We introduce the operators
\begin{eqnarray}
J_{ab}\left(\mathbf{p}_1,\mathbf{p}_2\right) &\equiv& \sum_{s} \, c^\dagger_{K\mathbf{k}^{(a)}+\mathbf{p}_1,s}\,c^\pdag_{K\mathbf{k}^{(b)}+\mathbf{p}_2,s}\nonumber \\
&=& \hat{J}\left(K\mathbf{k}^{(a)}+\mathbf{p}_1,K\mathbf{k}^{(b)}+\mathbf{p}_2\right).
\end{eqnarray}

\noindent
In terms of the $J_{ab}\left(\mathbf{p},\mathbf{p}'\right)$, the part of the flowing interaction $\tilde{U}\left(\mathbf{p}_1,\mathbf{p}_2,\mathbf{p}_3\right)$ that acts on states in the vicinity of the saddle points can be expanded as
\begin{equation}
\begin{split}
\ham{int} =& \frac{1}{N} \sum_{\mathbf{p}_1\cdots\mathbf{p}_4}  \delta\left(\mathbf{p}_1+\mathbf{p}_2-\mathbf{p}_3-\mathbf{p}_4\right)\, \\
&\times\;\Big\{ \tilde{u}_1\left(\mathbf{p_1},\mathbf{p_2},\mathbf{p}_3\right)\, J_{12}\left(\mathbf{p}_1,\mathbf{p}_3\right) J_{21}\left(\mathbf{p}_2,\mathbf{p}_4\right)  \\ &+ \tilde{u}_2\left(\mathbf{p_1},\mathbf{p_2},\mathbf{p}_3\right)\,J_{11}\left(\mathbf{p}_1,\mathbf{p}_3\right) J_{22}\left(\mathbf{p}_2,\mathbf{p}_4\right) \\
& + \frac{\tilde{u}_3}{2} \left(\mathbf{p_1},\mathbf{p_2},\mathbf{p}_3\right)\left[J_{21}\left(\mathbf{p}_1,\mathbf{p}_3\right) J_{21}\left(\mathbf{p}_2,\mathbf{p}_4\right) + J_{12}\left(\mathbf{p}_1,\mathbf{p}_3\right) J_{12}\left(\mathbf{p}_2,\mathbf{p}_4\right)\right] \\
&  + \frac{\tilde{u}_4}{2} \left(\mathbf{p_1},\mathbf{p_2},\mathbf{p}_3\right)\left[J_{11}\left(\mathbf{p}_1,\mathbf{p}_3\right) J_{11}\left(\mathbf{p}_2,\mathbf{p}_4\right) + J_{22}\left(\mathbf{p}_1,\mathbf{p}_3\right) J_{22}\left(\mathbf{p}_2,\mathbf{p}_4\right)\right]\Big\},
\end{split}
\label{eq:sp_h_int_momentum_space}
\end{equation}
where all momentum sums are restricted to $-\pi/M < p_i <\pi/M$ for some $M$ that depends on the renormalization scale as discussed above.
The spatial dependence of the interaction for scales less than $M$ is contained in the average values of the $\tilde{u}_i\left(\mathbf{p}_1,\mathbf{p}_2,\mathbf{p}_3\right)$, whereas the dependence on scales greater than $M$ is given by the $\mathbf{p}$ dependence of each $\tilde{u}_i$. In accordance with the usual RG scaling arguments \cite{shankar}, we find numerically that the $\mathbf{p}$-dependence of the interactions is always small compared to the $a$-dependence, so that the effective interactions are short ranged for length scales larger than the renormalization scale.

\subsubsection{Effective Hamiltonian in the Wilson-Wannier basis}

\noindent
Summarizing the results established above, we arrive at a model on a square lattice with lattice constant $2M$. There are two orbitals per site, labelled by the orbital index $a=1,2$.  The hopping Hamiltonian connects nearest neighbors only, and is determined by the two hopping terms $t$ and $t'$ of the original Hamiltonian. Interactions are local, and there are four coupling constants $u_1,\ldots,u_4$ that are related directly to the corresponding coupling constants in the two-patch model from Sec.~\ref{sec:rg}. All couplings scale like $M^{-2}$, so that it makes sense pull this factor out to arrive at an $M$-independent Hamiltonian. The Hamiltonian is given by
\begin{eqnarray}
\left(M\right)^2 \ham{eff} &=& \ham{hop} + \ham{int}
\nonumber \\&=&-\sum_{\mathbf{m},\mathbf{m}'}\sum_{\sigma}\, T^{ab}_{\mathbf{m},\mathbf{m}'}\, \gamma^\dagger_{\mathbf{m} a\sigma} \, \gamma^\pdag_{\mathbf{m'} a\sigma} \\ && + \sum_{\mathbf{m}} \Big[   u_1  \, J^{12}_{\mathbf{m}} \, J^{21}_{\mathbf{m}} +  u_2 \, J^{11}_{\mathbf{m}}\,J^{22}_{\mathbf{m}} \nonumber \\
&& + \frac{u_3}{2} \left( J_{\mathbf{m}}^{12}\, J_{\mathbf{m}}^{12} + J_{\mathbf{m}}^{21}\,J_{\mathbf{m}}^{21}\right) + \frac{u_4}{2} \left(J_{\mathbf{m}}^{11}\,J_{\mathbf{m}}^{11} + J_{\mathbf{m}}^{22} \, J_{\mathbf{m}}^{22}\right)\Big] \nonumber,
\label{eq:WW_Hamiltonian}
\end{eqnarray}
where $J^{ab}_{\mathbf{m}} = \sum_\sigma \, \gamma^\dagger_{\mathbf{m} a \sigma}\, \gamma^\pdag_{\mathbf{m} b \sigma}$.

\section{The local problem}
\label{sec:saddle_point_local}

\subsubsection{Local Hilbert space}

\noindent
The unperturbed Hamiltonian $\ham{int}$ is local, so that we begin the investigation with the local physics. The local Hamiltonian has a non-degenerate ground state for all values of the couplings $u_i$ considered here. The local ground state is given by
\begin{equation}
\ket{0}_{\text{loc}} = \frac{1}{\sqrt{2}}\left(\ket{\begin{array}{c}\uparrow\downarrow\\0 \end{array}} - \ket{\begin{array}{c} 0 \\ \uparrow\downarrow \end{array}} \right), 
\label{eq:sp_local_ground_state}
\end{equation}
with energy
\begin{equation}
E_0 = u_1 - u_3 + 2 u_4.
\end{equation}
There are eight degenerate single particle excitations with energy $E_p$ per site, corresponding to all combinations of the three quantum numbers orbital, spin, and charge (particle or hole). Furthermore, there are six bosonic excitations with relatively low energy, corresponding to the three order parameters for antiferromagnetism (three states, labelled by $\ket{A^i}$), $d$-wave superconductivity (two states $\ket{\Delta^\pm}$) and $d$CDW (one state $\ket{\phi}$). The states are
\begin{eqnarray}
\ket{A^i} &=& \left\{\ket{\begin{array}{c} \uparrow\\ \uparrow\end{array}},\; \frac{1}{\sqrt{2}}\left( \ket{\begin{array}{c} \uparrow\\ \downarrow\end{array}}+\ket{\begin{array}{c} \downarrow\\ \uparrow\end{array}} \right),\;\ket{\begin{array}{c} \downarrow\\ \downarrow\end{array}} \right\}\nonumber \\
\ket{\Delta^\pm} &=& \left\{ \ket{\begin{array}{c} \uparrow\downarrow \\ \uparrow\downarrow  \end{array}},\; \ket{\begin{array}{c} 0 \\ 0  \end{array}}\right\},\nonumber\\
\ket{\phi} &=& \frac{1}{\sqrt{2}}\left( \ket{\begin{array}{c} \uparrow\\ \downarrow\end{array}}-\ket{\begin{array}{c} \downarrow\\ \uparrow\end{array}} \right).
\end{eqnarray}
The remaining state is given by an $s$-wave pair on a WW site, and is always highest in energy. The energies of the fermionic and bosonic excited states (relative to the groundstate) are
\begin{eqnarray}
E_{f} &=& -\frac{u_1 + u_4}{2} + u_2 + u_3\nonumber \\
E_A &=& -u_1 + u_2 + u_3 - u_4 \nonumber \\
E_\Delta &=& -u_1 + 2 u_2 + u_3 \nonumber \\
E_\phi &=& u_1 + u_2 + u_3 - u_4.
\label{eq:local_gaps}
\end{eqnarray}

\noindent
At the beginning of the flow we have $u_i \propto U$, so that the excitation energy of the triplet states $\ket{A^i}$ vanishes exactly. This can be understood by performing a linear transformation on the two states $\ket{1} = \ket{ \mathbf{k}^{(1)}}$ and $\ket{2}=\ket{\mathbf{k}^{(2)}}$ (at fixed $\mathbf{m}$), such that
\begin{eqnarray}
\ket{a} &=& \frac{1}{\sqrt{2}}\left(\ket{1} + \ket{2}\right) \nonumber \\ \ket{b} &=& \frac{1}{\sqrt{2}}\left(\ket{1} - \ket{2}\right).
\end{eqnarray}
In terms of the new states, the Hubbard interaction is a local repulsive interaction for a two-orbital model, so that locally the two orbitals are decoupled, and the ground state degeneracy is two per orbital. All other excitations have gaps of order $U$ (corresponding to about $U/4M^2$ in the original units). The following section discusses how this picture changes when renormalization is taken into account.

\subsubsection{Influence of renormalization on the local problem}

\begin{figure}
\centering
\subfigure[$t'/t = 0.1$]{\includegraphics[width=7cm]{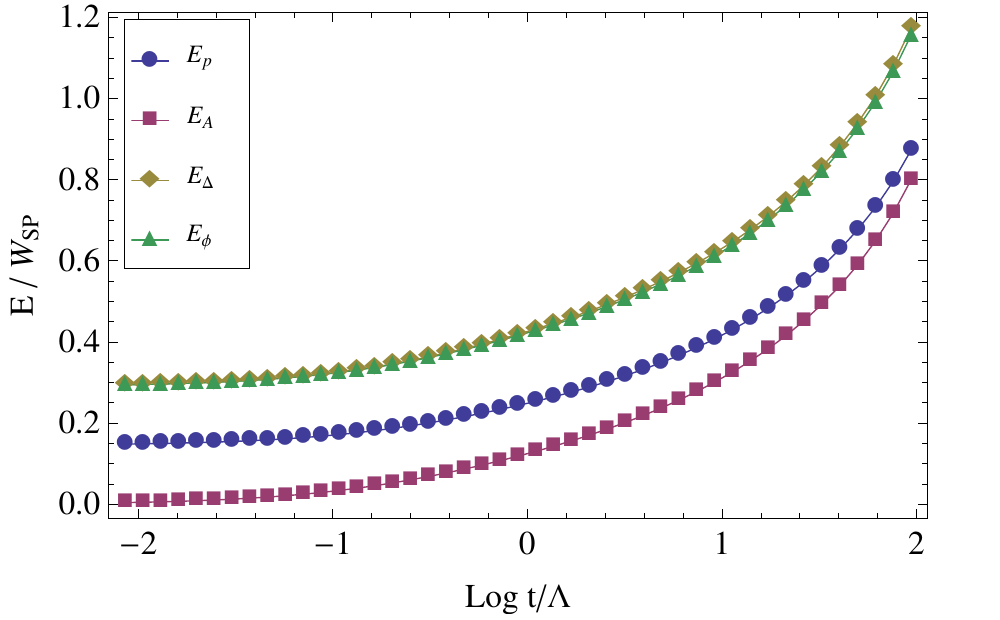}}
\subfigure[$t'/t = 0.2$]{\includegraphics[width=7cm]{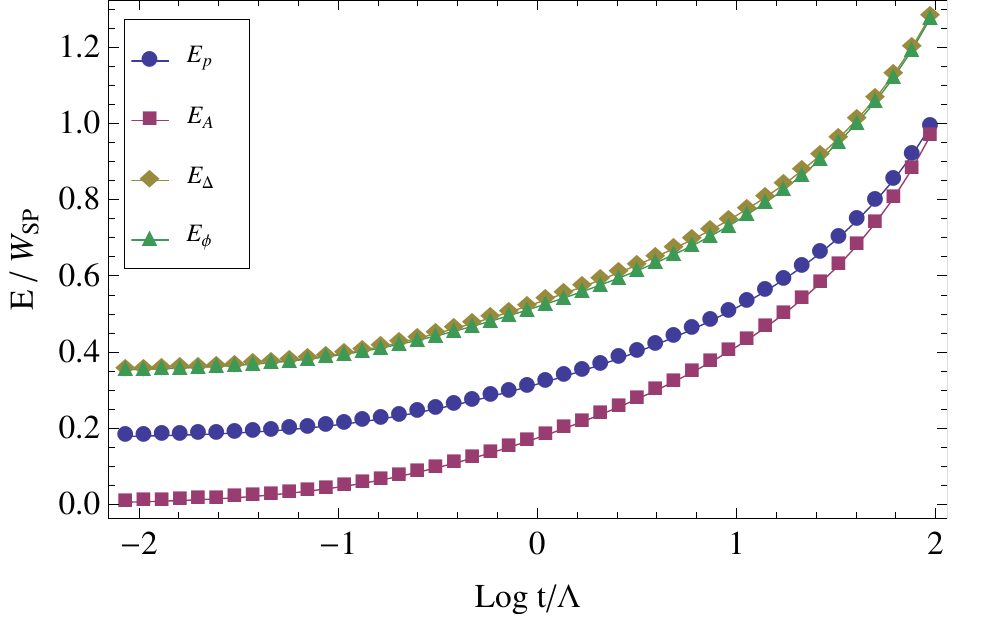}}
\subfigure[$t'/t = 0.3$]{\includegraphics[width=7cm]{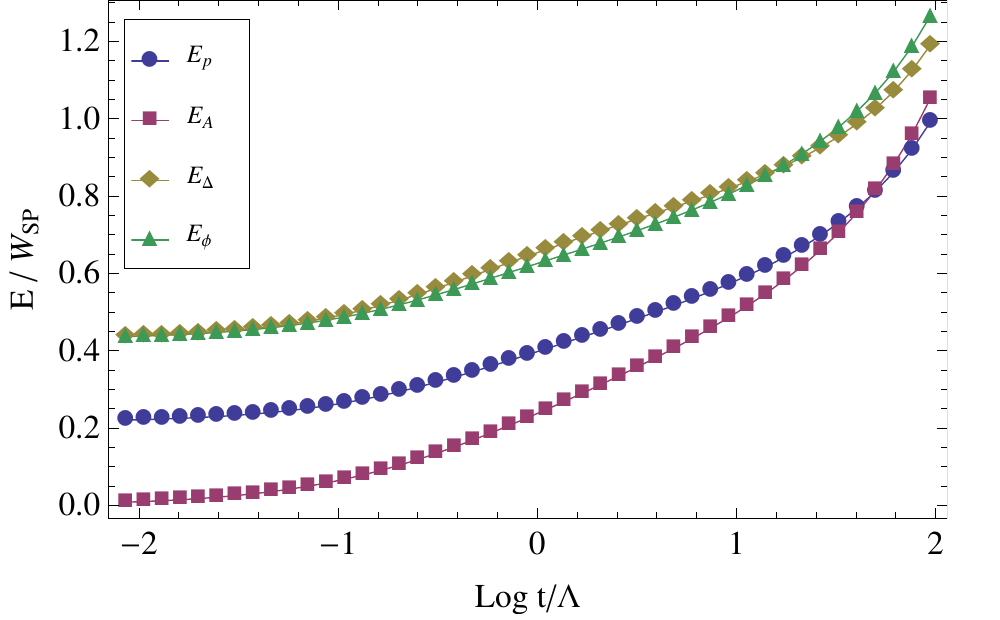}}
\caption{Flow of the excitation energies of the lowest excitations of the saddle point states for $t'/t=0.1,\,0.2,\,0.3$. Energies of single particle excitations ($E_p$), antiferromagnetic spin excitations ($E_A$), $d$-wave pair excitations ($E_\Delta$), and $d$-wave charge density excitations ($E_\phi$) are shown as functions of the logarithm of the RG scale $\Lambda$.}
\label{fig:flowing_gaps_loc}
\end{figure}
\noindent
In this section, we use the RG method discussed in Sec.~\ref{sec:rg_2d} together with the analysis of the local spectrum above to investigate the influence of the renormalization group flow on the saddle point states.
Instead of monitoring the flowing couplings $u_1,\ldots,u_4$, we use (\ref{eq:local_gaps}) and follow the flowing local energy gaps. Results for $U=3.7t$, $t'=0.3t$ are shown in fig. \ref{fig:flowing_gaps_loc}. Initially, the excitation energies for charge and (AF) spin excitations are separated: The spin excitation energy vanishes for the initial interaction, whereas the pair and $d$CDW energies are exactly twice the single particle energy $U$. However, in the course of the RG flow a spin gap starts to build up. At the end of the flow at $\Lambda \approx 0.1t$, the spin gap is comparable to the single particle gap. The energy of the $d$CDW excitations remains comparable to the energy to excite a $d$-wave pair, but the pair excitations are slightly lowered when $t'$ is increased. For $t'=0.3t$, the spin excitations are pushed up in energy, so that they lie above the single particle excitations and their energy is comparable to that of a pair excitation. The local behavior for $t'=0.3t$ is similar to the strong coupling fixed point of the RG for the two-patch model \cite{chubukov_two_patch} with particle-hole symmetry, which displays an emergent O$(6)$ symmetry. In this case, all three bosonic excitations are degenerate. However, the symmetry is realized only locally in the case at hand. Particle-hole symmetry violating terms show up in the non-local part of the Hamiltonian, in particular in the hopping term. In order to obtain a more complete picture of the local physics of the effective saddle point Hamiltonian we turn to the diagonalization of a plaquette in the next section.

\section{Diagonalization of small clusters}
\label{sec:sp_flow_cluster}

\noindent
In order to better estimate the interplay between kinetic energy and interaction at low energy scales, we diagonalize the effective Hamiltonian on a $2\times2$ plaquette. We also take into account the non-local part of the interaction. The non-local interactions are generally an order of magnitude smaller than their local counterparts, so that their overall effect is not large, but still noticeable. The main effect is to push pair- and spin-excitations down in energy with respect to single particle and $d$CDW excitations. We use periodic boundary conditions, so that bandwidth of the hopping operator is the same as for an infinite size system. The effect of renormalization on the saddle point degrees of freedom is discussed in the same way as before, by evaluating all quantities at different stages of the flow, characterized by the flow parameter $\Lambda$.

\begin{figure}
\centering
\includegraphics[width=7cm]{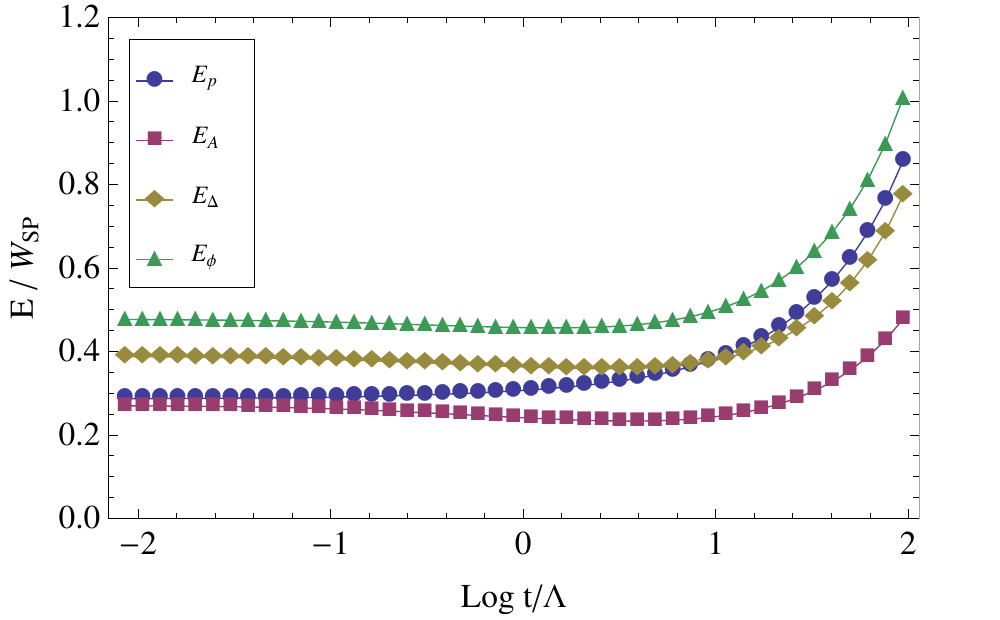}
\includegraphics[width=7cm]{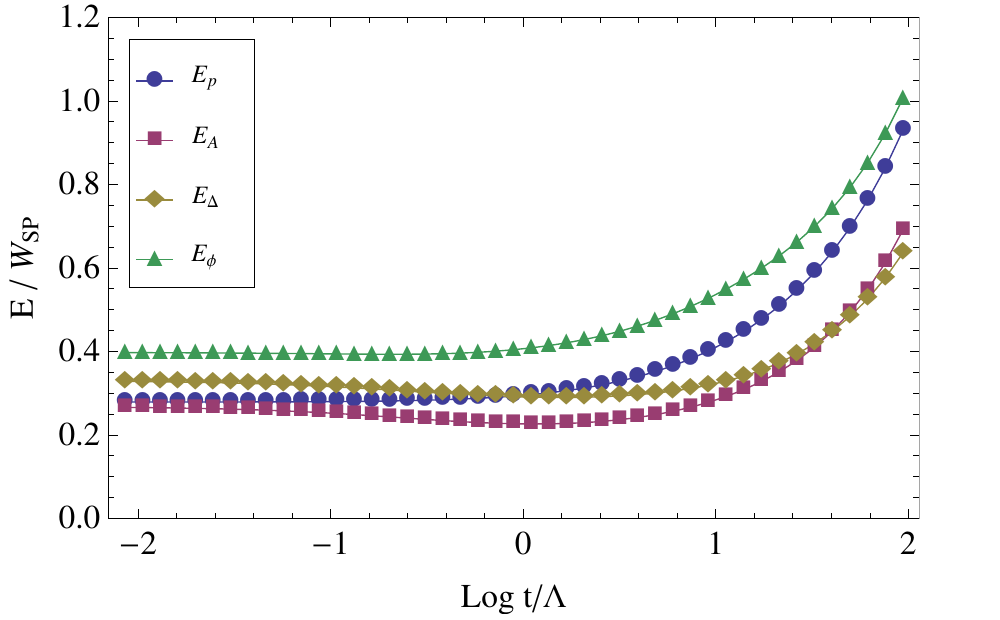}
\includegraphics[width=7cm]{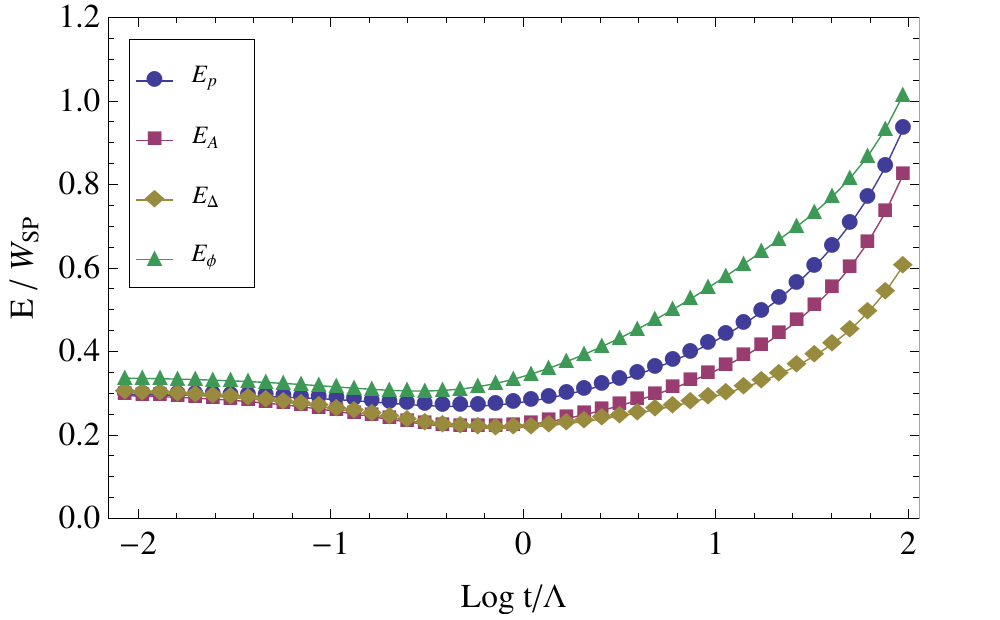}
\caption{Excitation energies for single particles ($E_p$), AF spin excitations ($E_A$), pairs ($E_\Delta$), and $d$CDW excitations ($E_\phi$) on a plaquette. The WW transform of the flowing RG interactions is used as discussed in Sec.~\ref{sec:effective_hamiltonian_saddle_points}.} 
\label{fig:flowing_gaps_plaq}
\end{figure}

\noindent
We compute three different kinds of observables: First, the excitation energies for the four types of interactions discussed above in Sec.~\ref{sec:saddle_point_local}. The flow of the excitation energies is shown in Fig. \ref{fig:flowing_gaps_plaq}. Compared to the local Hamiltonian above, we note that the $d$CDW excitations are raised in energy compared to the other excitations in the later stages of the flow. For the other bosonic excitations, we find that there is a crossover controlled by $t'$. For $t'=0.1t$, the AF spin excitations are lowest in energy, and are lowered in energy compared to the local Hamiltonian, so that they lie significantly below the single particle excitations. For $t'=0.3t$, on the other hand, the pair excitations have the lowest energy. At $t'=0.2t$, pair and spin excitations are approximately degenerate, and are both comparable to the single particle energy. Note that for all parameter values the bosonic excitations retain a large gap that is of the order of the fermion gap on the plaquette. However, it is clear that the system is very small, so that finite size effects are expected to be large. In order to gain a better understanding of the model, we now turn to the ground state properties of the plaquette Hamiltonian.

\begin{figure}
\centering
\includegraphics[width=7cm]{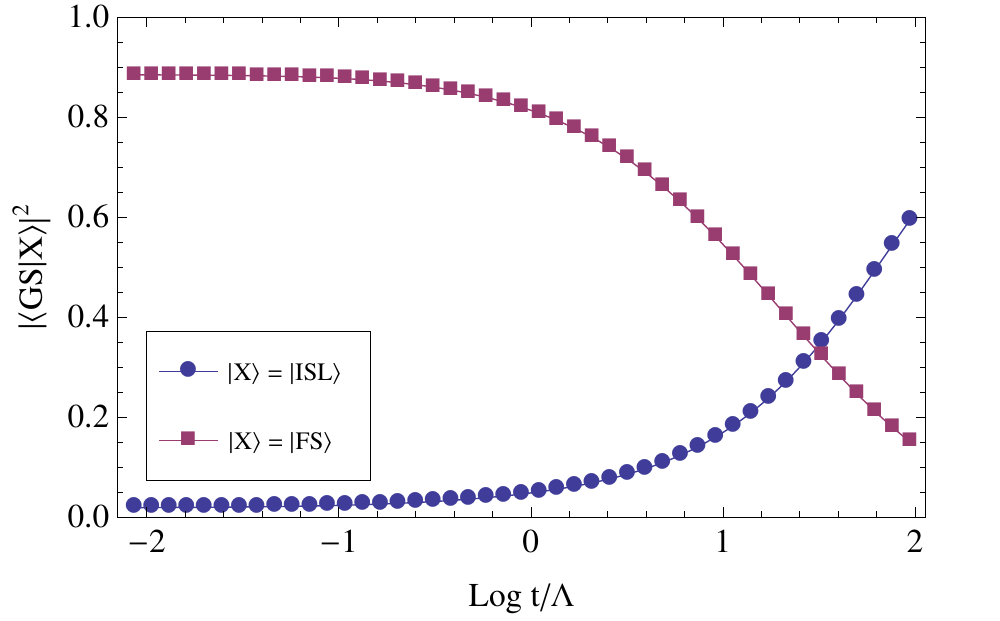}
\includegraphics[width=7cm]{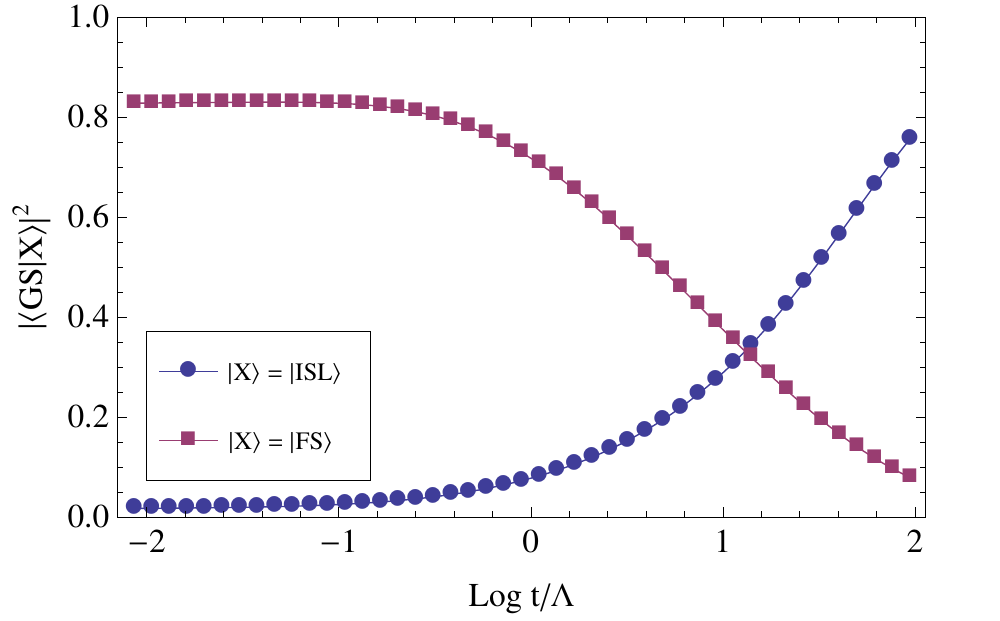}
\includegraphics[width=7cm]{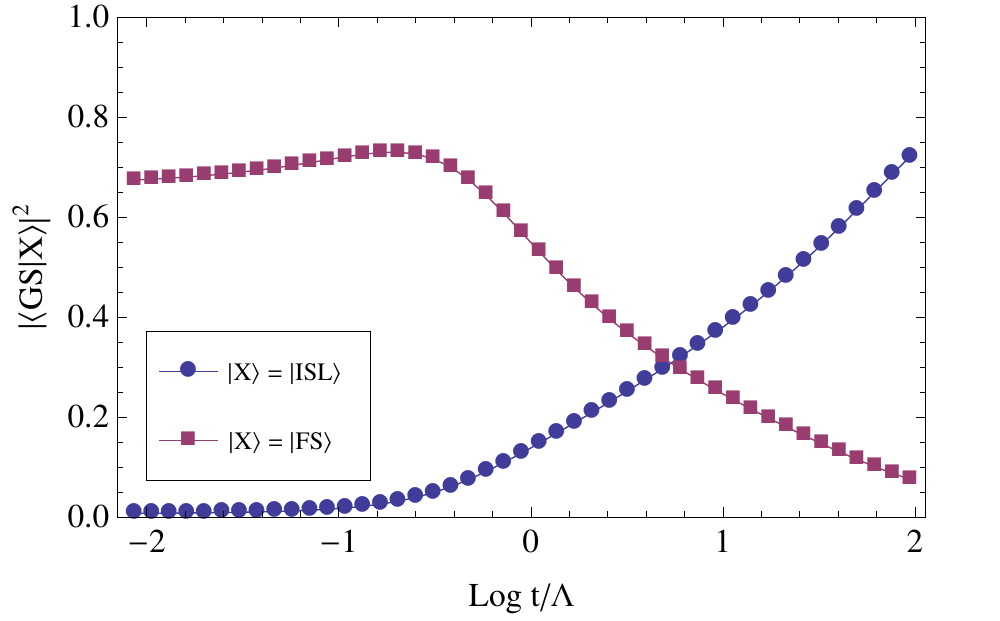}
\caption{Weight of the two trial ground states $\ket{\text{FS}}$ and $\ket{\text{ISL}}$ in the ground state of the saddle point model on a plaquette for different values of $\Lambda$, and $t'/t = 0.1, \, 0.2,\, 0.3$.}
\label{fig:sp_weight_flow}
\end{figure}

\noindent
The saddle point Hamiltonian consists of a local part, with a non-degenerate local ground state, and the non-local terms, which are dominated by the kinetic energy. The ground state of the kinetic energy is the half-filled Fermi sea. In order to characterize the behavior of the saddle point system, we compute the weight of two trial ground states in the plaquette ground state. The first state is the uncorrelated Fermi sea state
\begin{equation}
\ket{\text{FS}} = \prod_{|\mathbf{p}}^{E_{\mathbf{p}} <0} \prod_\sigma \gamma^\dagger_{\mathbf{p}\sigma} \ket{\text{vac}},
\end{equation}
where here the momenta $\mathbf{p}$ refer to the Fourier transform with respect to $\mathbf{m}$, and $E_{\mathbf{p}}$ is the single particle energy associated to the WW hopping operator $T^{ab}_{\mathbf{m}\mathbf{m}'}$. The second state is the tensor product of the local ground state $\ket{0}_{\text{loc}}$, defined in Eq.~(\ref{eq:sp_local_ground_state}):
\begin{equation}
\ket{\text{ISL}} = \prod_{\mathbf{m}} \ket{0}_{\text{loc},\mathbf{m}}
\end{equation}
The label $\text{ISL}$ is shorthand for insulating spin liquid, because the state is compatible with a charge gap and has short ranged spin correlation only. The weight of the two states in the $\Lambda$-dependent plaquette ground state is shown in Fig. \ref{fig:sp_weight_flow}.
In all three cases ($t'/t = 0.1,\,0.2,\,0.3$) there is a crossover during the RG flow. Initially, the FS state is always a better approximation to the plaquette ground state than the ISL state. Due to the renormalization of interactions, this situation is reversed in the later stages of the flow. As an interesting feature, we find that the weight of the ISL state at the end of the flow is maximal for $t'=0.2t$, which, by comparison with Fig. \ref{fig:flowing_gaps_plaq}, marks the transition between an AF and a SC dominated regime. In all three cases we find that the excitation energy of AF and pair excitations is comparable to or lower than the single particle gap. This motivates a mapping to a bosonic low-energy model, which is the subject of the next section.

\section{Effective quantum rotor model}
\label{sec:sp_quantum_rotor}

\noindent
The form of the excitation spectrum of the saddle point model on a plaquette suggests to reduce the model to an effective bosonic model, with either $d$-wave pairs, AF triplet excitations, or both as degrees of freedom. The mapping is performed using the contractor renormalization (CORE) \cite{morningstar} algorithm to reduce the local Hilbert space appropriately. 

\noindent
We first revisit the local Hilbert space and show that the low-energy part of the local Hilbert space is the same as that for a truncated O$(N)$ quantum rotor. The value of $N$ depends on the degrees of freedom that are kept, and is given by $N=2$ when only pairs are kept, $N=3$ for triplet excitations only, and $N=5$ when both are included. Note that the effective Hamiltonian is O$(N)$ symmetric for the case $N=3$ only, where the symmetry is the usual spin rotation invariance. 

\noindent
Afterwards we briefly discuss the CORE algorithm, and point out some problems we have encountered connected to the fact that the gaps of bosonic excitations are large. Finally, we introduce a variational coherent state for the effective model in order to overcome the limitations of small clusters.

\subsubsection{The quantum rotor subspace of the local Hilbert space}

\noindent
In order to elucidate the structure of the low-energy local Hilbert space, we discuss its properties and show that it has the same structure as a truncated quantum rotor.

\noindent
The local Hilbert space is 16-dimensional, and has already been discussed in Sec.~\ref{sec:saddle_point_local}. We keep the local ground state $\ket{0}$, and the pair and/or spin excitations $\ket{\Delta^\pm}$ and $\ket{A^i}$, respectively. The excited states are obtained from the ground state by applying pair- and AF-spin-operators:
\begin{eqnarray}
\Delta^{\pm} \ket{0} &=& \ket{\Delta^{\pm}} \nonumber \\
A^i\ket{0} &=& \ket{A^i},
\end{eqnarray}
where
\begin{eqnarray}
\Delta^+ &=& \frac{1}{\sqrt{2}} \left(\gamma^\dagger_{1\uparrow}\,\gamma^\dagger_{1\downarrow} - \gamma^\dagger_{2\uparrow}\,\gamma^\dagger_{2\downarrow}\right)
\nonumber \\ 
\Delta^- &=& \left[\Delta^+\right]^\dagger \nonumber \\
A^i &=& \frac{1}{2} \sigma^i_{\sigma\sigma'} \left(\gamma^\dagger_{2\sigma}\gamma^\pdag_{1\sigma'} + \gamma^\dagger_{1\sigma}\gamma^\pdag_{2\sigma'}\right).
\end{eqnarray}
We also introduce the total charge and spin operators for a WW site, that locally generate the corresponding O$(2)$ (equivalently U$(1)$) and O$(3)$ symmetries:
\begin{eqnarray}
n &=& \frac{1}{2}\left(\sum_a \sum_{\sigma}\gamma^\dagger_{a\sigma}\,\gamma^\pdag_{a\sigma}\right) - 2 \nonumber \\
S^i &=& \sum_a \sum_{\sigma,\sigma'} \gamma^\dagger_{a\sigma}\, \sigma^i_{\sigma\sigma'}\, \gamma^\pdag_{a\sigma'}.
\label{eq:local_symmetry_generators}
\end{eqnarray}
Note that the charge operator has zero eigenvalue at half-filling, and that the charge of a fermion is $\pm 1/2$ in this convention. Then it is straightforward to verify the commutation relations
\begin{eqnarray}
\left[n, \Delta^\pm\right] &=& \pm \Delta^\pm \nonumber \\
\left[\Delta^+, \Delta^-\right] &=& n \nonumber \\
\left[S^i, A^j\right] &=& i\epsilon_{ijk} A^k \nonumber \\
\left[A^i,A^j\right] &=& i\epsilon_{ijk} S^k.
\end{eqnarray}
The first and third line are identical to the defining commutation relations of the O$(2)$ and O$(3)$ quantum rotors, respectively \cite{sachdev}. The second and fourth line originate in the fact that the Hilbert space is finite, as opposed to the infinite Hilbert space of the quantum rotor. When both pairs and AF triplets are kept, the symmetry generators (\ref{eq:local_symmetry_generators}) can be augmented by the so-called $\pi$-generators
\begin{equation}
\pi^i = \frac{i}{2}\left(\sigma^y \sigma^i\right)_{\sigma\sigma'} \gamma_{1\sigma}\,\gamma_{2\sigma},
\end{equation}
and their hermitian conjugates. These operators transform $\Delta^{\pm}$ into $A^i$ and vice versa. In conjunction with the spin and charge operators, they generate the O$(5)$ algebra \cite{zhang98}. 

\noindent
The local ground state $\ket{0}$ is annihilated by all the generators, and is thus a singlet under O$(5)$ and its O$(2)$ and O$(3)$ subgroups, the five states $\ket{\Delta^\pm}$ and $\ket{A^i}$ transform under the vector representation of O$(5)$. Hence the bosonic state may be viewed as the first two levels of a quantum rotor system with the appropriate dimensionality (2,3, or 5). We stress, however, that the Hamiltonian is not O$(5)$ symmetric, and that the rotor model is only a convenient way to organize the low-energy sector of the local Hilbert space. The Hamiltonian is always O$(3)$ symmetric. The O$(2)$ charge symmetry holds when it is particle-hole symmetric, which can only be the case when $t'=0$.

\subsubsection{Contractor Renormalization}

\noindent
The CORE method is a real space renormalization scheme for lattice systems. It consists of two steps: First, one truncates the local Hilbert space to a subset of states that are the most relevant for the low energy behavior. The effective Hamiltonian for these local degrees of freedom is then found by diagonalization of small clusters using the linked cluster theorem, essentially by projecting low energy cluster states to the reduced Hilbert space followed by orthogonalization. Contributions from clusters with $n$ sites lead to $n$-site interactions in the effective model. $n$ is called the range of an interaction. When the reduction is good, the importance of interactions decays with their range. For a detailed discussion, we refer to \cite{morningstar} and to App.~\ref{sec:core}. In the following, we keep interactions up to range 3.

\noindent
Since the CORE algorithm is based on an application of the linked cluster theorem and it is controlled only when the renormalization due to high energy states is short ranged. For this reason, we apply it only to the Hamiltonian obtained at the end of the flow, where the weight of the RVB state in the ground state of the plaquette system is large, cf. Fig.~\ref{fig:sp_weight_flow}.

\noindent
We have encountered problems when applying the method to the spin excitations. The main reason is that the bosonic excitations are not well separated from the fermionic excitations, due to the insulating spin liquid nature of the local ground state. In a nutshell, the algorithm works by diagonalizing a small cluster, and projecting low-lying eigenstates to the tensor product of local bosonic states. When there are $\kappa$ states in the local projected Hilbert space, then $\kappa^2$ eigenstates have to be projected for the simplest case of a two-site cluster. However, even when these bosonic states are lower in energy locally than the fermions, the tensor product state contains high-energy states as well. For example, there are nine possible states with two spin excitations, with $S=0,1,2$. The state with $S=0$ has the same quantum numbers as the ground state, a state with two fermion excitations that form a singlet, or a state with two pair excitations of opposite charge. Hence there are many possibilities which Eigenstate should be chosen, and the algorithm tends to become unstable with respect to level crossings. 

\noindent
As a consequence, we limit ourselves to the case $t'/t=0.3$, and obtain an effective Hamiltonian for the pair states.

\subsubsection{Variational ground state for $d$-wave pair excitations}

\noindent
The effective quantum rotor Hamiltonian obtained from the CORE method is very complicated, since it contains interactions involving up to three sites in the present setup. Since our main interest here is to distinguish regimes with long range order from quantum disordered states, we use a family of coherent states in order to estimate the behavior of the effective model on longer length scales. 

\begin{figure}
\centering
\includegraphics[width=8cm]{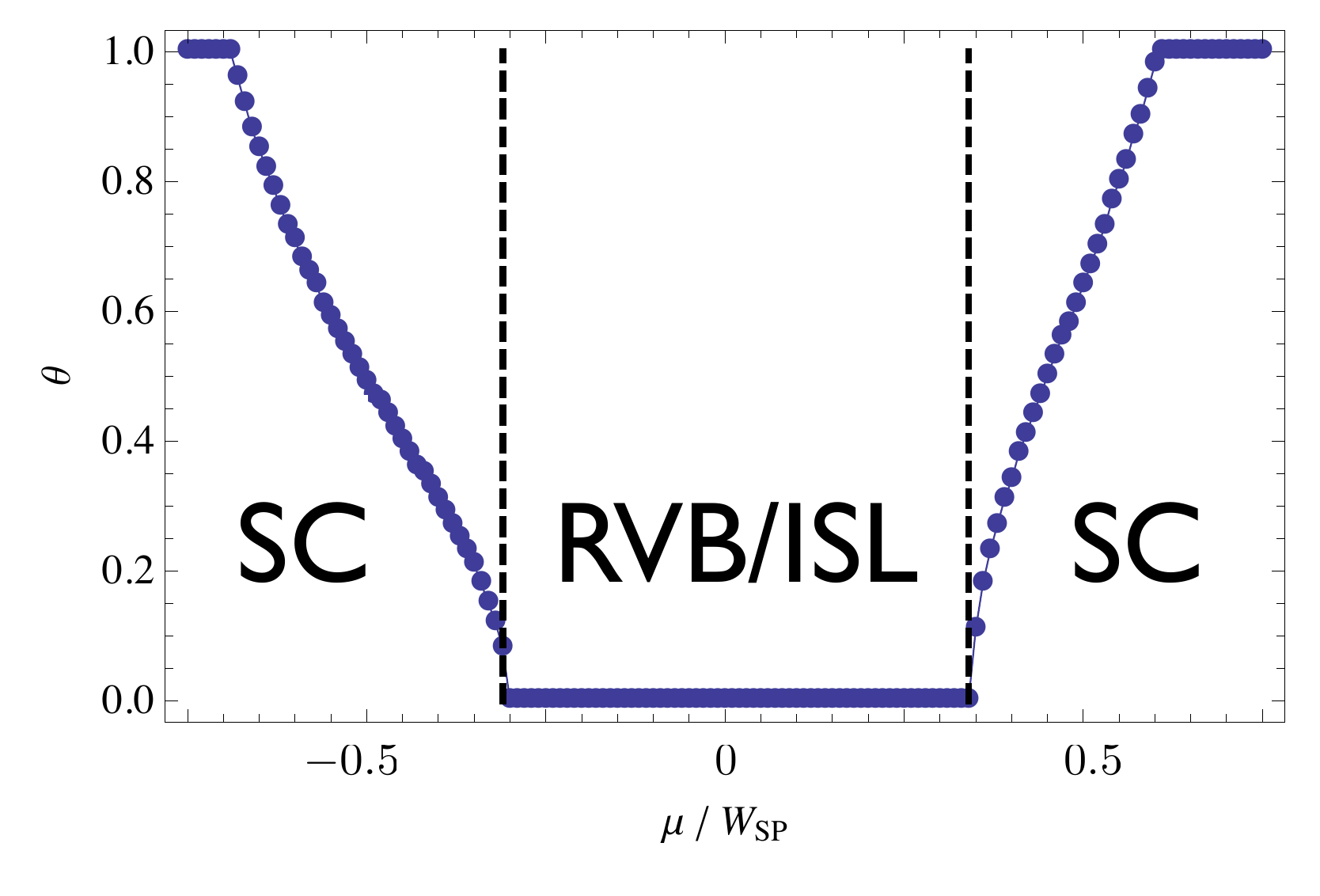}
\caption{Phase diagram for the effective pair model at $t'=0.3t$. The energy of the effective Hamiltonian obtained from the CORE method is evaluated for the variational coherent states \ref{eq:coherent_state_sc}. In order to estimate the size of the charge gap, we add a chemical potential term to the Hamiltonian. The figure shows the optimized value of the parameter $\theta$ as a function of the chemical potential $\mu$. $\theta =0$ corresponds to the insulating spin liquid (ISL) or RVB state. We find that the charge gap remains large for moderate values of $\mu$. Only at values of $\mu \approx \pm W_{\text{SP}}/4$, pairs begin to be doped into the system. Here $W_{\text{SP}}$ is the band width of the saddle point states.}
\label{fig:sc_isl}
\end{figure}

\noindent
For $t'=0.3t$, we reduce the model to the pair excitations (or O$(2)$ rotor model). We estimate the charge gap using the variational wave function
\begin{eqnarray}
\ket{\theta} &=& \prod_{\mathbf{m}} \ket{\theta}_{\mathbf{m}} \\
\ket{\theta}_{\mathbf{m}} &=& \cos \theta \, \ket{0}_{\mathbf{m}} + \sin\theta \, \ket{\Delta^\pm}_{\mathbf{m}},
\label{eq:coherent_state_sc}
\end{eqnarray}
where the sign of the charge carriers is selected according to the charge carrier type with the lower gap. The wave function $\ket{\theta}$ is a coherent state wave function, the local charge fluctuations depend on $\theta$. For $\theta = 0$, there are no charge fluctuations at the mean-field level, and the system is insulating. For finite $\theta$, the system is doped away from half-filling, and charges move freely. The charge gap is estimated by introducing a chemical potential term into the Hamiltonian. Results are shown in Fig. \ref{fig:sc_isl}. For small values of $\mu$, the system remains insulating. For large negative (positive) values, hole (particle) pairs are doped into the system, and it becomes superconducting, with a superfluid density that depends on the doping level. The size of the insulating region yields an estimate of the charge gap. We find that the charge gap remains large within this approach. It is clear, however, that the nature of the approximations made does not allow to make firm conclusions at present. 

\section{Conclusions}

\noindent
In summary, we have applied the wave packet approach to the saddle point regime of the two-dimensional Hubbard model. Based on arguments developed in earlier chapters, we have argued that a separation of length and energy scales takes place in this regime, allowing the states at the saddle points to become localized independently of the behavior of the nodal states, which extend over much larger length scales. We have then used the WW basis states to isolate the region around the saddle points, and have analyzed the flow of the effective saddle point Hamiltonian using exact diagonalization and a mapping to an effective bosonic model. We observed that the local Hilbert space at the saddle points is identical to the two-leg ladder, Ch.~\ref{ch:ww1}. Similarly, the local ground state is the same, even though the excitation spectrum depends on the doping, displaying a crossover between AF dominated and $d$SC dominated regimes as $t'$ (and therefore the doping) is increased. Correspondingly, the non-degeneracy of the local ground state and the fact that it is compatible with AF and $d$SC fluctuations make it very difficult to perturb. Within all methods we found that all excitations remain gapped in a finite range of chemical potentials.

\chapter{Conclusions and outlook}
\label{ch:conclusions}

\section{Summary}

\noindent
The unifying theme of this work is the investigation of the competition and mutual reinforcement of antiferromagnetism and superconductivity in one- and two-dimensional interacting electron systems, motivated by the phenomenology of cuprate superconductors. We have pursued a weak coupling approach, sacrificing the possibility to treat strong coupling effects, while gaining momentum space resolution. Throughout, we have employed the renormalization group which treats all particle-particle and particle-hole channels on an equal footing - a prerequisite for the study of the mutual influence of different possible instabilities. 

\subsubsection{Anisotropic scattering rates in the Hubbard model}

\noindent
We began our investigation with an experimentally motivated study of anisotropic quasi-particle scattering rates in the two-dimensional Hubbard model with parameters corresponding to the overdoped regime of Tl$_2$Ba$_2$CuO$_{6+x}$. We found that the strongly renormalized interactions lead to enhanced scattering for electrons in the anti-nodal direction. In conjunction with the scale-dependence of the renormalized vertex this was found to give rise to a highly anisotropic quasi-particle scattering rate at the Fermi surface. Moreover, the anisotropic part, which was shown to be peaked in the anti-nodal direction, was seen to have an almost linear temperature dependence down to very low temperatures, in qualitative agreement with experiment \cite{abdel}. We traced this behavior back to the simultaneous growth of correlations in the antiferromagnetic and $d$-wave superconducting channels in the vicinity of the saddle points. 

\noindent
In fact, it has been known for a long time that the Hubbard model exhibits other peculiar features in the so-called saddle point regime \cite{saddlepointregime}, where the Fermi surface lies in the vicinity of the saddle points (but not necessarily \emph{at} these points). In particular, the overlap between the antiferromagnetic and $d$-wave pairing channels   is large there, especially when the couplings are not too weak. The strong coupling phase does not appear to lead to a simple ordered phase, but to an insulating spin liquid phase with RVB correlations \cite{laeuchli, rice_rvb}.

\subsubsection{The wave packet approach to the saddle point regime}

\noindent
In order to investigate this phase we set out to try to obtain a better grasp of the low-energy physics in this regime in the remaining chapters. To this end, we introduced a novel tool, the use of the Wilson-Wannier basis functions \cite{wilsonbasis, daubechies, discretewilson, wilsonbasis2}, for the study of the strongly correlated low energy problem. The basis is generated from wave packet states with a fixed length scale $M$. It involves two coordinates, a coarse grained momentum coordinate that describes the physics on scales less than $M$, and a coarse grained real space coordinate, that describes scale larger than $M$.

\noindent
The basic idea was based on the fact that a gap in the single particle spectrum introduces a length scale into the problem that is given by the exponential decay of spatial correlations. This led us to expect that the description simplifies if a basis is chosen that reflects this length scale, the fermions should 'disappear' from the physics at larger length scales, leaving only effective degrees of freedom that can be determined from an analysis of the Hamiltonian. At the same time, the dependence on the wave packet momentum can be used to single out the low energy states close to the Fermi surface, and to incorporate effects of Fermi surface anisotropy.

\noindent
Since the approach is new, we spent some time developing the necessary formalism and useful approximation methods. These were used in order to highlight the influence of the (coarse grained) geometry of the Brillouin zone on Fermi surface instabilities. In particular, we have seen in Ch.~\ref{ch:ww_pairing} that a separation of length scales between nodal and anti-nodal states occurs whenever the Fermi surface lies in the vicinity of the saddle points. While it is well known that because of the van Hove singularity the low energy phase space tends to concentrate around the saddle points \cite{saddle1, saddle2}, our approach allowed us to estimate the anisotropy of the strength of correlations at a fixed length scale. We found that generically the states at the saddle points are much more correlated than the nodal states at the same length scale, so that they effectively decouple. 

\noindent
We proceeded by studying one-dimensional chains with quasi-long range order, and the two-leg ladder at half-filling, which is known to exhibit a RVB-like insulating spin liquid phase \cite{linbalents}. The main aim was to compare the results from the wave packet approach to exact solutions, and we found good qualitative agreement in all three cases despite of very simplistic approximations. We explained the agreement in terms of the separation of scales between fermionic and bosonic excitations at the length scale where the pairing occurs. 

\noindent
We used this separation of scales and computed and effective Hamiltonian for the anti-nodal states in isolation. We analyzed this Hamiltonian, and compared it to the two-leg ladder system that was treated in a similar approximation. We found that locally, the two models are very similar. Consequently, the states at the saddle point appear to be prone to localize in a state with strong (but short ranged) singlet correlations, resembling the RVB state of the ladder system with large gaps for all excitations at the pairing scale, in agreement with earlier calculations based on exact diagonalization \cite{laeuchli}. 

\noindent
We emphasize that all the results are only qualitative, and that the approximations made are quite drastic. Nevertheless, we think that the underlying physical arguments based on 
the separation of length scales on the one hand and the non-degeneracy of local ground states are sound. Clearly the WW basis breaks the underlying translational invariance so that one might think that it overestimates gap formation. While this is certainly true to some extent, the states involved can localize because they are coupled by umklapp scattering. Moreover we have seen that our approach does lead to quasi-long range order for one-dimensional chains, where the order parameter modes separate from the fermionic degrees of freedom. It is worth pointing out that it is not necessary for the Fermi surface to lie exactly at the van Hove points, since we consider singlet-pair formation on rather short scales (about 8 lattice constants), so that the pairs are too delocalized in momentum space to resolve the exact position of the Fermi surface.

\noindent
Finally, our results are compatible with other studies that start from the strong correlation limit. In particular, exact diagonalization studies for the $t-J$-model \cite{exactdiagonalization} find a cooperon mode with weight at the saddle point at finite energy for a lightly doped system. This is consistent with our results which naturally lead to such a mode. Similarly, recent cluster DMFT calculations indicate that strong short-ranged singlet correlations can lead to the formation of a pseudogap in the anti-nodal direction without long range ordering \cite{georges_dmft_cluster}. The authors attributed the opening of the single fermion gap to strong single correlations, similar to our observation.

\noindent
In summary, the wave packet approach to interacting fermions provides surprising insights to complicated problems already in its simplest implementation. Its main strength in our view is that it provides a relatively straightforward bridge between effective interactions and the geometry of the Brillouin zone on the one hand to effective models low energy physics on the other hand. However, the approach is still in its infancy, and much more work is needed in order to assess its merits and shortcomings.

\section{Outlook}

\noindent
We see several possible extensions to this work. First and foremost, we think it would be highly useful to improve on the approximations made in this thesis by using more WW orbitals (recall that we have truncated the basis to only one or two states per WW site). There is no problem of principle, and only the finite time horizon of this project has prevented us from pursuing this route so far. 

\noindent
From the point of view of flexibility it would be interesting to see whether similar constructions can be worked out for other lattice geometries, such as the honeycomb lattice. Note that the WW basis is in essence a one-dimensional construct, so that it can be used for all rectangular lattices directly, but not for lattices that are not tensor products of the one-dimensional chain. We hope that our group theoretical reformulation from Ch.~\ref{ch:wwbasis} may be helpful for this problem.

\noindent
The WW basis incorporates a single length scale, but many problems, such as the pseudogap problem, exhibit multiple length scales. In this work we invoked the separation of length scales in order to treat the anti-nodal states in isolation. However, it is clear that for a full account of the phenomenology the nodal states have to be dealt with. 
We see two possible approaches: First, in the sprit of two-length-scale expansions, one could use the effective Hamiltonian for the $d$-wave pairs at anti-nodal states and couple them to the nodal states. This is feasible since all couplings are known. Since the translational invariance breaking strongly distorts the Fermi surface, one might then take the continuum limit, effectively setting the length scale for the nodal states to infinity. This leads to a model of mobile fermions coupled to immobile pairs, similar to some phenomenological models \cite{geshkenbein}.

\noindent
On the other hand, one might try to develop a true multiscale approach, that can deal with several length scales at once, for example by applying the WW transformation for a second time to some of the WW orbitals of the first transformation. However, we can not judge at present the feasibility or usefulness of this idea.

\noindent
Finally, the model developed so far exhibits several features which should be elaborated from a phenomenological point of view: First, as the doping level is increased, a transition in the charge sector arises at the nodal points (cf. Fig.~\ref{fig:sc_isl}, and the hole pairs become mobile. Intuitively, this transition may be linked to the rapid rise of the superconducting dome at this point. Second, our model includes spin excitations with wave vector around $\left(\pi,\pi\right)$, which are expected to be coherent since they lie below the particle-hole continuum at the saddle points. These might offer a natural explanation of the so-called hourglass that is observed at optimal doping \cite{hourglass}.

\noindent
We are confident that more such extensions can be found, and that many routes are open to extend the very modest first steps presented in this work.

\appendix
\chapter{Construction of the window function}
\label{ch:windowfunction}

\section{Conditions on the window function}
\label{sec:gconditions}

\noindent
We derive the conditions that the window function $g(j)$ has to satisfy to make the Wilson basis orthonormal. The derivation closely follows the ones given in \cite{daubechies,discretewilson}. 

\noindent
The conditions on $g(j)$ that make the $\psi_{mk}(j)$ an orthogonal basis can be derived from the conditions 
\begin{equation}
\sum_{mk} \psi_{mk}(j_1)\,\psi_{mk}(j_2) = \delta_{j_1,j_2}  \label{eq:unity}.
\end{equation}
We consider only the case of a real window function $g(j)$, so that the $\psi_{mk}(j)$ are real, too. Condition (\ref{eq:unity}) then amounts to demanding that the matrix $\psi_{mk,j}$ with rows given by $\psi_{mk}(j)$ is an orthogonal matrix. This implies that the condition of orthonormality, $\sum_j \psi_{m_1 k_1}(j)\psi_{m_2 k_2}(j) = \delta_{m_1m_2}\,\delta_{k_1 k_2}$, is automatically fulfilled whenever (\ref{eq:unity}) is satisfied.

\noindent
We use the definition (\ref{eq:defpsi}), and write $\psi_{mk}(j)$ in terms of the window function $g(j)$. Moreover, we split the sum over $m$ into sums over even and odd $m$,
\begin{equation}
\begin{split}
\sum_{mk}\,\psi_{mk}(j_1)\,\psi_{mk}(j_2) =& \sum_{l=0}^{L/2-1} g(j_1-2M l) g(j_2-2Ml)\;\Big\{ 1 + (-1)^{j_1+j_2} \\&  + 2 \sum_{k=1}^{M-1} \cos\left[\frac{\pi}{M}kj_1-\phi_k\right]\cos\left[\frac{\pi}{M}k j_2 -\phi_k\right]\Big\} \\
& + \sum_{l=0}^{M/2-1} g(j-M(2l+1))\,g(j-M(2l+1)) \\ & \times2\sum_{k=1}^{M-1} \cos\left[\frac{\pi}{M} k j_1 - \phi_{k+1}\right]\cos\left[\frac{\pi}{M}k j_2 - \phi_{k+1}\right]
\end{split}
\end{equation}
Now we use 
\begin{equation*}
2\cos\left[\frac{\pi}{M}k j_1 -\phi\right]\cos\left[\frac{\pi}{M}k j_2 - \phi\right] = \cos\left[\frac{\pi}{M} k\left(j_1 - j_2\right)\right] + \cos\left[\frac{\pi}{M}k\left(j_1+j_2\right)-2\phi\right],
\end{equation*}
and notice that
\begin{displaymath}
2\phi_k = \pi k \mod 2\pi.
\end{displaymath}
Then
\begin{displaymath}
\cos\left[\frac{\pi}{M} k (j_1 + j_2) - 2\phi_{k+m}\right] = (-1)^m \cos\left[\frac{\pi}{M} k \left(j_1 + j_2 - M\right)\right].
\end{displaymath}
Hence we have
\begin{eqnarray}
\sum_{mk}\,\psi_{mk}(j_1)\,\psi_{mk}(j_2) &=& \sum_{l=0}^{L/2-1} g(j_1 - 2 M l) g(j_2-2Ml) \Bigg\{ \left(1+(-1)^{j_1+j_2}\right) \\
&&+ \sum_{k=1}^{M-1}  \cos\left[\frac{\pi}{M} k(j_1-j_2)\right] +  \cos\left[\frac{\pi}{M}k (j_1+j_2-M)\right]\Bigg\}\nonumber \\
&& + \sum_{l=0}^{L/2-1} g(j_1 - M(2l+1)) g(j_2- M(2l +1)) \nonumber \\
&& \times \Bigg\{\sum_{k=1}^{M-1} \left(\cos\left[\frac{\pi}{M} k(j_1-j_2)\right] - \cos\left[\frac{\pi}{M} k (j_1+j_2-M)\right]\right) \Bigg\}\nonumber
\end{eqnarray}
Because of the symmetry $\cos x = \cos\left(-x\right)$ we can transform the domain of the sums over $k$ from $\{1,\ldots, M-1\}$ to $\{-M+1,\dots,M\}$ as follows:
\begin{eqnarray}
\left(1+(-1)^{j_1 + j_2}\right) +  \sum_{k=1}^{M-1} \Big( \cos\left[\frac{\pi}{M} k(j_1-j_2)\right] +  \cos\left[\frac{\pi}{M}k (j_1+j_2-M)\right]\Big) && \nonumber \\= \frac{1}{2} \sum_{k=-M+1}^M \cos\left[\frac{\pi}{M} k(j_1-j_2)\right]  + \frac{1}{2} \sum_{k=-M+1}^M  \cos\left[\frac{\pi}{M}k (j_1+j_2-M)\right]
\end{eqnarray}
for terms even in $m$ and
\begin{eqnarray}
\sum_{k=1}^{M-1} \Big( \cos\left[\frac{\pi}{M}k (j_1 - j_2)\right] - \cos\left[ \frac{\pi}{M}k (j_1+j_2-M)\right] \Big) && \nonumber \\
= \frac{1}{2} \sum_{k=-M+1}^M \cos\left[\frac{\pi}{M} k(j_1-j_2)\right] - \frac{1}{2} \sum_{k=-M+1}^M \cos\left[\frac{\pi}{M} k(j_1+j_2-M)\right]
\end{eqnarray}
for the odd terms. Now the summation over $k$ can be performed using the orthogonality of exponential functions, yielding
\begin{equation}
\frac{1}{2} \sum_{k=-M+1}^M \cos\left[\frac{2\pi}{2 M} k a\right] = M \, \delta^{(2M)}_{a,0},
\end{equation}
where $a$ is any integer and $\delta^{(2M)}_{ij}$ is the Kronecker delta modulo $2M$. The final expression is then
\begin{equation}
\begin{split}
\sum_{mk} \psi_{mk}(j_1)\psi_{mk}(j_2) =& \;\;M \sum_{m=0}^{L-1} g(j_1-M m) g(j_2 - M m) \, \delta^{(2M)}_{j_1, j_2}  \\&\; + M \sum_{m=0}^{L-1} \, (-1)^m g(j_1-Mm)g(j_2-Mm) \, \delta^{(2M)}_{j_1+j_2,M}.
\end{split}
 \label{eq:sumpsi}
\end{equation}
Now we show that whenever $g(j) = g(-j)$, the second line vanishes identically. We set $j_2 = -j_1 + M(2l+1)$ (with $0 \leq l < L/2$) to satisfy the Kronecker delta. The resulting expression is
\begin{displaymath}
\sum_{m=0}^{L-1} (-1)^m g(j - M m) g(-j - M(m-2l-1)) = \sum_{m=0}^{L-1} (-1)^m g(j-Mm) g(j-(2l+1-m)),
\end{displaymath}
where we have used $g(j) = g(-j)$. To see that this sum vanishes we introduce the new summation variable $\tilde{m} = 2l+1-m$. Under this transformation the product of the window functions is invariant, but $(-1)^m = -(-1)^{\tilde{m}}$, and hence the sum has to vanish. 

\noindent
From (\ref{eq:sumpsi}) we then conclude that the conditions 
\begin{equation}
\sum_{m=0}^{L-1} g(j- M m) g(j - M (m+2l)) = \frac{1}{M}\, \delta_{l,0} \qquad 0\leq l < L/2
\label{eq:orthonormality_condition}
\end{equation}
guarantee orthonormality of the Wilson basis. These are the conditions stated above in eq. (\ref{eq:gconditions}). Since (\ref{eq:orthonormality_condition}) depends on $j \mod M$ only, the number of independent conditions is $N/2$.

\section{Zak transformation}
\label{sec:zaktransform}

\begin{figure}
\centering
\includegraphics[width=6cm]{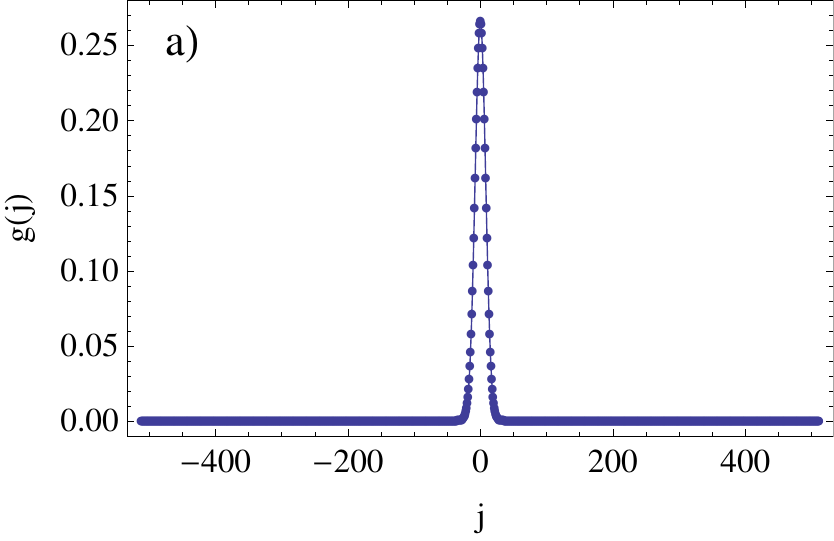}
\hspace{1cm}
\includegraphics[width=6cm]{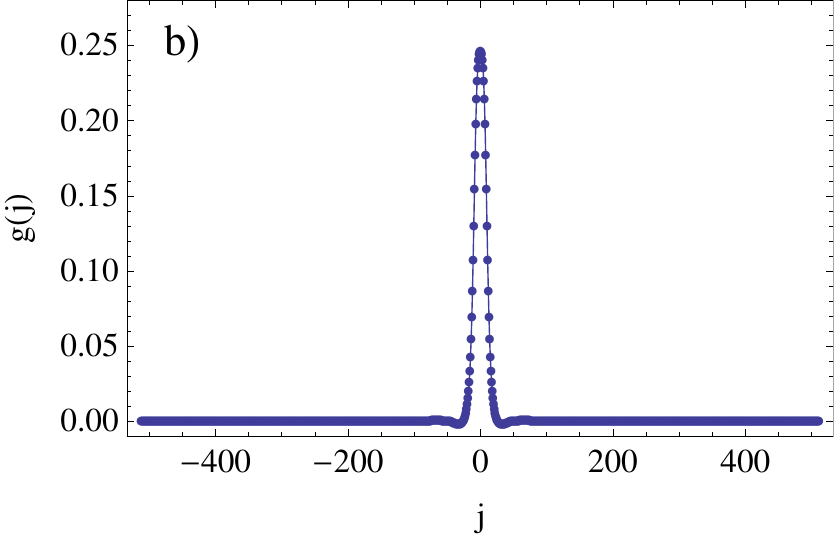}
\caption{Window functions in real space. a) shows the function $g_{0}(j)$ before the application of the transformation (\ref{eq:gtransformation}), in b) the function $g(j)$ that results from the transformation is shown.}
\label{fig:gplot}
\end{figure}

\noindent
To simplify these conditions it is convenient to introduce the so-called finite Zak transform $Z_g(j, p)$ of $g(j)$ \cite{zaktransform}. From the definition
\begin{eqnarray}
Z_g(j, p) &=& \frac{1}{L} \sum_{m=0}^{L-1} g(j + m M) e^{-\frac{2\pi i}{L} p m}\nonumber \\
g(j + m M) &=& \frac{1}{L} \sum_{p=0}^{L-1} Z_g(j, p) e^{\frac{2\pi i}{L} p m},
\end{eqnarray}
where $0 \leq j < L$ and $0 \leq p < M$, one can see that the Zak transform of a function is very similar to the Bloch representation of wave functions in a periodic potential. The lattice is split into unit cells of size $M$, and the degrees of freedom within one unit cell are represented in real space, but the change from unit cell to unit cell is Fourier transformed leading to the pseudo-momentum $p$. The Zak transform simplifies conditions (\ref{eq:gconditions}) because they are in the form of a convolution with the sum running over unit cells of size $M$. The convolution is turned into a multiplication, and the conditions on $Z_g$ are
\begin{equation}
\left|Z_g(j, p)\right|^2 + \left|Z_g(j, p + L/2)\right|^2 = C(j),
\label{eq:Zconditions}
\end{equation}
where $C(j)$ is independent of $p$. $C(j)$ is fixed by the condition
\begin{equation}
\sum_{p=0}^{L-1} \left|Z(j, p)\right|^2 = \frac{1}{M}.
\label{eq:zaknormalization}
\end{equation}
As a practical consequence of this simple condition, one can use (almost) any initial window function $g_0(j)$ as long as $Z_{g_0}(j, p)$ and $Z_{g_0}(j,p+L/2)$ do not vanish simultaneously for some $j$. A window function $g(j)$ that satisfies conditions (\ref{eq:Zconditions}) up to a constant prefactor can then be obtained via
\begin{equation}
Z_g(j, p) \propto \frac{Z_{g_0}(j,p)}{\sqrt{|Z_{g_0}(j,p)|^2 + |Z_{g_0}(j,p+L/2)|^2}}.
\label{eq:gtransformation}
\end{equation}
The normalization is fixed by demanding
\begin{equation}
\sum_j g^2(j) = 1
\end{equation}
for the modified window function. For finite lattices this procedure is easily implemented on a computer. A convenient choice of $g(j)$ is a Gaussian centered around $j=0$ with width $M$, but the qualitative results do not depend on the exact choice of $g(j)$. In particular, one one can show that salient features like the exponential decay are preserved \cite{daubechies}. To illustrate the preservation of rapid decay, we display the effect of the transformation (\ref{eq:gtransformation}) on a Gaussian initial window function in fig. \ref{fig:gplot}. 

\section{Conditions for band limited window functions}
\label{sec:gcondition_band_limited}

\noindent
The orthogonality conditions (\ref{eq:orthonormality_condition}) on the window function simplify when it is band limited in the sense that the Fourier transform $\tilde{g}(p)=\frac{1}{\sqrt{N}} \sum_j g(j) e^{-ipj}$ satisfies
\begin{equation}
\tilde{g}(p) = 0 \text{ for }  \left|p\right| \geq \frac{\pi}{M}.
\label{eq:bandlimit}
\end{equation}
In this case it is more useful to express (\ref{eq:orthonormality_condition}) in momentum space. We first express $g(j)$ in (\ref{eq:orthonormality_condition}) in terms of $\tilde{g}(p)$, which leads to
\begin{equation}
\frac{1}{N} \sum_{p,p'} \tilde{g}\left(p\right)\,\tilde{g}\left(p'\right) \, e^{i j \left(p+p'\right)} e^{-i2Ml p} \sum_{m=0}^{L-1} e^{-iMm\left(p+p'\right)} = \frac{1}{M}\delta_{l,0}. 
\label{eq:othocond_p1}
\end{equation}
Now we take the Fourier transform of (\ref{eq:orthonormality_condition}) with respect to $j$. The right hand side becomes
\begin{equation}
\frac{1}{N} \sum_j e^{-iqj} \frac{1}{M} \delta_{l,0} = \frac{1}{M} \delta_{q,0}\,\delta_{l,0}.
\end{equation}
The left hand side becomes
\begin{equation}
\frac{1}{N}\sum_{p,p'} \tilde{g}\left(p\right)\,\tilde{g}\left(p'\right) e^{-i2Mlp}\times\underbrace{\left[ \frac{1}{N} \sum_j e^{i j\left(p+p' - q\right)} \right]}_{=\delta_{p+p',q}} \times \underbrace{\left[\sum_{m=0}^{L-1} e^{-iM m\left(p+p'\right)}\right]}_{=L \sum_a \delta_{p+p', \frac{2\pi}{M} a}}.
\end{equation}
The first Kronecker delta can be used to eliminate via $p' = q-p$, and using $L=N/M$ we arrive at
\begin{equation}
\frac{1}{M} \sum_p  \tilde{g}\left(p\right)\,\tilde{g}\left(\frac{2\pi}{M} a-p\right) e^{-i2Mlp} = \frac{1}{M}\delta_{a,0}\,\delta_{l,0},
\end{equation}
where we have used the Kronecker deltas to replace $q$ by the integer $a$. From the condition (\ref{eq:bandlimit}) that $\tilde{g}(p)$ is band limited it follows that
\begin{equation}
\tilde{g}\left(p\right) \,\tilde{g}\left(\frac{2\pi}{M}a - p\right) = 0 \text{ for } a \neq 0,
\end{equation}
so that all conditions with $a\neq 0$ are identically satisfied, so that only the case $a=0$ has to be taken into account. Hence the orthogonality conditions become
\begin{eqnarray}
\delta_{l,0}&=& \sum_p \tilde{g}\left(p\right) \, \tilde{g}\left(-p\right) e^{-i2Mlp}  \nonumber \\
&=& \sum_p \left|\tilde{g}(p)\right|^2 e^{-i2Mlp},
\label{eq:orthocond_p2}
\end{eqnarray}
since $\tilde{g}(p) = \tilde{g}(-p)$ by assumption. The exponential on the right hand side is periodic in $p$ with period $\frac{\pi}{M}$, so that we can write
\begin{equation}
\sum_p \left|\tilde{g}(p)\right|^2 e^{-i2Mlp} = \sum_{p \in \text{rBZ}} e^{-i2Mlp} \sum_{a} \left|\tilde{g}\left(p+\frac{\pi}{M} a\right)\right|^2,
\end{equation}
where we have introduced the reduced Brillouin zone (rBZ) $\frac{-\pi}{M} \leq p < \frac{\pi}{M}$. Hence (\ref{eq:orthocond_p2}) states that
\begin{equation}
\sum_a \left|\tilde{g}\left(p + \frac{\pi}{M} a\right)\right|^2 = \frac{2M}{N} \; \forall\; p \in \text{rBZ}.
\end{equation}
The value of the constant on the right hand side is determined by the number of points in the reduced Brillouin zone, which is $\frac{N}{2M}$. This can be further simplified using the fact that $\tilde{g}(p)$ is band limited and symmetric. The band limitation implies that only $a=-1,0,+1$ contribute to the sum. Moreover, we must have $a p < 0$, otherwise the term vanishes. By virtue of the symmetry of $\tilde{g}(p)$, it is clear that $\tilde{g}\left(p - \frac{\pi}{M}\right) = \tilde{g}\left(-p + \frac{\pi}{M}\right)$, so that it is sufficient to consider the case $p\geq 0$. Hence we obtain the final form of the orthogonality conditions:
\begin{equation}
\left|\tilde{g}\left(p\right)\right|^2 + \left|\tilde{g}\left(\frac{\pi}{M} - p \right)\right|^2 = \frac{2M}{N} \;\text{ for } 0 \leq p \leq \frac{\pi}{M}.
\label{eq:orthonormality_condition_p}
\end{equation}
Note that these conditions can be trivially satisfied by fixing $\tilde{g}(p)$ to arbitrary values less than $\sqrt{2M/N}$ for $0 < p < \pi/2M$, and to use (\ref{eq:orthonormality_condition_p}) to infer the absolute value of $\tilde{g}(p)$ for $\pi/2M < p < \pi/M$. Note that $\tilde{g}(0) = \sqrt{2M/N}$ and $\tilde{g}\left(\pi/2M\right)=\sqrt{M/N}$ are fixed. When we choose $\tilde{g}(p) \geq 0$ everywhere, the window function is fully determined.


\chapter{Window function gymnastics}
\label{sec:window_function_gymnastics}

\noindent
This appendix summarizes some useful relations between window functions. In particular, it is shown how to replace the shifted window functions $g_{mk}(j)$ by their unshifted counterparts. All relations follow from the fact that $g(j)$ is symmetric around the origin and real for all $j$, i.e.
\begin{eqnarray}
g(j) &\in& \mathbb{R} \label{eq:g_prop_real} \\
g(j) &=& g(-j) \label{eq:g_prop_sym}
\end{eqnarray}

\noindent
We recall the definition (cf. Eq. (\ref{eq:gshift})) of the phase space shift of $g(j)$:
\begin{eqnarray}
g_{m,k}(j) &=& e^{i\frac{\pi}{M}k j} \, g\left(j-Mm\right)
\label{eq:gshift_app}
\end{eqnarray}
It follows directly that
\begin{equation}
g_{m,k}(j)^\ast = g_{m,-k}(j),
\end{equation}
where $g_{m,k}(j)^\ast$ is the complex conjugate of $g_{m,k}(j)$.

\noindent
For the Fourier transform $\tilde{g}_{m,k}(p)$ we find
\begin{eqnarray}
\tilde{g}_{m,k}(p) &=& \frac{1}{\sqrt{N}} \sum_j e^{-ipj} g_{m,k}(j) \nonumber \\
&=& \frac{1}{\sqrt{N}} \sum_j e^{-i \left(p-\frac{\pi}{M} k\right)} g\left(j-M m\right) \nonumber \\
&=&\left(-1\right)^{mk}\, e^{-iMmp} \frac{1}{\sqrt{N}} \sum_j e^{-i\left(p-\frac{\pi}{M} k\right)j} g(j) \nonumber \\
&=& \left(-1\right)^{mk}\,e^{-iMmp} \tilde{g}\left(p-\frac{\pi}{M}k\right),
\label{eq:gshift_p}
\end{eqnarray}
where we have used (\ref{eq:gshift_app}) in the second line, and shifted the summation variable $j\rightarrow j+Mm$ in the third line.

\noindent
Now we note that the properties (\ref{eq:g_prop_real}) and (\ref{eq:g_prop_sym}) imply that $\tilde{g}(p)$ is real, and that $\tilde{g}(p) = \tilde{g}(-p)$. For the complex conjugate $\tilde{g}_{m,k}(p)^\ast$ of $\tilde{g}_{m,k}(p)$ this leads to
\begin{eqnarray}
\tilde{g}_{m,k}(p)^\ast &=& \left(-1\right)^{mk}\,e^{+iMmp} \tilde{g}\left(p-\frac{\pi}{M}k\right)\nonumber \\
&=& \left(-1\right)^{mk}\,e^{+iMmp} \tilde{g}\left(-p+\frac{\pi}{M}k\right)\nonumber \\
&=& \tilde{g}_{m,-k}(-p),
\label{eq:g_ast_p}
\end{eqnarray}
where we have used (\ref{eq:gshift_p}) in the first line and $\tilde{g}(p) = \tilde{g}(-p)$ in the second line.

\noindent
The replacement of $\tilde{g}_{mk}(p)$ by $\tilde{g}(p)$, Eq. (\ref{eq:gshift_p}) can be used in order to simplify sums involving the shifted window function. Let $f(p)$ be an arbitrary function of the momentum, then
\begin{eqnarray}
\sum_p f(p) \tilde{g}_{mk}(p) &=& \left(-1\right)^{mk}\sum_p f(p)  e^{-iMmp}\tilde{g}\left(p-\frac{\pi}{M} k\right) \nonumber \\
&=& \left(-1\right)^{mk} e^{i\pi M m k} \,\sum_p f\left(p+\frac{\pi}{M}k\right) e^{-i Mmp} \tilde{g}\left(p\right)\nonumber \\
&=&\sum_p f\left(p+\frac{\pi}{M}k\right) e^{-i Mmp} \tilde{g}\left(p\right),
\label{eq:g_sum_simplification}
\end{eqnarray}
where we have used (\ref{eq:gshift_p}) in the first line, and shifted $p\rightarrow p+\frac{\pi}{M}k$ in the second line.

\noindent
Finally, it is worth mentioning that the phase space shift (\ref{eq:gshift_app}) could also be defined as
\begin{equation}
g_{m,k}(j) = e^{i\frac{\pi}{M} k \left(j-Mm\right)} \, g\left(j-Mm\right).
\label{eq:gshift_alternative}
\end{equation}
Both definitions can be used, since the difference amounts to a gauge transformation
\begin{equation}
\ket{m,k} \rightarrow \left(-1\right)^{mk} \ket{m,k}.
\end{equation}
With the alternative definition (\ref{eq:gshift_alternative}), the factor $(-1)^{mk}$ shows up in the real space representation of the shifted window function instead of the momentum space form, i.e.
\begin{eqnarray}
g_{m,k}(j) &=& \left(-1\right)^{mk}\, e^{i\frac{\pi}{M} k j} g\left(j-Mm\right),\nonumber \\
\tilde{g}_{m,k}(p) &=&  e^{-iMmp} \tilde{g}\left(p-\frac{\pi}{M}k\right).
\end{eqnarray}
In this work, we will stick to the definition (\ref{eq:gshift_app}), however, since it is more convenient for practial calculations.

\chapter{One loop RG equations from Wegner's flow equation}
\label{ch:wegner}

\noindent
We derive equations that are structurally similar to the one loop RG equations from Wegner's flow equation \cite{wegner}, also known as continuous unitary transformation (CUT, \cite{cut1,cut2}). In this method, a sequence of infinitesimal canonical transformations is applied to partially diagonalize the Hamiltonian. Since each transformation is infinitesimal, it can be approximated by linearizing in its generator $\eta$:
\begin{equation}
e^{\eta} \ham{} e^{-\eta} = \left[ \eta, \ham{}\right] + \ldots
\end{equation}
Introducing a parameter flow parameter $B$, the flow equation for the Hamiltonian is given by
\begin{equation}
\frac{d}{dB} \ham{}(B) = \left[\eta(B), \ham{}(B)\right].
\label{eq:flow_equation}
\end{equation}
One can show \cite{wegner} that the so-called canonical generator
\begin{equation}
\eta(B) = \left[ \ham{kin}(B), \ham{int}(B)\right]
\label{eq:def_canonical_generator}
\end{equation}
leads to a Hamiltonian that commutes with the kinetic energy in the limit $B\rightarrow \infty$. In general, single particle states with a kinetic energy difference $\Delta E_{\rm kin}$ start to decouple at $B \sim \left(\Delta E_{\rm kin}\right)^{-2}$. In particular, states with a kinetic energy greater than $B^{-1/2}$ decouple from the states at the Fermi surface. From the low energy point of view, the method is thus similar to the usual renormalization group approach of integrating out states \cite{shankar} that are far away from the Fermi surface, with the RG scale $\Lambda \sim B^{-1/2}$. However, it should be noted that no modes are integrated out, and thus the full information about the Hamiltonian and its spectrum is conserved in the flow equation approach. Moreover, whereas in the renormalization group an effective \emph{action} is obtained, the flow equation yields an effective \emph{Hamiltonian}, which is more suitable for our purposes. 

\noindent
In order to solve eq.~(\ref{eq:flow_equation}), approximations are necessary, since infinitely many operators are generated during the flow. We follow the usual practice of making an ansatz for the flowing Hamiltonian $\ham{}(B)$, and thus truncating the number of operators. More specifically, we use the ansatz
\begin{eqnarray}
\ham{}(B) &=& \sum_{p} \epsilon_p \, :J(p, p): \nonumber \\&&+ \frac{1}{2 N} \sum_{p_1\cdots p_4} U_{p_1p_2p_3}(B)\,\delta_{p_1+p_2,p_3+p_4}\, :J\left(p_1,p_3\right) \, J\left(p_2,p_4\right):,\nonumber\\
J\left(p_1,p_2\right) &=& \sum_\sigma \, c^\dagger_{p_1\sigma}\, c^\pdag_{p_2 \sigma},
\label{eq:ham_ansatz}
\end{eqnarray} 
where $: O :$ denotes normal ordering of the operator $O$ with respect to the Fermi sea. We omit the flow of the kinetic energy in the following, so that only the couplings $U_{p_1p_2p_3}\left(B\right)$ depend on $B$. From (\ref{eq:def_canonical_generator}), the canonical generator is then given by
\begin{eqnarray}
\eta\left(B\right) &=& \frac{1}{2N} \sum_{p_1\cdots p_4} U_{p_1p_2p_3}\left(B\right) D_{p_1 p_2 p_3} \delta_{p_1+p_2,p_3+p_4} \, :J\left(p_1,p_3\right) \, J\left(p_2,p_4\right):,\nonumber\\
D_{p_1p_2p_3} &=& \epsilon_{p_3} + \epsilon_{p_4} - \epsilon_{p_2} - \epsilon_{p_1}.
\label{eq:def_eta}
\end{eqnarray}
Plugging this into the flow equation (\ref{eq:flow_equation}), we see that there are two different contributions:
\begin{equation}
\frac{d}{dB}\ham{}(B) = \underbrace{\left[\eta(B), \ham{kin}\right]}_{O(U)} + \underbrace{\left[\eta(B), \ham{int}(B)\right]}_{O\left(U^2\right)}.
\label{eq:flow_equation_orders}
\end{equation}
The first term gives rise to a set of linear differential equations for the couplings $U_{p_1p_2p_3}(B)$, whereas the second term incorporates perturbative renormalization effects. We will first investigate the flow equation to $O(U)$. With the generator given by (\ref{eq:def_eta}) we obtain
\begin{eqnarray}
\frac{d}{dB}\ham{}(B) &=&  \left[\eta(B), \ham{kin}\right] + O\left(U^2\right)  \\
&=& -\frac{1}{2N} \sum_{p_1\cdots p_4} \delta_{p_1+p_2,p_3+p_4}\, D_{p_1p_2p_3}^2 U_{p_1p_2p_3}\left(B\right) :J\left(p_1,p_3\right)\,J\left(p_2,p_4\right):\nonumber
\end{eqnarray}
Comparing the coefficients of the operators  $:$$J\left(p_1,p_3\right) J\left(p_2,p_4\right)$: on both sides, we obtain an equation for the couplings $U_{p_1p_2p_3}(B)$:
\begin{equation}
\frac{d}{dB} U_{p_1p_2p_3}(B) = -D_{p_1p_2 p_3}^2 U_{p_1 p_2 p_3}(B) + O\left( U^2\right),
\end{equation}
which is solved by
\begin{equation}
U_{p_1 p_2 p_3}(B) = U_{p_1 p_2 p_3}(0)\,e^{-D_{p_1p_2p_3}^2 B}.
\end{equation}
From the solution one sees that interaction matrix elements that couple states with energy difference larger than $B^{-1/2}$ are suppressed exponentially, as announced above. In order to include the renormalization effects, the second order equation has to be solved. In order to satisfy the first order equation automatically, it makes sense to write
\begin{equation}
U_{p_1 p_2 p_3}(B) = e^{-D_{p_1p_2p_3}^2 B} \; F_{p_1 p_2 p_3}(B),
\label{eq:relation_U_F}
\end{equation}
and to use $F_{p_1 p_2 p_3}(B)$ as flowing coupling function. The flow equation for $F_{p_1p_2p_3}(B)$ is then given by
\begin{equation}
\frac{d}{dB} F_{p_1 p_2 p_3}(B) = e^{D^2_{p_1 p_2 p_3} B} D^2_{p_1 p_2 p_3} \, F_{p_1p_2 p_3}(B) + e^{D^2_{p_1 p_2 p_3} B} \frac{d}{dB} U_{p_1 p_2 p_3}(B).
\label{eq:ddB_F}
\end{equation}
The first term cancels the $O(U)$ term in the flow equation (\ref{eq:flow_equation_orders}) for $U_{p_1p_2p_3}(B)$, so that only the second order term contributes. In order to evaluate the commutator $\left[\eta(B), \ham{int}(B)\right]$, Wick's theorem for the product of normal ordered operators can be used. This allows to decompose a product of normal ordered operators into a sum of normal ordered operators. The decomposition is achieved by means of contractions of fermion operators. With the ansatz (\ref{eq:ham_ansatz}), only terms with two contractions are needed. When contractions are represented as Feynman diagrams, the standard second order diagrams for the renormalization of the interaction appear. 

\begin{figure}
\centering
\includegraphics[width=8cm]{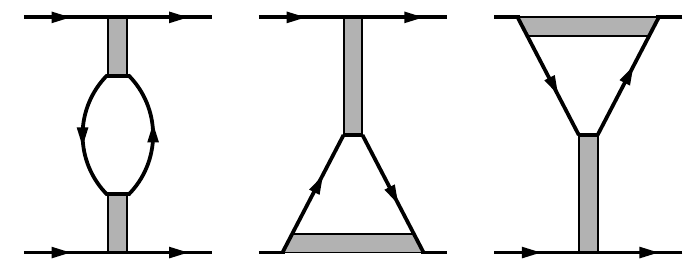}
\\
\includegraphics[width=4cm]{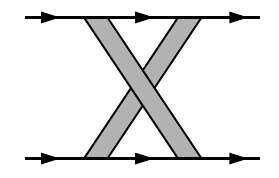}
\includegraphics[width=4cm]{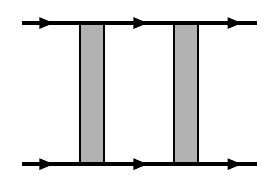}
\caption{One loop diagrams contributing to the renormalized vertex (Figure from \cite{tflow})}
\label{fig:one_loop_diagrams}
\end{figure}

\noindent
Applying Wick's theorem and comparing coefficients of operators on both sides of the flow equation yields the following equation for $U_{p_1 p_2 p_3}(B)$:
\begin{equation}
\frac{d}{dB} U_{p_1, p_2, p_3}(B) = \mathcal{T}^{(\text{phd})}_{p_1,p_2,p_3}(B) + \mathcal{T}^{\text{(phc)}}_{p_1,p_2,_3}(B) + \mathcal{T}^{\text{(pp)}}_{p_1,p_2,p_3}(B)
\label{eq:flow_equation_U_second_order}
\end{equation}

\noindent
The three terms in (\ref{eq:flow_equation_U_second_order}) correspond to different physical renormalization processes, via particle-hole excitations in the charge ($\mathcal{T}^{\text{(phd)}}$) and spin ($\mathcal{T}^{\text{(phc)}}$) channels, and through particle-particle excitations ($\mathcal{T}^{\text{(pp)}}$). The diagrams contributing to each term are shown in Fig.~\ref{fig:one_loop_diagrams}. The mathematical expressions look different from the usual one loop diagrams, but we will show in the following how they are related to them.
\begin{eqnarray}
\mathcal{T}^{(\text{phd})}_{p_1,p_2,p_3}(B) &=&\frac{1}{N}\sum_q \Big\{ 4 U_{p_1,q,p_3}\,U_{q + p_3-p_1,p_2,q} \,  D_{p_1,q,p_3} \left(n_{q}-n_{q+p_3-p_1}\right)  \label{eq:T_phd}\\
&& \phantom{\frac{1}{N}}+2 U_{q+p_3-p_1,p_1,q}\,U_{p_2,q,q+p_3-p_1} \, D_{q+p_3-p_1,p_1,q} \left(n_{q}-n_{q+p_3-p_1}\right) \nonumber \\
&& \phantom{\frac{1}{N}}+2 U_{q+p_3-p_1,p_1,q} \, U_{q,p_2,q+p_3-p_1}\,D_{q+p_3-p_1,p_1,q} \left(n_{q}-n_{q+p_3-p_1}\right) \Big\} \nonumber \\
\mathcal{T}^{(\text{phc})}_{p_1,p_2,p_3}(B) &=& -2\frac{1}{N}\sum_q  U_{q,p_2,p_3}\,U_{p_1,q+p_2-p_3,q} \, D_{q,p_2,p_3}\left(n_q - n_{q + p_2 - p_3}\right)\label{eq:T_phc}  \\
\mathcal{T}^{(\text{pp})}_{p_1,p_2,p_3}(B)&=& -\frac{1}{N}\sum_q \Big\{  U_{q,-q+p_1+p_2,p_3}\, U_{p_1,p_2,q} D_{q,-q+p_1+p_2,p_3} \left(1-n_q - n_{-q+p_1+p_2}\right) \nonumber \\
&&\phantom{\frac{1}{N}}- U_{p_1,p_2,q}\,U_{q,-q+p_1 + p_2,p_3} \, D_{p_1,p_2,q} \left(1-n_q - n_{-q+p_1+p_2} \right) \Big\},
\label{eq:T_pp}
\end{eqnarray}
where $n_p$ is the Fermi function evaluated at $\epsilon_p$. The terms containing Fermi functions correspond to the numerators in one loop diagrams, whereas the denominators of the diagrams are not explicitly visible in the present formulation. In the remainder we show how Eqns.~(\ref{eq:T_phd}-\ref{eq:T_pp}) relate to the usual one loop RG expressions. Since the the steps are essentially the same for all three types of diagrams, we will do the analysis for $\mathcal{T}^{(\text{phc})}$ only, and summarize the results for the other contributions in the end.

\noindent
First we substitute $F_{p_1p_2p_3}$ for $U_{p_1p_2p_3}$ in (\ref{eq:flow_equation_U_second_order}) and plug the result into the flow equation for $F_{p_1,p_2,p_3}$. We obtain
\begin{eqnarray}
\frac{d}{dB} F_{p_1,p_2,p_3}\Big|_{\text{phc}} &=& - e^{BD_{p_1,p_2,p_3}^2 } \, \frac{1}{N}\sum_q\Big\{ 2 F_{q,p_2,p_3}\,F_{p_1,q+p_2-p_3,q}\\&& \times  \;D_{q,p_2,p_3}e^{-B\left(D_{q,p_2,p_3}^2+D_{p_1,q+p_2-p_3,q}^2\right)} \left( n_q - n_{q+p_2-p_3}\right) \Big\} \nonumber,
\label{eq:flow_equation_F_phc}
\end{eqnarray}
where the subscript 'phc' indicates that we have suppressed phd and pp contributions. In analogy with the one loop RG we now distinguish between internal (or loop) lines and external legs. The external legs correspond to the momenta $p_1,\ldots, p_4$, whereas the internal lines are the momenta which are summed over, here $q$ and $q+p_2-p_3$. Since we are chiefly interested in the couplings in the vicinity of the Fermi surface, we set the kinetic energy of all external lines to zero in the $D_{p_1,p_2,p_3}$. This step is the analogue of evaluating diagrams at zero external frequency in the one loop RG. This simplifies the flow equation considerably. In (\ref{eq:flow_equation_F_phc}), this leads to the replacements
\begin{eqnarray}
D_{p_1,p_2,p_3} &\rightarrow& 0 \nonumber \\
D_{q,p_2,p_3} &\rightarrow& \epsilon_q - \epsilon_{q+p_3-p_1} \nonumber \\
D_{p_1,q+p_2-p_3,q} &\rightarrow& \epsilon_{q+p_2 - p_3} - \epsilon_q,
\end{eqnarray}
i.e. only the energies of the loop momenta $q$ and $q+p_2 - p_3$ remain. This leads to the approximation
\begin{eqnarray}
\frac{d}{dB} F_{p_1,p_2,p_3}\Big|_{\text{phc}} &\approx&  -\frac{1}{N}\sum_q\Big\{  F_{q,p_2,p_3}\,F_{p_1,q+p_2-p_3,q}\\&& \phantom{\frac{1}{N}}\times  \;2\left(\epsilon_q - \epsilon_{q+p_2-p_3}\right) e^{-2B\left(\epsilon_q - \epsilon_{q+p_2-p_3}\right)^2} \left( n_q - n_{q+p_2-p_3}\right) \Big\} \nonumber,
\label{eq:flow_equation_F_phc_zero_frequency}
\end{eqnarray}

\noindent
We now define the ph loop function
\begin{eqnarray}
L^{(\text{ph})}_{q, p}(B) &\equiv& R\left(\epsilon_q - \epsilon_{q+p}, B\right) \frac{n_q - n_{q+p}}{\epsilon_q - \epsilon_{q+p}}
\end{eqnarray}
which differs from the standard expression by the scale dependent prefactor
\begin{equation}
R\left(\epsilon, B\right) = 1-e^{-2 B \epsilon^2}
\end{equation}
that suppresses low-energy excitations with energy less than $B^{-1/2}$. Then we have
\begin{equation}
\frac{d}{dB} L^{(\text{ph})}_{q,p_2-p_3}(B) = 2 \left(\epsilon_q - \epsilon_{q+p_2-p_3}\right)e^{-2B\left(\epsilon_q - \epsilon_{q+p_2-p_3}\right)^2}\left(n_q - n_{q+p_2-p_3}\right),
\end{equation}
which is just the second line of (\ref{eq:flow_equation_F_phc_zero_frequency}). Thus we can write
\begin{equation}
\frac{d}{dB} F_{p_1,p_2,p_3}\Big|_{\text{phc}} =   -\frac{1}{N}\sum_q\, F_{q,p_2,p_3}\,F_{p_1,q+p_2-p_3,q} \;\frac{d}{dB} L^{(\text{ph})}_{q,p_2-p_3}(B).
\label{eq:flow_equation_F_phc_L_derivative}
\end{equation}
Finally, we can substitute the scale $\Lambda = B^{-1/2}/4$ for the flow parameter $B$, where $\Lambda$ is chosen such that the flow at scale $\Lambda$ receives contributions mainly from states with single particle energy $\Lambda$. Since the derivative with respect to $B$ appears on both sides, the form of (\ref{eq:flow_equation_F_phc_L_derivative}) is unchanged under this substitution.

\noindent
Similar manipulations on the other contributions to the flow equation lead to the final form
\begin{equation}
\frac{d}{d\Lambda} F_{p_1,p_2,p_3}(\Lambda) = \mathcal{T}^{(\text{phd})}_{p_1,p_2,p_3}(\Lambda) + \mathcal{T}^{(\text{phc})}_{p_1,p_2,p_3}(\Lambda) + \mathcal{T}^{(\text{pp})}_{p_1,p_2,p_3}(\Lambda),
\label{eq:flow_equation_F}
\end{equation}
with the contributions from the different channels given by
\begin{eqnarray}
\mathcal{T}^{(\text{phd})}_{p_1,p_2,p_3}(\Lambda) &=&-\frac{1}{N}\sum_q \Big( -2 F_{p_1,q,p_3}\,F_{q + p_3-p_1,p_2,q} + F_{q+p_3-p_1,p_1,q}\,F_{p_2,q,q+p_3-p_1}  \nonumber \\
&& \phantom{\frac{1}{N}}+ F_{q+p_3-p_1,p_1,q} \, F_{q,p_2,q+p_3-p_1}\Big)\frac{d}{d\Lambda} L^{(ph)}_{q,q+p_3-p_1}(\Lambda) \label{eq:T_phd_F} \\
\mathcal{T}^{(\text{phc})}_{p_1,p_2,p_3}(\Lambda) &=& -\frac{1}{N}\sum_q  F_{q,p_2,p_3}\,F_{p_1,q+p_2-p_3,q} \,\frac{d}{d\Lambda} L^{(ph)}_{q,q+p_2-p_3}(\Lambda) \label{eq:T_phc_F}  \\
\mathcal{T}^{(\text{pp})}_{p_1,p_2,p_3}(\Lambda)&=& -\frac{1}{N}\sum_q   F_{q,-q+p_1+p_2,p_3}\, F_{p_1,p_2,q} \,\frac{d}{d\Lambda} L^{(\text{pp})}_{q,p_1+p_2}(\Lambda) \label{eq:T_pp_F}.
\end{eqnarray}
The loop functions with $\Lambda$ as a flow variable are
\begin{eqnarray}
L^{(\text{ph})}_{q,p}(\Lambda) &=& R_\Lambda\left(\epsilon_q - \epsilon_{q+p}\right) \, \frac{n_q - n_{q+p}}{\epsilon_q - \epsilon_{q+p}} \\
L^{(\text{pp})}_{q,p}(\Lambda) &=& R_\Lambda\left(\epsilon_q + \epsilon_{-q+p}\right) \, \frac{1 - n_q - n_{-q+p}}{\epsilon_q + \epsilon_{-q+p}} \\
R_\Lambda(\epsilon) &=& 1 - e^{-\epsilon^2/8\Lambda^2}
\end{eqnarray}

\noindent
Once the flow equation is solved up to some scale $\Lambda$, the renormalized interaction Hamiltonian is obtained by transforming from $F_{p_1,p_2,p_3}(\Lambda)$ back to $U_{p_1,p_2,p_3}(\Lambda)$. Recalling (\ref{eq:relation_U_F}), we have
\begin{equation}
U_{p_1,p_2,p_3}(\Lambda) = e^{-D^2_{p_1,p_2,p_3}/16\Lambda^2} F_{p_1,p_2,p_3}(\Lambda),
\end{equation}
the physical interaction decouples states at the Fermi surface from states with single particle energy $|\epsilon| > \Lambda$. As a consequence, states above the cutoff may be neglected or treated perturbatively, so that only states below the cutoff remain.
\begin{figure}
\centering
\includegraphics[height=4cm]{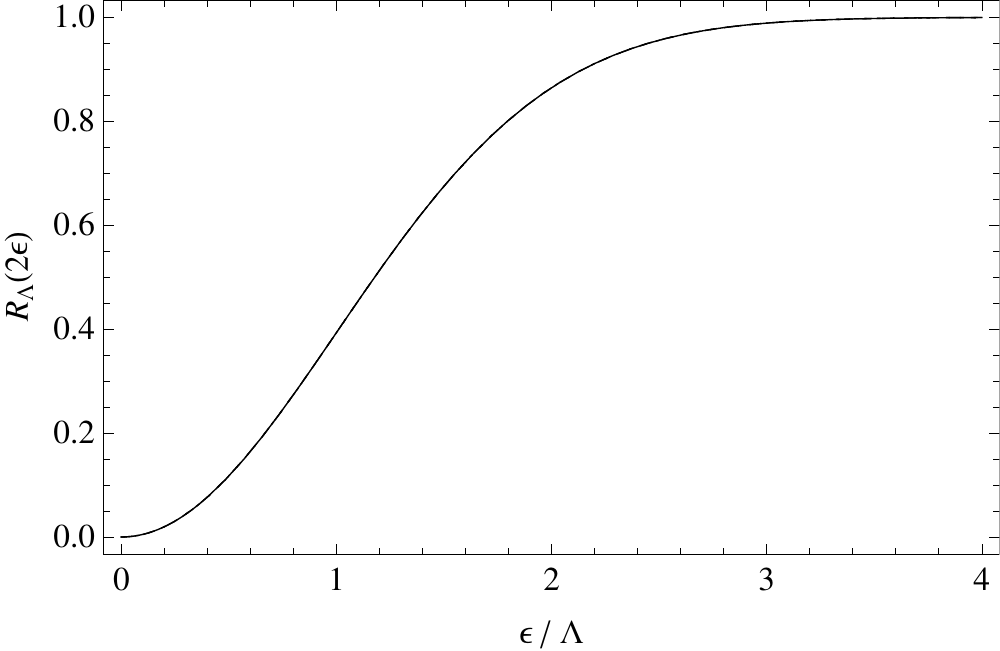}
\includegraphics[height=4cm]{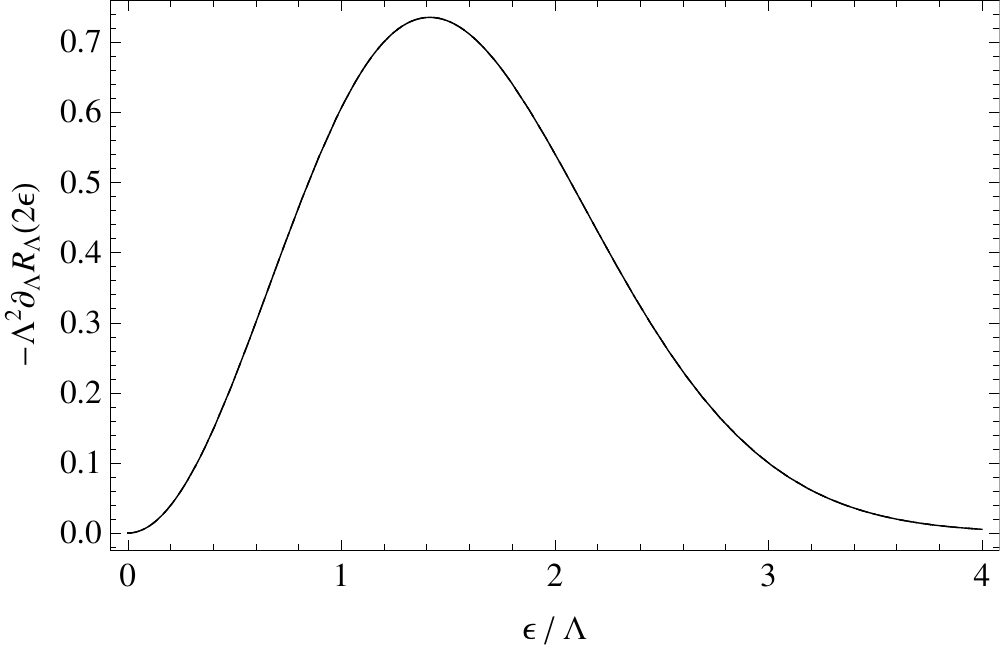}
\caption{Dependence of the cutoff function $R_\Lambda^{(n)}(2\epsilon)$ and its scale derivative on the average single-particle energy $\epsilon$.}
\label{fig:cutoff_function}
\end{figure}
\noindent

\noindent
Comparison of (\ref{eq:flow_equation_F}) with the flow equation derived from functional RG \cite{tflow} shows that the structure of the two equations is identical when the FRG equation is evaluated in the one-loop approximation, i.e.~omitting the frequency dependence and using bare propagators in all diagrams. In both equations, the flow is generated by second order perturbation theory. The bubble diagrams are scale dependent due to the presence of an infrared cutoff, and their contribution to the flowing couplings is given by the scale derivative of the bubble. However, in the present approach, the cutoff function acts on the energy of two-particle excitations (pp or ph), whereas in the functional RG the cutoff acts on single particle energies. From a practical point of view, however, this difference should be small in most situations, since the two types of cutoff only differ when the single particle energies of the two particles taking part in the excitation are very different. Due to kinematic restrictions in the presence of a Fermi surface this is not expected to matter in most situations.

\chapter{Contractor renormalization}
\label{sec:core}

In the following we review briefly the contractor renormalization (CORE) method \cite{morningstar} for correlated lattice systems.
The method is based on first choosing a truncated set of $\kappa$ states for the local Hilbert space of a lattice system. There is no general rule how to choose these states, so that the choice has to be based on physical insight. These $\kappa$ states are used as effective degrees of freedom per site. The dimension of the local Hilbert space is denoted by $\nu$. The couplings between sites are evaluated by diagonalizing connected clusters of lattice sites and projecting lowest states onto the tensor product of the truncated local basis. For an $N$-site cluster, one has to diagonalize an $\nu^N$-dimensional matrix and find its lowest $\kappa^N$ states. 

\noindent
We denote the $\kappa^2$ tensor products of projected single-site states by $|\alpha_i\rangle$. We need the $\kappa^2$ lowest eigenstates of the two-site cluster for the projection, and will denote the $n$-th such state by $|n\rangle$. The CORE algorithm consists of three steps:

\noindent
First the low-lying states of the cluster are projected onto the $|\alpha_i\rangle$,
\begin{equation}
|\psi_n\rangle = \sum_i \, |\alpha_i\rangle\langle \alpha_i|n\rangle.
\end{equation}
Second, the projected states $|\psi_n\rangle$ are orthonormalized, using the Gram-Schmidt method starting with the groundstate:
\begin{equation}
|\tilde{\psi}_n\rangle = \frac{1}{Z_n}\left(|\psi_n\rangle - \sum_{i<n} |\tilde{\psi}_i\rangle \langle \tilde{\psi}_i|\psi_n\rangle\right).
\end{equation}
The effective Hamiltonian for the $N$-site cluster is then given by
\begin{equation}
\mathcal{H}^{(N)} = \sum_n E_n |\tilde{\psi}_n\rangle\langle \tilde{\psi}_n|,
\end{equation}
where $E_n$ is the energy of state $|n\rangle$. Finally, one can write the total effective Hamiltonian as a sum over irreducible $N$-site operators, where the irreducible part of the $N$-site Hamiltonian for the cluster $C_N$ is defined by
\begin{equation}
\ham{irred.}(C_N) = \mathcal{H}^{(N)} - \sum_{N'<N} \sum_{C_{N'} \in C_{N}} \mathcal{H}(C_{N'})
\end{equation}
where $\sum_{C_{N'} \, \in \, C_N}$ denotes summation over all connected subclusters of $C_N$ of size $N'$. One can show \cite{morningstar} that the expansion 
\begin{equation}
\ham{eff} = \sum_{N=1}^{N_{\text{max}}} \sum_{C_N} \mathcal{H}(C_N)
\end{equation}
reproduces the lowest $\kappa$ energies of the original Hamiltonian in the limit $N_{\text{max}}\rightarrow\infty$. In practice, however, the expansion has to be truncated at some point, and here we will consider the simplest approximation only, and restrict ourselves to two-site interactions.

%

\backmatter
\vspace{-2cm}
\chapter*{Acknowledgements}

\noindent

During the time I spent working on this thesis, I had the pleasure of interacting with many people that contributed directly or indirectly. I would like to thank my supervisors Manfred Sigrist and T. Maurice Rice for giving me the opportunity to work on a project which for a long time resembled Schr\"odingers cat, oscillating between success and failure. 

The time I spent at my office was made much more fun by the people that shared it with me, Andreas R\"uegg, Hiroto Adachi, Kaiyu Yang, Sebastian Pilgrim, Yoshiki Imai, and Daniel M\"uller. I thank all of them for discussions, creating a friendly atmosphere, and in particular for enduring my frequent and lengthy rants, digressions, and 'brief' sketches of ideas. 

I would also like to thank the other members of the solid state theory groups, Jonathan Buhmann, Adrien Bouhon, Mark Fischer, Sarah Thaler, Ludwig Klam, Jun Goryo, Sebastian Huber, Fabian Hassler, Barbara Theiler, Alexander Thomann, Roland Willa, and David Oehri. I have learnt many things through fruitful discussions with them, and benefitted greatly from their kindness in many ways.

Finally, I am very grateful to Lena Hartmann for spending the last few years with me emotionally, and for her tolerance in spending them mainly without me physically.

\cleardoublepage

\end{document}